\documentclass[a4paper,12pt,times,numbered,index,custombib,twoside]{Classes/PhDThesisPSnPDF}

\input{Preamble/preamble}

\title{Aspects of Gravitational Collapse and the formation of Spacetime Singularities}
\author{Soumya Chakrabarti}

\degreetitle{\small Thesis submitted to IISER Kolkata 
\\
for the fulfilment of the requirements for the Degree of Doctor of Philosophy}

\renewcommand{\submissiontext}{\bf Supervisor: Prof. Narayan Banerjee}

\dept{ Department of Physical Sciences}

\university{ Indian Institute of Science Education and Research Kolkata}
\crest{\centering \includegraphics[width=0.25\textwidth]{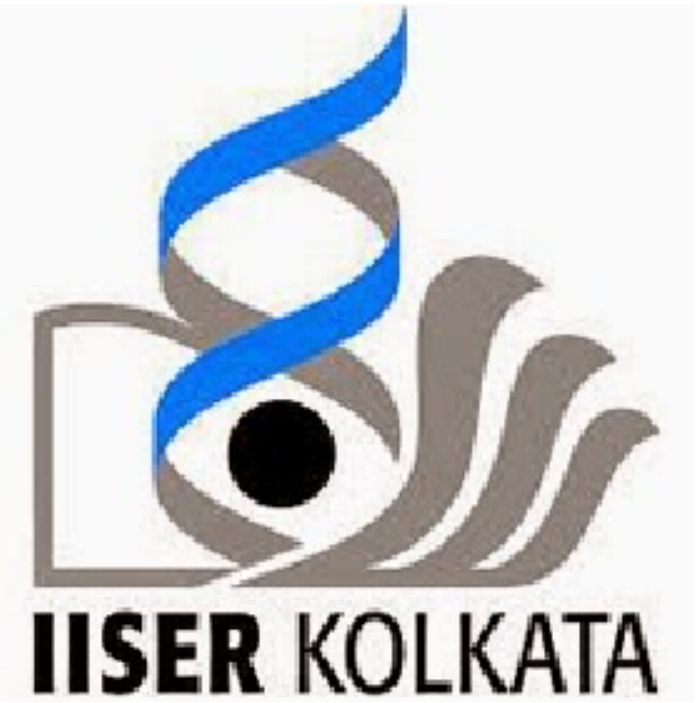}}
\subject{LaTeX} \keywords{{LaTeX} {PhD Thesis} {DPS} {IISER Kolkata}}

\ifdefineAbstract
\fi

\begin{document}

\frontmatter
\begin{titlepage}
  \maketitle
\end{titlepage}

% ******************************* Thesis Dedidcation ********************************

\begin{dedication} 

{\Large To

Mother...  }

\end{dedication}

% ******************************* Thesis Declaration ***************************

\begin{declaration}

I certify that this work contains no material which has been accepted for the award of any other degree or diploma in my name, in any university or other tertiary institution, to the best of my knowledge and belief, contains no material previously published or written by another person, except where due reference has been made in the text and acknowledgement of collaborative research. In addition, I certify that no part of this work will, in the future, be used in a submission in my name, for any other degree or diploma in any university or other tertiary institution without the prior approval of Indian Institute of Science Education and Research Kolkata, India and collaborators.

% Author and date will be inserted automatically from thesis.tex \author \degreedate

\end{declaration}

\begin{dedication}

\begin{center}

{ \Large \bf \underline{Certificate} }

\end{center}

\begin{flushleft}

{\centering This is to certify that the Ph.D. thesis entitled "{\bf Aspects of Gravitational Collapse and the formation of Spacetime Singularities}" submitted by {\bf Soumya Chakrabarti} is absolutely based upon his own work under the supervision of {\bf Prof. Narayan Banerjee} at the Indian Institute of Science Education and Research, Kolkata (IISER Kolkata) and that neither this thesis nor any part of it has been submitted for either any degree/diploma or any other academic award anywhere before.}

\end{flushleft}

~~
~~
~~

\begin{flushright}
\vskip 3cm 
\noindent\rule{6cm}{0.4pt}

{\bf Prof. Narayan Banerjee}

Professor

Department of Physical Sciences

IISER Kolkata
\end{flushright}
\end{dedication}

% ************************** Thesis Acknowledgements **************************

\begin{acknowledgements}      

It is indeed a pleasure on my part to express my gratitude to my supervisor, Professor Narayan Banerjee. I would not be in my position today, without his support, guidance and his belief in me. At many times his insights and patience were the only things that came to my aid; academically or else.  \\ 

I owe my gratitude to Dr. Rituparno Goswami (University of KwaZulu Natal) for his ideas and helpful comments which helped me towards a deeper understanding of my work. My special thanks to Dr. Golam Mortuza Hossain (IISER Kolkata) for his help and valuable suggestions.  \\

During my stay at IISER Kolkata, it was a memorable experience with all my friends and colleagues. I was blessed to have seniors like Arghya Da, Nandan Da and Gopal Da from whom I learnt a lot through numerous 'interactive' tea sessions. I thank all my juniors, Anushree, Chiranjib, Subhajit, Avijit, Srijita, Sachin; especially Chiranjib and Subhajit for making our days full of laughter and pratical jokes. I must thank all the other members of the Department Of Physical Sciences, IISER Kolkata, for their support and help.  \\
A very special gratitude to Debmalya who was a wonderful companion during this journey and at times was like an elder brother to me. I hope that we can continue to have even more refreshing conversations as we grow old and experienced.  \\
I would like to thank my batch-mate Ankan for being helpful in many difficult times. I wish him a lot of success in life.  \\

I am extremely indebted to my mother Sarbari Chakrabarti, who was my very first teacher and mentor, and never stopped dreaming for me. I wish that someday I would be a better son and give her a lot more peace, happiness and joyous memories and make her proud.  \\
I cannot end without thanking my grandfather, for encouraging my interest in mathematics when I was a child. I thank my sister and express my deepest affection. I wish her a lot of happiness in life.  \\
To my father, I understand it is extremely difficult to grow up to become a man like you were. A day does not go by when I don't miss you.  \\

I owe my heartiest gratitude and many colourful memories to Torsa, who never loses faith in me, being as loving, patient and charming as only she can be; to whom I dedicate these words... "Grow old with me, the best is yet to be..."

\end{acknowledgements}

% ************************** Thesis Abstract *****************************
% Use `abstract' as an option in the document class to print only the titlepage and the abstract.
\begin{abstract}
Possibilities emerging out of the dynamical evolutions of collapsing systems are addressed in this thesis through analytical investigations of the highly non-linear Einstein Field Equations. 		\\

Studies of exact solutions and their properties, play a non-trivial role in general relativity, even in the current context. Finding non-trivial solutions to the Einstein field equations requires some reduction of the problem, which usually is done by exploiting symmetries or other properties. Exact solutions of the Einstein’s field equations describing an unhindered gravitational collapse are studied which generally predict an ultimate singular end-state. In the vicinity of such a spacetime singularity, the energy densities, spacetime curvatures, and all other physical quantities blow up. Despite exhaustive attempts over decades, the famous conjecture that the formation of a singularity during stellar collapse necessarily accompanies the formation of an event horizon, thereby covering the central singularity, still remains without a proof. Moreover, there are examples of stellar collapse models with reasonable matter contribution in which an event horizon does not form at all, giving rise to a naked singularity from which both matter and radiation can fall in and come out. These examples suggest that the so-called “cosmic censorship” conjecture may not be a general rule. Therefore one must embark upon analysis of realistic theoretical models of gravitational collapse and gradually generalizing previous efforts.  \\

Viable $f(R)$ models are quite successful in providing a geometrical origin of the dark energy sector of
the universe. However, they possess considerable problems in some other significant sectors, such as, difficulty to find exact solutions of the field equations which are fourth order differential equations in the metric components. Moreover, a recent proposition that homogeneous collapsing stellar models (e.g. Oppenheimer-Snyder-Datt model of a collapsing homogeneous dust ball with an exterior Schwarzschild spacetime) of General Relativity can not be viable models in $f(R)$ theories, heavily constrict the set of useful astrophysical solutions. In this thesis, we address some collapsing models in $f(R)$ gravity such that at the comoving boundary of the collapsing star, the interior spacetime matches smoothly with an exterior spacetime. The presence and importance of spatial inhomogeneity is duely noted and discussed. The ultimate spacetime singularity remains hidden or exposed to an exterior observer depending on initial conditions from which the collapse evolves.   \\

The study of collapsing solutions of the Einstein equations with a scalar field as the matter contribution owes special importance, because one would like to know if cosmic censorship is necessarily preserved or violated in gravitational collapse for fundamental matter fields, which are derived from a suitable Lagrangian. In this thesis we have studied some models of gravitational collapse under spherical symmetry, with a self-interacting scalar field minimally coupled to gravity along with a fluid description. The field equations are solved under certain significant symmetry assumption at the outset (for instance, conformal flatness, self-similarity) without assuming any particular equation of state for the matter contribution. The relevance of such investigations stems from the present importance of a scalar field as the dark energy {\it vis-a-vis} the fluid, whose distribution still remains unknown apart from the general belief that the dark energy does not cluster at any scale below the Hubble scale. The study of collapse of scalar fields, particularly in the presence of a fluid may in some way enlighten us regarding the possible clustering of dark energy. The collapsing models are studied in this thesis for certain popular and physically significant forms of the self-interaction potential, for example, a power-law or an exponential dependence over the scalar field. The end-state of the collapse is investigated by analyzing the apparent horizon curve and existence of radial null geodesics emanating from the spacetime singularity.

\end{abstract}

\begin{dedication}

\textbf{\Large List of Publications}

\vskip 2.0 cm  

\begin{flushleft}

1. Soumya Chakrabarti and Narayan Banerjee, {\it "Spherical Collapse in vacuum f(R) gravity"}, Astrophys. Space Sci. {\bf 354} (2014) no.2, 2118; Erratum: Astrophys. Space Sci. {\bf 359} (2015) no.1, 36. ({\it Not included in the thesis})

\vskip 1.0 cm  

2. Soumya Chakrabarti and Narayan Banerjee, {\it "Spherically symmetric collapse of a perfect fluid in f(R) gravity"}, Gen. Relativ. Gravit. {\bf 48} : 57 (2016).

\vskip 1.0 cm  

3. Soumya Chakrabarti and Narayan Banerjee, {\it "Gravitational collapse in f(R) gravity for a spherically symmetric spacetime admitting a homothetic Killing vector"}, Eur. Phys. J. Plus {\bf 131} : 144 (2016).

\vskip 1.0 cm  

4. Soumya Chakrabarti and Narayan Banerjee, {\it "Scalar field collapse in a conformally flat spacetime"}, Eur. Phys. J. C. {\bf 77} no.3, 166 (2017).

\vskip 1.0 cm  

5. Narayan Banerjee and Soumya Chakrabarti, {\it "Self-similar scalar field collapse"}, Phys. Rev. D. {\bf 95}, 024015 (2017).

\vskip 1.0 cm  

6. Soumya Chakrabarti, {\it "Scalar Field Collapse with an exponential potential"}, Gen. Relativ. Gravit. {\bf 49} : 24 (2017).

\end{flushleft}

\end{dedication}

% *********************** Adding TOC and List of Figures ***********************

\tableofcontents

%\listoffigures

%\listoftables

% \printnomencl[space] space can be set as 2em between symbol and description
%\printnomencl[3em]

\printnomencl

% ******************************** Main Matter ********************************* 
\mainmatter

%*******************************************************************************
%*********************************** First Chapter *****************************
%*******************************************************************************

\chapter{Introduction}\label{intro}  %Title of the First Chapter

\ifpdf
    \graphicspath{{Chapter1/Figs/Raster/}{Chapter1/Figs/PDF/}{Chapter1/Figs/}}
\else
    \graphicspath{{Chapter1/Figs/Vector/}{Chapter1/Figs/}}
\fi

%********************************** %First Section  **************************************

\section{Gravitational collapse and spacetime singularity}\label{gcss}
One of the most important aspects of General Relativity is the formation of spacetime singularities and horizons as the end results of a gravitational collapse of a massive object. A massive star, in general, replenishes the heat that radiates from its hot surface into the depths of interstellar space by steadily burning nuclear fuel in its deep interior. The nuclear fuel supply for any star is limited. Eventually, after some billions of years of its lifetime, the star exhausts its nuclear fuel, and dies \cite{thorne, shapiro}.   \\

An investigation of such massive distributions, where density of matter and strength of the gravitational field is expected to be extreme, requires the theory of General Relativity which provides a description of gravity as a geometric property of spacetime; formulated by Einstein in $1915$ using the equivalence between gravitation and inertia successfully. The relation between curvature of spacetime and the energy-momentum of whatever matter or radiation are present, is well defined by the famous Einstein Field Equations (in the units $G = 1$, $c = 1$) as
\begin{equation}
G_{\mu\nu} = 8 \pi T_{\mu\nu},
\end{equation}
where $G_{\mu\nu}$ is the Einstein tensor, $T_{\mu\nu}$ is the energy-momentum tensor. It is precisely these equations that are thought to govern different aspects of gravitational physics of which we now have observational evidences; such as the expansion of the evolution of the Universe, the behavior of black holes, the propagation of gravitational waves, and many other phenomena involvng gravity. Some predictions of general relativity differ significantly from classical newtonian physics, and such predictions are confirmed in all experiments and observations to date. For instance, general relativity predicts a precession in planetary orbits, which was derived in case of mercury by Einstein when he \cite{einstein1, einstein2, einstein3} published his vacuum field equations in $1915$. Experimental findings like the Lense-Thirring gravitomagnetic precession and the gravitational deflection of light by the Sun (as measured in $1919$ during a Solar eclipse by Eddington) were in well agreement with theoretical predictions of the theory.   \\

What is then, the final outcome of the gravitational collapse of a massive star which has exhausted its nuclear fuel? Detailed studies of a gravitational collapse have been conducted within the context of general relativity and the generic conclusions emerging from these studies are striking. While the collapse is generally expected to produce curvature generated singularities characterized by diverging curvatures and densities, a trapped surface may not develop early enough to always shield this process from an outside observer. Depending on the nature of the initial data from which the collapse evolves, either a black hole or a naked singularity results as the final outcome of the collapse.  \\

In $1916$, Schwarzschild \cite{sch} discovered the exact solution to the Einstein field equations representing the external gravitational field of a static, spherically symmetric body. This solution allows a very simple derivation of the perihelion advance and was the first black hole solution to Einstein’s equations; though at that time the concept of a black hole was not introduced. However, dynamics of a gravitatonal contraction as governed by the field equations were not addressed until in $1930$s, when Chandrasekhar \cite{chandra} and Landau \cite{landau} suggested an upper limit (the Chandrasekhar Limit) to the mass of a massive astronomical body. Any massive object (stars, remnants, supernovae) having a mass above the limit cannot continue to support itself against the inward pull of gravity. In $1935$, Eddington \cite{eddi} discussed that any star having a mass exceeding the Chandrasekhar limit, shall contract indefinitely producing an end state from which even light could not escape. A complete general relativistic treatment of a star undergoing gravitational collapse to zero volume was given in $1939$ by Oppenheimer and Snyder \cite{os} and these works now serve as the paradigm of a gravitational collapse.   \\
 
Gravity is a nonlinear phenomena and its predictability lies within the nonlinear field equations. Rigorous study of the field equations by exploiting different symmetry assumptions can indeed throw light into the physical process of gravitational collapse, not only in general relativity, but in different viable modifications of gravity as well. In this thesis, we discuss some reasonable examples of self-gravitating systems undergoing an unhindered gravitational collapse.

\subsection{Basic features of a Gravitational Collapse}
During the $1960$s and early $1970$s, a first understanding of the classical processes involved in gravitational collapse were established after rigorous studies; concept of a black hole was introduced as well \cite{wald, penrose, hp}. To define in a very simple manner, the contraction of an astronomical massive body due to the influence of its own gravity is known as a Gravitational Collapse.            \\

Gravitational collapse is the fundamental mechanism thought to be responsible for structure formation in the universe. For example, a star is born through the gradual gravitational collapse of a cloud of interstellar matter. The outward thermal pressure generated from a thermonuclear fusion occurring at the central region of the star gradually halts the contraction, balancing the inward gravitational pull, forming a state of dynamic equilibrium. When a star has burned out its fuel supply, it will undergo a contraction that can be halted only if it reaches a new state of equilibrium. Depending on the mass during it's lifetime, these equilibrium states can take certain forms.

\begin{itemize}
\item{ White dwarfs, in which gravity is opposed by the electron degeneracy pressure.
}
\item{ Neutron stars, in which gravity is opposed by the neutron degeneracy pressure and short-range repulsive neutron–neutron interactions mediated by the strong force.
}
\item{ Unhindered gravitational collapse to zero volume; according to Einstein's theory, for massive stars stars above the Tolman–Oppenheimer–Volkoff \cite{ov} limit, no known form of internal pressure can balance the inward pull due to gravity. Hence, the contraction continues until a zero volume is reached with nothing to stop it.
}
\end{itemize}

\subsection{End-state of an unhindered Gravitational Collapse: Blackholes}
Newtonian physics says that only matter can be a source of gravitation. However, general relativity allows any form of energy, including gravitational energy to contribute to the field as well. Thus we have the strange possibility of a gravitational field that feeds upon itself, becoming infinitely strong even where there is no matter \cite{shapiro}. If Einstein's conception of gravity introduces this concept of singularity, however, it also offers a possible escape clause. Whenever simple configurations of matter collapse according to the rules of general relativity, the collapsed region always seems to be enveloped in a Black Hole before a singularity forms, provided an event horizon develops prior to the formation of the singularity. Neither matter nor radiation nor information can cross the event horizon. Inside the event horizon, the presence of the singularity renders all the known laws of physics quite useless; one cannot predict what will happen there. But because there can be no communication from inside the black hole to the rest of the universe, life can go on undisturbed elsewhere \cite{hp}.              \\

The formation of horizon is governed purely by general relativity. However, applying the theory to stellar collapse still remains a formidable task since Einstein's equations are notoriously non-linear, and solving them requires simplifying assumptions. Oppenheimer and Snyder \cite{os} and independently, Datt \cite{datt} made an initial attempt in the late $1930$s. They considered perfectly spherical stars, consisting of a gas of homogeneous density and no gas pressure and found that under idealized conditions, a collapsing cloud of matter with zero pressure will necessarily trap all light and matter inside an event horizon, thereby ending up in a blackhole. However, real stars are more complicated to deal with. Their density is inhomogeneous, the gas in them may exert pressure, and they can assume different shapes. Whether or not every sufficiently massive collapsing star turn into a black hole, remains an intriguing question even today. The first definitive attempt to close the question, came in $1969$, when Penrose conjectured that \cite{penrose1} the formation of a singularity during stellar collapse necessarily accompanies the formation of an event horizon, thereby covering the central singularity. Penrose's conjecture is termed as the Cosmic Censorship Hypothesis. If the cosmic censorship hypothesis holds, then general relativity remains a safe and reliable theory of the structure of the universe. Its only area of inadequacy, however, remains in describing the bizarre environment inside a black hole. 

\begin{figure}[h]
\centering
\includegraphics[width=0.5\textwidth]{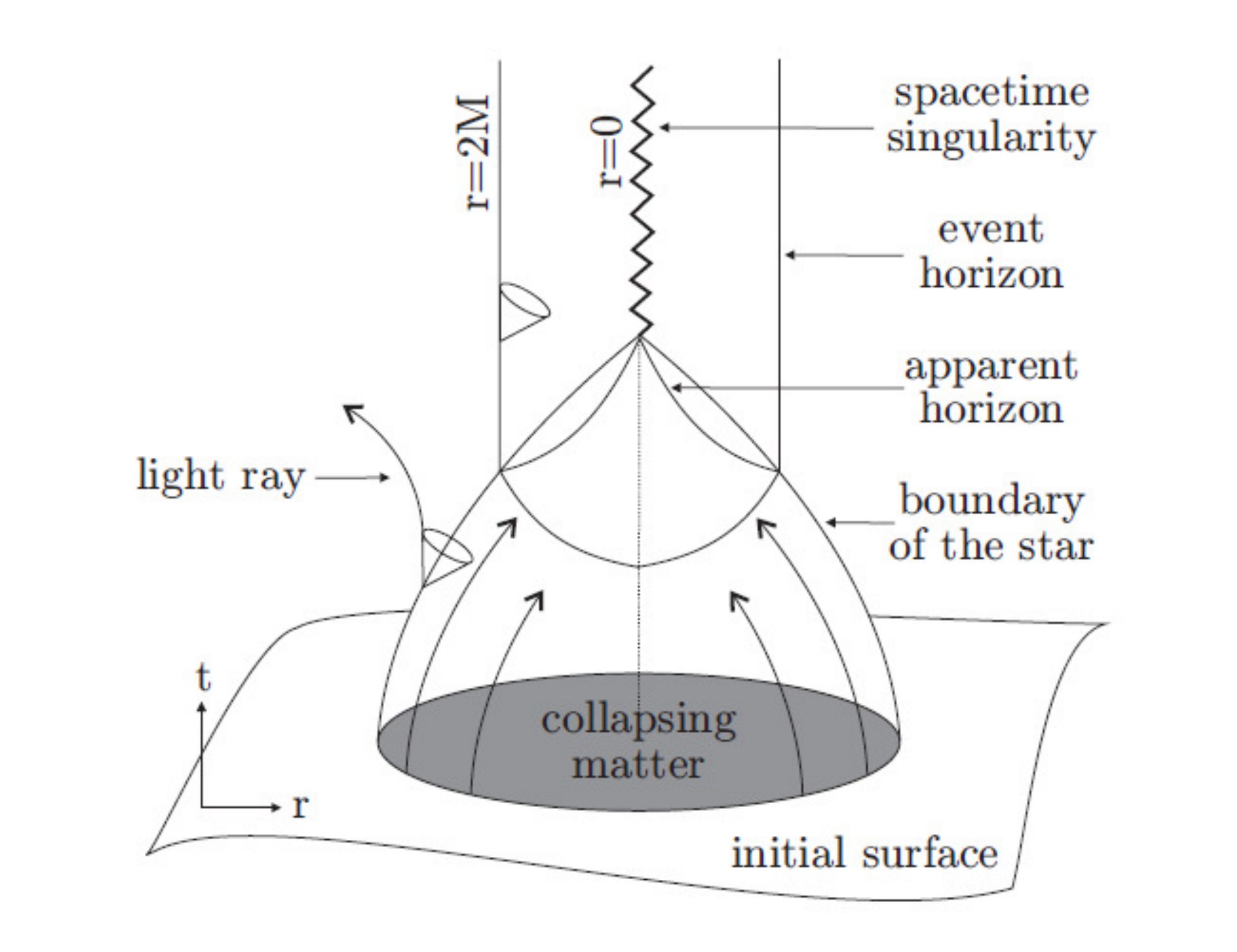}
\includegraphics[width=0.5\textwidth]{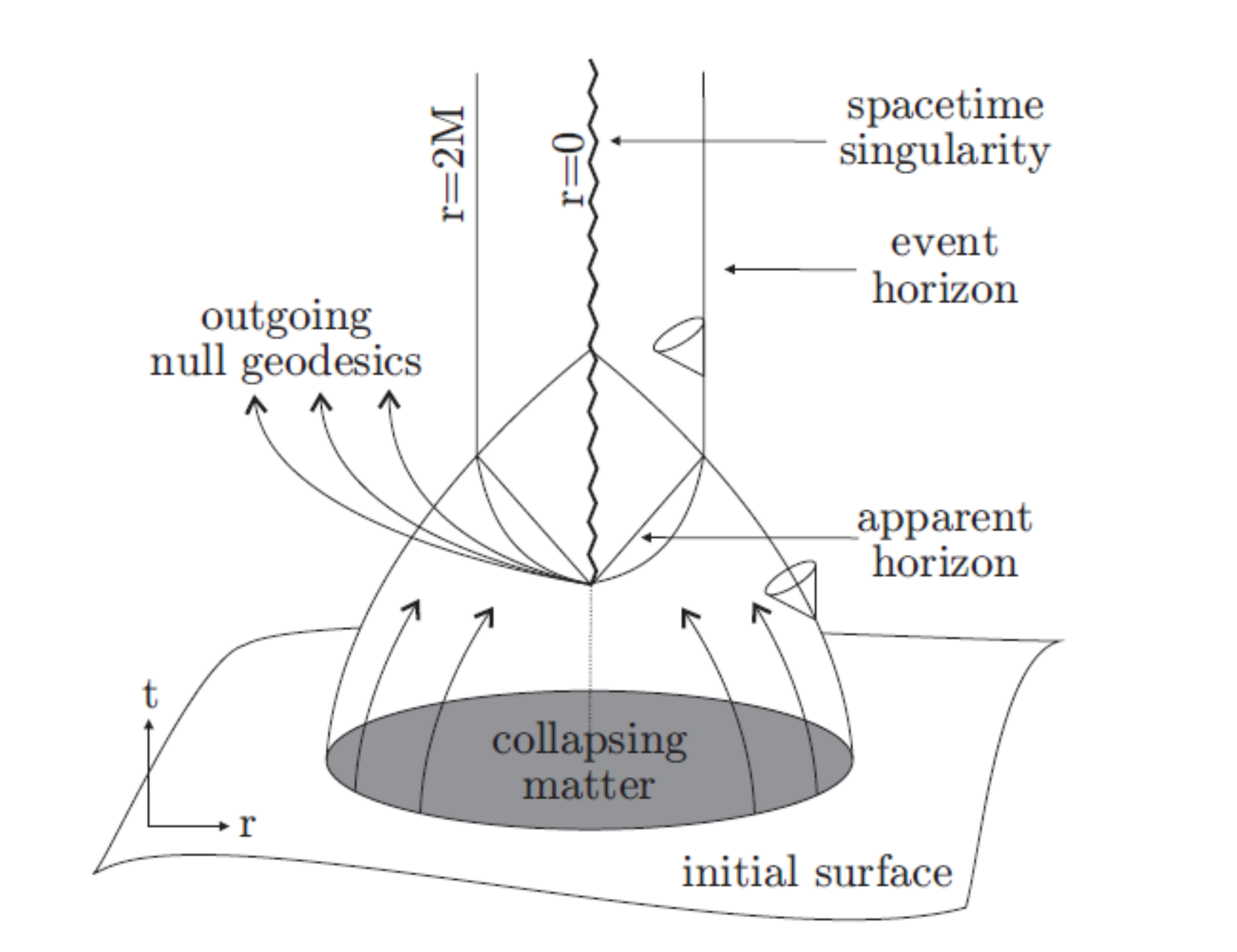}
\caption{Schematic idea of a Homogeneous dust cloud collapse. If the event horizon forms prior to the singularity, creates a blackhole as the collapse endstate. For a globally visible singularity, the outgoing rays reach the boundary of the cloud and can reach faraway external observers}
\label{fig:coll1}
\end{figure}

\subsection{Naked Singularities}
Determining the validity of cosmic censorship is perhaps one of the most outstanding problem in the study of general relativity. Indeed, there are examples of stellar collapse models in which an event horizon does not form at all, so that the singularity remains exposed to the view of an exterior observer. This gives rise to the concept of a naked singularity which, unlike a blackhole, can allow exchange of both matter and radiation with the exterior. These are indeed, counterexamples to Penrose's conjecture and they suggest that the cosmic censorship is not a general rule. Moreover, there is no direct proof of censorship as yet that applies under all conditions. Therefore one must embark upon analysis of realistic theoretical models of gravitational collapse one by one gradually generalising previous efforts. For instance, Yozdis, Seifert and their colleagues considered inhomogeneity in the collapsing configuration and showed that layers of collapsing matter can create momentary singularities that are not covered by horizons \cite{seif1, seif2}. Eardley and Smarr \cite{smarr} performed a numerical simulation of a collapsing star with a reasonable density profile. An exact analytical treatment of the similar situation was undertaken by Christodoulou \cite{christozero}. In both of these models it was shown that the star could shrink to zero proper volume producing a naked singularity.         \\ 

In the beginning, naked singularities were thought to be weak in nature \cite{newman}, but that may not be a general statement. Joshi and his collaborators \cite{joshi1, joshi2, joshidwivedi1, joshidwivedi2, joshidwivedi3} discussed possible scenarios of inhomogeneous gravitational collapse leading to strong curvature singularities yet remaining visible to external observers. Waugh and Lake \cite{waughlake1, waughlake2} discussed the strength of shell-focussing naked singularities considering evolution of a gravitational collapse from regular initial conditions, considering self-similarity in the space-time. In the early $1990$s, the effects of gas pressure in the models of realistic gravitational collapse were considered. Numerical simulations carried out by Ori and Piran \cite{ori1, ori2} and analytical studies of Joshi and Dwivedi \cite{joshidwivedi1, joshidwivedi2, joshidwivedi3} manifested that massive stars can in fact, collapse to naked singularities. Magli et. al. \cite{magli1, magli2, magli3} and Nakao \cite{nakao} considered different forms of pressure contribution within a collapsing star and showed that in a wide variety of situations, collapse can end in a naked singularity. These studies analyzed perfectly spherical stars, which is not as severe a limitation as it might appear, because most stars in nature are very close to this shape. Physicists have been exploring nonspherical collapse as well, for example, Shapiro and Teukolsky \cite{shapiro1, shapiro2} presented numerical simulations in which rotating spheroids could collapse to a naked singularity.    \\

These scenarios, with a gravitational collapse ending in a so-called {\bf naked singularities} open some important physical questions. These may not be just some counter-examples to the cosmic censorship conjecure. Indeed the formation of a naked singularity may imply that fundamental processes occurring near the singularities are not always forbidden from the exterior.

\subsection{Physical implications of a Naked Singularity}
For a realistic study one should consider general matter fields so as to include important physical features in the collapse, such as inhomogeneities in matter distribution, non-vanishing pressure, different forms for the equations of state of the collapsing matter, and other such aspects. The nature of the outcome of a collapse generally depends on the initial configuration from which the collapse evolves, and the allowed dynamical evolutions in the spacetime, as permitted by the Einstein equations \cite{joshi1}. Recent results and different aspects regarding the dynamics of a continued gravitational collapse is summerized by Joshi \cite{joshi1, joshi2}.
\\

Gravity is a complicated phenomenon involving not only a force of attraction, other physical attributions of the matter distribution can play significant role in the allowed dynamics as well. For example, presence of a non-zero shear means different layers of material are shifted laterally with different velocities. In a region where a high density massive collapsing star can trap light, non-zero shear and inhomogeneities may create escape routes. It was discussed by Joshi, Dadhich and Maartens \cite{maartens1, maartens2}, Joshi, Goswami and Dadhich \cite{joshiritudadhi} that shearing of the collapsing material close to the singularity, can indeed set off powerful shock waves that can eject matter and light.
\\

If a homogeneous star collapses, gravity increases in strength and bends the paths of moving objects severely until light can no longer propagate away from the star. Since for a homogeneous star, the density varies only with time, the entire star is crushed to the singular point simultaneously as discussed by Joshi, Goswami and Dadhich \cite{joshiritudadhi}. The trapping of light occurs well before this moment, so the singularity remains hidden. Due to inhomogeneity the density decreases with distance from the center. The denser inner regions feel a stronger pull of gravity and collapse faster than the outer ones. Therefore the entire star does not collapse to the singularity simultaneously. The innermost shells collapse first, and then the outer shells pile on gradually and as argued by Joshi, Goswami and Dadhich \cite{joshiritudadhi}, the resulting delay can postpone the formation of an event horizon in principle. If density decreases with distance too rapidly, these shells may not constitute enough mass to trap light. In such cases the singularity, when it forms, will be naked. Intuitively one can guess that perhaps there is a threshold: if the degree of inhomogeneity is very small, below a critical limit, a black hole will form; with sufficient inhomogeneity, a naked singularity may result.
\\

The rapidity of collapse can perhaps be another contributing factor in the black hole-naked singularity conundrum, specially in models where stellar gas has converted fully to radiation, first considered by Vaidya \cite{vaidya1, vaidya2}. The radiation effects are important during the later stages of gravitational collapse when the star is perhaps throwing away considerable mass as radiation. Such a non-static distribution is then be surrounded by an ever-expanding zone of radiation. When the collapse is fast enough, the horizon forms well before the singularity as to fully cover it, and a blackhole forms as the collapse endstate. However, for a slow enough collapse the horizon and trapped surface formation is delayed and a naked singularity develops as the final state of collapse \cite{joshi1}. However, not much is known today on the actual equation of state for the collapsing star in the very late stages of its gravitational collapse. These considerations in fact show a few possibilities within the framework of general relativistic gravitational collapse, subject to various physical conditions, such as the validity of energy conditions and evolution from regular initial data.                 \\

Although spherical symmetry is mathematically elegant, most of the stellar bodies do show departure from sphericity. One may approach the problem in two different ways, either looking for exact non-spherical models or try a perturbative approach where a departure from the spherical symmetry in collapse scenario is considered. In terms of analytic calculations, not much is known about a non-spherical gravitational collapse. There are indications, however, in view of the properties of the Kerr geometry, that to avoid naked singularity formation the collapsing object should not be rotating very fast, discussed by Miller and Sciama \cite{millersciama}. In Kerr geometry, the metric is characterized in terms of the mass of the particle and its angular momentum. If the angular momentum is larger than the mass, then a naked singularity naturally forms in the spacetime. It should be noted, however, that there is no interior metric yet known for a Kerr exterior solution. The ideal situation would be the existence of a non-spherical collapsing cloud, the exterior of which is given by the Kerr solution, and then to investigate the final state of the collapse. There are, however, a few examples of non-spherical collapse in literature. For an extensive review, the reader is referred to the work of Miller and Sciama \cite{millersciama}. Bronnikov and Kovalchuk made an extensive study of various non-spherical collapse with a pressureless dust distribution \cite{bronni1} and then generalized the same to include a charged dust in planar symmetry \cite{bronni2} and later an electromagnetic field for various non-spherical distributions \cite{bronni3}. Ganguly and Banerjee extended the work, investigating some exact non-spherical models with a minimally coupled scalar field in planar, cylindrical, toroidal and pseudo-planar symmetries \cite{gangulybanerjee}.

\subsection{Black Holes vs Naked Singularities}
It is perhaps obvious that the two possible outcomes of a complete gravitational collapse of massive astronomical bodies, black hole and naked singularity, must be very different from each other physically, and must have different observational signatures \cite{joshi3}. A naked singularity may present opportunities to observe the physical effects near the ultra dense regions that form in the very final stages of a collapse. In a black hole scenario, such regions are necessarily hidden within the event horizon.              \\

There have been attempts to explore physical applications and implications of naked singularities (we refer to the works of Joshi and Malafarina \cite{joshimala1}, Joshi, Malafarina and Narayan \cite{joshimalanarayan1} and references therein). An interesting recent proposal in this connection is presented by Patil, Joshi and Malafarina, to consider the naked singularities as possible particle accelerators \cite{patiljoshi1, patiljoshi2}. Among the most prominent features of how a singularity can affect incoming particles are either in the form of light bending, such as in gravitational lensing \cite{sahupatiljoshi}, particle collisions close to the singularity \cite{patiljoshi3, patiljoshi4, joshimala2}, or properties of accretion disks \cite{joshimalanarayan2}. The accretion discs around a naked singularity are regions where the matter particles are attracted towards or repelled away from the singularity with great velocities. These regions can provide an excellent arena for predictions of important observational signatures. In a scenario where the central ultra-high density region around the formation of singularity is exposed while the outer shells are still falling towards the center, shockwaves emanating from the superdense region and collisions of particles near the horizon can have significant effects on the outer layers. These effects must be considerably different from those appearing during the formation of a black hole.              \\

To summarize, recent studies on formation of naked singularities as collapse end states have manifested some of the most intriguing basic questions as follows:
\begin{itemize}
\item{ can the ultra-dense regions forming in a physically realistic collapse of a massive star remain visible forever?
}
\item{ Are there really any observable astrophysical consequences of this phenomena?
}
\item{ What is the causal structure of space-time in the vicinity of a singularity as decided by the dynamics of collapse which evolves from a regular initial collapsing profile?
}
\item{ How early or late the horizons may actually develop in a physically realistic gravitational collapse?
}
\end{itemize}

A continuing study of the collapse phenomena within a general framework may be the only way to have answers on some of these issues. This could lead us to novel physical insights and possibilities emerging out of the intricacies of gravitational force and nature of gravity, as emerging from examining the dynamical evolutions as allowed by Einstein equations.

\section{Modified theory of gravity}
A viable relativistic theory of gravity must include well-defined foundation requirements, as summarised by Faraoni \cite{faraoni}, Clifton, Ferreira and Padilla \cite{clifton}, Sotiriou \cite{soti1}, Sotiriou and Faraoni \cite{soti2}. For example, it must pass the tests imposed by Solar System and terrestrial experiments on relativistic gravity such as different observations involving the propagation of light and the orbits of massive bodies. General Relativity has been widely accepted as a complete and fundamental theory of gravity considering these foundation requirements. However, alternative theories do exist for some reasons which include a need to describe the recently observed accelerated expansion of the universe, and to fill in for the so called dark energy component. 

\subsection{Motivation for modifying Gravity}
Theories of gravity, alternative to Einstein's General Relativity, have been proposed to cure the problems of
the standard cosmological model. Motivation for studying an alternative theory arise also from the perpective of quantizations of gravity. The alternative gravitational theories constitute an attempt to formulate a semi-classical scheme in which general relativity and its most successful features can be recovered. One of the most fruitful approaches thus far has been that of Extended Theories of Gravity, which have become a paradigm in the study of the gravitational interaction. The paradigm consists, essentially, of adding higher order curvature invariants and/or minimally or non-minimally coupled scalar fields to the dynamics; these corrections emerge from the effective action of quantum gravity (for a useful discussion, we refer to the monograph by Buchbinder, Odintsov and Shapiro \cite{buchbinder}). Further motivation to modify gravity arises from the problem of fully implementing Mach's principle in a theory of gravity, which leads one to contemplate a varying gravitational coupling. Mach's principle states that the local inertial frame is determined by the average motion of distant astronomical objects \cite{bondibook}. An independent motivation for extending gravity comes from the fact that every unification scheme of the fundamental interactions, such as Superstring, Supergravity or Grand Unified Theories exhibit effective actions containing non-minimal couplings to the geometry or higher order terms in the curvature invariants. Specifically, this scheme was adopted in the study of quantum field theory on curved spacetime and it was found that interactions between quantum scalar fields and background geometry, or gravitational self-interactions, yield such corrections to the Einstein-Hilbert Lagrangian \cite{birrell}. Moreover, it has been realized that these corrective terms are inescapable in the effective action of quantum gravity close to the Planck energy as discussed by Vilkovisky \cite{vilkov}. Of course, all these approaches do not constitute a full quantum gravity theory, but are needed as a working schemes toward it.      \\

In summary, higher order terms in the invariants of the Riemann tensor, such as $R^2$, $R_{ab} R^{ab}$ or $R_{abcd} R^{abcd}$ and non-minimal coupling terms between scalar fields and geometry, have to be incorporated to the effective gravitational Lagrangian when quantum corrections are introduced. These terms occur also in the effective Lagrangian of string or Kaluza-Klein theories when a mechanism of compactification of extra spatial dimensions is used as discussed by Gasperini and Veneziano \cite{gasperini}. Besides fundamental physics motivations, all these theories have been the subject of enormous attention in cosmology due to the fact that they naturally exhibit an inflationary behaviour which can overcome the shortcomings of the general relativity based standard cosmological model. A bird's eye view of such relevant issues and possibilities has been nicely summarized in the review of Capozziello, De Laurentis and Faraoni \cite{birdseye}.

Recent developments of observational cosmology suggests that the universe has experienced two phases of cosmic acceleration. The first one is called inflation (we refer to the discussions of Starobinsky \cite{starob}, Kazanas \cite{kazanas}, Guth \cite{guth} and Sato \cite{sato}), which is believed to have occurred prior to the radiation dominated era (we refer to the works of Liddle and Lyth \cite{liddle}, Lyth and Riotto \cite{lyth}, Bassett, Tsujikawa and Wands \cite{bassette} for extensive reviews). This phase is required to solve the flatness and horizon problems of big-bang cosmology, as well as to explain a nearly flat spectrum of temperature anisotropies observed in Cosmic Microwave Background (CMB) \cite{smoot}. The second accelerating phase begins well within the matter domination era and an unknown component is believed to be behind this late-time cosmic acceleration, popularly known as the dark energy (detailed reviews regarding various aspects of the dark energy component can be found in the works of Huterer and Turner \cite{huterer}, Sahni and Starobinsky \cite{sahni}, Padmanabhan \cite{paddy}, Peebles and Ratra \cite{peebles}, Copeland, Sami and Tsujikawa \cite{copeland}). These two phases of cosmic acceleration cannot be explained by considering standard matter in the form of a fluid alone. Some component of an effective negative pressure is further required to realize the acceleration.             \\

However, the simplest possible model based on cosmological constant has its own problems such as its theoretically predicted energy scale is way too large to be compatible with the dark energy density, assuming that it originates from the vacuum energy, as discussed by Weinberg \cite{weinberg}. A scalar field with a slowly varying potential serves as a competent candidate for inflation as well as for dark energy. However, with many scalar field potentials for inflation being constructed, the CMB observations still do not show particular evidence to favor any one of such models. On the other hand, while scalar field models of inflation and dark energy correspond to a modification of the energy-momentum tensor in Einstein equations, there is another approach to explain the acceleration of the universe. Without thinking of an exotic matter component responsible for a negative pressure, modifying the theory of gravity itself may be an alternative approach where there is an effective energy-momentum tensor with purely geometrical origin due to the modification of the action \cite{cruz, clifton}.

\subsection{$f(R)$ Gravity}
The instigation of looking into a 'perhaps more general' theory of gravity came as early as in $1918$, only a few years after the first papers on General Relativity by Einstein. In that sense, fourth-order theories of gravity have a long history. These theories generalise the Einstein-Hilbert action by adding additional scalar curvature invariants to the action. Specific attention has been given to theories where general relativity is modified by making the action a more general analytic function $f(R)$ of the Ricci scalar $R$. These are generically referred to as $f(R)$ theories of gravity. The merits and demerits of these theories have been thoroughly studied in literature. These theories can lead to a period of accelerating expansion early in the history of the Universe, without the need to consider any exotic matter component. More recently, these theories have been of considerable interest as a possible explanation for the observed late-time accelerating expansion of the Universe as well (for detailed discussions on motivations, successes and challenges of $f(R)$ theories of gravity we refer to the reviews by Clifton, Ferreira, Padilla and Skordis \cite{clifton}, Sotiriou \cite{soti1}, Sotiriou and Faraoni \cite{soti2}).      \\

The $f(R)$ modifications of general relativity are derived from an action of the form
\begin{equation}\label{EHfr}
S_{EH} = \frac{1}{16 \pi G} \int d^{4}x \sqrt{-g}f(R).
\end{equation}
For $f(R) = R$, one gets the standard Einsteins' gravity. The field equations, giving rise to the modified Einstein field equations can be derived from the least action principle by using different variational approach. Two such variational principles have been mainly considered in existing literature.

\begin{enumerate}
\item{The standard metric formalism, where the the connections are considered to be metric dependent and therefore the only fields present are those coming from the metric tensor.
}
\item{The so-called Palatini variational principle where metric and connection are assumed to be independent fields. In this case the action is varied with respect to both of them.
}
\end{enumerate}
For an action linear in $R$ both formalisms lead to the same field equations \cite{cruz}. For an exhaustive review on different formalisms of modified theory of gravity we refer the reader to the the work of Sotiriou and Liberati \cite{soti3} . In this thesis, we shall restrict ourselves to the metric formulation. For that purpose, we shall assume the connection to be the usual Levi-Civita connection given by
\begin{equation}\label{christof}
\Gamma^{\alpha}_{\mu\nu} = \frac{1}{2}g^{\alpha\gamma}\Bigg(\frac{\partial g_{\gamma\nu}}{\partial x^{\mu}}+\frac{\partial g_{\mu\gamma}}{\partial x^{\nu}}-\frac{\partial g_{\mu\nu}}{\partial x^{\gamma}}\Bigg),
\end{equation}
where, as in the rest of the work, Einstein’s convention for implicit summation is assumed. The convention to be used for the metric signature will be $(+,-,-,-)$. The Riemann tensor is defined as
\begin{equation}\label{Riemann}
{\it R}^{\mu}_{\nu\alpha\beta} = \frac{\partial \Gamma^{\mu}_{\nu\alpha}}{\partial x^{\beta}}-\frac{\partial \Gamma^{\mu}_{\nu\beta}}{\partial x^{\alpha}} + \Gamma^{\mu}_{\sigma\beta}\Gamma^{\sigma}_{\nu\alpha} - \Gamma^{\mu}_{\nu\alpha}\Gamma^{\mu}_{\nu\beta}.
\end{equation}

From expression (\ref{Riemann}), the corresponding Ricci tensor and scalar curvature can be written respectively as
\begin{equation}\label{Ricci}
{\it R}_{\mu\nu} = {\it R}^{\alpha}_{\mu\alpha\nu} ~~ {\it R} = {\it R}^{\alpha}_{\alpha}.
\end{equation}

\subsection{$f(R)$ gravity in metric formalism}
Adding a matter term to the action (\ref{EHfr}), the total action for $f(R)$ gravity becomes
\begin{equation}\label{action}
A=\int\Bigg(\frac{f(R)}{16\pi G}+L_{m}\Bigg)\sqrt{-g}~d^{4}x,
\end{equation}
where $L_{m}$ is the Lagrangian for the matter distribution. The variation of the action (\ref{action}) with respect to the metric tensor leads to the following fourth order partial differential equation,
\begin{equation}
F(R)R_{\mu\nu}-\frac{1}{2}f(R)g_{\mu\nu}-\nabla_{\mu}\nabla_{\nu}F(R)+g_{\mu\nu}\Box{F(R)}=-8\pi G T^{m}_{\mu\nu},
\end{equation}
where $F(R)=\frac{df}{dR}$. Writing this equation in the form of Einstein tensor, one obtains
\begin{equation}
G_{\mu\nu}=\frac{8 \pi G}{F}(T^{m}_{\mu\nu}+T^{C}_{\mu\nu}),
\end{equation}
where
\begin{equation}\label{curvstresstensor}
T^{C}_{\mu\nu}=\frac{1}{8 \pi G}\Bigg(\frac{f(R)-RF(R)}{2}g_{\mu\nu}+\nabla_{\mu}\nabla_{\nu}F(R)-g_{\mu\nu}\Box{F(R)}\Bigg).
\end{equation}

$T^{C}_{\mu\nu}$ represents the contribution of the curvature. This may formally be treated as an {\bf effective stress-energy tensor} with a purely geometrical origin. It must be noted that this energy-momentum tensor does not necessarily obey the strong energy condition which holds in ordinary fluids (dust matter, radiation, etc.).

\subsection{Criteria for Viability}
It must be stressed that $f(R)$ theories ar investigated mainly as toy models, as proofs of the principle that modified gravity can explain the observed acceleration of the universe without the need for a dark energy component. Indeed, such examples in the domain of $f (R)$ gravity do exist that pass all the observational and theoretical constraints, for example the Starobinsky model \cite{starob}, where $f(R)$ is given by $f(R) = R + \lambda R_{0} \Bigg[\frac{1}{\Big(1+\frac{R^2}{R_{0}^2}\Big)^{n}-1}\Bigg]$. To summarise, $f (R)$ theories have helped our understanding of the peculiarities of general relativity in the broader spectrum of relativistic theories of gravity, and have taught us about important aspects of its simple generalisations. They even constitute viable alternatives to dark energy models in explaining the cosmic acceleration, although at present there is no definite prediction that sets them apart once and for all from dark energy and other models. However, there are well defined criteria for viability of modified theories of gravity which one must exercise with additional caution. Referring to the comprehensive description by Faraoni \cite{faraoni}, Sotiriou and Faraoni \cite{soti2} and references therein, we review the relevant issues in brief in this section.  \\

Certain metric $f (R)$ models have been found to suffer from a non-linear instability, which makes it difficult to construct relativistic stars in the presence of strong gravity because of a singularity developing at large curvature. For example, Dolgov and Kawasaki discovered an instability in the prototype model $f (R) = R - \frac{\mu^4}{R}$ \cite{dolgov}, famously known as the Dolgov-Kawasaki instability, which manifests itself on an extremely short time scale. This result was confirmed by Nojiri and Odintsov \cite{nojiodi1, nojiodi2}. They also showed that adding an extra $R^2$ term to this specific $f (R)$ model removes the instability. The instability was rediscovered by Baghram, Farhang and Rahvar \cite{bagh} for a specific form of the function $f (R)$ and was later generalized to arbitrary $f (R)$ theories in the metric formalism by Faraoni \cite{faraoni, faraoni1}, Cognola and Zerbini \cite{cognola}. \\
A general stability analysis of $f(R)$ models can be done by parametrizing the deviations from general relativity by assuming $f(R)$ of the form
\begin{equation}
f(R) = R + \epsilon \chi(R),
\end{equation}
where $\epsilon$ is a small positive constant with the dimensions of a mass squared and the function $\chi$ is dimensionless. For a step by step calculation we refer the reader to the reviews of Sotiriou and Faraoni \cite{soti2, faraoni, faraoni2}.                                    \\
The ultimate conclusion is that the theory is stable if and only if $\frac{d^2 f(R)}{dR^2} \geq 0$ and unstable if $\frac{d^2 f(R)}{dR^2} < 0$. The instability of stars made of any type of matter in theories with $\frac{d^2 f(R)}{dR^2} < 0$ (for example, $f(R) = \frac{1}{R}$) was confirmed with a generalized variational principle approach by Seifert \cite{mdseifert}. The stability condition $\frac{d^2 f(R)}{dR^2} \geq 0$ was readressed by Sawicki and Hu through a detailed study of cosmological perturbation \cite{husawi}.         \\
The criterion  $\frac{d^2 f(R)}{dR^2} \geq 0$ can be interpretated in a simple manner following the arguments given by Faraoni \cite{faraoni2}. Remembering that the effective gravitational coupling is $G_{eff} = \frac{G}{F(R)}$, if $\frac{dG_{eff}}{dR} = -\frac{G \frac{d^2 f}{dR^2}}{(\frac{df}{dR})^2}  > 0$ (which corresponds to $\frac{d^2 f}{dR^2} < 0$), then $G_{eff}$ increases with $R$ and a large curvature causes gravity to become stronger, which in turn causes a larger $R$, in a positive feedback mechanism driving the system away. There is no stable ground state if a small curvature grows rapidly without limit and the system runs away. If instead the effective gravitational coupling decreases when $R$ increases, which is achieved when $\frac{dG_{eff}}{dR} \geq 0$, then a negative feedback damps the increase in the gravitational coupling strength and there is no running away of the solutions.    \\

A well-posed initial value problem is necessary in order for a physical theory to have a predictability. The initial value problem for metric $f (R)$ gravity is well-posed in vacuum and also for reasonable forms of matter; as in the context of general relativity, as discussed by Salgado \cite{salgado1}, Salgado, Martinez del Rio, Alcubierre and Nunez \cite{salgado2}. A comparative study of the cauchy problem for metric $f (R)$ gravity and for general relativity can be found in the works of Noakes \cite{noakes}, Cocke and Cohen \cite{cocke}.               \\

Ghosts are massive states of negative norm which cause lack of unitarity and are common when trying to generalize Einstein’s gravity. More general theories of the form $f(R, R_{ab} R^{ab}, R_{abcd} R^{abcd}$ may contain ghost fields. A possible exception can be the case in which the extra terms appear in the Gauss-Bonnet combination, for instance $G = R^2 - 4R_{ab} R^{ab} + R_{abcd} R^{abcd}$ \cite{nunez}, but in general $f (R)$ gravity is ghost-free.     \\

A proper cosmological dynamics must include an early inflationary era followed by a radiation era and a matter era during which structure formation took place, followed by the present accelerated era. The transitions between consecutive eras must be smooth. While describing the exit from radiation era some models of $f(R)$ cosmology lead to inconsistency. Problems regarding a smooth exit from radiation era in $f(R)$ cosmological models have also been thoroughly addressed in literature (we refer to the descriptions by Faraoni \cite{faraoni}, Amendola, Polarski and Tsujikawa \cite{amendola1} and Amendola, Gannouji, Polarski and Tsujikawa \cite{amendola2}. The allowed functional form of the function $f (R)$ is not unique but rather, depends heavily on the expansion history of the universe. It is not expected to assume a simple form of the Ricci scalar as discussed by Capozziello, Stabile and Troisi \cite{capo1} and Amendola et. al. \cite{amendola2} who derived the conditions for the cosmological viability of $f (R)$ dark energy models.   \\
Structure formation of the universe depends heavily on the theory of gravity. Song, Hu, and Sawicki \cite{hu, soko, faulk} assumed an expansion history typical of the $\Lambda CDM$ model and found that various modes of structure formation are not affected by $f (R)$ corrections to Einstein gravity. The condition $\frac{d^2 f}{dR^2} \geq 0$ for the stability of perturbations are also well-established in this regards, in agreement with the arguments discussed above. Overall, the study of structure formation in modified gravity is still a 'work in progress', and often is performed within the context of specific models, some of which are already in trouble because they do not pass the weak-field limit or the stability constraints.     \\

A viable $f(R)$ model needs to be close to General Relativity in local regions for consistency with local gravity constraints. In the weak-limit approximation, all these classes of theories should be expected to reproduce general relativity which, in any case, is experimentally tested most successfully in this limit \cite{willbook}. This fact is matter of debate since extended theories of gravity do not reproduce exactly Einstein results in the Newtonian approximation but, in some sense, generalize them. In fact, as it was firstly noticed by Stelle \cite{stellegrg}, a $R^2$ theory gives rise to Yukawa like corrections to the Newtonian potential which could have interesting physical consequences. There are claims for explaining the flat rotation curves of galaxies by using such terms by Sanders \cite{sanders1, sanders2, sanders3}. Various issues regarding Newtonian limit of extended theories of gravity has been nicely summarized by Capozziello \cite{caponewton}.    \\

The weak-field limit of metric $f (R)$ gravity was studied by Sotiriou and Faraoni \cite{soti2}, following early work on particular models by Chiba et. al. \cite{chiba} and Olmo \cite{olmo}. These investigations confirm the existence of a chameleon effect in $f (R)$ gravity models. It is analogous to the chameleon mechanism of quintessence scalar field models with potentials $V(\phi) \sim \frac{1}{\phi^{\alpha}}$ (with $\alpha > 0$) (for an idea regarding the chameleon mechanism of scalar field cosmology we refer to the work of Sokolowski \cite{soko}, Faulkner, Tegmark, Bunn and Mao \cite{faulk} and to the review of Faraoni \cite{faraoni}). Clifton and Barrow investigated the cosmological and weak-field properties of models extending general relativity by means of a Lagrangian proportional to $R^{1+\delta}$. They discussed that in order to be compatible with the local astronomical tests like the perihelion shift, an $f (R)$ model should be very close to General Relativity; for instance, for $f (R) = R^{1+\delta}$, $\delta$ could at most be of the order of $10^{-19}$ \cite{cliftonbarrowweakfield}.

\subsection{$f(R)$ models in the context of the present accelerated expansion of the universe}
Lots of $f(R)$ models have been proposed and scrutinized from various physical perspectives over the years. To modify the standard Einstein gravity at low curvature, the simplest possibility is to consider a $\frac{1}{R}$ term in the Einstein-Hilbert action. Such a theory maybe suitable to derive cosmological models with a late time acceleration. However, it is now realized that inclusion of such terms in the Einstein-Hilbert action leads to instabilities \cite{clifton, paul}.           \\

The model with $f (R) = R + \alpha R^{2} (\alpha > 0)$ can lead to the accelerated expansion of the Universe because of the presence of the $\alpha R^{2}$ term. In fact, this was the first model of inflation proposed by Starobinsky in $1980$ \cite{starob}. This model is well consistent with the temperature anisotropies observed in Cosmic Microwave Background as discussed by Clifton et. al. \cite{clifton}, Paul, Debnath and Ghose \cite{paul}, Starobinsky \cite{starob}, Sotiriou and Faraoni \cite{soti2}. The discovery of dark energy in $1998$ also stimulated the idea that cosmic acceleration today may originate from some modification of gravity to General Relativity and that led to extensive studies of different possible forms of $f (R)$ to explain the dark energy and to realize the late-time acceleration. For example, the model with a Lagrangian density $f (R) = R - \frac{\alpha}{R^n} (\alpha > 0, n > 0)$ was proposed for dark energy in the metric formalism by Paul, Debnath and Ghose \cite{paul}. However, this model carries a matter instability as well as few difficulties to satisfy local gravity constraints; moreover it does not possess a standard matter-dominated epoch because of a large coupling between dark energy and dark matter. Further, addition of an $R^2$ term or an $ln R$ term to the action leads to consistent modified theory of gravity which may pass satisfactorily the solar system tests, free from instability problems as shown by Nojiri and Odintsov \cite{nojiodi1, nojiodi2}.         \\

It is only logical to explore a theory which could accommodate both an inflationary scenario at the early universe and an accelerating phase of expansion at late time followed by a matter dominated phase. From this idea, modified theories of gravity which contains both positive and negative powers of the Ricci Scalar $R$ are studied. A general form of $f(R)$ can be written as
\begin{equation}
f(R) = R + \alpha R^m + \frac{\beta}{R^n},
\end{equation}
where $\alpha$, $\beta$ represent coupling constants, and $m$ and $n$ are positive constants. The $R^m$ term dominates and it permits power law inflation if $1 \leq m \leq 2$, in the large curvature limit \cite{paul}.            \\

However, no definite physical criteria has been found so far to select a particular kind of theory capable of matching the data at all scales. For example, simple $R^m$ models do permit a matter dominated universe but fails to connect to the late accelerating phase \cite{paul}. Observational signature of $f(R)$ dark energy models that satisfy cosmological and local gravity constraints fairly well were discussed by Copeland, Sami and Tsujikawa \cite{copeland}. Das, Banerjee and Dadhich \cite{dasbanerjee} showed that it is quite possible for an $f (R)$ gravity to drive a smooth transition from a decelerated to an accelerated phase of expansion at a recent past.

\subsection{Exact solutions and Gravitational Collapse in $f(R)$ Gravity}
It is indeed true that viable $f (R)$ models are quite successful in providing a geometrical origin of the dark energy sector of the universe. However, they pose considerable difficulties in some other significant sectors. It is extremely difficult to find exact solutions as the field equations are fourth order differential equations in the metric components. Nevertheless, significant attention has been dedicated in finding spherically symmetric solutions in $f (R)$ theories \cite{nzioki}, including the collapsing solutions \cite{cembra, ghosh}.

\subsubsection{Static Spherically Symmetric Solutions}
A great deal of attention has been devoted to tatic spherically symmetric (SSS) solutions of the gravitational field equations in $f(R)$ gravities \cite{capo2, multamaki}. Solutions in vacuum have been found by simplifying assumptions like considering relations among metric coefficients or imposing a constant Ricci curvature scalar.
\begin{itemize}
\item{ Static spherically symmetric exact solutions are obtained by Capozziello, Nojiri, Odintsov and Troisi for the case where Ricci scalar is a function of radial coordinate only \cite{capo1}.
}
\item{ SSS black-hole solution is obtained for a positive constant curvature scalar by Cognola et. al. \cite{cognola1}, and a black hole solution is discussed by requiring a negative constant curvature scalar by Cruz-Dombriz, Dobado and Maroto \cite{delacruz}.
}
\item{ There are some black hole solutions in literature for nonconstant curvature scalar \cite{sebastiani, berg, agha} as well and for gravity coupled to Yang-Mills field \cite{moon, mazha1} and coupled to nonlinear electrodynamics \cite{mazha2, lobo}.
}
\item{ Hendi et al. \cite{hendi1, hendi2} further generalized the static spherically symmetric $f(R)$ black hole solution to include conformally invariant Maxwell sources. The $f(R)$ Maxwell black hole solutions has been analyzed in detail by Moon et al. \cite{moon}.
}
\item{ A new covariant formalism for static spherically symmetric solutions in $f(R)$ theories is introduced by Nzioki, Carloni, Goswami and Dunsby quite recently which shows that Schwarzschild solution is not a unique solution in these theories \cite{nzioki}. Thus, the Birkhoff theorem of General Relativity may not carry over to $f(R)$ gravity \cite{troisi, vale, saezgomez} in general.
}
\end{itemize} 

\subsubsection{Non static collapsing solutions}
Only a few exact solutions in $f(R)$ gravity are known which can describe the time evolution of massive astronomial bodies. Hence, it is of immense importance to look for non-static solutions in $f(R)$ gravity and discuss its difference from the conventional picture of gravitational collapse. Though field equations of $f(R)$ gravity are notoriously non-linear, there are some attempts in recent literature to study the time evolution of a spherical gravitational collapse.

\begin{itemize}
\item{ Bamba, Nojiri and Odintsov \cite{bamba} investigated the curvature singularity appearing in stellar collapse process in $f (R)$ theories.
}
\item{ The time scale of the appearance of singularity in exponential $f(R)$ gravity was studied by Arbuzova and Dolgov \cite{arbuzo} and they found that, explosive phenomena in a finite time may appear in systems with time dependent increasing mass density.
}
\item{ Santos \cite{esantos} investigated neutron stars in $f (R)$ theories and showed that $f (R)$ theory allows neutron stars to equilibriate with arbitrary baryon number, independent of their size.
}
\item{ Charged black holes in $f(R)$ gravity can have a new type of singularity due to higher curvature corrections, the so-called $f(R)$-induced singularity, although it is highly model-dependent, as discussed by Hwang, Lee and Yeom \cite{hwang}.
}
\item{ There has been a recent investigation using numerical simulations by Borisov, Jain and Zhang \cite{bori}. Guo, Wang and Frolov \cite{guo} investigated spherical collapse in $f(R)$ gravity numerically in double-null coordinates in Einstein frame, where the nonlinear contribution of the curvature in the action is reduced to a nonminimally coupled scalar field via a conformal transformation.
}
\item{ Kausar and Noureen \cite{kaus} worked on the effect of an anisotropy and dissipation in the fluid distribution of a collapsing sphere in a particular $f (R)$ model given by $f(R) = R + \alpha R^n$ and studied the effect of electromagnetic field on the instability range of gravitational collapse. The instability ranges in the Newtonian and post-Newtonian regimes of an adiabatic anisotropic collapsing sphere in the context of Palatini $f(R)$ gravity was studied in detail by Sharif and Yousaf \cite{sharifyousaf1}, by constructing the collapse equation with the help of the contracted Bianchi identities of the effective as well as the usual energy-momentum tensor. The dynamical instability of the charged expansion-free evolution of a spherical collapse in $f(R)$ gravity was also addressed by Sharif and Yousaf \cite{sharifyousaf2}.
}
\item{ In a very recent work Cembranos, Cruz-Dombriz and Nunez \cite{cembra} investigated the gravitational collapse in $f (R)$ gravity theories in detail for a general $f(R)$ model with uniformly collapsing cloud of self-gravitating dust particles.
}
\end{itemize}

\subsubsection{Matching conditions and spherically symmetric collapsing bodies in $f(R)$ gravity}
For any massive self-gravitating body undergoing gravitational collapse, the spacetime of the interior has to be matched smoothly with the exterior spacetime. This is not a straight-forward task in $f(R)$ theories since the fourth order field equations lead to extra matching conditions between two spacetime beyond the usual Israel-Darmois \cite{darmo, israel} conditions in General Relativity. The extra conditions arising from the matching of the Ricci scalar and it’s normal derivative across the matching surface, heavily constrict the set of useful astrophysical solutions \cite{cliftondunsby, ganguly}.                                 \\

In a very recent extensive work by Goswami, Nzioki, Maharaj and Ghosh, based on the existence of extra matching conditions, an exact inhomogeneous solution in the Starobinski model with $f (R) = R + \alpha R^2$ was found, where the collapsing stellar matter has anisotropic pressure and heat flux \cite{ritu1}. It is quite interesting to note that this solution mimics the Lemaitre-Tolman-Bondi dust solution \cite{lemaitre, tolman, bondi} in General Relativity. A general conclusion from the work of Goswami et. al. \cite{ritu1} can be summerized that for any non-linear function $f(R)$, homogeneous dynamic stars with non-constant Ricci scalar cannot be matched to a static exterior spacetime across a fixed boundary. This result has remarkable consequence such as no stellar collapse model mimicking the Oppenheimer-Snyder-Datt \cite{os} dust collapse (A collapsing homogeneous dust ball with an exterior Schwarzschild spacetime) can be admissible in $f (R)$ theories anymore; investigations regarding gravitational collapse of homogeneous matter in modified gravity (for example \cite{cembra}) also become redundant from this point of view. However, it still remains to be seen the results of collapsing evolution in other $f(R)$ models.           \\

Assuming any physically realistic star with spatial inhomogeneity should be matched with a static vacuum solution (which, in the case of spherical symmetry is the Schwarzschild solution) we briefly review here the matching of two spacetimes $\nu^{\pm}$ across the boundary surface denoted by $\Sigma$, mainly following the arguements of Goswami et. al. \cite{ritu1}. The boundary surface must be the same in $\nu^{+}$ and $\nu^{-}$, which implies continuity of both the metric and the extrinsic curvature of $\Sigma$ as in General Relativity \cite{darmo, israel}. Moreover, in $f(R)$-theories of gravity, continuity of the Ricci scalar across the boundary surface and continuity of its normal derivative must also be taken care of. For a mathematical detail we refer to the works of Senovilla \cite{seno}, Clifton, Dunsby, Goswami and Nzioki \cite{cliftondunsby}, Deruelle, Sasaki and Sendouda \cite{deru}.   \\
Writing the metric of the interior and exterior spacetime locally (near the matching surface) in terms of the Gaussian coordinates
\begin{equation}
\label{gaussmetric}
ds^2 =g_{ab} d\xi^{a} d\xi^{b} = d\tau^2 + \gamma_{ij} d\xi^{i} d\xi^{j},
\end{equation}
where $\xi^{i}, i = 1,2,3$ are the intrinsic coordinates to $\Sigma$, $\gamma_{ij}$ is the intrinsic metric (first fundamental form) of $\Sigma$ and the boundary is located at $\tau =0$. Given \ref{gaussmetric}, together with the extrinsic curvature (second fundamental form) of the boundary surface defined by
\begin{equation}
K_{ij} = -\frac{1}{2} \partial_\tau \gamma_{ij},
\label{exrinsic}
\end{equation}
the Ricci scalar can be written as
\begin{equation}
\label{R}
R = 2 \partial_\tau K - \tilde{K}_{ij} \tilde{K}^{ij} - \frac{4}{3} K^2 + \cal{R},
\end{equation}
where $\cal{R}$ is the Ricci curvature constructed from the 3-metric $\gamma_{ij}$, $K$ is the trace part of the extrinsic curvature and $\tilde{K}_{ij}$ is the trace-free part.       \\

The continuity requirements at the boundary lead to the following junction conditions in $f(R)$-theories
\begin{eqnarray}
\label{junction1}
\left[\gamma_{ij} \right]^{+}_{-} = 0, \\
\label{junction2}
\frac{d^2 f(R)}{dR^2} \left[ \partial_\tau R \right]^{+}_{-}  = 0, \\
\label{junction3}
\frac{d f(R)}{dR}  \left[ \tilde{K}_{ij} \right]^{+}_{-} = 0, \\
\label{junction4}
\left[ K \right]^{+}_{-} = 0, \\
\label{junction5}
\left[ R \right]^{+}_{-} = 0,
\end{eqnarray}
provided $\frac{d^2 f(R)}{dR^2} \neq 0$. The conditions (\ref{junction2}) and (\ref{junction5}) are the extra conditions that arise in $f(R)$-theories with non-linear function $f(R)$. These extra conditions are necessary for the continuity of the field equations across the boundary hypersurface and they impose considerable constraints on the viability and consistency of the solutions.             \\

For a non-static collapsing solution, the interior must have a well-defined time evolution. Thus the dynamic homogeneous spacetime interior to the boundary hypersurface has non-constant Ricci scalar. Therefore one has a conundrum on the cards since the Ricci scalar will evolve with time on one side of the boundary, whereas on the other side of the boundary the Ricci scalar remains constant for a static exterior spacetime. Hence the junction condition (\ref{junction5}) can never be satisfied for all epochs. This immediately nullifies the existence of homogeneous dynamic stars with non-constant Ricci scalar for example, for collapsing dustlike matter as in Oppenheimer-Snyder-Datt model \cite{os, datt}. The only homogeneous collapsing stars that can be matched to a static exterior are those which have a constant Ricci scalar in the interior. Thus the modification in the theory of gravity restricts the allowed structure and the thermodynamic properties of the collapsing star. However, alternatively the exterior of a collapsing star can be assumed as non-static, though the solar system experiments seem to suggest otherwise. The matching of the Ricci scalar and the normal derivative across the surface of a homogeneous star is still not possible unless one allows a jump in the curvature terms of the field equations across the boundary \cite{nziokithesis, tclifton}. This will lead to surface stress energy terms, that are purely generated by the dynamic curvature. Those terms on a realistic collapsing stars must have observational signatures and can perhaps be established via experimental evidences \cite{ritu1}.             \\

In summary, one must ask the question whether it is at all possible to find a physically realistic inhomogeneous collapsing solution in $f (R)$ gravity such that the collapsing stellar matter obeys all the energy conditions and at the comoving boundary of the collapsing star, the interior spacetime matches smoothly with a Schwarzschild exterior spacetime. Although highly non-linear and non-trivial to solve, under suitable symmetry assumptions the field equations can be written in integrable form to extract exact collapsing solutions in $f(R)$ gravity. In this thesis a few simple models of gravitational collapse in $f(R)$ gravity are discussed which predict an unhindered contraction of inhomogeneous fluid distribution ending in either a black hole or a naked singularity, depending on initial collapsing profiles and functional forms of $f(R)$.

\section{Introduction to Scalar Fields}
As already discussed in section $1.2.1$, the recent observation of type $Ia$ supernova ($SNIa$) (\cite{riess1, riess2} and \cite{perl}) indicated that the present universe is undergoing an accelerated expansion. The phase of cosmic acceleration of the universe cannot be explained by considering standard matter in the form of a fluid alone. An additional component is needed to explain the accelerated expansion of the universe, generating a pressure that works in the opposite way of gravity. This component is popularly called the dark energy.             \\

The most simple model to create an acceleration is the Cosmological Constant, $\Lambda$ in the Einstein equations. However, this cosmological constant suffers a serious fine tuning problem \cite{weinberg}, thw observationally required value is unfathomably smaller than the theoretically predicted value of the vacuum energy density, which actually plays the role of $\Lambda$. Therefore it is better to assume that the vacuum energy density is zero and that something else causes the negative pressure in the universe. From this motivation, time dependent scalar fields $\phi = \phi(t)$ with a self-interaction potential are studied as a likely candidate for the driver of the late-time acceleration such that the Lagrangian can lead to a dynamical equation of state parameter $w = w(t)$, which can evolve to pick up a value less than $-\frac{1}{3}$ so as to drive the present accelerated expansion.         \\

The dark energy could not have been dominant in the early universe, because in such a case structure formation would be an impossibility. Therefore it is preferrable to describe models in a manner such that dark energy naturally becomes dominant at late times independent of initial conditions. From this perspective, models with a tracker behavior were first proposed, in which the dark energy density closely tracks the radiation density until the radiation-matter equality is reached; after this epoch, a scalar field starts behaving as dark energy, eventually dominating the universe. Zlatev, Wang and Steinhardt \cite{zlatev1} introduced the concept of a tracker field. They showed that the evolution of tracker field is blind to a very wide range of initial conditions and rapidly converge to a common evolution track; eventually the scalar field energy density overtaking the matter density. It was also argued that a sufficiently stiff interaction potential can lead to tracking behaviour \cite{zlatev2}. Peebles and Ratra \cite{ratra} showed that exponential potentials can lead to stable tracking solutions. Liddle and Scherrer \cite{liddlescher} studied a classification of scalar field potentials with cosmological scaling solutions in details, analysing exact solutions and their stability properties, over a range of possible cosmological applications. These considerations motivated a search for a dynamical dark energy model caused by an exotic scalar field. Theoretically, in particle physics and string theory, scalar fields arise in a natural way, to give masses to standard-model fields without breaking gauge symmetry, for example as the Higgs particle, the dilaton field and tachyons. Scalar fields also arise in the low-energy limit of higher-dimensional theories of gravity \cite{barrowcot, whit, wand}.                \\               
A very popular choice includes a scalar field minimally coupled to the gravitational field, for which the relevant action is given by 
\begin{equation}
\textit{A}=\int{\sqrt{-g}d^4x[R + \frac{1}{2} \partial^{\mu}\phi \partial_{\mu}\phi - V(\phi) + L_{m}]},
\end{equation}
where $V(\phi)$ is the interaction potential and $L_{m}$ is the Lagrangian density for the fluid distribution. From this action, the contribution to the energy-momentum tensor from the scalar field $\phi$ can be written as
\begin{equation}
T^\phi_{\mu\nu}=\partial_\mu\phi\partial_\nu\phi-g_{\mu\nu}\Bigg[\frac{1}{2}g^{\alpha\beta}\partial_\alpha\phi\partial_\beta\phi-V(\phi)\Bigg]. 
\end{equation}
where $R$ is the curvature scalar, $\phi$ is the scalar field, $V\left( \phi \right) $ is the self-interaction potential and it is standard to use natural units. For a flat FRW scalar field dominated Universe with the line element
\begin{equation}
ds^{2}=dt^{2}-a^{2}(t)\left( dx^{2}+dy^{2}+dz^{2}\right) ,
\end{equation}
where $a(t)$ is the scale factor, the evolution of a cosmological model is determined by the system of the field equations
\begin{eqnarray}
3H^{2} &=&\rho _{\phi }=\frac{\dot{\phi}^{2}}{2}+V\left( \phi \right) ,
\label{H} \\
2\dot{H}+3H^{2} &=&-p_{\phi }=-\frac{\dot{\phi}^{2}}{2}+V\left( \phi \right),
\label{H1}
\end{eqnarray}
and the evolution equation for the scalar field
\begin{equation}
\ddot{\phi}+3H\dot{\phi}+V^{\prime }\left( \phi \right) =0,  \label{phi}
\end{equation}
where $H = \dot{a}/a>0$ is the Hubble expansion rate, the overhead dot denotes the derivative with respect to the cosmic time $t$, while the prime denotes the derivative with respect to the scalar field $\phi$, respectively.                    \\
In the next section we review the existing cosmological models based on scalar fields in very brief. 

\subsection{A brief review of Cosmological models based on scalar fields}
\subsubsection*{Inflation}
In order to overcome a number of cosmological problems such as flatness and horizon problems which plagues in the standard big-bang scenario, it is required to consider an epoch of accelerated expansion in the early
universe, i.e., inflation. From this point of view, homogeneous scalar fields, dubbed inflaton, are considered as possible candidates whose potential energy leads to an exponential expansion of the universe.  In general, the investigations go under the assumption of a single scalar field, with some underlying potential. The usual strategy includes a treatment of the system under the slow-roll approximation \cite{copeland, liddle, liddle1}, through simplifying the classical inflationary dynamics of expansion, ignoring the contribution of the kinetic energy of the inflation to the expansion rate. This has been a vastly popular approach to address the issue of inflation and there are lots of relevant investigations in the literature (for example, Copeland, Sami and Tsujikawa \cite{copeland}, Harko, Lobo and Mak \cite{harko1}, Liddle and Lyth \cite{liddle}, Caldwell, Dave and Steinhardt \cite{cald}, Bassett, Tsujikawa and Wands \cite{bas}, Caldwell and Linder \cite{linder}, Barreiro, Copeland and Nunes \cite{barre}, Maleknejad, Sheikh-Jabbari and Soda \cite{malek}); the respective potentials allowing a graceful exit have also been well classified \cite{schunk}. The basic ideas of inflation were originally proposed by Guth \cite{guth} and Sato \cite{sato} independently in $1981$. A revised version was proposed by Linde \cite{linde}, Albrecht and Steinhardt \cite{alb} in $1982$, which is dubbed as new inflation. In $1983$ Linde \cite{chaoticlinde} considered chaotic inflation, in which initial conditions of scalar fields were chaotic. By far the most useful property of inflation is that it generates both density perturbations and gravitational waves. There are different widely accepted methods of such measurements, for example, the analysis of microwave background anisotropies, velocity flows in the Universe, clustering of galaxies and the abundance of gravitationally bound objects of various types as discussed by Liddle\cite{liddle1}. Many kinds of inflationary models have been constructed over the past couple of decades (we refer to the work of Kolb \cite{kolb}, Linde \cite{linde, chaoticlinde, hybridlinde}, Albrecht and Steinhardt \cite{alb}, Bassett, Tsujikawa and Wands \cite{bas}, Copeland et. al. \cite{wandetal} for a classification). In particular, the recent trend is to construct consistent models of inflation based on superstring or supergravity models (for a review in such a direction we refer to the work of Lyth and Riotto \cite{lyth}).  

\subsubsection*{The Dark Energy component}
A large number of cosmological models which can account for the dark energy component of the universe are there in the literature. All of them have their merits, but none perhaps has a firm precedence over the others.

\begin{enumerate}
\item{ Quintessence (originates from a greek word quinta essentia, the fifth element after air, earth, fire and water, a sublime perfect substance.) is a dynamical alternative to cosmological constant. In quintessence scalar field models of dark energy an ordinary scalar field $\phi$ is minimally coupled to gravity. The potential of the Quintessence field is assumed to be dominant with respect to its kinetic energy. For a detailed review we refer the reader to the monographs by Sami \cite{booksami} and Tsujikawa \cite{tsuji} and references therein.
}
\item{ It is also possible to have models of dark energy where the kinetic energy drives the acceleration. Originally kinetic energy driven acceleration was introduced to describe inflation of the early universe and this model was named as k-inflation \cite{arme1}. Chiba, Okabe and Yamaguchi \cite{chiba} first introduced this idea to describe late time acceleration, later generalized by Armendariz-Picon, Mukhanov and Steinhardt \cite{arme2, arme3} and called the K-essence models (abbreviation of Kinetic Quintessence) of dark energy.
}
\item{ Atypical scalar field models include Phantom scalar field models and Galileons. Scalar fields $\phi$ that are minimally coupled to gravity and carry a negative kinetic energy, are known as phantom fields. They were first introduced by Caldwell \cite{caldphantom}. The phantom fields accelerate the universe to a stage of infinite volume and an infinite expansion rate at a finite future. The properties of phantom cosmological models, the phenomenon of the phantom divide line crossing in the scalar field models were investigated by Carrol, Hoffman and Trodden \cite{carrol} and Kamenshchik \cite{kamen}. Galileons on the other hand, include non-linear derivative interaction. In fact, the standard model Higgs boson may act as an inflaton due to Galileon-like interaction as discussed by Kamada, Kobayashi, Yamaguchi and Yokoyama \cite{kamada}. Generalized Galileons as a framework to develop single-field inflation models were studied by Kobayashi, Yamaguchi and Yokoyama \cite{kobaya}. However, some of these theories admit solutions violating the null energy condition and have no obvious pathologies.
}
\end{enumerate}

However, it must be mentioned that despite of exhaustive attempts over the years, the distribution of the dark energy vis-a-vis the fluid is not known. It is generally believed that the dark energy does not cluster at any scale below the Hubble scale. The study of collapse of scalar fields, particularly in the presence of a fluid may in some way enlighten us regarding the possible clustering of dark energy. In the next section we briefly summerise existing and ongoing possibilities coming out of a Scalar Field Collapse; collapsing solutions of the Einstein field equations in the spherically symmetric case with a massless or massive scalar field as matter contribution. So scalar fields considered as the matter description is indeed quite significant.  \\
Apart from its role in cosmology, scalar fields have significance in system undergoing gravitational collapse by its own right, since a scalar field with a variety of interaction potential can mimic the evolution of many a kind of matter distribution; for instance, Goncalves and Moss \cite{goncalves} showed that the collapse of a spherically symmetric self-interacting scalar field can be formally treated as a collapsing dust ball. Many reasonable matter distribution can be modelled with power law interaction potentials, for example, a quadratic potential, on the average, mimics a pressureless dust whereas the quartic potential exhibits radiation like behavior \cite{booksami}.

\subsection{Scalar Field Collapse}
The Cosmic Censorship question is one of the fundamental open problems in general relativity. It is believed that the study of Einstein field equations under realistic simplifying assumptions, like spherical or axial symmetry, may ultimately lead to a possible resolution. Special importance is attached to the investigation of collapse for a scalar field. This is because one would like to know if cosmic censorship is necessarily preserved or violated in gravitational collapse for fundamental matter fields, which are derived from a suitable Lagrangian.                 \\

The investigation of collapsing solutions of Einstein equations under spherical symmetry with a massless scalar field as matter contribution began with the pioneering works of Chritodolou \cite{christo1, christo2, christo3}. It is commonly expected that such a gravitational collapse will lead to a black hole end-state only if the initial field is strong enough. It was confirmed by Goldwirth and Piran \cite{piran} who discussed the collapse of a massless scalar field, using a characteristic numerical method. A similar behavior was observed by Choptuik \cite{chop} who solved, independently, the same problem using numerical formalism and finite-differencing techniques.

\subsubsection{Critical Phenomena in massless scalar field collapse}
The now famous critical phenomena in gravitational collapse was discovered by Choptuik in $1993$ \cite{chop},  who examined numerically the spacetime evolution of massless scalar field minimally coupled to gravity. The solutions exhibit critical behaviour; if $\alpha$ is the parameter which characterizes the solutions, then for $\alpha > 0$ the scalar field collapses, interacts and disperses leaving behind nothing but flat space. The exactly critical case is given by $\alpha = 0$ and corresponds to a spacetime which is asymptotically flat, containing a null, scalar-curvature singularity at $r = 0$. The remaining case, $\alpha < 0$, corresponds to a black hole formation. The model studied by Choptuik was the simplest one so that numerical studies are accurate enough. The conclusions are intriguing and potentially a rich store for further investigations.
\begin{enumerate}
\item{ The first is the mass scaling law in the resulting black hole mass $M$ as
\begin{equation}
M \simeq k (p - p_{\ast})^{\gamma},
\end{equation}
where $p$ is a parameter in a one-parameter family of initial data which is varied to give different initial conditions. While constant $k$ and critical value $p_{\ast}$ depend on the particular one-parameter family, the exponent $\gamma$ is universal.
}
\item{ The second finding is universality. For a finite time in a finite region of space, the spacetime generated by all initial data takes the same form as long as they are close to the so called critical conditions. The commonly approached solution is called the Critical Solution.
}
\item{ The critical solution has an amazing property, the third phenomenon, called scale-echoing. In the model studied by Choptuik, the critical solution is scale invariant by a factor $e^{\delta}$ as
\begin{equation}
\phi^{\ast}(r,t) = \phi^{\ast}(e^{\delta} r,e^{\delta} t),
\end{equation}
where $\delta \simeq 3.44$.
}
\end{enumerate}

Following Choptuiks' findings, a lot of other matter models were studied, and similar critical phenomena were discovered (for example, Brady, Chambers and Goncalves \cite{brady} and Gundlach \cite{gund1, gund2}). It is now clear that the critical phenomena in gravitational collapse are in fact common features.               \\

There are actually two kinds of critical phenomena, type $I$ and type $II$, named after analogy to critical phase transitions in statistical mechanics.
\begin{enumerate}
\item{ A critical solution with temporal periodicity instead of self-similarity or scale invariance is related to type $I$ phenomena where the resulting black hole mass $M$ is always finite when a black hole is formed.
}
\item{ A critical solution with scale invariance is related to type $II$ critical phenomena, such as the case studied by Choptuik \cite{chop}.
}
\end{enumerate}
Analytical investigations of the critical collapse problem, and a search for a theoretical explanation of the behaviour discovered by Choptuik were carried out by Brady \cite{brady} under the assumption that the collapse is self-similar, i.e. there exists a vector field, $\xi$, such that the spacetime metric, $g$, satisfies $L_{\xi} g = 2g$, where $L_{\xi}$ denotes the Lie derivative with respect to $\xi$. Such a scenario leads to a one parameter family of solutions representing scalar field collapse. These solutions were discussed in details by Roberts \cite{roberts} in the context of an investigation of counter-examples to cosmic censorship. There are recent analytical investigations consisting of increasingly generalised setup \cite{oli, frolov, soda}. Critical collapse in the context of primordial black hole initial mass function was addressed by Green and Liddle \cite{green}. A general framework for understanding and analyzing critical behaviour in gravitational collapse was given by Hara, Koike and Adachi \cite{hara} adopting the method of renormalization group, providing a natural explanation for various types of universality and scaling observed in numerical studies. In particular, universality in initial data space and universality for different models were understood in a unified way. For a detailed review on the Critical Phenomena in Gravitational Collapse we refer to the monograph by Gundlach \cite{gund2}.     \\

During the last decade, the problem of critical phenomena in a gravitational collapse of massless scalar field has been addressed rigorously under various interesting setups. Critical collapse of a massless scalar field with angular momentum in spherical symmetry was studied by Olabarrieta, Ventrella, Choptuik and Unruh \cite{olabari}. Hawley and Choptuik \cite{hawley} found type $I$ critical solutions dynamically by tuning a one-parameter family of initial data consisting of a boson star and a massless real scalar field, and numerically evolving the data, they showed that boson stars are unstable to dispersal in addition to black hole formation. A numerical study of the critical regime at the threshold of black hole formation in the spherically symmetric, general relativistic collapse of collisionless matter was carried out by Olabarrieta and Choptuik \cite{olabari1}. Ventrella and Choptuik investigated numerically \cite{ventrella} a collapse of spherically symmetric, massless spin$-\frac{1}{2}$ fields at the threshold of black hole formation and found strong evidence for a Type $II$ critical solution at the threshold between dispersal and black hole formation. A numerical study of critical gravitational collapse of axisymmetric distributions of massless scalar field was presented by Choptuik, Hirschmann, Liebling and Pretorius \cite{pretorius}. Critical collapse models of neutron stars in spherical symmetry were extensively studied in the past decade as well. Noble and Choptuik \cite{noble1} investigated type$-II$ critical phenomane in neutron star models. Radice, Rezzolla and Kellermann \cite{radice1, radice2} studied the critical evolution of a family of linearly unstable isolated spherical neutron stars under the effects of very small, perturbations and found that the system exhibits a type I critical behaviour. Quite recently Noble and Choptuik \cite{noble2} addressed the critical collapse of initially stable neutron star models that are driven to collapse by the addition of either an initially ingoing velocity profile, or a shell of minimally coupled, massless scalar field that falls onto the star.

\par Using the hyperbolic formulations of Einstein’s equations, Akbarian and Choptuik \cite{akbarchop} studied type II critical collapse of a massless scalar field in spherical symmetry and generalized the Baumgarte-Shapiro-Shibata-Nakamura formulation, very recently. Adopting standard dynamical gauge choices, they evolved the initial data sufficiently close to the black hole threshold to $(1)$ unambiguously identify the discrete self-similarity of the critical solution, $(2)$ determine an echoing exponent consistent with previous calculations, and $(3)$ measure a mass scaling exponent, also in agreement with prior computations.       

\subsubsection{Massive scalar field collapse}
Gravitational collapse with a scalar field minimally coupled to gravity have been studied extensively in literature, but usually consisting of a massless scalar field. There is only a limited amount of work on massive scalar field collapse and that too in a very restricted scenario, which we summerize in brief below. 

\begin{itemize}
\item{ Giambo addressed the conditions under which gravity coupled to self interacting scalar field determines singularity formation and showed that under a suitable matching with an exterior solution,the collapsing scalar field may give rise to a naked singularity \cite{massivegiambo}.
}
\item{ A spherically symmetric collapse of a real, minimally coupled, massive scalar field in an asymptotically Einstein–de Sitter spacetime was studied by Goncalves \cite{massivegoncalves}; using an eikonal approximation for the field and metric functions, a simple analytical criterion was found involving the physical size and mass scales of the initial matter configuration data to collapse to a black hole.
}
\item{ Goswami and Joshi constructed a class of collapsing scalar field models with a non-zero potential, which resulted in a naked singularity as the end state of collapse \cite{massiveritu1, massiveritu2}.
}
\item{ It was shown by Ganguly and Banerjee \cite{massiveganguly} that a scalar field, minimally coupled to gravity, may have collapsing modes even when the energy condition is violated, and discussed the significance of the result in the context of possible clustering of dark energy.
}
\item{ Quite recently Baier, Nishimura and Stricker \cite{baiernishi} proved that a scalar field collapse, along with a negative cosmological constant, can lead to the formation of a naked singularity.
}
\item{ It must be noted that non-spherical models of scalar field collapse are there in literature as well, for example, self-similar scalar field solutions to the Einstein equations in cylindrical symmetry symmetry by Condron and Nolan \cite{condron1, condron2}, scalar field collapse with planar as well as toroidal, cylindrical and pseudoplanar symmetries by Ganguly and Banerjee \cite{nonsphericalganguly}, however, the more popular approach being the spherically symmetric collapse.
}
\end{itemize}
            
\section{Summary of the Present Work}
In this thesis, the issue of spherically symmetric Gravitational Collapse is addressed. Exact solutions of the Einstein's field equations in closed form are studied in this regards and spacetime singularities are noted to form as an end state. Visibility of such a singular end-state depends on initial collapsing profiles from which the collapse evolves and in some cases, on the parameters of the theory itself.        \\

The Chapters $2$ and $3$, present some simple examples of exact collapsing solutions in $f(R)$ theories of gravity.

\begin{enumerate}
\item{ In {\bf Chapter $2$} a simple collapsing scenario for a general $f(R)$ model is studied. An exact solution is obtained under the assumption of separability of metric coefficients and pressure isotropy. The interior solution is matched smoothly with a vacuum exterior solution and it is discussed that the domain of validity of the collapsing solution is conditional, for instance, for $f (R) = R + \alpha R^2$, one has to choose the parameter $\alpha$ and the constant $R_0$ such that $1 + 2\alpha R_0 = 0$; where the Ricci scalar is of the form $R = R_{0} + \psi(r,t)$. The visibility of the ultimate shell-focussing curvature singularity depends on constants of integration coming out from the boundary matching condition and on the particular form of $f(R)$ (for example, the parameter $\alpha$ for $f (R) = R + \alpha R^2$).
}
\item{ In {\bf Chapter $3$}, an attempt has been made to look for collapsing solutions for a class of theories where $f(R)$ is given as a power law function of $R$, i.e., $f(R) \sim R^n$ , for as general a value of $n$ as possible. Imposing the existence of a homothetic Killing vector at the outset, the metric tensor can be simplified which has nicely been summarized and utilized by Wagh, Govinder \cite{wagh1} and Wagh, Saraykar, Muktibodh, Govinder \cite{wagh2}. A spacetime admitting a homothetic Killing vector is called self-similar, where one can have repetitive structures at various scales. Self-similar solutions have their own significance in various physical systems describing dynamics or equilibrium structure formations. To name a few examples, dynamics of strong explosions and thermal waves exhibit self-similarity. We study the collapsing solutions for such a space-time and find that for certain scenarios the collapsing fluid indeed hit shell-focussing curvature singularity, but they are always covered with an apparent horizon. Some interesting cases are observed as well where the collapsing body, after an initial collapsing era, eventually settles to a constant radius at a given value of the radial coordinate rather than crushing to singularity.
}
\end{enumerate}

In Chapters $4$, $5$ and $6$, the collapse of a self-interacting scalar field along with a fluid distribution is investigated. The scalar field is minimally coupled to gravity. Generally, no equation of state for the fluid is assumed at the outset. The relevance of such investigations stems from the present importance of a scalar field as the dark energy as already discussed. Since the distribution of the dark energy vis-a-vis the fluid is not known except the general belief that dark energy does not cluster at any scale below the Hubble scale. The study of collapse of scalar fields under different formalisms, particularly in the presence of a fluid may in some way enlighten us regarding the possible clustering of dark energy.

\begin{enumerate}
\item{ In {\bf Chapter $4$}, the collapse of a massive scalar field along with a distribution of perfect fluid is investigated for a conformally flat spacetime. The potential is taken to be a power law ($V(\phi) \sim \phi^{(n+1)}$ ) where $n$ can take a wide range of values. Adopting a completely different strategy, we exploit the integrability conditions for an anharmonic oscillator equation and transform the scalar field evolution equation into an integrable form. The integrability criterion itself leads to a second order equation for the scale factor and some general conclusions regarding the outcome of the collapse can be extracted. It is found that a central singularity results which is covered by an apparent horizon for all $n > 0$ and $n < -3$.
}
\item{ The aim of {\bf Chapter $5$} is to look at the collapse of a massive scalar field along with a fluid distribution which is locally anisotropic and contains a radial heat flux. The potential is taken to be a power law function of the scalar field or suitable combinations of power-law terms. The existence of a homothetic Killing vector implying a Self-Similarity in the spacetime is assumed at the outset. Anisotropic fluid pressure and dissipative processes are quite relevant in the study of compact objects and considerable attention has been given to this in existing literature. Particularly when a collapsing star becomes too compact, the size of the constituent particles can no longer be neglected in comparison with the mean free path, and dissipative processes can indeed play a vital role, in shedding off energy so as to settle down to a stable system. Collapsing modes leading to a final singular state are found. Whether the singularity of a zero proper volume occurs at a finite future or the modes are forever collapsing without practically hitting the singular state, depends on the potential as well as the initial conditions. Some of the singularities are found not covered by an apparent horizon. This perhaps indicates the fact that the scalar field contribution can violate the energy conditions. Anisotropy of the fluid pressure and the heat flux (departure from the perfect fluid) can also contribute towards this existence of Naked Singularities.
}
\item{ In {\bf Chapter $6$}, an analogue of the Oppenheimer-Synder collapsing model is treated analytically, where the matter source is a scalar field with an exponential potential. The Klein-Gordon equation describing the dynamics of the scalar field is simplified into a first order non-linear differential equation. The end state of the collapse is predicted to be a finite time shell-focusing singularity. The evolution of the system is found to be independent of different parameters defining the self-interacting potential. The collapse is simulteneous and results in a singularity which acts as a sink for all the curves of the collapsing congruence. An apparent horizon is always expected to form before the formation of zero proper volume singularity, which therefore remains hidden forever.
}
\end{enumerate}

%\begin{figure}[t]
%\centering
%\includegraphics[width=0.802\textwidth]{confi.pdf}
%\caption{Allowed and disallowed hoping processes in the fully projected space. (a) and (d) are the only two processes that do not increase or decrease the number of double occupancies during the hopping process. (b) increases the number of double occupancy and (c) decreases it. Though (d) does not increase the double occupancy, the initial state in the process (d) already has double occupancy and is not considered in the fully projected basis. So, the only allowed process is (a). We will see in section \ref{partialdouble} that in partial projected states, (d) is also considered.
%}
%\label{fig:kinallowed}
%\end{figure}

%*******************************************************************************
%****************************** Third Chapter *********************************
%*******************************************************************************
\chapter[Spherically symmetric collapse of a perfect fluid in f(R) gravity]{Spherically symmetric collapse of a perfect fluid in f(R) gravity\footnote[1]{The results of this chapter are reported in {\it General Relativity and Gravitation {\bf 48}:57 (2016)}}}\label{GRG1}

\ifpdf
    \graphicspath{{Chapter2/Figs/Raster/}{Chapter2/Figs/PDF/}{Chapter2/Figs/}}
\else
    \graphicspath{{Chapter2/Figs/Vector/}{Chapter2/Figs/}}
\fi

In this chapter the gravitational collapse of a perfect fluid in a general $f(R)$ gravity model is investigated. For a general $f(R)$ theory, through an investigation of the field equations it is shown that a collapse is quite possible. The singularity formed as a result of the collapse is found to be a curvature singularity of shell focussing type. The possibility of the formation of an apparent horizon hiding the central singularity depends on various initial conditions and a smooth matching of the collapsing interior to a static exterior solution.                      \\

As already discussed in sections $1.2.2$ and $1.2.3$, replacing the Ricci curvature $R$ by any analytic function of $R$ in the Einstein-Hilbert action is generically called an $f(R)$ theory of gravity. In fact, every different function $f = f(R)$ leads to a different theory. Besides the explanation ''why not?'', the primary motivation was to check whether such an $f(R)$ theory, particularly for $f(R) = R^{2}$, can give rise to an inflationary regime in the early universe \cite{starob, kerner}. The implications of an $f(R)$ theory in the context of cosmology was investigated by Barrow and Ottewill \cite{barrow1}. In the context of the discovery that the universe is undergoing an accelerated expansion at the present epoch, $f(R)$ theories find a rejuvenated interest so as to provide a possibility of a curvature driven late time acceleration where no exotic matter component has to be put in by hand. After the intial work by Capozziello {\it et al}\cite{capo} and Caroll {\it et al}\cite{carroll1}, a lot of work has been done where a late time acceleration of the universe has been sought out of inverse powers of $R$ in the Einstein-Hilbert action. Das, Banerjee and Dadhich \cite{dasbanerjee} showed that it is quite possible for an $f(R)$ gravity to drive a smooth transition from a decelerated to an accelerated phase of expansion at a recent past. Various $f(R)$ theories and their suitability in connection with various observations have been dealt with in detail by Amendola {\it et al} \cite{amendola1, amendola2}, Felice and Tsujikawa \cite{felice}, Nojiri and Odintsov \cite{nojiriphysrep}. There are well defined criteria for vialbility of modified theories of gravity which one must look at carefully.   \\

For a local distribution of mass, a gravitational collapse may lead to different interesting possibilities. For example, if the end product of a collapse is a singularity, then the question arises whether the singularity is hidden from the exterior by an event horizon or is visible for an observer. Basic ideas of such an unhindered gravitational collapse and the possibility of different end-states is discussed in section $1.1$. Although various aspects of $f(R)$ gravity has been investigated, the outcome of a collapse has hardly been addressed. The motivation of this chapter is to investigate the possibility of the formation of a black hole or a naked singularity as a result of a perfect fluid collapse in an $f(R)$ gravity model. A simple form of the metric is assumed to start with, so in that sense it may not very general, but an exact collapsing solution can be found which is valid for a fairly large domain of the theory. The density and pressure of the collapsing fluid are found to be inhomogeneous throughout the evolution, so the result is in accord with that obtained by the recent extensive work of Goswami {\it et al} \cite{ritu1} who concluded that for any non-linear function $f (R)$, homogeneous dynamic stars with non-constant Ricci scalar cannot be matched to a static exterior spacetime across a fixed boundary. The possibility of the formation of a black hole, i.e. an apparent horizon or a naked singularity as the end product of the collapse is found to be dependent on the initial conditions.

\section{Collapsing model and formation of singularity}

\subsection{Field Equations}
In $f(R)$ theories, the Einstein-Hilbert action of General Relativity is modified by using a general analytic function $f(R)$ instead of $R$. The action is given by
\begin{equation}\label{actionc2}
A=\int\Bigg(\frac{f(R)}{16\pi G}+L_{m}\Bigg)\sqrt{-g}~d^{4}x,
\end{equation}

where $L_{m}$ is the Lagrangian for the matter distribution. We take up the standard metric formulation where the action is varied with respect to $g_{\mu\nu}$ as opposed to a Palatini variation where both of the metric and the affine connections are taken as the arguments of variation. The variation of the action (\ref{actionc2}) with respect to the metric tensor leads to the following fourth order partial differential equation as the field equation,
\begin{equation}
F(R)R_{\mu\nu}-\frac{1}{2}f(R)g_{\mu\nu}-\nabla_{\mu}\nabla_{\nu}F(R)+g_{\mu\nu}\Box{F(R)}=-8\pi G T^{m}_{\mu\nu},
\end{equation}
where $F(R)=\frac{df}{dR}$ and $T^{m}_{\mu\nu}$ is the fluid contribution to the energy momentum tensor. Writing this equation in the term of Einstein tensor, one obtains
\begin{equation}\label{gmunuc2}
G_{\mu\nu}=\frac{\kappa}{F}(T^{m}_{\mu\nu}+T^{D}_{\mu\nu}),
\end{equation}
where
\begin{equation}\label{curvstresstensorc2}
T^{C}_{\mu\nu}=\frac{1}{\kappa}\Bigg(\frac{f(R)-RF(R)}{2}g_{\mu\nu}+\nabla_{\mu}\nabla_{\nu}F(R)-g_{\mu\nu}\Box{F(R)}\Bigg).
\end{equation}

$T^{D}_{\mu\nu}$ represents the contribution of the curvature in addition to Einstein tensor and $\kappa = 8 \pi G$. This may formally be treated as an effective stress-energy tensor $T^{D}_{\mu\nu}$ with a purely geometrical origin. The stress-energy tensor for a perfect fluid is given by $T^{m}_{\mu\nu}=(\rho+p)v_{\mu}v_{\nu}-pg_{\mu\nu}$. Here $\rho$ and $p$ are the density and isotropic pressure of the fluid respectively and $v^{\mu}$ is the velocity four-vector of the fluid particles, which, being a timelike vector, can be normalized as $v^{\mu}v_{\mu} = 1$.

\subsection{Exact Solution and time evolution of the collapsing sphere}
The metric is taken to be Lemaitre-Tolman-Bondi \cite{lemaitre, tolman, bondi} type with separable metric components,
\begin{equation}\label{metricc2}
ds^2=dt^2-B(t)^2X(r)^2dr^2-{r^2B(t)^2}d\Omega^2,
\end{equation}
where $d\Omega^2$ is indeed the metric on a unit two-sphere. The $(0,1)$ component of the effective Einstein's equation (\ref{gmunuc2}) can be written from equations (\ref{metricc2}) and (\ref{curvstresstensorc2}) as
\begin{equation}\label{FRc2}
\frac{\dot{F'}}{F'}=\frac{\dot{B}}{B},
\end{equation}
which readily integrates to give,
\begin{equation}\label{FandB1c2}
F'={k_0(r)}B
\end{equation}
where $k_0(r)$ is an arbitrary function of $r$, which comes from the integration with respect to $t$. Overhead dot and prime represent differentiations with respect to time $t$ and $r$ respectively.              \\
This equation can be written in the form 
\begin{equation}\label{FandB2c2}
F=B\int{k_0(r)}dr=B(t)k_1(r),
\end{equation}
where we have written $\int{k_0(r)}dr=k_1(r)$. \\
Equation (\ref{FandB2c2}), along with the condition of pressure isotropy, yields
\begin{equation}\label{eqforBc2} 
2\frac{\ddot{B}}{B}-2\frac{\dot{B}^2}{B^2}=\frac{1}{B^2}\Bigg(\frac{1}{X^2r^2}-\frac{1}{r^2}+\frac{X'}{rX^3}+\frac{k_1'}{rk_1X^2}-\frac{k_1''}{k_1X^2}+\frac{k_1'X'}{k_1X^3}\Bigg).
\end{equation}
Multiplying both sides by $B^2$, one can easily see that LHS of the resulting equation is a function of time whereas RHS is a function of $r$ only. Therefore both sides must be equal to a constant. Since we are mainly interested in the time evolution of the collapsing system, we concentrate on the time dependent part of the equation,
\begin{equation}\label{b1c2}
2\frac{\ddot{B}}{B}-2\frac{\dot{B}^2}{B^2}+\frac{\lambda}{B^2}=0,
\end{equation}
where $\lambda$ is the separation constant and is positive.       \\
This yields a first integral as
\begin{equation}\label{b2c2}
\dot{B}^2={\beta}B^2+\frac{\lambda}{2},
\end{equation}
where $\beta$ is a constant of integration. Since we are interested in a collapsing situation we shall henceforth be using the negative root, i.e., $\dot{B}<0$.
With this assumption equation (\ref{b2c2}) is integrated to yield a simple solution.
\begin{equation}\label{b-intc2}
B(t)=\frac{1}{2}e^{\sqrt{\beta}(t_0-t)}-\frac{\lambda}{4\beta}e^{-\sqrt{\beta}(t_0-t)}.
\end{equation}

The $r$-dependent part of equation (\ref{eqforBc2}) works as a constraint over the choice of initial data and gives a relation between $k_1(r)$ and $X(r)$ as
\begin{equation}\label{k1-Xc2}
\frac{1}{X^2r^2}-\frac{1}{r^2}+\frac{X'}{rX^3}+\frac{k_1'}{rk_1X^2}-\frac{k_1''}{k_1X^2}+\frac{k_1'X'}{k_1X^3}+\lambda=0.
\end{equation}

\subsection{Evolution of Density, Pressure, Curvature invariants and formation of a shell-focussing Curvature Singularity}
Using equation (\ref{b2c2}) for the $G^{0}_{0}$ and $G^{1}_{1}$ components of the field equations, the evolutions of density and pressure in terms of $f(R)$, $F(R)$ and the metric coefficients can be written as
\begin{equation}\label{densityc2}
\rho=-(4\beta k_1+\frac{f}{2F})B-\frac{k_1}{B}\Bigg(2\lambda+\frac{2}{r^2}+\frac{k_1''}{k_1X^2}+\frac{2}{r^2X^2}+\frac{2k_1'}{rk_1X^2}-\frac{X'k_1'}{k_1X^3}\Bigg),
\end{equation}
\begin{equation}\label{pressurec2}
p=(4\beta k_1+\frac{f}{2F})B+\frac{k_1}{B}\Bigg(\frac{7\lambda}{2}+\frac{2}{r^2X^2}+\frac{2X'}{rX^3}+\frac{2k_1'}{rk_1X^2}\Bigg).
\end{equation}

Using the definition of Misner-Sharp mass function, which can be written as the total energy contained by the sphere \cite{sharp},
\begin{equation}\label{misnerc2}
m(r,t)=\Bigg[\frac{C}{2}(1+\frac{\dot{C}^2}{A^2}-\frac{C'^2}{B^2})\Bigg], 
\end{equation}
for a general spherically symmetric spacetime given by $ds^2=A^2(t,r)dt^2-B^2(t,r)dr^2-C^2(t,r)d\Omega^2$. Using the first integral (\ref{b2c2}) we can write 
\begin{equation}\label{misner1c2}
m(r,t)=\Bigg[\frac{rB}{2}(1+r^2{\beta}B^2+r^2\frac{\lambda}{2}-\frac{1}{X^2})\Bigg],
\end{equation}
for the present case. \\

One can see from equation (\ref{b-intc2}) that when $t=t_0-\frac{1}{2\sqrt{\beta}}\ln(\frac{\lambda}{2\beta})$, $B(t)$ goes to zero, hence the collapsing fluid crushes to a singularity of zero proper volume whose measure is $\sqrt{-g} = B^{3}X$. Equations (\ref{densityc2}), (\ref{pressurec2}) show that both the fluid density and pressure diverge to infinitely large values when $B(t) \rightarrow 0$, i.e. at the singularity of zero proper volume. As these physical quantities, like density and pressure, are all functions of $r$, the fluid distribution is not spatially homogeneous. The collapse is thus different from the Oppenheimer-Snyder collapse in general relativity \cite{os}.          \\

The Ricci Scalar $R = R^{\alpha}_{\alpha}$ for the metric (\ref{metricc2}) is given by
\begin{equation}
\label{riccic2}
R=-12\beta-\frac{2}{B^2}\Big(3\frac{\lambda}{2}+\frac{2X'}{rX^3}+\frac{1}{r^2}-\frac{1}{r^2X^2}\Big),
\end{equation}
and the Kretschmann scalar is given by
\begin{equation}\label{kretsc2}
K=6\beta^2+\frac{1}{B^4}\Bigg[4\frac{(rX^3\dot{B}^2+X')^2}{r^2X^6}+2\frac{(-1+X^2+r^2X^2\dot{B}^2)^2}{r^2X^4}\Bigg].
\end{equation}

One can easily note that both these scalars blow up to infinity at $t=t_{s}$ where $B\Rightarrow 0$. Thus this is indeed a curvature singularity. In the present case, at the singularity $t = t_{0}$, $g_{\theta\theta} = 0$. This ensures that the singularity is a shell focusing singularity and not a shell crossing singularity \cite{seif1, seif2, waughlake3}.   \\
Existence of a singularity in a spacetime can be proved by considering gravitational focusing caused by the spacetime curvature in congruences of timelike and null geodesics. This turns out to be the main cause of the existence of singularity in the form of non-spacelike incomplete geodesics in spacetime. The issue of physical nature of a spacetime singularity is very important. There are many types of singular behaviors possible for a spacetime and some of these could be regarded as mathematical pathologies in the spacetime rather than having any physical significance. This will be especially so if the spacetime curvature and similar other physical quantities remain finite along an incomplete nonspacelike geodesic in the limit of approaching the singularity. For instance, while studying the properties of space-time in the vicinity of the Schwarzschild black-hole singularity, Lukash and Strokov \cite{lukash} came to the notion of an integrable singularity that is, in a sense, weaker than the conventional singularity and allows the (effective) matter to pass to a white-hole region. A singularity will only be physically important when there is a powerful enough curvature growth along singular geodesics, and the physical interpretation and implications of the same are to be considered.        \\
In this chapter, the curvature singularity encountered is indeed a physically significant non-integrable singularity, indicating an occurrence of nonspacelike geodesic incompleteness.    \\
An interesting point to be noted here is that $F(R)$ also turns out to be separable as functions of $r$ and $t$. The Ricci scalar for the metric (\ref{metricc2}) can be calculated as $R = -12\beta - \frac{2}{B^2}\Big(3\frac{\lambda}{2}+\frac{2X'}{rX^3}+\frac{1}{r^2}-\frac{1}{r^2X^2}\Big),$ and can be written in a simplified form as 
\begin{equation}
R = R_0 + \psi (r) \chi (t),
\end{equation}
where $R_0$ is a constant. Therefore, the requirement of separability of $F(R)=\frac{df(R)}{dR}$ may be difficult to impose for any general functional form of $f(R)$ without any restrictions or special cases. For instance, for $f(R) = R + \alpha R^{2}$, one has to choose the parameter $\alpha$ and the constant $R_0$ such that $(1+ 2 \alpha R_0) = 0$. In that sense, this model works along with certain restrictions. 

\section{Matching of the collapsing sphere with an exterior vacuum spacetime}
The parameters $\lambda$, $\beta$ and $t_0$ can be estimated from suitable matching of the interior collapsing fluid with a vacuum exterior geometry. Generally, in general relativistic collapsing models, the interior is matched with a vacuum Schwarzschild exterior, which implies continuity of both the metric and the extrinsic curvature on the boundary (\cite{darmo, israel}). However, in $f(R)$ theories of gravity, continuity of the Ricci scalar across the boundary surface and continuity of its normal derivative are also required (\cite{tclifton, deru, seno, nziokithesis}), as already discussed in section $1.2.6$.    \\

Developing a new covariant formalism to treat spherically symmetric spacetimes in metric $f(R)$ theories of gravity the general equations for a static and spherically symmetric metric in a general $f(R)$ gravity was derived by Nzioki, Carloni, Goswami and Dunsby \cite{nzioki}; these equations were used to show that validity of the Schwarzschild metric as the only vacuum solution is conditional in $f(R)$ theories. Nzioki, Goswami and Dunsby proved a Jebsen-Birkhoff-like theorem for $f(R)$ theories of gravity and found the necessary conditions required for the existence of the Schwarzschild solution in these theories \cite{nzioki2}; so that it can act as the stable limit of certain $f(R)$ models \cite{nzioki, nzioki2}. In connection with the Schwarzchild limit in $f(R)$ gravity we also refer to the recent work by Ganguly {\it et al}\cite{ganguly}. As the non-Schwarzschild counterparts are obtained mainly for $\frac{1}{R}$ theories which will hardly fall into the scheme of a separable $\frac{df}{dR}$ models, we can match our solutions to an exterior Schwarzschild solution.  \\
Matching of the first and second fundamenal form across the boundary hypersurface $\Sigma$ yields:
\begin{equation}
m(t,r)_{\Sigma} = M,
\end{equation}

\begin{equation}
X(r)_{\Sigma} = \frac{1}{\Big(1+\lambda\frac{r^2}{4}\Big)^{1/2}}
\end{equation}
where $m(t,r)$ is the Misner-Sharp mass as defined in the reference (\cite{sharp}) and $M$ is the Schwarzschild mass. The problem of matching Ricci Scalar and its' normal derivative was studied in detail by Deruelle, Sasaki and Sendouda (\cite{deru}). They generalized the Israel junction conditions (\cite{israel}) for this class of theories by direct integration of the field equations. It was utilised by Clifton et. al. (\cite{tclifton}) and Goswami et. al. (\cite{ritu1}) quite recently. Following these investigations, we match the Ricci scalar and its spatial derivative across the boundary hypersurface.       \\

For a spherical geometry where the time-evolution is governed by (\ref{b-intc2}), a smooth boundary matching of the Ricci scalar requires that the scalar can be taken in a general functional form 
\begin{equation}
R=T(t)+\frac{f_{1}(r)}{f_{2}(t)}.
\end{equation}
Here $T(t)$, $f_{1}(r)$ and $f_{2}(t)$ are defined in terms of $t$, $r$ and parameters such as $\lambda$, $\beta$ etc.
Therefore, at the boundary $r=r_{\Sigma}$, by an inspection of the continuity of $R'$, one can write
\begin{equation}
\frac{2X'}{rX^3}+\frac{1}{r^2}-\frac{1}{r^2X^2}{=^\Sigma} \Lambda_1,
\end{equation}
where $\Lambda_1$ is a constant which can be estimated in terms of the parameter $\lambda$.

\section{Visibility of the central singularity}
Whether or not the central shell-focussing singularity is visible, depends on the formation of an Apparent Horizon, which is the surface surrounding a black hole on which outgoing light rays are just trapped, and cannot expand outward. The apparent horizon thus satisfies a stronger condition than that of the event horizon, and the apparent horizon always lies inside the event horizon, or coincides with it \cite{wald, joshi1, millersciama}. The condition for the formation of such a surface is given by
\begin{equation}
g^{\mu\nu}Y,_{\mu}Y,_{\nu}=0,
\end{equation}
where $Y$ is the proper radius of the two-sphere. So $Y = rB(t)$ in the present case. Thus the relevant equation reads as
\begin{equation}\label{apphor1c2}
{r^2}{\dot{B}}^2-\frac{1}{X^2}=0.
\end{equation}

Taking advantage of the fact that $B$ and $X$ are functions of single variables, namely $t$ and $r$ respectively, one can write,
\begin{equation}\label{apphor2c2}
\dot{B}^2=\frac{1}{r^2X^2}=\delta^2,
\end{equation}
where $\delta$ is a constant. \\

Using equations (\ref{b2c2}) and (\ref{apphor2c2}), one can find, by some simple algebra, the time ($t_{ap}$) of formation of the apparent horizon as

\begin{equation}\label{t-appc2}
t_{ap}=t_0-\frac{1}{\sqrt{\beta}}\ln\Bigg(\sqrt{\frac{\delta^2}{\beta}}{\pm}\sqrt{\frac{\delta^2-\frac{\lambda}{2}}{\beta}\Bigg)}.
\end{equation}

This immediately yields the condition for the formation of the apparent horizon as ${\delta}^{2}\geq \frac{\lambda}{2}$. \\

From equation (\ref{b-intc2}), the time ($t_s$) of formation of singularity ($B=0$) is given by
\begin{equation}\label{t-s}
t_s=t_0-\frac{1}{2\sqrt{\beta}}\ln(\frac{\lambda}{2\beta}).
\end{equation}

Depending on $\lambda$ and $\delta$, the visibilty of the central singularity is determined. From the last two equations, 
one has
\begin{equation}
\label{t-appcondc2}
t_{s} - t_{ap} = \frac{1}{\sqrt{\beta}} \ln \Bigg[\frac{(\delta {\pm} \sqrt{{\delta}^{2} - \frac{\lambda}{2}})}{\sqrt{\lambda}}\Bigg].
\end{equation}

As the singularity is independent of $r$, this scenario is essentially a non-central one and appears at all points simultaneously. It was discussed by Goswami and Joshi\cite{gosjoshi}, Joshi, Goswami and Dadhich\cite{joshiritudadhi} that in such a case, there is no possibility of a naked singularity.

\section{Discussion}

With a simple metric where the metric components are separable as products of functions of time and the radial coordinate, a spherically symmetric gravitational collapse in a framework of a general $f(R)$ theory is discussed in this chapter. $F = \frac{df}{dR}$ also happens to be a separable function of $r$ and $t$ coordinates which defines certain conditional domain of validity for this theory. It is shown that the collapse necessarily leads to a singularity of zero proper volume, and the physical quantities like density, pressure and the scalar curvature etc. diverge to infinity. The question of the formation of apparent horizon depends on the relative values of $\lambda$ and $\delta$, both of which are separation constants. These constants may be fixed either by matching the collapsing solution to the exterior metric or by initial conditions. In $f(R)$ gravity, a stable Schwarzschild analogue is not guaranteed, however, there are examples of such an analogue for quite general classes of $f(R)$ theories \cite{nzioki, nzioki2, ganguly}. Assuming the existence of a stable Schwarzchild solution for the exterior, matching at the boundary is discussed.         \\

It deserves mention that one can find the condition for a vacuum collapse by simply setting $p=\rho=0$. This puts a condition on the arbitrary function of $r$
\begin{equation}
k_1 = \frac{r}{X(r)} e^{\int(\frac{1}{r}-\frac{\lambda r}{2})X(r)^2dr},
\end{equation}
which is a result of the simplication of the expressions for the density and pressure in equations (\ref{densityc2}, \ref{pressurec2}).         \\

The conclusion that the the density and pressure remains inhomogeneous, strongly supports the result obtained by Goswami {\it et al}, which is the only extensive work in $f(R)$ collapse \cite{ritu1}. The advantage of the model presented in this chapter is that this is a simple solution, and thus can be useful for any further study. In particular, these simple models can potentially serve a secondary purpose. There is no significant knowledge regarding the possible clustering of dark energy, it is more or less granted that it does not cluster at any scale smaller than the Hubble scale. The investigation regarding collapse may also indicate the possibilities in this connection in a modified theory of gravity.

%We believe that the stronger inhomogeneity in the IMT method looses the LDOS modulations easily, and peaks smear out and become obscured in the large background noise compared to the GIMT findings.
%We conclude that, while the renormalization of effective disorder plays a significant role to differentiate the GIMT and IMT results, there are more to it, likely related to the complex interplay of different order-parameters through a complete self-consistency.

\chapter[Gravitational collapse in f(R) gravity for a spherically symmetric spacetime admitting a homothetic Killing vector]{Gravitational collapse in f(R) gravity for a spherically symmetric spacetime admitting a homothetic Killing vector\footnote[1]{The results of this chapter are reported in Eur. Phys. J. Plus (2016) {\bf 131}: 144}}

% **************************** Define Graphics Path**************************

\ifpdf
    \graphicspath{{Chapter3/Figs/Raster/}{Chapter3/Figs/PDF/}{Chapter3/Figs/}}
\else
    \graphicspath{{Chapter3/Figs/Vector/}{Chapter3/Figs/}}
\fi

The motivation of this chapter is to look at a collapsing scenario in $f(R)$ gravity in a spherically symmetric spacetime. Although various aspects of $f(R)$ gravity theories have been studied quite extensively, interest in collapsing models in such theories came into being quite recently. In the second chapter, a collapsing model was discussed for a general $f(R)$ model (without choosing any particular form) where a condition of separability of the metric tensor was assumed at the outset so as to deal with the non-linearity of the equation system. In this chapter the discussion starts assuming a general form of $f(R)$, but eventually attempts to look for solutions for $f(R) \sim R^{n}$, for as general a value of $n$ as possible. Relevance of such $f(R)$ models has been discussed in brief in section $1.2.5$. The equation system is treated after imposing the existence of a homothetic Killing vector at the outset. The existence of a homothetic Killing vector results in a simplification of the metric tensor. This perhaps restricts the geometry, but keeps alive the dependence on the radial coordinate $r$ which is indeed crucial in keeping the option open for a formation of horizon, in case an ultimate curvature singularity develops as a result of the collapse.

\section{Conformal symmetry in general relativity}
Symmetry analysis provides a very useful tool in providing new exact solutions to the field equations. Amongst a large number of treatise on spacetime symmetries, we refer to only a few, namely the reviews of Choquet-Bruhat et. al. \cite{bruhat}, Stephani et. al. \cite{stephani} and Hall \cite{hall}.     \\
Conformal Symmetries have the geometric property of preserving the structure of the null cone by mapping null geodesics to null geodesics. With a conformal symmetry the angle between two curves remains the same and only the distance between two points are scaled by a factor depending on the spacetime points. There are extensive studies of conformal symmetries in general relativity from various perspectives, for example, conformal geometry has been studied in a Robertson-Walker spacetime by Maartens and Maharaj \cite{maartens1c3} and Keane and Barrett \cite{keane1}. A detailed analysis of conformal vectors has been undertaken by Maartens and Maharaj \cite{maartens2c3} and Keane and Tupper \cite{keane2} in $pp$-wave spacetimes. Tupper et. al. \cite{tupper} considered the existence of conformal vectors in null spacetimes. General conformal equations in plane symmetric static spacetimes were studied by Saifullah and Yazdan \cite{saifullah}. The full conformal structure of a spherically symmetric static spacetime was found by Maartens, Maharaj and Tupper \cite{maartens3c3}. The conformal geometry of nonstatic spherically symmetric spacetimes was analysed by Moopanar and Maharaj \cite{moopanar} without specifying any form of the matter content. Chrobok and Borzeszkowski \cite{chrobok} modelled irreversible thermodynamic processes close to equilibrium and Bohmer et al \cite{bohmer} modelled wormhole structures with exotic matter. Mak and Harko \cite{mak} studied charged strange stars with a quark equation of state, Esculpi and Aloma \cite{esculpi} generated anisotropic relativistic charged fluid spheres with a linear barotropic equation of state and Herrera, Di Prisco, Ibanez \cite{herrera1} studied reversible dissipative processes and Landau damping in stellar systems. Shear-free relativistic fluids which are expanding and accelerating are important for describing gravitational processes in inhomogeneous cosmological models (for a overview we refer to the monograph by Krasinski \cite{krasinski}) and radiating stellar spheres as discussed by Herrera and Santos \cite{herrera2}. For a general form of matter content, the complete conformal geometry of shear-free spacetimes in spherical symmetry was studied by Moopanar and Maharaj \cite{moopanar1}.             \\ 

A vector field $X$ is called a Killing vector field if it satisfies the Killing equation given as:
\begin{equation}\label{killingc3}
L_{X}g_{ab}=g_{{ab},c}X^c+g_{cb}X^c,a+g_{ac}X^c,b=0,
\end{equation}
where $L$ is the Lie derivative operator along the vector field $X$. The existence of a Killing vector signifies a symmetry and hence a conserved quantity. For example, if a timelike Killing vector exists for a particular metric, the energy is conserved.          \\

If the right hand side of equation (\ref{killingc3}) is not zero but proportinal to $g_{ab}$, 
\begin{equation}\label{homothetyc3}
L_{x}g_{ab}=2\Phi g_{ab},
\end{equation}
$X$ is called conformal Killing vector. Here $\Phi$ is a function, called the conformal factor. The space time admits a conformal symmetry if the equation (\ref{homothetyc3}) admits a solution for $X$.

\subsection{Homothetic Killing Vector}
For a Homothetic Killing vector (HKV), $\Phi$ must be a constant and is thus a special case of a conformal killing vector. Such vectors scale distances by the same constant factor and also preserve the null geodesic affine parameters.        \\

A Self-Similar Spacetime is characterized by the existence of a Homothetic Killing vector. Spherically symmetric self-similar solutions of the Einstein field equations for a perfect fluid was studied in extensive details by Cahill and Taub \cite{cahilltaub}. Any spherically symmetric spacetime is self-similar if it admits a radial area coordinate $r$ and an orthogonal time coordinate $\tau$ such that for the metric components $g_{\tau\tau}$ and $g_{rr}$, the following relations hold
\begin{eqnarray}
g_{\tau\tau}\left(\kappa \tau, \kappa r\right) &=& g_{\tau\tau}\left(\tau, r\right), \\
g_{rr}\left(\kappa \tau, \kappa r\right) &=& g_{rr}\left(\tau, r\right), 
\end{eqnarray}
for all constants $\kappa > 0$. For a self-similar spacetime, the Einstein field equations, a set of partial differential equations, reduce to ordinary differential equations with the metric components being functions of a single arguement $z = \frac{\tau}{r}$.              \\
Importance of self-similar spacetimes in cosmological context has been discussed extensively in the literature (we refer to the work of Cahill and Taub \cite{cahilltaub} and the monograph by Joshi \cite{joshi1}). While self-similarity is a strong restriction of geometry, it has been successfully exploited in various physical scenarios (for a recent survey on the importance of self-similarity in General Relativity we refer to the summary by Carr and Coley \cite{carrcoley}). In the next sub-section it is discussed that the existence of a homothetic Killing vector field for a spherically symmetric spacetime automatically implies the separability of the spacetime metric coefficients in terms of the co-moving coordinates and that the metric can be written in a simplified unique form (following the work of Wagh and Govinder \cite{waghgovi}).

\subsection{Spherically Symmetric, Self-Similar Spacetimes}
We assume that a spherically symmetric spacetime admits a Homothetic Killing vector of the form 
\begin{equation}\label{vec1}
X^{a} = (0,f(r,t),0,0).
\end{equation}
Typically, a homothetic Killing vector is written in the form
\begin{equation}\label{vec2}
{X}^a = (T,R,0,0) .
\end{equation} 
However, any vector of the form (\ref{vec1}) can be transformed into the form (\ref{vec2}) via a coordinate transformation
\begin{equation} \label{sstrans}
R = l(t) \exp\left(\int f^{-1} dr\right) \qquad T = k(t) \exp\left(\int f^{-1} dr\right), 
\end{equation} 
without loosing any generality.         \\

If a general spherically symmetric line element given by
\begin{equation}\label{met1} 
ds^2 = e^{2\nu(t,r)} dt^2 - e^{2\lambda(t,r)} dr^2 - Y^{2}(t,r)(d\theta^2 + \sin^2 \theta d\phi^2)
\end{equation}
admits a Homothetic Killing vector of the form (\ref{vec1}), the expression (\ref{homothetyc3}) reduces to the system of four equations given by
\begin{eqnarray}
f(r,t) \frac{\partial \nu}{\partial r} &=& \Phi \label{phinu} \\
\frac{\partial f(r,t)}{\partial t} &=& 0 \label{fcr}\\
\left( \frac{1}{Y} \frac{\partial Y}{\partial r} - \frac{\partial \nu}{\partial r} \right) f(r,t) &=& 0 \\
\left( \frac{\partial \lambda}{\partial r} - \frac{\partial \nu}{\partial r} \right) f(r,t) + \frac{\partial f(r,t)}{\partial r} &=&0,
\end{eqnarray}
where $\Phi$ is a constant.        \\

Solving the above system of equations, one can obtain
\begin{eqnarray}
f &=& F(r), \label{feqn} \\
Y &=& g(t) \exp\left(\int\frac{\Phi}{F(r)}d r\right), \\
\lambda &=& \int\frac{\Phi}{F(r)}d r - \log F(r) + h(t), \\
\nu &=& \int\frac{\Phi}{F(r)}d r + k(t),  
\end{eqnarray} 
so that the spacetime metric becomes separable and is given by 
\begin{equation} \label{finalmet}
ds^2 = k^2(t)\exp\left(2\int\frac{\Phi}{F(r)}dr\right) \left[dt^2 - \frac{h^2(t)}{F^2(r)} dr^2 + g^2(t)\left(  d\theta^2 + \sin^2{\theta} d\phi^2 \right)\right].
\end{equation}

The explicit forms of the metric coefficients depend on the Einstein Field Equations for the choice of specific energy-momentum tensor for the matter in the spacetime. This metric can be written as (Wagh et. al.\cite{waghh})
\begin{equation} \label{metgen}
ds^2 = y^2(r)A^2(t)dt^2 - \gamma^2 B^2(t) \left(\frac{dy}{dr}\right)^2 dr^2 + y^2(r) R^2(t) \left[d\theta^2 + \sin^2{\theta} d\phi^2 \right],
\end{equation}
where $y(r)$ is an arbitrary function of $r$ and $y'(r) = \frac{dy(r)}{dr}$. One can easily check that the metric of the form (\ref{metgen}) admits a Homothetic Killing vector given by
\begin{equation}\label{methkv}
X = \frac{y}{\Phi y'} \frac{\partial}{\partial r}.
\end{equation}

One important point to note here is that the general metric admitting (\ref{vec2}) has metric coefficients which are functions of $\frac{t}{r}$. If written in terms of the transformed coordinates $R$ and $T$, the metric will not be diagonal. The imposition of diagonality of the metric will require a relationship between $l(t)$ and $k(t)$. Such a relation can always be imposed as discussed by Wagh and Govinder \cite{waghgovi}. However, the equivalence of the spacetimes only holds if the transformation (\ref{sstrans}) is non-singular. For example, in the case of the Robertson-Walker spacetime, the transformation is singular as the function $F(r) = 0$.   \\

The requirement of self-similarity of a spherically symmetric spacetime uniquely fixes the metric to the form (\ref{metgen}). The spacetime (\ref{metgen}) is radiating and shearing. There are attempts to describe gravitational collapse under such a configuration, by suitable matching with an appropriate exterior spacetime, for instance by Ori and Piran\cite{ori1}, Wagh et. al.\cite{waghh}. In the next section, we study the evolution of a gravitational collapse in $f(R)$ gravity for a metric of the form (\ref{metgen}).

\section{Exact Collapsing Solution for $f(R)=\frac{R^{(n+1)}}{(n+1)}$}
As in the previous chapter, we work in the standard metric formalism of $f(R)$ gravity where the action is varied with respect to $g_{\mu\nu}$. The variation of the action with respect to the metric tensor leads to the following partial differential equations (already discussed in sections $1.2.3$ and $2.1.1$) as the field equations,
\begin{equation}\label{fec3}
F(R)R_{\mu\nu}-\frac{1}{2}f(R)g_{\mu\nu}-\nabla_{\mu}\nabla_{\nu}F(R)+g_{\mu\nu}\Box{F(R)}=-8\pi G T^{m}_{\mu\nu},
\end{equation}
where $F(R)=\frac{df}{dR}$.
Writing this equation in the form of Einstein tensor, one obtains
\begin{equation}\label{fefec3}
G_{\mu\nu}=\frac{\kappa}{F}(T^{m}_{\mu\nu}+T^{C}_{\mu\nu}),
\end{equation}
where $T^{m}_{\mu\nu}$ is the stress energy tensor for the matter distribution defined by $L_{m}$ and 
\begin{equation}\label{curvstresstensorc3}
T^{C}_{\mu\nu}=\frac{1}{\kappa}\Bigg(\frac{f(R)-RF(R)}{2}g_{\mu\nu}+\nabla_{\mu}\nabla_{\nu}F(R)-g_{\mu\nu}\Box{F(R)}\Bigg).
\end{equation}    
$T^{C}_{\mu\nu}$ represents the contribution of the curvature and may formally be treated as an effective stress-energy tensor with a purely geometrical origin. The stress-energy tensor for the matter part is taken to be that of a perfect fluid which is given by $T^{m}_{\mu\nu}=(\rho+p)v_{\mu}v_{\nu}-pg_{\mu\nu}$. Here $\rho$ and $p$ are the density and pressure of the fluid respectively and $v^{\mu}$ is the velocity four-vector of the fluid particles, which, being a timelike vector, can be normalized as $v^{\mu}v_{\mu} = 1$. This theory is essentially a nonminimally coupled theory, the curvature related term $F(R)$ couples with the matter sector $T^{m}_{\mu\nu}$ nonminimally.          \\

Following the discussion of the last subsection, the metric is assumed to be in the form 
\begin{equation}\label{metricc3}
ds^2=y^2(r)dt^2-2B^2(t)\Bigg(\frac{dy}{dr}\Bigg)^2dr^2-y^2(r)B^2(t)d\Omega^2,
\end{equation}

The Ricci Scalar for the metric (\ref{metricc3}) can be calculated to be
\begin{equation}\label{riccic3}
R=\frac{1}{y^2(r)}\Bigg(\frac{1}{B^2}-6\frac{\dot{B}^2}{B^2}-6\frac{\ddot{B}}{B}\Bigg),
\end{equation}
where an overhead dot and a prime indicate differentiation with respect to time $t$ and the radial coordinate $r$ respectively. It is interesting to note that the Ricci scalar also turns out to be separable in functions of $t$ and $r$ for this class of spacetime line elements.   \\

From (\ref{metricc3}) and (\ref{fec3}), the $G_{01}$ equation yields
\begin{equation}\label{g01c3}
2\frac{\dot{B}}{B}\frac{y'}{y}=\frac{\dot{F}'}{F}-\frac{y'}{y}\frac{\dot{F}}{F}-\frac{\dot{B}}{B}\frac{F'}{F}.
\end{equation}

From equation (\ref{riccic3}), the derivative of the Ricci Scalar with respect to $r$ can be written in the form
\begin{equation}\label{trickc3}
R'=-2\frac{y'}{y}R.
\end{equation}

Equations (\ref{g01c3}) and (\ref{trickc3}) can be combined to yield
\begin{equation}\label{trick2c3}
2\frac{\dot{B}}{B}=\Bigg[\frac{2R\frac{d^2F}{dR^2}+3\frac{dF}{dR}}{2R\frac{dF}{dR}-F}\Bigg]\dot{R}=\Lambda(R)\dot{R},
\end{equation}
where $\Lambda(R)$ depends solely on the choice of $f(R)$. \\

In what follows, a particular form of $f(R)$, namely, $f(R)=\frac{R^{(n+1)}}{(n+1)}$ ($n \neq -1$) is chosen, so that $F(R)=R^n$. With this choice, equation (\ref{trick2c3}) simplifies to the form

\begin{equation}
\label{trick3c3}
\frac{\dot{B}}{B}=\frac{n(2n+1)}{2(2n-1)}\frac{\dot{R}}{R}=m\frac{\dot{R}}{R},
\end{equation}
where $m=\frac{n(2n+1)}{2(2n-1)}$.
Hence, it is straightforward to write
\begin{equation}\label{BandRc3}
B=\delta(r)R^m,
\end{equation}
where $\delta(r)$ is an arbitray function of $r$ arising out of the integration over time.       \\

From the metric (\ref{metricc3}) and the field equations (\ref{fec3}) one can write the condition for isotropy of the fluid pressure as 
\begin{equation}\label{isopressc3}
2\frac{\ddot{B}}{y^2B}=2\frac{\dot{B}\dot{F}}{y^2BF}+\frac{3F'}{2yy'B^2F}-\frac{F''}{2B^2y'^2F}-\frac{F'}{2B^2y'^2F}\Bigg(2\frac{y'}{y}-\frac{y''}{y'}\Bigg).
\end{equation}

Using equation (\ref{trickc3}) and (\ref{BandRc3}) in (\ref{isopressc3}), one can derive a non-linear second order differential equation for the scale factor $B(t)$ as
\begin{equation}\label{timeevolutionc3}
2\ddot{B}B-4\frac{(2n-1)}{(2n+1)}\dot{B}^2+2n(n+1)=0,
\end{equation}

where $\alpha=4\frac{(2n-1)}{(2n+1)}$. This equation can be readily integrated to yield a first integral as
\begin{equation}\label{1stintc3}
\dot{B}^2=\lambda B^{\alpha}+\frac{2n(n+1)}{\alpha},
\end{equation}
where $\lambda$ is a constant of integration and sensitive to initial collapsing profile. The general solution of (\ref{1stintc3}) can be written in terms of Gauss' Hypergeometric Function,
\begin{equation}\label{exactsolc3}
\frac{B}{\sqrt{\frac{2n(n+1)}{\alpha\lambda}}}{_2}F_{1}\Bigg[\frac{1}{2},\frac{1}{\alpha};\Big(1+\frac{1}{\alpha}\Big);-\frac{\alpha\lambda B^\alpha}{2n(n+1)}\Bigg]=\sqrt{\lambda}(t_0-t).
\end{equation}
For a real solution $\Big(1+\frac{1}{\alpha}\Big) > 0$, which obviously imposes some restriction over the possible choices of $n$, i.e. the choice of the form of $f(R)$.  One can note that for $n=0$, one goes back to Einstein gravity, for $n=\frac{1}{2}$, equation (\ref{trick3c3}) is not valid and for $n=-\frac{1}{2}$, $B$ is constant so there is no evolution at all. To have a real time-evolution one must impose that $n\notin [-1,-\frac{1}{2}]\cup[0,1)$.

\section{Analysis of the solution; study of the formation of an Apparent Horizon}

\begin{figure}[t]
\begin{center}
\includegraphics[width=0.5\textwidth]{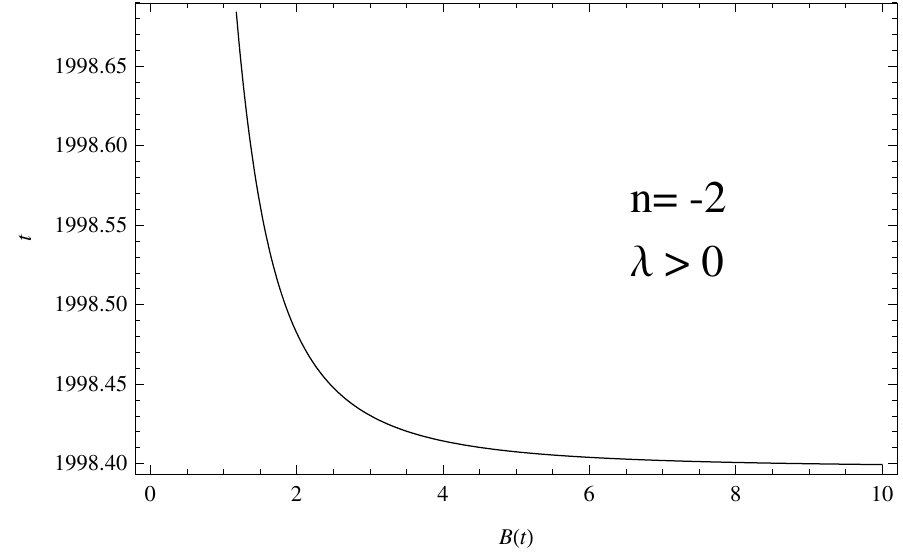}
\includegraphics[width=0.5\textwidth]{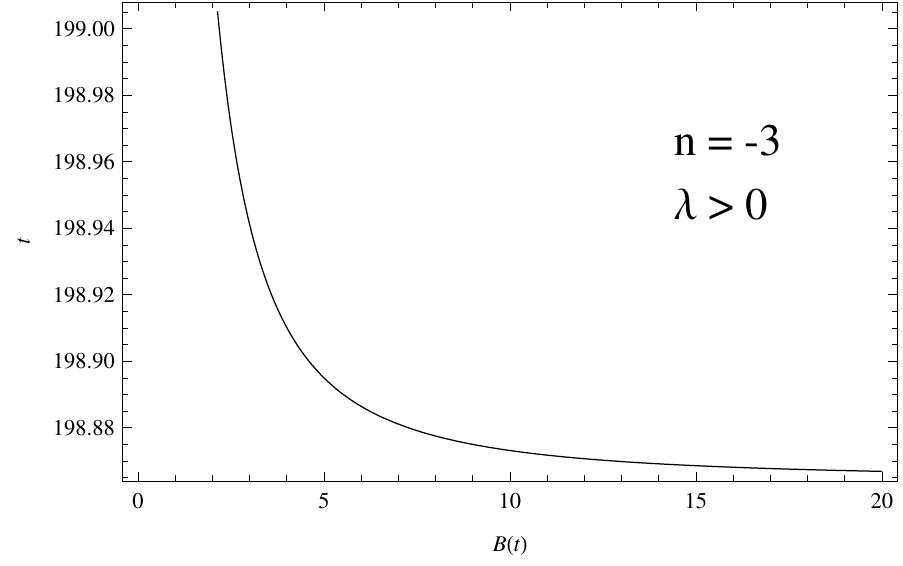}
\caption{Time evolution for $\lambda > 0$ for $f(R)=-\frac{1}{R}$ and $f(R)=-\frac{1}{2R^2}$.}
\end{center}
\label{fig:ss12}
\end{figure}

Generally, it is not easy to invert (\ref{exactsolc3}) to write $B(t)$ explicitly as a function of $t$. So in what follows we try to look at the collapsing modes with the help of plots of $t$ vs $B(t)$ for different initial conditions, using the expression (\ref{exactsolc3}). Figures $3.1$ shows the behaviour of the collapse for $\lambda > 0$, for $n=-2$ and $n=-3$. For both $n=-2$ and $n=-3$ the scale factor $B$ and hence the volume decreases with time, but the rate of collapse slows down, eventually, with the sphere asymptotically settling down to a minimum non-zero volume. 

\begin{figure}[h]
\begin{center}
\includegraphics[width=0.5\textwidth]{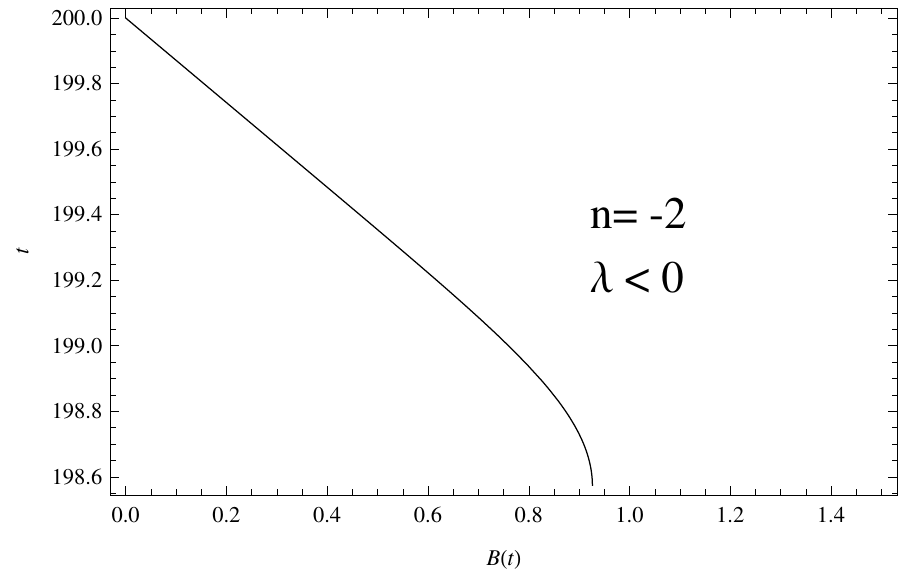}
\includegraphics[width=0.5\textwidth]{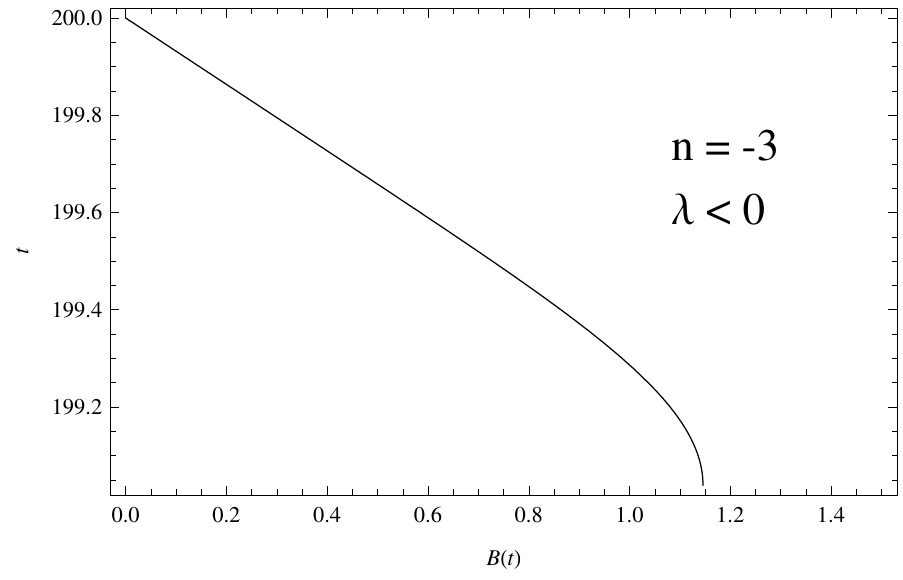}
\caption{Time evolution for $\lambda < 0$ for $f(R)=-\frac{1}{R}$ and $f(R)=-\frac{1}{2R^2}$.}
\end{center}
\label{fig:ss34}
\end{figure}

However, for $\lambda < 0$, the radius of the two sphere goes to zero quite rapidly and reaches zero at a finite future for $n=-2$ and $n=-3$ as shown by figure $3.2$. The figures also show that the nature of collapse hardly depends on the particular choice of $n$, only the time of reaching the singularity changes. For other negative values of $n$ allowed by the model, plots of exactly similar nature can be obtained.

\begin{figure}[h]
\begin{center}
\includegraphics[width=0.5\textwidth]{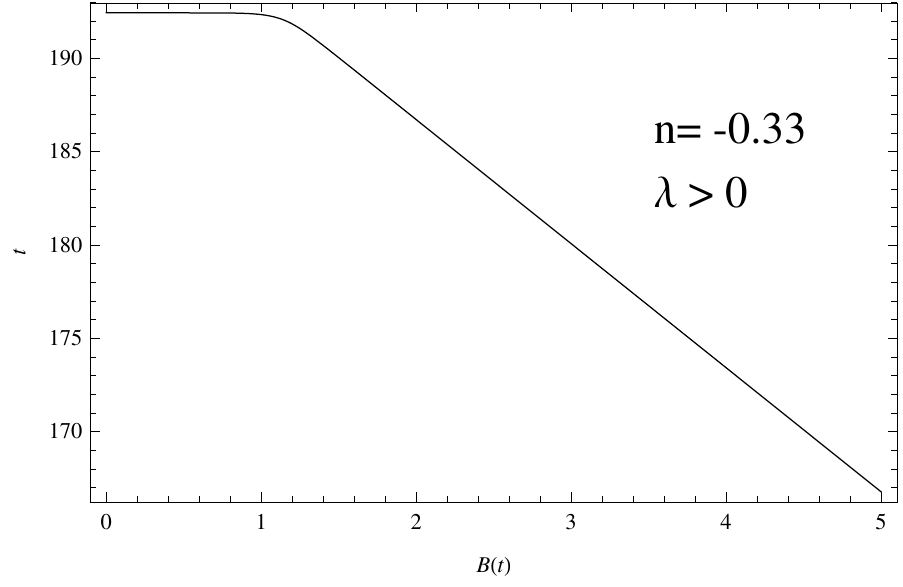}
\includegraphics[width=0.5\textwidth]{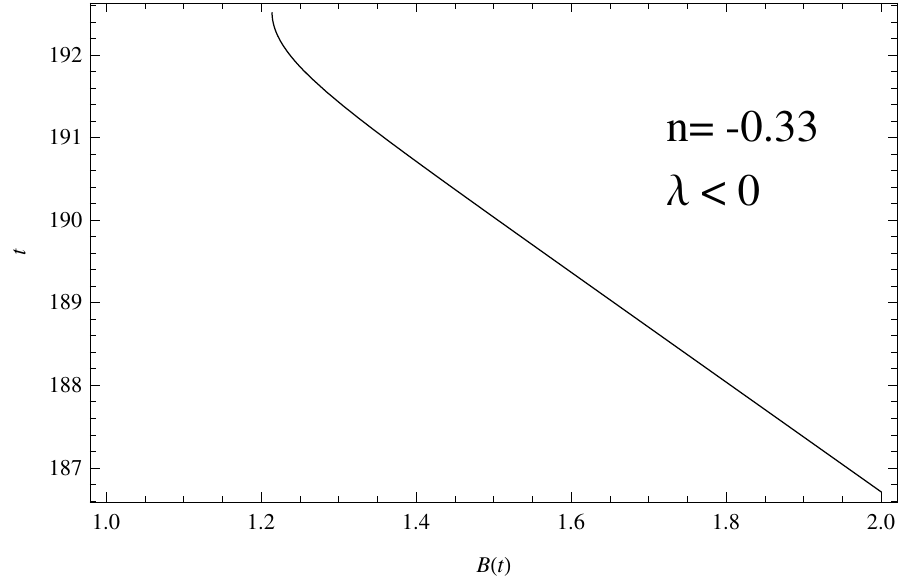}
\caption{Time evolution for different signatures of initial condition ($\lambda$) and $f(R) = \frac{3R^{2/3}}{2}$.}
\end{center}
\label{fig:ss78}
\end{figure}

For $-\frac{1}{2} < n <0$, the collapsing scenario as given in Figure $3.3$ is quite different. For a positive $\lambda$, the sphere contracts at a steady manner to begin with, but at a particular time the radius suddenly hurries towards zero. For $\lambda < 0$, the sphere collapses steadily to a certain volume, and apears to equilibriate itself at a finite volume.    

\begin{figure}[h]
\begin{center}
\includegraphics[width=0.5\textwidth]{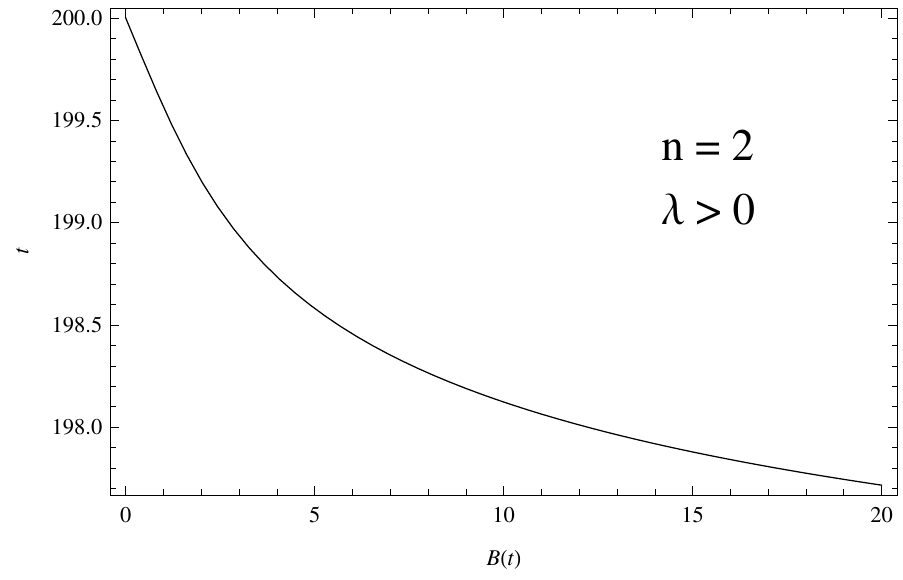}
\includegraphics[width=0.5\textwidth]{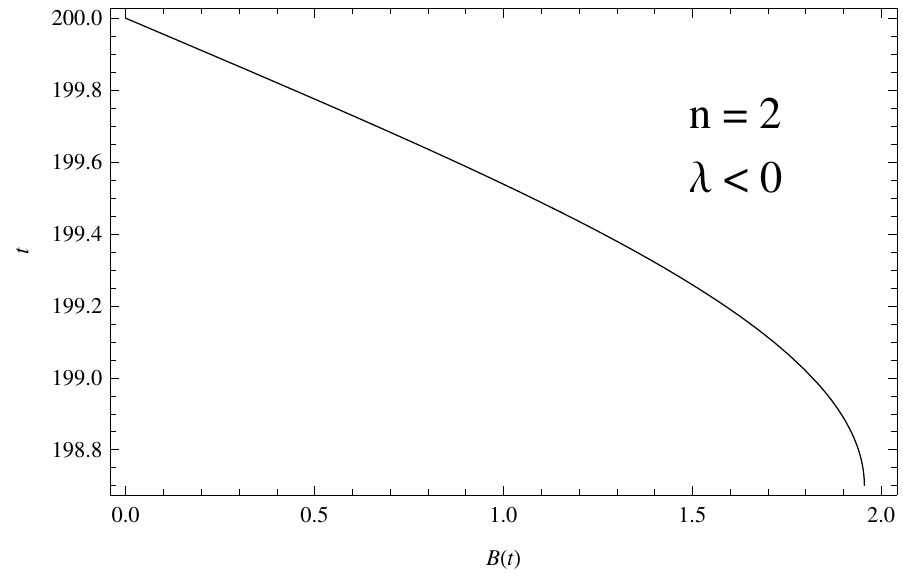}
\caption{Time evolution for different signatures of initial condition ($\lambda$) and $f(R)=\frac{R^3}{3}$.}
\end{center}
\label{fig:ss910}
\end{figure}

Figures $3.4$ and $3.5$ depict the scenario for positive values of $n$ in the allowed domain. The models show unhindered collapse, the radius goes to zero at a finite future. We have chosen two values of $n$ as examples, $n=2$ (Figure $3.4$) and $n=1$ (Figure $3.5$). Both cases show that for a positive $\lambda$, the process of collapse slows down towards the end, whereas for a negative $\lambda$ the radius shrinks zero rather rapidly. For any other positive values of $n$, it is easy to check that the scenario is exactly similar qualitatively. 

\begin{figure}[h]
\begin{center}
\includegraphics[width=0.4\textwidth]{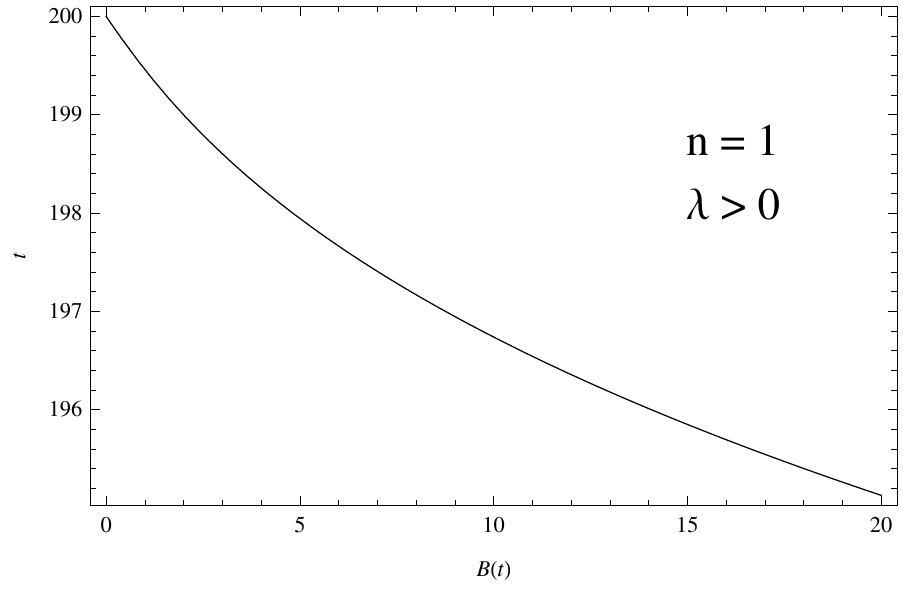}
\includegraphics[width=0.4\textwidth]{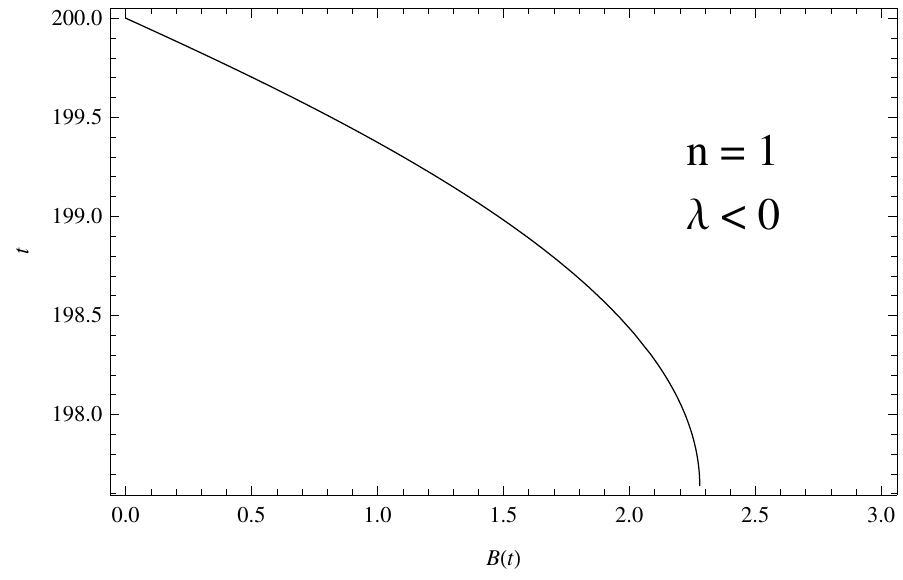}
\caption{Time evolution for different signatures of initial condition ($\lambda$) and $f(R)=\frac{R^2}{2}$.}
\end{center}
\label{fig:ss1112}
\end{figure}

It deserves mention, however, that these plots are not numerical plots, but rather plotted using the analytical expression (\ref{exactsolc3}). \\
The condition for the formation of an {\bf Apparent Horizon} is given by
\begin{equation}
\label{app-horc3}
g^{\mu\nu}R,_{\mu}R,_{\nu}=0,
\end{equation}
where $R(r,t)$ is the proper radius of the two-sphere given by $y(r)B(t)$ in this case. For the metric in the present case, this yields a simple result as 
\begin{equation}
\label{app-hor-constc3}
\dot{B}^2=\frac{1}{2}.
\end{equation}

With the help of the equation (\ref{1stintc3}), this condition yields the value of $B$ corresponding to the apparent horizon in terms of $n$ as 

\begin{equation} \label{app-hor_nc3}
B_{ap}=\Bigg[-\frac{1}{4\lambda}\Bigg(\frac{(2n^3+3n^2-n+1)}{(2n-1)}\Bigg)\Bigg]^{\frac{(2n+1))}{4(2n-1)}}.
\end{equation}

Amongst the examples worked out in this section, the value of $B_{ap}$ from equation (\ref{app-hor-constc3}) comes out to be $(-\frac{5}{4\lambda})^{\frac{3}{4}}$ and $(-\frac{9}{4\lambda})^{\frac{5}{12}}$ for the last two cases respectively. Figures $3.4$ and $3.5$ indicates that only for a negative $\lambda$ one has a finite time curvature singularity. So the singularities are well covered by a horizon. For a positive $\lambda$, there is no respectable horizon, but there is no requirement of that either as the volume goes to zero only asymptotically. In a similar way, in all the three other examples, one has a horizon when it is relevant. In the present model, as it happens, the singularity is independent of $r$, and thus is not a central singularity. This kind of singularity is expected to stay hidden by a horizon \cite{gosjoshi, maartens2, joshiritudadhi}. So the existence of the apparent horizon in all the relevant cases is quite a consistent result.

\section{Nature of the singularity}
From the metric (\ref{metricc3}), one can calculate the {\bf Kretschmann Curvature Scalar} as
\begin{equation}
{\textit{K}}={R}_{abcd}{R}^{abcd}=\frac{1}{y^4}\Bigg[\frac{3}{B^4}-4\frac{\dot{B}^2}{B^4}+12\frac{\dot{B}^4}{B^4}-8\frac{\ddot{B}}{B^3}+12\frac{\ddot{B}^2}{B^2}\Bigg].
\end{equation}
Though an explicit expression of $B(t)$ as a function of time could not be written, one can assess the nature of the Kretschmann Scalar from the first integral (\ref{1stintc3}).
\begin{eqnarray}\nonumber \label{Kretc3}
&& K = \frac{1}{y^4}\Bigg[\frac{3\alpha^2-8n\alpha(n+1)+48n^2(n+1)^2}{\alpha^2B^4}+\frac{48n\lambda(n+1)-4\alpha^2\lambda-4\alpha\lambda}{B^{4-\alpha}} \\ && +\frac{3\lambda^2\alpha^2+12\lambda^2}{B^{4-2\alpha}}\Bigg].
\end{eqnarray}

It is clearly seen that for $B \rightarrow 0$, at least the first term in the expression for $K$ will always blow up, indicating that the singularity that one obtains is indeed a Curvature Singularity in nature. The second and third terms within the parenthesis are proportional to $B^{(\alpha -4)}$ and $B^{(2\alpha -4)}$ respectively and may or may not diverge at zero proper volume, for all choices of $f(R)$, depending on the values of $n$ and $\alpha$. For example, for $n = 1$, i.e. for $f(R) \sim R^{2}$, $\alpha = \frac{4}{3}$, and all the terms diverge when $B(t) \rightarrow 0$. However $n = -\frac{3}{2}$, i.e., for $f(R) \sim R^{-\frac{1}{2}}$, $\alpha = 8$, and the second term and the third term are proportional to $B^4$ and $B^{12}$ respectively. Naturally, these terms go to zero as the zero proper volume singularity is approached. However, the first term, being always proportional to $B^{-4}$, diverges for all choices of $n$ or $\alpha$.    \\
From the field equations (\ref{fec3}) and (\ref{fefec3}), one can write the evolution of density and pressure respectively as
\begin{eqnarray}\nonumber \label{exp-densityc3}
&& \rho = \frac{1}{\kappa} \Bigg[\frac{1}{y(r)^2}\Bigg(3+\frac{n}{m}\Bigg)\Bigg(\lambda B^{\alpha}+\frac{2n(n+1)}{\alpha}\Bigg)\frac{B^{\frac{n}{m}-2}}{\delta(r)^{\frac{n}{m}}}+\frac{1}{2}\frac{n}{(n+1)}\Bigg(\frac{B}{\delta(r)}\Bigg)^{\frac{n+1}{m}} \\ && +\frac{n(2n-1)}{y(r)^2}\frac{B^{\frac{n}{m}-2}}{\delta(r)^{\frac{n}{m}}}\Bigg],
\end{eqnarray}

and, 

\begin{eqnarray}\nonumber \label{exp-pressc3}
&& p = \frac{1}{\kappa} \Bigg[\frac{1}{y(r)^2}{\Bigg(\frac{n}{m}+\frac{n^2}{m^2}-1\Bigg)}\Bigg(\lambda B^{\alpha}+\frac{2n(n+1)}{\alpha}\Bigg)\frac{B^{\frac{n}{m}-2}}{\delta(r)^{\frac{n}{m}}}+\frac{n(2n-1)+\frac{1}{2}}{y(r)^2}\frac{B^{\frac{n}{m}-2}}{\delta(r)^{\frac{n}{m}}} \\ &&-\frac{1}{2}\frac{n}{(n+1)}\Bigg(\frac{B}{\delta(r)}\Bigg)^{\frac{n+1}{m}}-\frac{n \alpha \lambda}{m y(r)^2}\frac{B^{\alpha+\frac{n}{m}-2}}{\delta(r)^{\frac{n}{m}}}\Bigg],
\end{eqnarray}

where, $m=\frac{n(2n+1)}{2(2n-1)}$ and $\alpha=4\frac{(2n-1)}{(2n+1)}$. As these physical quantities are all functions of $r$, it must be concluded that the collapsing fluid distribution is not spatially homogeneous. The collapse is thus different from the Oppenheimer-Snyder collapse in general relativity\cite{os}.    \\

For both $\rho$ and $p$, the powers involved for $B(t)$ are $(\frac{n}{m}-2)$, $(\alpha+\frac{n}{m}-2)$ and $(\frac{n+1}{m})$. For all $n > -\frac{1}{2}$, at least one of them is negative making both pressure and density proportional to inverse powers of $B(t)$ in these cases, and they do diverge when a zero proper volume is reached. In the examples discussed, for $n=1$ and $n=2$, $\frac{n}{m} -2$ is negative, and for $n=-0.33$, two of them, $(\frac{n}{m}-2)$ and $(\alpha+\frac{n}{m}-2)$ are negative. This is indeed the expected behaviour of density at a singularity.    \\

However, for $n< -\frac{1}{2}$, all these powers are positive, which leads to the peculiar result that the density and pressure both vanish at $B(t)=0$, i.e., at the singularity of zero proper volume. This is an intriguing result. If the matter sector has a conservation for itself, it really cannot happen. But since $f(R)$ theories are nonminimally coupled theories, this indicates that there must be an exchange of energy between the matter and the curvature sector, and the latter drains the energy from the former during collapse, and diverges at the singularity. This is evident from the expression for the Kretschmann scalar in equation (\ref{Kretc3}), which diverges anyway regardless of the choice of $n$.     

\section{Matching of the collapsing fluid with a Schwarzschild exterior}
The arbitrary function $\delta(r)$ in (\ref{BandRc3}) can be estimated from a suitable matching of the solutions for the collapsing fluid with that of a vacuum exterior solution at the boundary. Generally, in collapsing models, the interior is matched at a boundary hypersurface with a vacuum Schwarzschild exterior, which requires the continuity of both the metric and the extrinsic curvature on the boundary hypersurface (\cite{darmo, israel}). However, in $f(R)$ theories of gravity, continuity of the Ricci scalar across the boundary surface and continuity of its normal derivative are also required, (\cite{tclifton, deru, seno, nziokithesis, ritu1, ganguly}) as discussed in section $2.2$. The rationale behind matching with a Schwarzchild solution was discussed and utilised by Goswami et. al. \cite{ritu1}. Schwarzchild solution can be written as

\begin{equation}
\label{schwarzc3}
 ds^{2} = (1-\frac{2M}{r})dt^{2}-(1-\frac{2M}{r})^{-1}dr^{2}-r^{2}(d{\theta}^{2}+\sin^{2}\theta d{\phi}^{2}),
\end{equation}
where $M$ is the total mass contained by the interior. For an interior metric given by 

\begin{equation}
\label{met-intc3}
 ds^2=A^2(t,r)dt^{2}-N^2(t,r)dr^{2}-C^2(t,r)(d\theta^{2}+\sin^2\theta d\phi^{2}),
\end{equation}
the matching with the metric (\ref{schwarzc3}) yields (using the matching of the second fundamental form or the extrinsic curvature)
\begin{eqnarray}\nonumber \label{extrinscurvc3}
&&2\Bigg[\left(\frac{\dot{C'}}{C}-\frac{\dot{C}A'}{CA}-\frac{\dot{N}C'}{NC}\right)\Bigg]_{\Sigma}
=
-\frac{N}{A}\left[\frac{2\ddot{C}}{C}-\left(\frac{2\dot{A}}{A}-
\frac{\dot{C}}{C}\right) \frac{\dot{C}}{C}\right]\\ &&
+\frac{A}{N}\left[\left(\frac{2A'}{A}+\frac{C'}{C}\right)\frac{C'}{C}
-\left(\frac{N}{C}\right)^2\right],
\end{eqnarray}
where $\Sigma$ is the boundary.

If $A, N, C$ of the metric (\ref{met-intc3}) are replaced by the solutions obtained in this chapter given by (\ref{1stintc3}), one obtains
\begin{equation}
\Bigg[\frac{\ddot{B}}{B}+\frac{\dot{B}^2}{B^2}-\frac{1}{2B^2}-\frac{\sqrt{2}\dot{B}}{B^2}\Bigg]_{\Sigma} = 0.
\end{equation}
The arbitrary constant $\lambda$ from (\ref{1stintc3}) can be estimated from this matching condition with the help of an explicit form of $B(t)$ as function of $t$.
The Misner and Sharp mass function\cite{sharp}, defined as 
\begin{equation}
\label{MS-mfc3}
m(t,r)=\frac{C}{2}(1+g^{\mu\nu}C_{,\mu}C_{,\nu})=\frac{C}{2}\left(1+\frac{\dot{C}^2}{A^2}
-\frac{C'^2}{N^2}\right),
\end{equation}
yields the mass contained by the surface defined by the radial coordinate $r$ in the present case as
\begin{equation}
\label{MS-mf2c3}
m(t,r)=\frac{yB}{2}(\frac{1}{2}+\dot{B}^2).
\end{equation}
The Scwarzschild mass $M$, i.e., the total mass contaiined by the collapsing fluid is given by the right hand side of equation (\ref{MS-mf2c3}) calculated at the boundary.   \\

The continuity requirements found from the matching of Ricci Scalar and its normal derivative yield
\begin{equation}
R_{\Sigma} = R'_{\Sigma} = 0.
\end{equation}

From the $G_{01}$ equation we already have
\begin{equation}\label{BandRc3p}
B=\delta(r)R^m.
\end{equation}
Since the Ricci scalar is separable, following the arguements of Goswami et. al.\cite{ritu1} and Clifton et. al.\cite{tclifton} one can see that it should rather be of the form 
\begin{equation}\label{riccimatchc3}
R = (r_{\Sigma}^2-r^2)^2 g(r) T(t),
\end{equation}
where $g(r)$ is a well-defined function of radial cordinate $r$ and $T(t)$ describes the time evolution of the scalar. The relevance of chosing this type of functional form for a spherical star in $f(R)$ gravity was elaborated by Goswami et. al. \cite{ritu1}. From (\ref{BandRc3p}) and (\ref{riccimatchc3}), one can write
\begin{equation}
\delta(r)=\Bigg(r_{\Sigma}^2-r^2\Bigg)^{\frac{n(2n+1)}{(1-2n)}}g(r)^{\frac{n(2n+1)}{2(1-2n)}},
\end{equation}
which describes the radial profile of the spherical collapsing body.

\section{Discussion}
A spherical collapse in $f(R)$ gravity for a particular form of $f(R)$ is investigated analytically in this chapter. With the assumption of the existence of a homothetic Killing vector, exact solutions for the collapsing models are obtained for $f(R)\sim R^{n}$. The solutions are not readily invertible in the form of the proper radius as a function of time. So the approach to the singularity of a zero proper volume is investigated with the help of numerical plots of time against proper radius. Some of the collapsing modes indeed succumb to a curvature singularity at a finite future. In some cases, where $f(R)$ varies as positive powers of $R$, the singularity is reached only asymptotically. Singularities forming at a finite future are always covered by an apparent horizon. The investigations presented in the chapter, though a simple one, predicts different possibilities while a collapsing evolution in $f(R)$ gravity is concerned; this posibilities, certainly depends on the initial conditions, determined by $\lambda$ and the choice of $f(R)$. For example, when $f(R)$ varies as negative powers of $R$, the collapsing object might equilibriate at a constant proper radius and never hit the singularity.  \\
The discussion in this chapter is only for $f(R) \sim R^{n}$. Recently it was shown by Goheer, Larena and Dunsby\cite{goheer} that within the class of $f(R)$ gravity theories, $FLRW$ power-law perfect fluid solutions only exist for $R^{n}$ gravity. However, this configuration may also serve as a platform to study exact or approximate solutions for different choices of $f(R)$. The equation (\ref{trick2c3}) can be written elaborately for different $\Lambda(R)$ which solely depends on the choice of $f(R)$. One such example can perhaps be $f(R) = R + \alpha R^{n}$ which has its own importance in describing the late time accelerated expansion of the universe.   \\
Due to the nonlinearity of the field equations, an exact solution in $f(R)$ gravity is not always guarranteed. Both chapter $2$ and $3$ demonstrates collapsing solutions in $f(R)$ gravity, however, the main difference in this chapter being the assumption of the existence of a homothetic killing vector. It results in certain simplifications, but one has to restrict oneself to a particular form of $f(R)$ to investigate the field equations.

\chapter[Scalar field collapse in a conformally flat spacetime]{Scalar field collapse in a conformally flat spacetime\footnote[1]{The results of this chapter are reported in The European Physical Journal C, (2017), {\bf 77} : 166.}}

% **************************** Define Graphics Path **************************
\ifpdf
    \graphicspath{{Chapter4/Figs/Raster/}{Chapter4/Figs/PDF/}{Chapter4/Figs/}}
\else
    \graphicspath{{Chapter4/Figs/Vector/}{Chapter4/Figs/}}
\fi

Scalar fields, albeit having no pressing motivation from particle physics theory more often than not, have been of great interest in theories of gravity for various reasons such as fitting in superbly for cosmological requirements such as in the role of the driver of the past or even the present acceleration of the universe. Moreover, as already discussed in section $1.3$, a scalar field with a variety of potential can serve as an excellent fit to study evolution of self-gravitating evolution of many a kind of realistic matter distribution \cite{goncalves, booksami}.           \\

The subject of zero mass scalar field collapse has been addressed rigorously over the years and the possibility of different end-products in terms of a black-hole or a naked singularity has been discussed in detail. Christodoulou established the global existence and uniqueness of the solutions of Einstein-scalar field equations \cite{christo1, christo2}. A sufficient condition for the formation of a trapped surface in the evolution of a given initial data set was also studied by the same author \cite{christo3}. Goldwirth and Piran showed that a scalar field collapse leads to a singularity which is cut-off from the exterior observer by an event horizon \cite{piran}, therefore ending up in a black-hole. Throughout the last couple of decades, numerical investigations have provided useful insights into black hole formation in a scalar field collapse. In their pioneering works, Choptuik \cite{chop}, Brady \cite{brady} and Gundlach \cite{gund1, gund2} considered the numerical evolution of collapsing profiles characterised by a single parameter $(p)$ and showed that the behaviour of the resulting families of solutions depends on the value of p; there is a critical evolution with $p = p^{\ast}$, which signals the transition between complete dispersal and black hole formation.      \\

Gravitational collapse with a scalar field minimally coupled to gravity have been studied extensively in literature, but usually consisting of massless scalar fields. There is only very limited amount of work on massive scalar field collapse and that too in very restricted scenarios. For example, Giambo showed that depending on suitable boundary matching with an exterior solution, a collapsing scalar field may give rise to a naked singularity \cite{massivegiambo}. A spherically symmetric collapse of a real, minimally coupled, massive scalar field in an asymptotically Einstein–de Sitter spacetime was studied by Goncalves \cite{massivegoncalves}. Goswami and Joshi constructed a class of collapsing scalar field models with a non-zero potential, which resulted in a naked singularity as the collapse end state \cite{massiveritu1, massiveritu2}. It was shown by Ganguly and Banerjee \cite{massiveganguly}, that a scalar field, minimally coupled to gravity, may have collapsing modes even when the energy condition is violated, and discussed the significance of the result in the context of possible clustering of dark energy. Quite recently Baier, Nishimura and Stricker \cite{baiernishi} proved that a scalar field collapse, along with a negative cosmological constant, can lead to the formation of a naked singularity. It must be noted that non-spherical models of scalar field collapse are there in literature as well (for example self-similar scalar field solutions to the Einstein equations in cylinder symmetry by Condron and Nolan \cite{condron1, condron2}, scalar field collapse with planar as well as toroidal, cylindrical and pseudoplanar symmetries by Ganguly and Banerjee \cite{nonsphericalganguly}).                       \\
The aim of this chapter is to discuss the collapse of a self-interacting scalar field along with a distribution of perfect fluid. In section $1.3.2$, significance of such a massive scalar field collapse has been reviewed in brief with proper references.            \\
The self-interaction potential of the scalar field is taken to be a power law ($V\sim {\phi}^{n}$) where $n$ can take a wide range of values. The assumption of conformal flatness is made at the outset. Conformally flat gravitational collapse of a fluid with heat flux has been investigated in the context of general relativity, for instance, a recent work of Sharma, Das and Tikekar \cite{sharmadastike}. The present work, in a way, is therefore a scalar field generalization of their work.  \\

The nonlinear field equations are studied in this chapter by invoking the integrability conditions for a classical anharmonic oscillator equation, developed by Euler \cite{euler1, euler2} and utilized by Harko, Lobo and Mak \cite{harkolobomak}. This leads to an integrable second order equation for the scale factor. Thus some general comments, regarding the possibility and nature of a wide range of power law potentials can be made. 

\section{Conformally flat metric and a scalar field collapse}
A conformally flat space-time is considered to discuss a relativistic model of a spherically symmetric matter source, whose collapse is accompanied with dissipation in the form of radial heat flux following the work of Santos \cite{santosc4}. The spacetime metric is chosen so as to have a vanishing Weyl tensor implying its conformal flatness. Conformally flat spacetimes, in the context of radiating fluid spheres, were first studied by Som and Santos \cite{som}. Later, the most general class of conformally flat solutions for a shear-free radiating star were obtained and examined by Maiti \cite{maiti}, Modak \cite{modak}, Banerjee, Dutta Chowdhury and Bhui \cite{bhui}, Patel and Tikekar \cite{patel}, Schafer and Goenner \cite{schafer}, Ivanov \cite{ivanov}. Herrera et. al. \cite{herreraetal} examined models of shear-free collapsing fluids accompanied by a dissipation of heat on a space-time background admitting a vanishing Weyl tensor.       \\
The metric can be written as 
\begin{equation}
\label{metricc4}
ds^2=\frac{1}{{A(r,t)}^2}\Bigg[dt^2-\frac{dr^2}{1-kr^2}-r^2d\Omega^2\Bigg],
\end{equation}
for which the Weyl tensor components vanish. Here, $k$ is a constant which can pick up values from $-1, 0, +1$. The energy momentum tensor is taken to be that of a perfect fluid given by
\begin{equation}
\label{em-tensorc4}
T^{\mu}_{\nu} = (\rho + p)u^{\mu}u_{\nu} - p\delta^{\mu}_{\nu},
\end{equation}
where $\rho$ is the energy density, $p$ is the isotropic fluid pressure, $u_{\mu}$ is the 4-velocity of the fluid. The scalar field contribution is defined as a scalar field $\phi$ minimally coupled to gravity and the relevant action is given by 
\begin{equation}
{\cal A} = \int{\sqrt{-g}d^4x[R+\frac{1}{2} \partial^{\mu}\phi \partial_{\mu}\phi - V(\phi) + L_{m}]},
\end{equation}

where $V(\phi)$ is the scalar potential and $L_{m}$ is the Lagrangian density for the fluid distribution. From this action, the contribution to the energy-momentum tensor from the scalar field $\phi$ can be written as
\begin{equation}
T^\phi_{\mu\nu}=\partial_\mu\phi\partial_\nu\phi-g_{\mu\nu}\Bigg[\frac{1}{2}g^{\alpha\beta}\partial_\alpha\phi\partial_\beta\phi-V(\phi)\Bigg]. 
\end{equation}

Einstein field equations $G_{\mu\nu} = -8\pi G T_{\mu\nu}$ (in the units $8\pi G = 1$) can thus be written as
\begin{eqnarray}\nonumber \label{fe1c4}
&& 3kA^2+3\dot{A}^2-3(1-kr^2)A'^2+2(1-kr^2)AA''+\frac{2(2-3kr^2)}{r}AA' \\&&
=\rho+\frac{1}{2}A^2\dot{\phi}^2-\frac{1}{2}A^2(1-kr^2)\phi'^2 +V(\phi),
\end{eqnarray}

\begin{eqnarray}\nonumber
\label{fe2c4}
&& -kA^2+2\ddot{A}A-3\dot{A}^2+3(1-kr^2)A'^2-\frac{4}{r}(1-kr^2)AA' \\&&
=p+\frac{1}{2}{\phi'}^2A^2(1-kr^2)+\frac{1}{2}A^2\dot{\phi}^2-V(\phi),
\end{eqnarray}

\begin{eqnarray}\nonumber
\label{fe3c4}
&& -kA^2+2\ddot{A}A-3\dot{A}^2+3(1-kr^2)A'^2-\frac{2}{r}(1-2kr^2)AA'-2(1-kr^2)AA'' \\&&
=p-\frac{1}{2}{\phi'}^2A^2(1-kr^2)+\frac{1}{2}A^2\dot{\phi}^2-V(\phi),
\end{eqnarray}

\begin{equation}
\label{fe4c4}
\frac{2\dot{A}'}{A}=\dot{\phi}\phi' .
\end{equation}

\par The wave equation for the scalar field is given by
\begin{equation}
\label{wavec4}
\Box\phi+\frac{dV}{d\phi}=0.
\end{equation}

For the sake of simpliciy, $\phi(r,t)$ is assumed to be a function of time $t$ alone. Consequently, from equation (\ref{fe4c4}), one can see that $A(r,t)$ can also be written as a function of time alone, and this is consistent with the rest of the equations of the system. With this, equations (\ref{fe2c4})and (\ref{fe3c4}) become identical. So effectively an Oppenheimer-Snyder type collapse scenario \cite{os} in the presence of a minimally coupled scalar field is considered here.         \\

With the assumptions $A=A(t)$ and $\phi=\phi(t)$, the field equations simplify as

\begin{equation}
\label{fenew1c4}
3kA^2+3\dot{A}^2=\rho+\frac{1}{2}A^2\dot{\phi}^2+V(\phi),
\end{equation}

\begin{equation}
\label{fenew2c4}
-kA^2+2\ddot{A}A-3\dot{A}^2=p+\frac{1}{2}A^2\dot{\phi}^2-V(\phi),
\end{equation}

and the wave equation looks like
\begin{equation}
\label{wave2c4}
\ddot{\phi}-2\frac{\dot{A}}{A}\dot{\phi}+\frac{1}{A^2}\frac{dV}{d\phi}=0.
\end{equation}

Now we have three equations (\ref{fenew1c4} - \ref{wave2c4}) to solve for four unknowns, namely $A(t), \phi(t), \rho$ and $p$. $V(\phi)$ of course is given as a function of $\phi$. One can close the system by a choice of an equation of state $p = p(\rho)$. We shall, however, would not try to close the system, but adopt a different strategy to check for the collapsing modes.

\section{Integrability of anharmonic oscillator equation}
A nonlinear anharmonic oscillator with variable coefficients and a power law potential can be written in a general form as

\begin{equation}
\label{genc4}
\ddot{\phi}+f_1(t)\dot{\phi}+ f_2(t)\phi+f_3(t)\phi^n=0,
\end{equation}

where $f_i$ are functions of $t$ and $n \in {\cal Q}$ is a constant. An overhead dot represents a differentiation with respect to $t$. Using Euler’s theorem on the integrability of the general anharmonic oscillator equation \cite{euler1, euler2} and recent results given  by Harko {\it et al} \cite{harkolobomak}, this equation can be integrated under certain conditions. The essence can be written in the form of the following theorem \cite{euler1, harkolobomak}.

\textbf{Theorem} An equation of the form of equation (\ref{genc4}) can be transformed into an integrable form for $n\notin \left\{-3,-1,0,1\right\} $ if and only if the coefficients of Eq. (\ref{genc4}) satisfy the differential condition
\begin{eqnarray}\nonumber \label{int-genc4}
&& \frac{1}{n+3}\frac{1}{f_{3}(t)}\frac{d^{2}f_{3}}{dt^{2}%
}-\frac{n+4}{\left( n+3\right) ^{2}}\left[ \frac{1}{f_{3}(t)}\frac{df_{3}}{dt%
}\right] ^{2}+ \frac{n-1}{\left( n+3\right) ^{2}}\left[ \frac{1}{f_{3}(t)}%
\frac{df_{3}}{dt}\right] f_{1}\left( t\right) \\ &&
+ \frac{2}{n+3}\frac{df_{1}}{dt}%
+\frac{2\left( n+1\right) }{\left( n+3\right) ^{2}}f_{1}^{2}\left( t\right)=f_{2}(t).
\end{eqnarray} 

Introducing a pair of new variables $\Phi$ and $T$ given by 
\begin{eqnarray}
\label{Phic4}
\Phi\left( T\right) &=&C\phi\left( t\right) f_{3}^{\frac{1}{n+3}}\left( t\right)
e^{\frac{2}{n+3}\int^{t}f_{1}\left( x \right) dx },\\
\label{Tc4}
T\left( \phi,t\right) &=&C^{\frac{1-n}{2}}\int^{t}f_{3}^{\frac{2}{n+3}}\left(
\xi \right) e^{\left( \frac{1-n}{n+3}\right) \int^{\xi }f_{1}\left( x
\right) dx }d\xi ,
\end{eqnarray}%

where $C$ is a constant, equation (\ref{genc4}) can be written as 

\begin{equation}
\label{Phi1c4}
\frac{d^{2}\Phi}{dT^{2}}+\Phi^{n}\left( T\right) =0.
\end{equation}

In what follows, this integrability condition is used accordingly in order to extract information from the scalar field equation (\ref{wave2c4}) for some given forms of the potential $V=V(\phi)$.

\section{Power-law potential}
In the first example we assume that the potential is a power function of $\phi$, $V(\phi) = \frac{V_{0}{\phi}^{(n+1)}}{n+1} $ such that 
\begin{equation}
\label{power-lawc4}
\frac{dV}{d\phi}= V_{0}\phi^n,
\end{equation}
where $n \in {\cal Q}$. While the potential with a postive power of $\phi$, where $\frac{d^{2}V}{d{\phi}^{2}}$ evaluated at $\phi = 0$ gives the mass of the field, is quite well addressed, potentials with inverse powers of $\phi$ are also quite useful in cosmological contexts, particularly as tracking quintessence fields. Ratra and Peebles \cite{ratrac4} used a potential of the form $V = \frac{M^{4+\alpha}}{\phi^{\alpha}}$, where $M$ is the Planck mass. Similar potential had later been used as a tracker field by Steinhardt, Wang and Zlatev \cite{zlatev2} where $M$ loses the significance as the Planck mass and is rather used as a parameter to be fixed by observations.

\section{Integrability of the scalar field equation}
\subsection{Solution for the metric}
With this power law potential, the scalar field equation (\ref{wave2c4}) becomes

\begin{equation}
\label{wave-pwrlwc4}
\ddot{\phi}-2\frac{\dot{A}}{A}\dot{\phi}+\frac{V_{0}}{A^2}\phi^n=0,
\end{equation}

which can be written in a more general form of second order ordinary differential equation with variable coefficients as

\begin{equation}
\label{wave-pwrlw1c4}
\ddot{\phi}+f_1(t)\dot{\phi}+f_3(t)\phi^n=0,
\end{equation}

where $f_{i}(t)$ are functions of time, determined by $A(t)$ and its derivatives. Equation (\ref{wave-pwrlw1c4}) is easily identified to be a special case of equation(\ref{genc4}) with $f_1(t)=-2\frac{\dot{A}}{A}$, $f_{2} = 0$ and $f_3(t)=\frac{V_{0}}{A^2}$. Hence, the integrability condition as in equation (\ref{int-genc4}) yields a second order differential equation of $A(t)$ in the form

\begin{equation}
\label{evolutionAc4}
-\frac{6}{(n+3)}\frac{\ddot{A}}{A}+\frac{18(n+1)}{(n+3)^2}\frac{\dot{A}^2}{A^2}=0.
\end{equation}

This can be integrated to find an exact time-evolution of $A(t)$ as
\begin{equation}
\label{Ac4}
A(t)=\Bigg[\frac{2n\sqrt{\lambda}}{(n+3)}(t_{0}-t)\Bigg]^{-(\frac{n+3}{2n})},
\end{equation}
where $\lambda$ is a constant of integration coming from the first integral and is a positive real number. It is interesting to note that the conformal factor is independent of the choice of $V_{0}$. \\
As the theorem is valid for  $n\notin \left\{-3,-1,0,1\right\} $, we exclude these in the subsequent discussion. The radius of a two-sphere for the metric (\ref{metricc4}) is given by $rY(t)$ where 
\begin{equation}
\label{Yc4}
Y(t) = \frac{1}{A(t)} = \Bigg[\frac{2n\sqrt{\lambda}}{(n+3)}(t_{0}-t)\Bigg]^{(\frac{n+3}{2n})}.
\end{equation}

From (\ref{Ac4}) and (\ref{Yc4}), the time evolution of the collapsing fluid can be discussed for different domains of $n$, i.e. for different choices of the potential.
\begin{itemize}
\item {For both $n > 0$ (excluding {$n=1$}) and for $n < -3$, $(\frac{n+3}{2n})$ is strictly positive. Let us write $(\frac{n+3}{2n}) = {n_{0}}^2$. Then from equation (\ref{Yc4}) one can write the radius of the two-sphere as 
\begin{equation}\label{scale1c4}
r Y(t) = r \Bigg[\frac{\sqrt{\lambda}}{{n_{0}}^2}(t_{0}-t)\Bigg]^{{n_{0}}^2}.
\end{equation} 
It is straightforward to note that $rY(t)$ goes to zero when $t \rightarrow t_0$. Thus, for all $n > 0$ and for $n < -3$, the collapsing sphere reaches a singularity of zero proper volume at a finite time defined by $t_0$. }

\item {However, for $0 > n > -3$ ($n \neq -1$), $(\frac{n+3}{2n})$ is negative and it can be written as $(\frac{n+3}{2n}) = -{m_{0}}^2$. For this domain of $n$, the scale factor $Y$ can be written as
\begin{equation}\label{scale2c4}
r Y(t) = r \Bigg[\frac{\sqrt{\lambda}}{{m_{0}}^2}(t-t_{0})\Bigg]^{-{m_{0}}^2}.
\end{equation}
Clearly, we have a collapsing solution as it is easy to check that $\dot{Y(t)} < 0$. However, the collapsing fluid reaches the zero proper volume only when $t \rightarrow \infty$. This indicates the system is collapsing for ever rather than crushing to singularity at a finite time.}

\item {From (\ref{scale1c4}) and (\ref{scale2c4}), one can check that $\frac{dY(t)}{dt} < 0$ for all relevant cases, provided $\sqrt{\lambda} > 0$. On the other hand, a negative $\sqrt{\lambda}$, turns collapsing solutions into expanding solutions. This is quite consistent with the fact that equation (\ref{evolutionAc4}) is invariant under $t \rightarrow -t$, and therefore another solution can be construced simply by flipping the sign of $t$ and $t_{0}$ in equation (\ref{Ac4}).}
\end{itemize}
       
\subsection{Solution for the scalar field}            
Using the transformation equations (\ref{Phic4}) and (\ref{Tc4}), one can write the general solution for the scalar field $\phi$ as,
\begin{equation}
\label{phigenc4}
\phi\left( t\right) =\phi_{0}\left[ C^{\frac{1-n}{2}}\int^{t}f_{3}^{\frac{2}{n+3}%
}\left( \xi \right) e^{\left( \frac{1-n}{n+3}\right) \int^{\xi }f_{1}\left(
x \right) dx }d\xi -T_{0}\right] ^{\frac{2}{1-n}}f_{3}^{-\frac{1}{n+3}%
}\left( t\right) e^{-\frac{2}{n+3}\int^{t}f_{1}\left( x \right) dx
},
\end{equation}
where $\phi_{0}$ and $T_0$ are constants of integration and $C$ comes from the definition of the point transformations (\ref{Phic4}) and (\ref{Tc4}). Both $\phi_{0}$ and $C$ must be non-zero. Since the integrability criteria produces an exact time evolution of $A(t)$ as given in (\ref{Ac4}), equation (\ref{phigenc4}) can be simplified in the present case as 
\begin{equation}
\label{phi-plwc4}
\phi\left( t\right) =\phi_{0}{{V_{0}}^{-\frac{1}{(n+3)}}}\Bigg(\frac{2n\sqrt{\lambda}}{n+3}\Bigg)^{-\frac{1}{n}}(t_{0}-t)^{-\frac{3}{n}}\Bigg[C^{\frac{1-n}{2}}{{V_{0}}^{\frac{2}{(n+3)}}}{\frac{n}{3}}\Bigg(\frac{2n\sqrt{\lambda}}{n+3}\Bigg)^{\frac{2}{n}}\Bigg((t_{0}-t)^{\frac{3}{n}}+\delta\Bigg)-T_0\Bigg]^{\frac{2}{(1-n)}},
\end{equation}

where $\delta$ comes as a constant of integration. One can clearly see that at $t=t_0$, when the volume element goes to zero, the scalar field diverges for $n>0$ and $n<-3$. A simple example for the evolution of the scalar field can be obtained where the integration constants $\delta$ and $T_0$ are put to zero. The time evolution then can be written as $\phi(t) \sim (t_0-t)^{\frac{3(1+n)}{n(1-n)}}$, which is consistent with the solution for scalar field one obtains from equation (\ref{wave-pwrlwc4}).         \\

\subsection{Divergence of Density, Pressure and Kretschmann Scalar}
From the field equations, one can write the expressions for the density and the pressure in terms of $A(t)$ and $\phi(t)$ as
\begin{equation}
\label{den-plwc4}
\rho=3kA^2+3\dot{A}^2-\frac{1}{2}A^2\dot{\phi}^2-\frac{{V_{0}}\phi^{n+1}}{n+1},
\end{equation}
\begin{equation}
\label{press-plwc4}
p=-kA^2+2\ddot{A}A-3\dot{A}^2-\frac{1}{2}A^2\dot{\phi}^2+\frac{{V_{0}}\phi^{n+1}}{n+1}.
\end{equation}

Both pressure and density diverge as $t$ goes to $t_0$ for $n > 0$ and $n < -3$. The expression for density indicates that if the scalar field part goes to infinity faster than the rest, the fluid density may go to a negative infinity close to the singularity. For a simple case, where $n=3$, i.e., the potential is defined as $V(\phi)=\frac{{V_{0}}\phi^4}{4}$, one can write $\phi = {\phi_0}{C_{1}}{\lambda}^{-\frac{1}{2}}(t_0-t)^{-2}$, for the arbitrary integration constants $T_0=\delta=0$ ($C_{1}$ is a constant depending on the values of $C$ and $V_{0}$). In this case, the scalar field part, contributing negatively, blows up much quicker ($\sim (t_0 - t)^{-8}$) as $t \rightarrow t_{0}$ than the rest, which go to infinity as  $\sim (t_0 - t)^{-2}$ and $\sim (t_0 - t)^{-4}$. However, the strong energy condition, $(\rho+3p) > 0$, can still be satisfied. From (\ref{den-plwc4}) and (\ref{press-plwc4}), one can write

\begin{equation}
(\rho+3p)=6\ddot{A}A-6\dot{A}^2-2A^2\dot{\phi}^2+\frac{2{V_{0}}\phi^{(n+1)}}{(n+1)}, 
\end{equation}

which can indeed remain positive. For the particular example of $n=3$, this can be simplified into
\begin{equation}
(\rho+3p) \sim \frac{6}{\lambda(t_0-t)^4}+\frac{{\phi_0}^2{C_{1}}^2}{2\lambda^2(t_0-t)^8}(V_{0}{\phi_0}^2{C_{1}}^2-16).
\end{equation}
Here, $\lambda$ is always positive as discussed earlier and so is ${\phi_0}^2{C_{1}}^2$. Near the singularity, as $t\rightarrow t_0$, the second term on the RHS becomes more dominating than the first term; $(\rho+3p)$ must be greater than zero to satisfy the energy condition, and to avoid any possible negativity near $t=t_0$, the condition $V_{0}{\phi_0}^2{C_{1}}^2 > 16$ must be satisfied.      \\

It is intersting to note that in the Hawking radiation process, the stress energy tensor is known to behave in such a peculiar manner, such as the breakdown of weak energy condition $T^{\mu\nu}u_{\mu}u_{\nu} > 0$ in the classical sense, meaning a negative energy density \cite{romanc4}. It should also be noted that at the apparent horizon, which covers the singularity in all cases, the terms in the expression for the density do not diverge, and it is quite possible to ensure a positive $\rho$ by fixing the constants.       \\ 

The Kretschmann scalar can be calculated for the metric (\ref{metricc4}) ($k = 0$) as
\begin{equation}
\label{kretschc4}
K=R_{\alpha\beta\gamma\delta}R^{\alpha\beta\gamma\delta}=6\dot{A}^4+6\Bigg(\frac{\ddot{A}}{A}-\frac{\dot{A}^2}{A^2}\Bigg)^2A^4,
\end{equation}
which, in view of the solution (\ref{Yc4}) yields \\

                               $$ K \sim (t_{0} - t)^{-4(n_{0}^{2} + 1)}$$, \\

for $n>0$ and $n<-3$. Since the kretschmann scalar clearly diverges as $t\rightarrow t_{0}$, one indeed has a curvature singularity as a result of the collapse. As already discussed in Section $2.1.3$, existence of such a singularity in a spacetime is in general understood in the form of non-spacelike incomplete geodesics. If the spacetime curvature and similar other physical quantities remain finite along an incomplete nonspacelike geodesic in the limit of approaching the singularity, it could be regarded as mathematical pathology in the spacetime rather than having any physical significance (for example, an integrable singularity, weaker than the conventional singularity and allows the (effective) matter to pass to a white-hole region; Lukash and Strokov \cite{lukash}). A singularity will only be physically important when there is a powerful enough curvature growth along singular geodesics, and the physical interpretation and implications of the same are to be considered.        \\
In this chapter, the curvature singularity encountered is indeed a physically significant non-integrable singularity, indicating an occurrence of nonspacelike geodesic incompleteness. 

\section{Boundary Matching with an exterior Vaidya spacetime}
For a complete and consistent analysis of gravitational collapse, proper junction conditions are to be examined carefully which allow a smooth matching of an exterior geometry with the collapsing interior. Any astrophysical object is immersed in vacuum or almost vacuum spacetime, and hence the exterior spacetime around a spherically symmetric star is well described by the Schwarzschild geometry. Moreover it was extensively shown by Goncalves and Moss \cite{goncalves} that any sufficiently massive collapsing scalar field can be formally treated as collapsing dust ball. From the continuity of the first and second differential forms, the matching of the sphere to a Schwarzschild spacetime on the boundary surface, $\Sigma$, is extensively worked out in literature as already discussed in the earlier chapters (see also \cite{santosc4, chanc4, kolla, maharajc5}).         \\

However, conceptually this may lead to an inconsistency since the treatment allowed for a dust collapse may not be valid for a scalar field in general. For instance, since Schwarzschild has zero scalar field, such a matching would lead to a discontinuity in the scalar field, which means a delta function in the gradient of
the scalar field. As a consequence, there will appear square of a delta function in the stress-energy, which is definitely an inconsistency. In modified theories of gravity an alternative scenario is discussed sometimes where the exterior is non-static. However, the solar system experiments constrain heavily such a scenario. Another possible way to avoid such a scenario can perhaps be allowing jump in the curvature terms in the field equations. Such cases must result in surface stress energy terms, which in collapsing models must have observational signatures and can be established via experimental evidences \cite{ritu1}.        \\
Following the arguements of Goswami and Joshi \cite{massiveritu1, massiveritu11}, Ganguly and Banerjee \cite{massiveganguly}, we match the spherical ball of collapsing scalar field to a Vaidya exterior across a boundary hypersurface defined by $\Sigma$. The metric just inside $\Sigma$ is
\begin{equation}
\label{metric2c4}
ds^2=\frac{1}{{A(t)}^2}\Bigg[dt^2-dr^2-r^2d\Omega^2\Bigg],
\end{equation}
and the Vaidya metric is given by
\begin{equation}\label{vaidyac4}
ds^2=\Bigg[1-\frac{2m(v)}{R}\Bigg]dv^2+2dvdR-R^2d\Omega^2.
\end{equation}
The quantity $m(v)$ represents the Newtonian mass of the gravitating body as measured by an observer at infinity. The metric (\ref{vaidyac4}) is the unique spherically symmetric solution of the Einstein field equations for radiation in the form of a null fluid.  The necessary conditions for the smooth matching of the interior spacetime to the exterior spacetime was presented by Santos \cite{santosc4} and also discussed in detail by Chan \cite{chanc4} in context of a radiating gravitational collapse. Following their work, The relevant equations matching (\ref{metric2c4}) with (\ref{vaidyac4}) can be written as                         
\begin{equation}
\frac{r}{A(t)}_{\Sigma} = R,
\end{equation}

\begin{equation}
r(r B')_{\Sigma} = \Bigg[R A \Big(1-\frac{2m(v)}{R}\Big)\dot{v} + RA\dot{R}\Bigg],
\end{equation}

\begin{equation}
m(v)_{\Sigma} = \frac{r^3}{2A^3}\Bigg(\dot{A}^2-A'^2+\frac{A'A}{r}\Bigg),
\end{equation}
and
\begin{equation}\label{prqc4}
p_{\Sigma} = \frac{q}{A(t)}_{\Sigma} = 0,
\end{equation}                        
where $\Sigma$ is the boundary of the collapsing fluid and $q$ denotes any radial heat flux defined in the interior of the collapsing scalar field. For a non-viscous shear-free fluid with pressure anisotropy and heat conduction the condition (\ref{prqc4}) gives $p_{r_{\Sigma}} = \frac{q}{A(t)}$.           \\

The relation as in equation (\ref{prqc4}) yields a nonlinear differential condition between the conformal factor and the scalar field to be satisfied on the boundary hypersurface $\Sigma$ as
\begin{equation}\label{bconditionc4}
\Bigg[2\frac{\ddot{A}}{A} - 2\frac{\dot{A}^2}{A^2} - \frac{1}{2}\dot{\phi}^2 + \frac{V_{0}}{A^2}\frac{\phi^{(m+1)}}{(m+1)}\Bigg]_{\Sigma} = 0.
\end{equation}
Using the time evolution of the conformal factor and the scalar field, i.e. equations (\ref{Ac4}) and (\ref{phi-plwc4}), one can simplify this expression and establish some constraints over the choices of parameters such as $V_{0}$, $n$, $\lambda$, $\phi_{0}$, $\delta$ and $T_{0}$. Therefore, the validity of the present models is established along with certain constraints. Using the field equations in the boundary matching conditions, it is straightforward to obtain \cite{olic4} a functional dependence between the retarded time $v$ and $t$ given as
\begin{equation}
\dot{v}_{\Sigma} = \frac{1}{(A - 2r\dot{A})}.
\end{equation}

An interesting feature is observed if the interior solution is matched with a Schwarzschild exterior. On the boundary hypersurface $\Sigma$, the matching of extrinsic curvature gives
\begin{equation}
\Bigg[\frac{2{n_{0}}^2-{n_{0}}^4}{(t-t_0)^2}\Bigg]_{\Sigma} = 0,
\end{equation}
which means ${n_{0}}^{2} = 2$ and it is easy to note that the resulting metric corresponds to the Oppenheimer-Snyder model for the marginally bound case.  \\
However, it must be noted that $(\frac{n+3}{2n}) = n_{0}^2 = 2$ implies that $n = 1$, which does not fall in the domain of validity of the theorem employed in this work.

\section{Visibility and nature of the singularity}
Whether the curvature singularity is visible to an exterior observer or not, depends on the formation of an apparent horizon. The condition for such a surface is given by
\begin{equation}
\label{app-horc4}
g^{\mu\nu}R,_{\mu}R,_{\nu}=0,
\end{equation}
where $R$ is the proper radius of the two-sphere, given by $\frac{r}{A(t)} = r Y(t)$ in the present case. The relevant cases in the present work are certainly the ones for $n > 0$ and $n < -3$. Using the explicit time evolution of $A$ from equation (\ref{Ac4}), equation (\ref{app-horc4}) yields a simple differential equation,
\begin{equation}
\label{app-hor1c4}
r^2\dot{Y}^2-(1-kr^2)Y^2=0,
\end{equation}
which, in view of equations (\ref{Yc4}) and (\ref{scale1c4}), yields the algebraic equation at $t=t_{app}$
\begin{equation}
\label{app-hor2c4}
\frac{\dot{Y}}{Y} = \frac{n_{0}^{2}}{t_{app}-t_{0}}.
\end{equation}

Since the present interest is in a collapsing solution, scale factor must be a monotonically decreasing function of time. $\dot{Y}$ is negative and $Y$, being the scale factor, must always be positive. Thus from equation (\ref{app-hor2c4}), the condition is consistent if and only if $t_{app} < t_{0}$. This clearly indicates that the apparent horizon forms before the formation of the singularity, for all relevant cases. Thus, the curvature singularity is always covered from an exterior observer by the formation of an apparent horizon. It deserved mention that there is no apriori compulsion of the formation of the horizon ahead of the formation of a singularity as this is a central singularity. Had the formation of singularity been independent of $r$, indicating that it forms everywhere in the collapsing distribution instanteneously, the compulsion of the existence of the horizon would have been ensured \cite{joshiritudadhi}. At the singularity in the present case, one has $Y=0$ and $\dot{Y}\neq 0$. Equation (\ref{app-hor1c4}) indicates that this is consistent only with $r=0$ at the singularity. Thus the singularity is strictly a central singularity which could have been a naked singularity as well. \\

The standard analysis shows that the present singularity is a shell-focusing one (for which $g_{\theta\theta}=0$) and not a shell-crossing one (for which  $\frac{dg_{\theta\theta}}{dr}=0$, $g_{\theta\theta}\neq 0$ and $r>0$)\cite{seif1, seif2, waughlake1, waughlake2}.

\section{Potential as a combination of the form $V(\phi) = \frac{1}{2}{\phi}^{2} + \frac{\phi^{n+1}}{n+1}$}

For a very simple combination of two powers of $\phi$,
\begin{equation}
 \label{polyc4}
V(\phi) = \frac{1}{2}{\phi}^{2} + \frac{\phi^{n+1}}{n+1},
\end{equation}
the method of integrability of anharmonic oscillators can lead to some interesting informations about the behaviour of the collapse. With equation (\ref{polyc4}), one can write 
\begin{equation}
\label{poly3c4}
\frac{dV}{d\phi}=\phi+\phi^n.
\end{equation}

The scalar field equation (\ref{wavec4}), with the same metric (\ref{metricc4}), becomes
\begin{equation}
\label{wave-polyc4}
\ddot{\phi}-2\frac{\dot{A}}{A}\dot{\phi}+\frac{\phi}{A^2}+\frac{\phi^n}{A^2}=0,
\end{equation}
which can be written in a general form 
\begin{equation}
\label{wave-poly1c4}
\ddot{\phi}+f_1(t)\dot{\phi}+f_2(t)\phi+f_3(t)\phi^n=0,
\end{equation}
in a similar way as was done earlier for a simple power law potential. It is easy to recognize $f_{i}$'s as $f_{1} = -2\frac{\dot{A}}{A}, f_{2} = f_{3} = \frac{1}{A^{2}}$. Equation (\ref{int-genc4}), now reduces to

\begin{equation}
\label{wave-poly2c4}
\frac{\ddot{A}}{A}-3\frac{(n+1)}{(n+3)}\frac{\dot{A}^2}{A^2}+\frac{(n+3)}{6A^2}=0,
\end{equation}
where in order to ensure an integrability of the wave equation (\ref{wave-polyc4}), one must restrict the choice of $n$ such that $n\notin \left\{-3,-1,0,1\right\}$.
This differential equation yields a straightforward first integral given as
\begin{equation}
\dot{A}^2-{\lambda}A^{\frac{6(n+1)}{(n+3)}}-\frac{(n+3)^2}{18(n+1)}=0,
\end{equation}
where $\lambda$ comes as a constant of integration. This can be written in a simpler form,
\begin{equation}\label{wave-poly3c4}
\dot{A}=\Big({\lambda}A^p+q\Big)^\frac{1}{2},
\end{equation}
where, $p=\frac{6(n+1)}{(n+3)}$ and $q=\frac{(n+3)^2}{18(n+1)}$. The general solution of equation (\ref{wave-poly3c4}) can in fact be given in the form of Gauss' Hypergeometric function,

\begin{equation}
\label{sol-A-polyc4}
\frac{A}{\sqrt{q}}{_2}F{_1}\Bigg[\frac{1}{2},\frac{1}{p};(1+\frac{1}{p});-\frac{\lambda A^p}{q}\Bigg]=t-t_0,
\end{equation}
where $t_{0}$ is a constant of integration. \\

It is very difficult to invert the equation (\ref{sol-A-polyc4}) and write $A(t)$ as a function of $t$ explicitly. However, since we are interested in a regime of space-time, where the volume is very small, an approximate analysis of this equation can be given, assuming $A(t)\rightarrow\infty$, meaning the proper radius ($\sim \frac{1}{A}$) is very small. Then from the series expansion of the Hypergeometric function, one can write 
\begin{equation}
\label{large-xc4}
{_2}F_{1}(a,b;c;x)=\frac{\Gamma(b-a)\Gamma(c)}{\Gamma(b)\Gamma(c-a)}(-x)^{-a}\Bigg[1+\textit{O}\Bigg(\frac{1}{x}\Bigg)\Bigg]+\frac{\Gamma(a-b)\Gamma(c)}{\Gamma(a)\Gamma(c-b)}(-x)^{-b}\Bigg[1+\textit{O}\Bigg(\frac{1}{x}\Bigg)\Bigg]
\end{equation}
for $|x|\rightarrow\infty$ and $a\neq b$.  \\

Using (\ref{large-xc4}), the expression for $A$ can be written from equation (\ref{sol-A-polyc4}) as
\begin{eqnarray}\nonumber \label{large-Ac4}
&& \frac{A^{1-p/2}}{\lambda^{1/2}}\frac{\Gamma(1/p-1/2)\Gamma(1+1/p)}{\Gamma(1/p)\Gamma(1/p+1/2)}\Bigg[1+\textit{O}\Bigg(-\frac{q}{\lambda A^p}\Bigg)\Bigg] \\ &&
+\frac{q^{1/p-1/2}}{\lambda^{1/p}}\frac{\Gamma(1/2-1/p)\Gamma(1+1/p)}{\Gamma(1/2)\Gamma(1)}\Bigg[1+\textit{O}\Bigg(-\frac{q}{\lambda A^p}\Bigg)\Bigg]=t-t_0.
\end{eqnarray}

A careful study of equation (\ref{large-Ac4}) reveals that, for all $(1-\frac{p}{2})<0$, $A(t)\rightarrow\infty$, which implies that the scale factor ${\frac{1}{A(t)}\rightarrow 0}$, for a negative $(1-\frac{p}{2})$, at the time 
\begin{equation}
t=t_0+\Bigg[\frac{q^{1/p-1/2}}{\lambda^{1/p}}\Bigg]\Bigg[\frac{\Gamma(1/2-1/p)\Gamma(1+1/p)}{\Gamma(1/2)\Gamma(1)}\Bigg].
\end{equation}
It should be noted that as this would require $(1-\frac{p}{2})$ to have a negative value, $n$ is either positive or $n<-3$. The latter however will lead to imaginary solutions for the scale factor and will not be considered in the subsequent discussion. \\

For the sake of completeness, we should mention that the general solution for the scalar field equation can be written as 
\begin{equation}
T=T_{0}+\frac{\epsilon }{C_{0}}\Phi\sqrt{\frac{C_{0}\left( n+1\right) -{\Phi}^{n+1}}{%
2\left( n+1\right) }}\,_{2}F_{1}\left[ 1,\frac{n+3}{2\left( n+1\right) };%
\frac{n+2}{n+1};\frac{{\Phi}^{n+1}}{C_{0}(n+1)}\right] , \qquad   n\neq -1,
\end{equation}
where $T_0$ and $C_0$ are arbitrary constants of integration and  $\Phi$ and $T$ are defined by equations (\ref{Phic4}) and (\ref{Tc4}) respectively.

\begin{figure}[h]
\begin{center}
\includegraphics[width=0.5\textwidth]{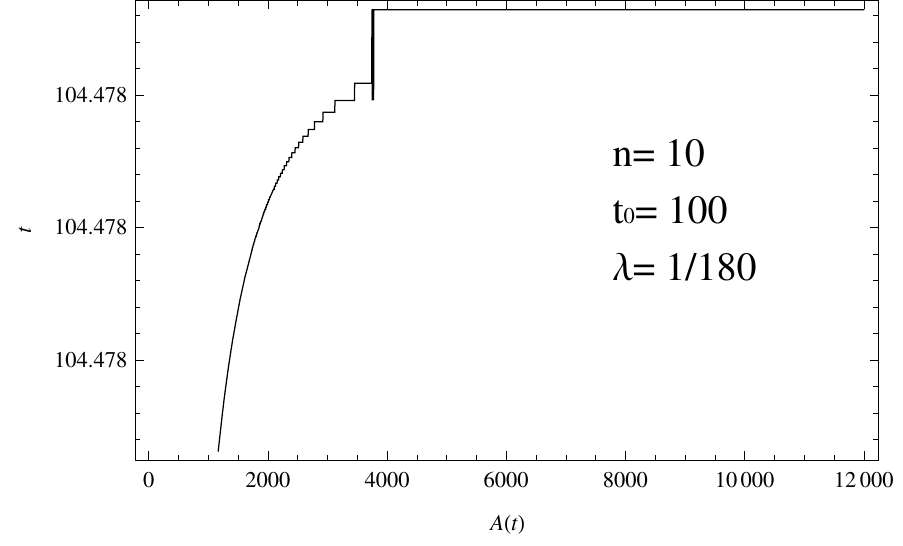}
\caption{Plot of $t$ vs $A(t)$ for $n = 10$ and a positive $\lambda$.}
\end{center}
\label{fig:hyp1}
\end{figure}

\begin{figure}[h]
\begin{center}
\includegraphics[width=0.5\textwidth]{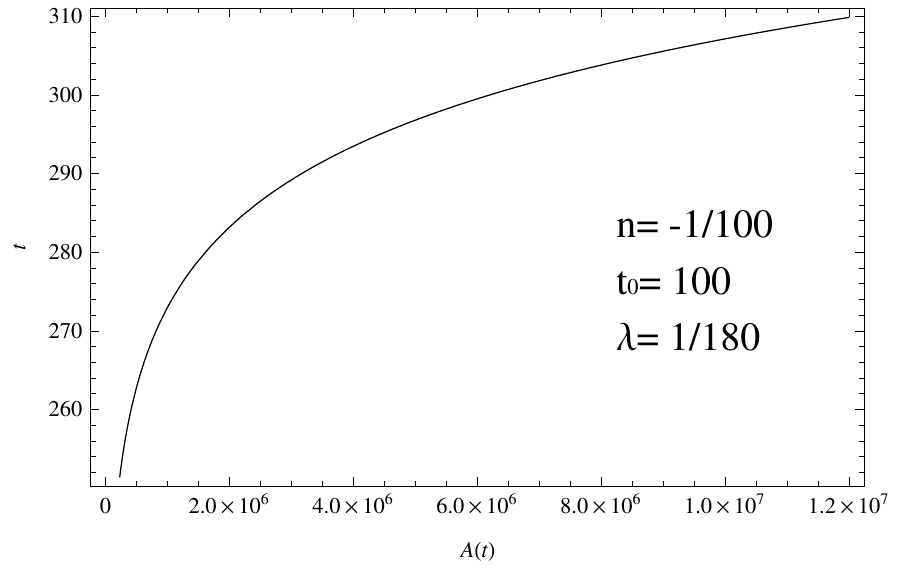}
\caption{Plot of $t$ vs $A(t)$ for $n=-\frac{1}{100}$ and a positive $\lambda$.}
\end{center}
\label{fig:hyp1}
\end{figure}

\begin{figure}[h]
\begin{center}
\includegraphics[width=0.5\textwidth]{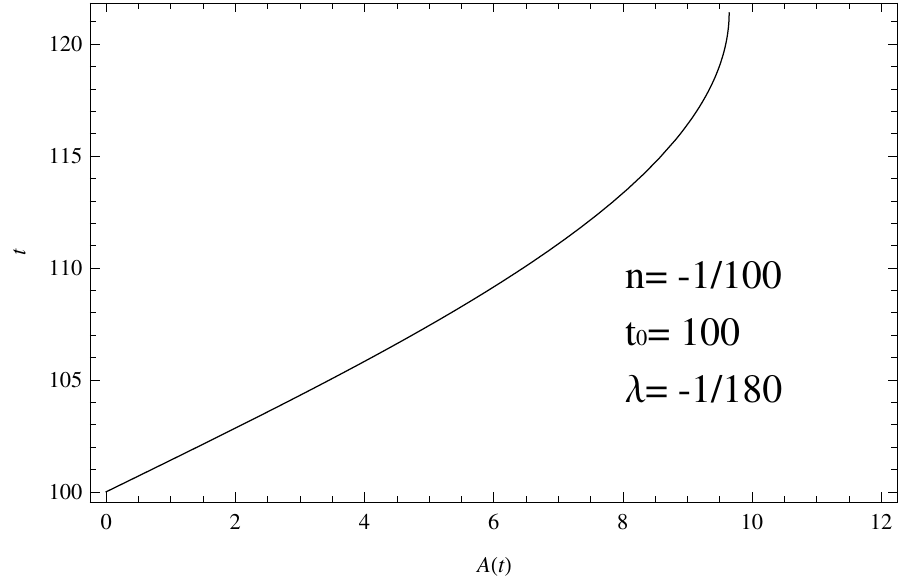}
\caption{Plot of $t$ vs $A(t)$ for negative $\lambda$.}
\end{center}
\label{fig:hyp1}
\end{figure}

We shall discuss a few examples with some values of the constants $n, \lambda$ and $t_{0}$ with the help of numerical plots. Figure $4.1$ shows that for $n=10$ and a positive $\lambda$, A increases very fast to an indefinitely large value at a finite value of time $t$ indicating that the proper radius ($\frac{1}{A}$) and hence the proper volume indeed crushes to a singularity. Figure $4.2$ shows that for a small negative value of $n$, namely $n=-\frac{1}{100}$, one has a collapsing situation, but the rate of collapse dies down and the singularity is not reached at a finite time.  This is quite consistent with the inference drawn from equation (\ref{large-Ac4}) that for a collapse to reach a singular state at a finite time, one would require a positive value of $n$. The behaviour is also sensitive to the initial conditions. For example, for a negative $\lambda$, the same small negative value of $n$ would lead to a situation where the distribution will not collapse beyond a certain constant finite volume at a finite time after the collapse begins, as shown in figure $4.3$.

\section{Discussion}
In the present chapter a massive scalar field collapse with a power law potential ($V \sim {\phi}^{n+1}$) is discussed in a very general situation which includes a very wide range of the values of $n$.

In order to study the problem analytically, a strategy of dealing with the integrability condition for the scalar field equation is adopted. The recently developed technique of solving anharmonic oscillator problem by Euler \cite{euler1, euler2} (see also \cite{harkolobomak}) has been utilized. It is interesting to note that the conclusions drawn from these calculations are independent of the choice of any equation of state for the fluid distribution. This is because the scale factor is calculated straightaway from the integrability condition. The field equations can be utilized in the determination of the fluid density and pressure as functions of $A$ and $\phi$ and hence as a function of $t$ (equations (\ref{den-plwc4}) and (\ref{press-plwc4})).  \\

The general result is that it is indeed possible to have a collapsing situation which crushes to the singularity of zero proper volume and infinite curvature. The singularity is hidden by an apparent horizon which forms before the formation of the singularity, thus the singularity is never visible. This situation is observed for potentials of the form $V\sim {\phi}^{n+1}$ where $n<-3$ or $n>0$. However, for $0>n>-3$, the model collapses for ever, reaching the singularity only at an infinite future. Therefore, for a continuous gravitational collapse of a massive scalar field with potential of the form $V(\phi) \sim \phi^{(n+1)}$, whenever one has a singularity at a finite future, it is necessarily covered by a horizon. This is completely consistent with the theorem proved by Hamid, Goswami and Maharaj \cite{hgmc4}, that for a continuous gravitational collapse in a conformally flat spacetime the end product is always a black hole.      \\
A quadratic potential is of a primary interest in scalar field theories. But this form of potential is out of the domain of validity of the theorem used (the method does not work for $n=1$, which corresponds to a quadratic potential). A discussion on a potential containing two terms, one of which is a quadratic in ${\phi}$, is included as well so as to include the effect of a quadratic potential albeit not independently. Although an elaborate discussion in this case has not been possible, quite a few interesting results from the asymptotic behaviour of the solution has been noted with the help of numerical plots. Depending on the initial conditions, there are many interesting possibilities, where the singularity is reached only at infinite time, and even a situation where the collapsing object settles down to a finite size rather than crushing into a singular state. This phenomena can perhaps be argued to be like a scalar field analogue of a white dwarf or a neutron star where the collapsing star equilibriates as a finite object when the degenerate fermion pressure is able to halt the gravitational collapse. Apparently a scalar field with a potential which is a power law of the field $\phi$ with a small negative exponent along with a ${\phi}^{2}$ term can also do the trick. \\

It deserves mention that method utilized works only for the cases when the scalar field equation is integrable, and does not work for certain values of $n$, but appears to be an extremely powerful tool, and should find extensive application in the physics of scalar field collapse.

\chapter[Self-similar scalar field collapse]{Self-similar scalar field collapse\footnote[1]{The results of this chapter are reported in Physical Review D (2017) {\bf 95}, 024015}}

% **************************** Define Graphics Path**************************

\ifpdf
    \graphicspath{{Chapter5/Figs/Raster/}{Chapter5/Figs/PDF/}{Chapter5/Figs/}}
\else
    \graphicspath{{Chapter5/Figs/Vector/}{Chapter5/Figs/}}
\fi

This chapter contains an investigation of the collapse of a massive scalar field along with a fluid distribution which is locally anisotropic and contains a radial heat flux. This vastly general matter distribution is analysed at the expense of a high degree of symmetry in the spacetime, the presence of both conformal flatness and self similarity. The self-interaction potential is taken to be a power law function of the scalar field or suitable combinations of power-law terms. Power law potentials are in fact quite relevant and well-studied in cosmological contexts as many realistic matter distribution can be modelled with power law potentials. For instance, a quadratic potential, on the average, mimics a pressureless dust whereas the quartic potential exhibits a radiation like behavior \cite{booksami}. A potential with a power less than unity results in an inverse power-law self interaction as in the wave equation, the $\frac{dV}{d\phi}$ term introduces a term with inverse power of $\phi$. Inverse power law potentials were extensively used for the construction of tracker fields\cite{zlatev2, johri1, johri2}. The scalar field evolution equation is studied extensively in this chapter for some reasonable choices of potentials so that examples from both positive and negative self interactions can be taken care of.           \\

In the previous chapter, an exact solution for such a scalar field collapse with a spatial homogeneity was discussed and it was shown that such a massive scalar field collapse without any apriori choice of an equation of state, necessarily leads to the formation of a black hole. Under the assumption that the scalar field evolution equation is integrable, the integrability conditions for a general anharmonic oscillator equation (developed by Euler \cite{euler1, euler2} and utilized by Harko, Lobo and Mak\cite{harkolobomak}) was extensively utilised in the last chapter, and is the main tool used in this chapter as well. The assumption of homogeneity is dropped and also the fluid content has a local anisotropy and a radial heat flux, thus making the system under consideration a lot more general than the one described in last chapter. However, an additional symmetry, the existence of a homothetic Killing vector, is assumed at the outset, implying a self-similarity in the spacetime.            \\

The assumption of self-similarity imposes a restriction on the metric tensor, but it is not really unphysical and there are lots of examples where self-similarity is indeed observed. In non-relativistic Newtonian fluid dynamics, self-similarity indicates that the physical variables are functions of a dimensioness variable $\frac{x}{l(t)}$ where $x$ and $t$ are space and time variables and $l$, a function of $t$, has the dimension of length. Existence of self-similarity indicates that the spatial distribution of physical variables remains similar to itself at all time. Such examples can be found in strong explosions and thermal waves. In general relativity also, self-similarity, characterized by the existence of a homethetic Killing vector, finds application. We refer to the work of Carr and Coley\cite{carrcoley} for a review on the implications of self-similarity in general relativity.      

\section{Conformally flat Scalar field and a fluid with pressure anisotropy and heat flux}
The space-time metric is chosen to have a vanishing Weyl tensor implying a conformal flatness. The metric can be written as
\begin{equation}
\label{metricc5}
ds^2=\frac{1}{{A(r,t)}^2}\Bigg[dt^2-dr^2-r^2d\Omega^2\Bigg],
\end{equation}
where $A(r,t)$ is the conformal factor and governs the evolution of the sphere. The fluid inside the spherically symmetric body is assumed to be locally anisotropic along with the presence of heat flux. Thus the energy-momentum tensor is given by
\begin{equation}\label{EMTc5}
T_{\alpha\beta}=(\rho+p_{t})u_{\alpha}u_{\beta}-p_{t}g_{\alpha\beta}+ (p_r-p_{t})\chi_{\alpha} \chi_{\beta}+q_{\alpha}u_{\beta}+q_{\beta}u_{\alpha},
\end{equation}
where $q^{\alpha}=(0,q,0,0)$ is the radially directed heat flux vector, $\rho$ is the energy density, $p_{t}$ the tangential pressure, $p_r$ the radial pressure, $u_{\alpha}$ the four-velocity of the fluid and $\chi_{\alpha}$ is the unit four-vector along the radial direction. The vectors $u_{\alpha}$ and $\chi_{\alpha}$ are normalised as
\begin{equation}
\label{normc5}
u^{\alpha}u_{\alpha}=1,\quad\chi^{\alpha}\chi_{\alpha}=-1,\quad\chi^{\alpha}u_{\alpha}=0.
\end{equation}

A comoving observer is chosen, so that $u^{\alpha} = A{\delta}^{\alpha}_{0}$ and the normalization equation (\ref{normc5}) is satisfied ($A = \frac{1}{\sqrt{g_{00}}}$). It is to be noted that there is no assumption of an isotropic fluid pressure to begin with. The radial and transverse pressures are different. Anisotropic fluid pressure is quite relevant in the study of compact objects and considerable attention has been given to this in existing literature. Comprehensive reviews can be found in the works of Herrera and Santos \cite{herresantoc5}, Herrera et. al.\cite{herreetal2}. It has also been shown by Herrera and Leon that on allowing a one-parameter group of conformal motions, a smooth matching of interior and exterior geometry is possible if and only if there is an pressure anisotropy in the fluid description \cite{herreraleon}.      \\

A dissipative process, namely heat conduction is also included in the system. In the evolution of stellar bodies, dissipative processes are of utmost importance. Particularly when a collapsing star becomes too compact, the size of the constituent particles can no longer be neglected in comparison with the mean free path, and dissipative processes can indeed play a vital role, in shedding off energy so as to settle down to a stable final system. For more of relevant details we refer to the works of Herrera et. al. \cite{herreetal2}, Kazanas and Schramm \cite{kaza}.        \\

When a scalar field $\phi=\phi(r,t)$ is minimally coupled to gravity, the relevant action is given by 
\begin{equation}
{\cal A} = \int{\sqrt{-g}d^4x[R + \frac{1}{2}\partial^{\mu}\phi \partial_{\mu}\phi - V(\phi) + L_{m}]},
\end{equation}

where $V(\phi)$ is the potential and $L_{m}$ is the Lagrangian density for the fluid distribution. From this action, the contribution to the energy-momentum tensor from the scalar field $\phi$ can be  written as
\begin{equation}
T^\phi_{\mu\nu}=\partial_\mu\phi\partial_\nu\phi-g_{\mu\nu}\Bigg[\frac{1}{2}g^{\alpha\beta}\partial_\alpha\phi\partial_\beta\phi-V(\phi)\Bigg]. 
\end{equation}

Einstein field equations (in the units $8 \pi G = 1$) can thus be written as
\begin{equation}
\label{fe1c5}
3\dot{A}^2-3A'^2+2AA''+\frac{4}{r}AA'=\rho+\frac{1}{2}A^2\dot{\phi}^2-\frac{1}{2}A^2\phi'^2 + V(\phi),
\end{equation}

\begin{equation}
\label{fe2c5}
2\ddot{A}A-3\dot{A}^2+3A'^2-\frac{4}{r}AA'=p_{r}+\frac{1}{2}{\phi'}^2A^2+\frac{1}{2}A^2\dot{\phi}^2-V(\phi),
\end{equation}

\begin{equation}
\label{fe3c5}
\ddot{A}A-3\dot{A}^2+3A'^2-\frac{2}{r}AA'-2AA''=p_{t}-\frac{1}{2}{\phi'}^2A^2 +\frac{1}{2}A^2\dot{\phi}^2-V(\phi),
\end{equation}

\begin{equation}
\label{fe4c5}
\frac{2\dot{A}'}{A}=-\frac{q}{A^3}+\dot{\phi}\phi'.
\end{equation}
The wave equation for the scalar field is given by
\begin{equation}
\label{wavec5}
\Box\phi+\frac{dV}{d\phi}=0,
\end{equation}
which, for the present metric (\ref{metricc5}), translates into
\begin{equation}
\label{wave2c5}
\ddot{\phi}-\phi''-2\frac{\dot{A}}{A}\dot{\phi}-2\frac{\phi'}{r}+2\frac{\phi'A'}{A}+\frac{1}{A^2}\frac{dV}{d\phi}=0.
\end{equation}
In this system of equations, there are $6$ unknowns, namely, $\rho$, $p_{r}$, $p_{t}$, $q$, $A$ and $\phi$, whereas only four equations $(6-8)$ to solve for them. Rather than the usual strategy of assuming any specific equation of state to close the system of equation, we probe the system under the assumption that the scalar field equation (\ref{wave2c5}) is integrable; which facilitates the appearence of an additional differential condition on the conformal factor, as discussed in the following sections. 

\section{Self Similarity and exact solution}
A self-similar solution is one in which the spacetime admits a homothetic killing vector $\xi$, which satisfies the equation
\begin{equation}
L_{\xi}g_{ab}=\xi_{a;b}+\xi_{b;a}=2g_{ab},
\end{equation}
where $L$ denotes the Lie derivative. In such a case, one can have repeatative structures at various scales. With a conformal symmetry the angle between two curves remains the same and the distance between two points are scaled depending on the spacetime dependence of the conformal factor ($\frac{1}{A}$ in the present case). For a self-similar space-time, by a suitable transformation of coordinates, all metric coefficients and dependent variables can be put in the form in which they are functions of a single independent variable, which is a dimensionless combination of space and time coordinates; for instance, in a spherically symmetric space this variable is $\frac{t}{r}$.      \\

It deserves mention that a homothetic Killing vector (HKV) is a special case of a conformal Killing vector (CKV) $\eta$ defined by $L_{\eta}g_{ab}=\eta_{a;b}+\eta_{b;a}=\lambda g_{ab}$, where $\lambda$ is a function. Clearly HKV is obtained from a CKV when $\lambda$ attains a constant value. For a brief but useful discussion on CKV, we refer to the works by Maartens and Maharaj \cite{maartens1c3, maartens2c3}, Maartens, Maharaj and Tupper \cite{maartens3c3}.   \\

Writing $A(r,t)=rB(z)$, the derivatives are transformed into derivatives with respect to $z$ and the scalar field equation (\ref{wave2c5}) is written as
\begin{equation}\label{wave3c5}
\phi^{\circ\circ}-\phi^{\circ}\Bigg[2\frac{B^{\circ}}{B}+\frac{2z}{1-z^2}\Bigg]+\frac{\frac{dV}{d\phi}}{B^2(1-z^2)}=0.
\end{equation}          
Here, an overhead $\circ$ denotes a derivative with respect to $z=\frac{t}{r}$.                      \\

Equation (\ref{wave3c5}) can be formally written as a general classical anharmonic oscillator equation with variable coefficients as 
\begin{equation}
\label{genc5}
\phi^{\circ\circ}+f_1(z)\phi^{\circ}+ f_2(z)\phi+f_3(z)\frac{dV}{d\phi}=0,
\end{equation}
where $f_i$'s are functions of $z$ only.          

\section{A note on the integrability of an anharmonic oscillator equation}
A nonlinear anharmonic oscillator with variable coefficients and a power law potential can be written in a general form as

\begin{equation}
\label{gennc5}
\ddot{\phi}+f_1(u)\dot{\phi}+ f_2(u)\phi+f_3(u)\phi^n=0,
\end{equation}

where $f_i$'s are functions of $u$ and $n \in {\cal Q}$, is a constant. Here $u$ is the independent variable. An overhead dot represents a differentiation with respect to $u$. Using Euler’s theorem on the integrability of the general anharmonic oscillator equation \cite{euler1, euler2} and recent results given  by Harko {\it et al} \cite{harkolobomak}, this equation can be integrated under certain conditions. This can be given in the form of a theorem as \cite{euler1, harkolobomak},

\textbf{Theorem} For $n\notin \left\{-3,-1,0,1\right\} $, if and only if the coefficients of Eq.(\ref{gennc5}) satisfy the differential condition
\begin{eqnarray}\nonumber
\label{int-genc5}
&&\frac{1}{(n+3)}\frac{1}{f_{3}(u)}\frac{d^{2}f_{3}}{du^{2}%
}-\frac{n+4}{\left( n+3\right) ^{2}}\left[ \frac{1}{f_{3}(u)}\frac{df_{3}}{du%
}\right] ^{2} + \frac{n-1}{\left( n+3\right) ^{2}}\left[ \frac{1}{f_{3}(u)}%
\frac{df_{3}}{du}\right] f_{1}\left( u\right) + \frac{2}{n+3}\frac{df_{1}}{du} \\&&
+\frac{2\left( n+1\right) }{\left( n+3\right) ^{2}}f_{1}^{2}\left( u\right)=f_2(u),
\end{eqnarray} 
equation (\ref{gennc5}) is integrable. \\

If one introduces a pair of new variables $\Phi$ and $U$ given by 
\begin{eqnarray}
\label{Phic5}
\Phi\left( U\right) &=&C\phi\left( u\right) f_{3}^{\frac{1}{n+3}}\left( u\right)
e^{\frac{2}{n+3}\int^{u}f_{1}\left( x \right) dx },\\
\label{Uc5}
U\left( \phi,u\right) &=&C^{\frac{1-n}{2}}\int^{u}f_{3}^{\frac{2}{n+3}}\left(
\xi \right) e^{\left( \frac{1-n}{n+3}\right) \int^{\xi }f_{1}\left( x
\right) dx }d\xi ,
\end{eqnarray}%
where $C$ is a constant, equation (\ref{gennc5}) can be written as 

\begin{equation}
\label{Phi1c5}
\frac{d^{2}\Phi}{dU^{2}}+\Phi^{n}\left( U\right) =0.
\end{equation}

\section{Power Law Potential : $V(\phi)=\frac{\phi^{m+1}}{(m+1)}$}
\subsection{Solution for the Scale Factor}
As already discussed in Section $1.3.1$, a  scalar field with a variety of interaction potential can mimic the evolution of many a kind of matter distribution; for instance, Goncalves and Moss \cite{goncalves} showed that the collapse of a spherically symmetric self-interacting scalar field can be formally treated as a collapsing dust ball. Many reasonable matter distribution can be modelled with power law interaction potentials, for example, a quadratic potential, on the average, mimics a pressureless dust whereas the quartic potential exhibits radiation like behavior \cite{booksami}. We choose to study a simple power law potential, for which $\frac{dV}{d\phi}=\phi^m$, such that one can study different possible outcomes of the gravitational collapse of such a self-interacting scalar field for different choice of $m$, i.e., for different definitions of the self-interaction. One can write equation (\ref{wave3c5}) as
\begin{equation}\label{wave4c5}
\phi^{\circ\circ}-\phi^{\circ}\Bigg[2\frac{B^{\circ}}{B}+\frac{2z}{1-z^2}\Bigg]+\frac{\phi^m}{B^2(1-z^2)}=0.
\end{equation}           \\
Comparing with (\ref{gennc5}), it is straightforward to identify $f_{1}(z)=-\Bigg[2\frac{B^{\circ}}{B}+\frac{2z}{1-z^2}\Bigg]$, $f_{2}(z)=0$ and $f_{3}(z)=\frac{1}{B^2(1-z^2)}$. The integrability criteria as mentioned in the last section, gives a second order non-linear differential condition on $B(z)$ as
\begin{equation} \label{solbasicc5}
\frac{B^{\circ\circ}}{B}-3\frac{(m+1)}{(m+3)}\Bigg(\frac{B^{\circ}}{B}\Bigg)^2-\frac{8(2m+3)}{6(m+3)}\frac{B^{\circ}z}{B(1-z^2)}-\frac{2(1+m)z^2-2(m+3)}{6(m+3)(1-z^2)^2}=0.
\end{equation}
Here $[m\neq{-3,-1,0,1}]$ as mentioned before.

\begin{figure}[h]\label{1stfig}
\begin{center}
\includegraphics[width=0.42\textwidth]{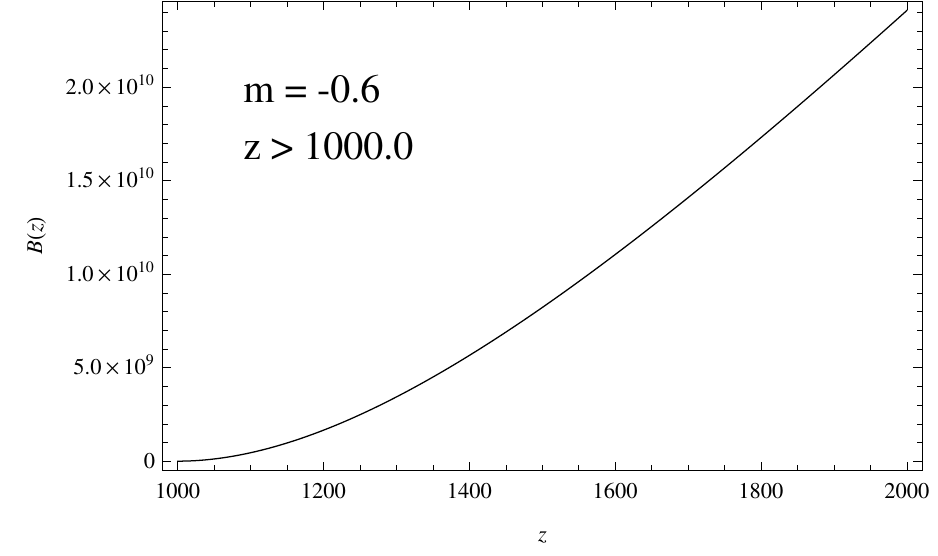}
\includegraphics[width=0.40\textwidth]{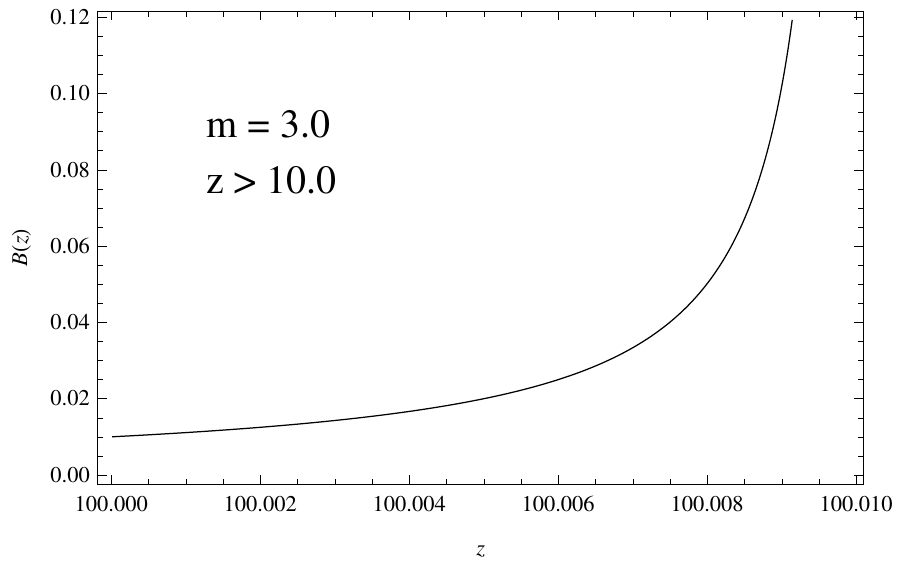}
\caption{Evolution of the conformal factor $B(z)$ with respect to $z$ for $z > 10$. The potentials are taken to be $V(\phi)= \frac{5\phi^{2/5}}{2}$ (graph on the left side) and $V(\phi)= \frac{\phi^4}{4}$ (graph on the right side).}
\end{center}
\end{figure}

It is indeed difficult to handle the equation (\ref{solbasicc5}) analytically, but it can be treated numerically so as to examine the nature of $B$ as a function of $z$. Our aim is to understand whether the system reaches any singularity of zero proper volume at any finite future. Figure $5.1$ shows the numerical evolution of $B(z)$ with respect to $z$ for $m=-\frac{3}{5}$ and $m=3$. Keeping in mind the fact that the scale factor is proportional to $\frac{1}{B(z)}$; it appears that for $m=-\frac{3}{5}$, i.e. $V(\phi)= \frac{5\phi^{2/5}}{2}$, the system undergoes a gravitational collapse, but the rate of collapse dies down eventually, and a zero proper volume singularity is reached only for $z \rightarrow \infty$, i.e. for finite $t = t_s$ but $r \rightarrow 0$. However, for $V(\phi)= \frac{\phi^4}{4}$, the system shrinks to zero rather rapidly, where $B(z) \rightarrow \infty$, at a finite value of $z$ and this singularity is not necessarily a central singularity.    \\

These plots, though represent the general evolution of the spherical body, further analysis like formation of an apparent horizon or the nature of curvature scalar may not be quite straightforward, without any solution for $B(z)$ in a closed form. With this in mind, we now investigate whether the collapsing system reaches any central singularity at a future given by $t \rightarrow t_{s}$ and $r \rightarrow 0$. For that purpose we now look for an approximate but analytical expression for $B(z)$ for $z >> 1$. In this domain the last term on the LHS of (\ref{solbasicc5}), $\frac{2(1+m)z^2-2(m+3)}{(m+3)^2 (1-z^2)^2}$ is of the order of $\sim \frac{1}{z^2}+\frac{1}{z^4}$ and hence can be ignored with respect to the other terms. This approximation does not affect the nature of the evolution as shall be seen in the subsequent analysis. Thus we write the effective equation governing the collapsing fluid for $z >> 1$ as
 
\begin{equation}\label{sol2c5}
\frac{B^{\circ\circ}}{B}-3\frac{(m+1)}{(m+3)}\Bigg(\frac{B^{\circ}}{B}\Bigg)^2-\frac{8(2m+3)}{6(m+3)}\frac{B^{\circ}z}{B(1-z^2)}=0.
\end{equation}

A solution of equation (\ref{sol2c5}) can be written in term of Gauss' Hypergeometric function as
\begin{equation}\label{hyperc5}
\frac{B^{1-\alpha}}{(1-\alpha)}= {_2}F{_1}\Bigg[\frac{1}{2},\beta;\frac{3}{2};z^2\Bigg]\varepsilon z + \varepsilon_0, 
\end{equation}

where $\alpha$ and $\beta$ are defined in terms of $m$ as $\alpha=3\frac{(m+1)}{(m+3)}$, $\beta=\frac{8(2m+3)}{6(m+3)}$ and  $\varepsilon$ and $\varepsilon_0$ are constants of integration. Equation (\ref{hyperc5}) describes the evolution of the spherical body as a function of the self-similarity variable $z$ and the exponent of the self interaction, $m$. One can expand equation (\ref{hyperc5}) in a power series in the limit $z \rightarrow \infty$ and write $B(z)$ explicitly as a function of $z$. However, in order to present some simple examples we choose three values of $m$ such that the parameter $\beta$ has the values $1$, $2$ and $0$ respectively; namely, $m=-\frac{3}{5}$, $m=3$ and $m= -\frac{3}{2}$. Therefore the potentials are effectively chosen as $V(\phi)= \frac{5\phi^{2/5}}{2}$, $V(\phi)= \frac{\phi^4}{4}$ and $V(\phi)= -\frac{2}{\phi^{1/2}}$. A case where $m = -\frac{3}{2}$ is also taken up so as to include an example for an inverse power law potential as well.

\begin{enumerate}
\item{{\bf Case $1$ : } {\bf $m=-\frac{3}{5}$, $\beta = 1$ and $\alpha = \frac{1}{2}$} \\
Putting these values in (\ref{sol2c5}), a solution can be written as
\begin{equation}\label{exact-solc5}
B(z)=C_{2}\Bigg[C_{1}+\frac{1}{2} ln\Big|z+(z^2-1)^\frac{1}{2}\Big|\Bigg]^2,
\end{equation}            
where $C_{1}$ and $C_{2}$ are constants of integration. Therefore the inverse of the conformal factor can be written explicitly as a function of $r$ and $t$ as
\begin{equation}\label{exact-sol2c5}
A(z)=rC_{2}\Bigg[C_{1}+\frac{1}{2} ln\Bigg|\frac{t}{r}+\Bigg(\frac{t^2}{r^2}-1\Bigg)^\frac{1}{2}\Bigg|\Bigg]^2.
\end{equation} 

\begin{figure}[h]\label{2ndfig}
\begin{center}
\includegraphics[width=0.42\textwidth]{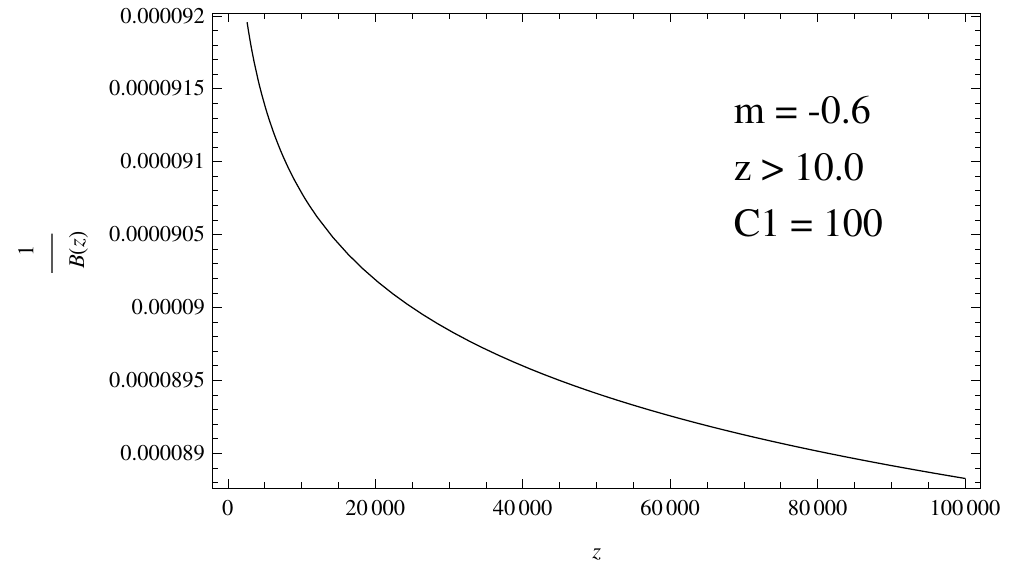}
\includegraphics[width=0.40\textwidth]{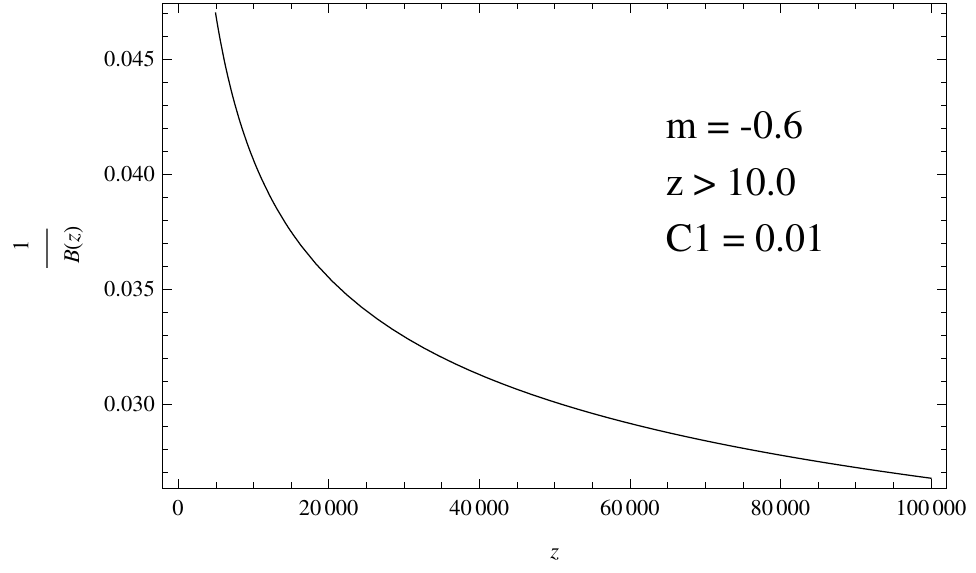}
\includegraphics[width=0.40\textwidth]{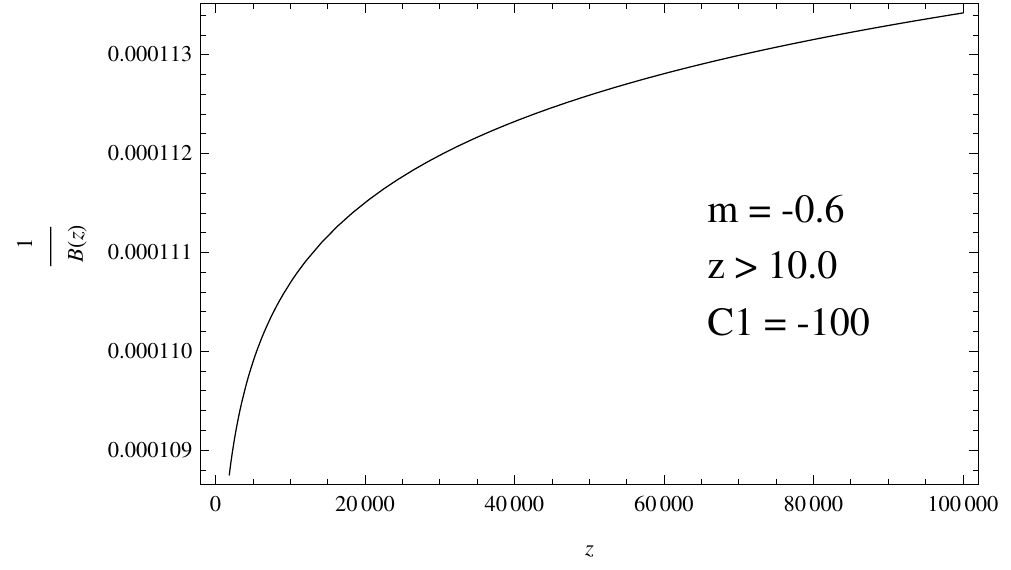}
\caption{Evolution of $B(z)$ with respect to $z$ for $V(\phi)= \frac{5\phi^{2/5}}{2}$, for different choices of the parameter $C_1$.}
\end{center}
\end{figure}

The evolution with respect to $z$ and therefore formation of a singularity will depend on the integration constants and of course, the choice of potential does play a crucial role. Without any loss of generality we assume $C_{2}=1$. It must be mentioned, that for all $C_{2}<0$, one has either a negative volume or no real evolution at all. Excluding those cases, now we examine equation (\ref{exact-solc5}) near the central singularity. It is evident that, the singularity, characterised by $B \rightarrow \infty$, appears only when $z \rightarrow \infty$, i.e, either when $t_s \rightarrow \infty$, or when $t \rightarrow t_s$, $r \rightarrow 0$. However, for all negative values of $C_1$, the system is not collapsing at all; rather, it settles down asymptotically at a finite volume after a period of steady expansion with respect to $z$. Figure $5.2$ shows the evolution of $B(z)$ with respect to $z$.
}
\item{{\bf Case $2$ : } {\bf $m=3$, $\alpha = \beta = 2$}. \\
With these values of the parameters, equation (\ref{sol2c5}) can be solved to write $B(z)$ as
\begin{equation}\label{exact-sol3c5}
B(z)= \frac{C_2}{ln{\Big[\frac{C_1(1+z)}{(1-z)}\Big]}}.
\end{equation}
Here, $C_2$ can again be chosen to be unity. For the solution to be valid in the region $z >> 1$, $C_1$ must be a non-zero negative number such that $ln{\Big[\frac{C_1(1+z)}{(1-z)}\Big]}$ is real.

\begin{figure}[h]\label{3rdfig}
\begin{center}
\includegraphics[width=0.40\textwidth]{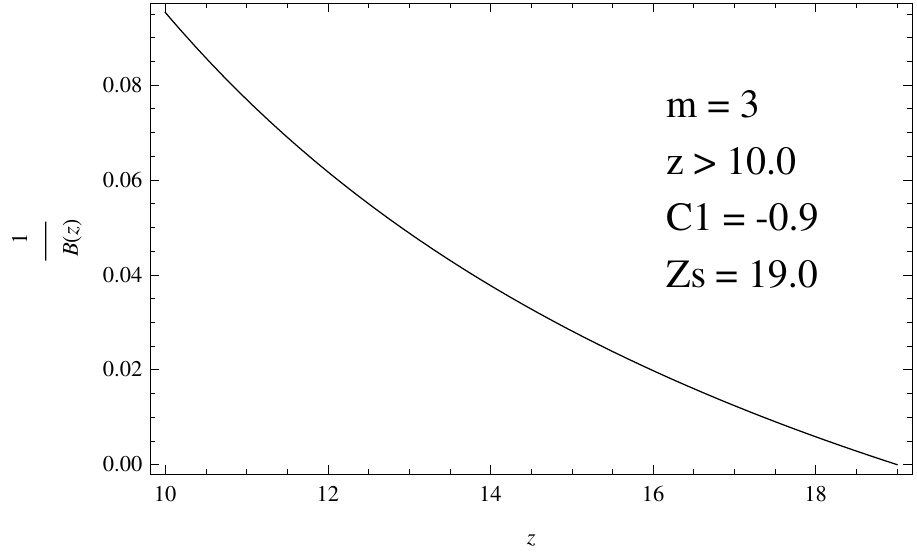}
\includegraphics[width=0.42\textwidth]{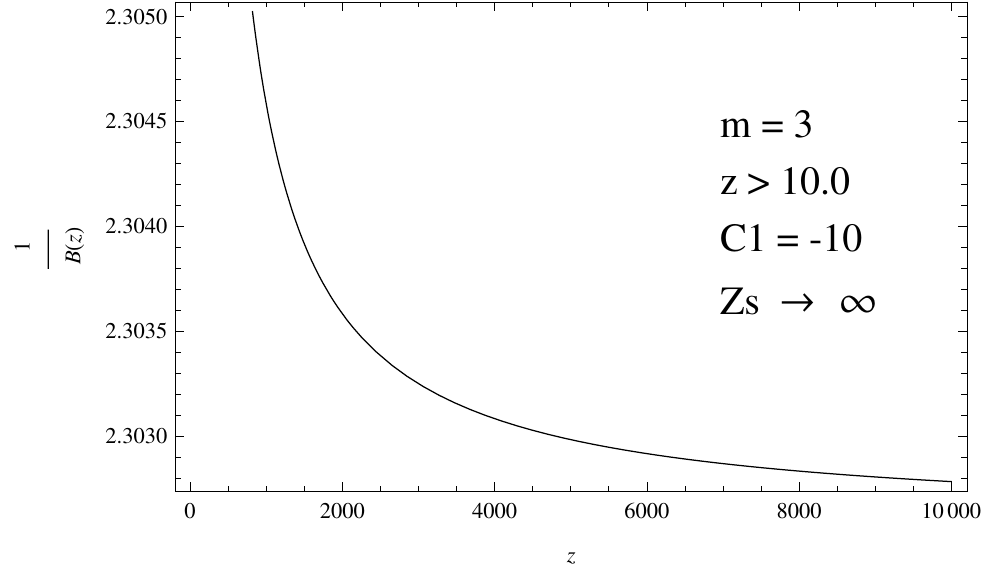}
\caption{Evolution of $B(z)$ with respect to $z$ for $V(\phi)= \frac{\phi^4}{4}$ for different negative values of the parameter $C_1$.}
\end{center}
\end{figure}

From (\ref{exact-sol3c5}) one can find that the system reaches zero proper volume singularity at a finite value of $z$ given by
\begin{equation}
z_s= \frac{1-C_1}{1+C_1}.
\end{equation}
However, for a collapsing model, the scale factor must be a decreasing function throughout. This means $\frac{d}{dz} ln\Big[\frac{(1+z)C_1}{(1-z)}\Big] < 0$. On simplification, this yields $\frac{2}{(1-z^2)} < 0$, which holds true if and only if $z > 1$. So $z_s$, defined by $z_s = \frac{1-C_1}{1+C_1}$ is greater than $1$, only if $-1 < C_1 < 0$, which is, therefore consistent with the requirement of a negative ${C_{1}}$. The discussion is supported graphically in figure $5.3$.
}

\item{{\bf Case $3$ : $m= -\frac{3}{2}$} {\bf $\alpha= -1$ and $\beta= 0$.} \\
Equation (\ref{sol2c5}) is simplified significantly and solution of $B(z)$ may be written as
\begin{equation}\label{exact-sol4c5}
B(z)= [2 C_1 (z+z_0)]^{\frac{1}{2}}.
\end{equation} 
Here, $C_1 (> 0)$ and $z_0$ are constants of integration. The scale factor is defined as $Y(r,t) = \frac{1}{[2 C_1 (\frac{t}{r}+z_0)]^{\frac{1}{2}}}$. 
\begin{figure}[h]\label{3rdfig}
\begin{center}
\includegraphics[width=0.40\textwidth]{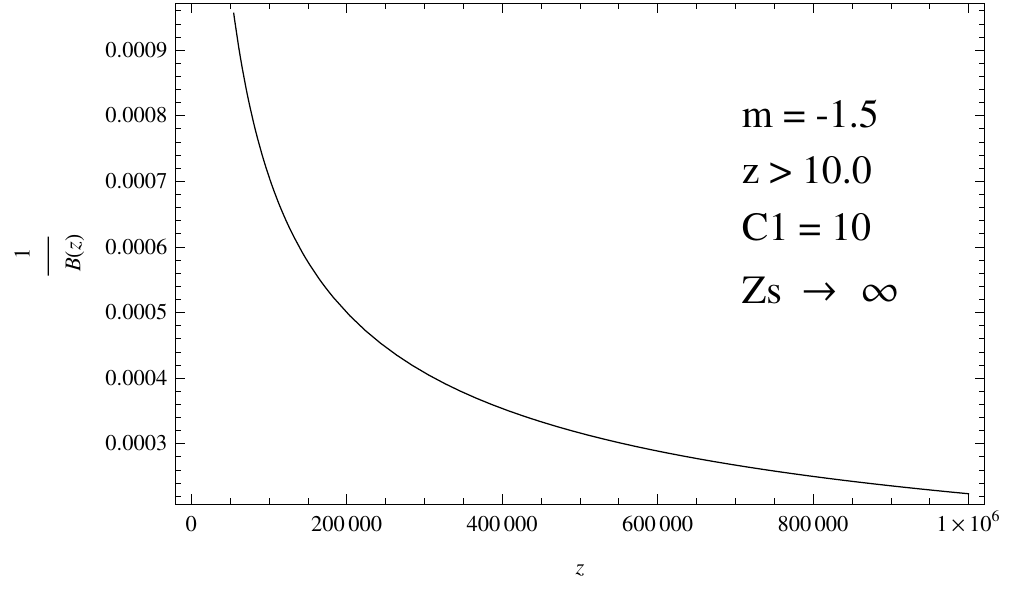}
\includegraphics[width=0.42\textwidth]{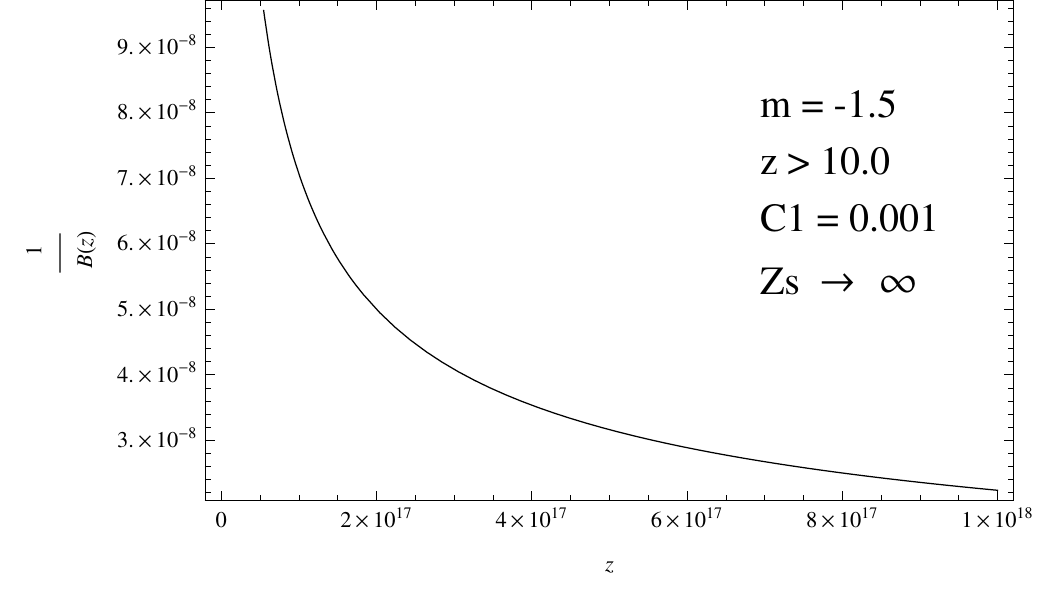}
\caption{Evolution of $B(z)$ with respect to $z$ for $V(\phi)= -\frac{2}{\phi^{1/2}}$ for different positive values of $C_1$. The qualitative behaviour remains the same over different choices of the parameter.}
\end{center}
\end{figure}
Now, $Y \rightarrow 0$ only when $z \rightarrow \infty$, i.e., $t \rightarrow t_s$, $r \rightarrow 0$. Thus the system reaches a central singularity ar $t \rightarrow t_s$, as shown in figure $5.4$ for different choices of initial conditions defined by the choice of $C_1$.         
}
\end{enumerate}

\subsection{Expressions for scalar field, physical quantities and curvature scalar}
Point transforming the scalar field equation using (\ref{Phic5}) and (\ref{Uc5}), equation (\ref{wave3c5}) can be written in an integrable form as (\ref{Phi1c5}). Using the exact form of the coefficients $f_{1}(z)$ and $f_{3}(z)$, after some algebra, we express the scalar field as a function of $z$ as
\begin{equation} \label{scalarfieldc5}
\phi(z)=\phi_{0}B^{\frac{6}{(m+3)}}(1-z^2)^{-\frac{1}{(m+3)}}\Bigg[C^{\frac{(1-m)}{2}}{\int B^{\frac{2(m-3)}{(m+3)}}(1-z^2)^{-\frac{(m+1)}{(m+3)}}dz}-\Phi_0\Bigg].
\end{equation}
Here, $C$ is an integration constant coming from the definition of the point transformation and $\phi_{0}$ is defined in terms of $C$. $\Phi_0$ is a constant coming from integration over $z$. For all choices of $m$ discussed in the previous section, $B^{\frac{6}{(m+3)}}$ is an increasing function w.r.t $z$, and diverges when $B(z) \rightarrow \infty$. Thus it is noted that the scalar field diverges at the singularity.       \\

From (\ref{fe1c5}), (\ref{fe2c5}), (\ref{fe3c5}) and (\ref{fe4c5}), the density, radial and tangential pressure and the heat flux can be expressed generally as
\begin{equation} \label{denstc5}
\rho=3\dot{A}^2-3A'^2+2A''A+\frac{4}{r}A'A-\frac{1}{2}A^2\dot{\phi}^2+\frac{1}{2}A^2\phi'^2-\frac{\phi^{m+1}}{(m+1)}.
\end{equation}

\begin{equation} \label{radpressc5}
p_{r}=2\ddot{A}A-3\dot{A}^2+3A'^2-\frac{4}{r}A'A-\frac{1}{2}\phi'^2A^2-\frac{1}{2}\dot{\phi}^2A^2 +\frac{\phi^{m+1}}{(m+1)}.
\end{equation}

\begin{equation} \label{tangpressc5}
p_{t}=2\ddot{A}A-3\dot{A}^2+3A'^2-2A''A-\frac{2}{r}A'A+\frac{1}{2}\phi'^2 A^2 -\frac{1}{2}\dot{\phi}^2A^2+\frac{\phi^{m+1}}{(m+1)}.
\end{equation}

\begin{equation}\label{heatfc5}
q=-2\dot{A}'A^2+\dot{\phi}\phi'A^3.
\end{equation}                               

One can use the approximate solution for $B$ from equations (\ref{exact-sol2c5}), (\ref{exact-sol3c5}) and (\ref{exact-sol4c5}) and use the fact that $A(r,t) = rB(z)$ in order to check the nature of the fluid variables at singularity. It is found that when the sphere shrinks to zero volume, all these quantities diverge to infinity, confirming the formation of a singularity. To comment on the nature of singularity we write the Kretschmann curvature scalar from (\ref{metricc5}) as 

\begin{eqnarray}\nonumber \label{kretc5}
&&K=-7\dot{A}'^2A^2+4(\dot{A}^2-A'^2 +\frac{AA'}{r}+A''A)^2+ 2(\dot{A}^2-A'^2+A''A -\ddot{A}A)^2 \\&&
+4(\dot{A}^2-A'^2+\frac{AA'}{r}-\ddot{A}A)^2 +(\dot{A}^2-A'^2+2\frac{AA'}{r})^2.
\end{eqnarray}

For any given time-evolution, the Kretschmann scalar diverges anyway when $r\rightarrow 0$, which indicates that the central shell focussing ends up in a curvature singularity. We note that a singularity may also form in relevant cases where $r \neq 0$. To investigate such a singularity, we write \ref{kretc5} as a function of $z$ given by

\begin{eqnarray}\nonumber \nonumber
&&K(z)= -7{B^{\circ}}^2z^2B^2 + 4[{B^{\circ}}^2-(B-B^{\circ}z)^2 + B(B-B^{\circ}z)-B^{\circ\circ}Bz^2]^2 + 2[{B^{\circ}}^2 \\&& \nonumber
-(B-B^{\circ}z)^2+ B^{\circ\circ}B -B^{\circ\circ}Bz^2]^2 + 4[{B^{\circ}}^2-(B-B^{\circ}z)^2 + B(B-B^{\circ}z)- B^{\circ\circ}B]^2 \\&& 
+[{B^{\circ}}^2-(B-B^{\circ}z)^2 + 2B(B-B^{\circ}z)]^2.
\end{eqnarray}
From this expression we note that the first term on the RHS of Kretschmann scalar ${B^{\circ}}^2z^2B^2$ diverges anyway when $B(z) \rightarrow \infty$. Thus, the singularity always turns out to be a curvature singularity.

\section{Apparent Horizon}
Visibility of the central singularity depends on the formation of an apparent horizon, the surface on which outgoing light rays are just trapped, and cannot escape outward. The condition of such a surface is given by
\begin{equation}\label{appdefc5}
g^{\mu\nu}R,_{\mu}R,_{\nu}=0,
\end{equation}
where $R(r,t)$ is the proper radius of the collapsing sphere, which is $rB(z)$ in the present work. Writing the derivative in terms of $z= \frac{t}{r}$, one can express (\ref{appdefc5}) as
\begin{equation}\label{appselfsimilarc5}
(1-z^2){B^{\circ}}^2=0.
\end{equation}
We investigate the formation of an apparent horizon in all the relevant cases discussed, for $V(\phi)= \frac{5\phi^{2/5}}{2}$, $V(\phi)= \frac{\phi^4}{4}$ and $V(\phi)= -\frac{2}{\phi^{1/2}}$, with the assumption $C_2 = 1$ and $z >> 1$.

\begin{enumerate}
\item{{\bf Case $1$ : $m=-\frac{3}{5}$} \\
The equation (\ref{appselfsimilarc5}) gives the condition
\begin{equation}
z+(z^2-1)^{1/2}=e^{-2(C_1-\delta_0^{1/2})}= e^{\gamma},
\end{equation}
where we have defined $\gamma = -2(C_1-\delta_0^{1/2})$ and $\delta_0$ is a constant of integration. The above equation can be simplified to find the time of formation of apparent horizon as
\begin{equation}
t_{ap} = \frac{re^{-\gamma}}{2}(e^{2\gamma}-1)= rsinh{\gamma}.
\end{equation} 
}
\item{{\bf Case $2$ : $m=3$} \\
The equation (\ref{appselfsimilarc5}) is simplified to yield the condition
\begin{equation}
C_{1}\Big(\frac{1+z}{1-z}\Big) = e^{1/{\Psi_0}}.
\end{equation}
Here $\Psi_0$ is a constant of integration over $z$. One can further simplify to write the time of formation of apparent horizon as
\begin{equation}
t_{ap} = r\Bigg(\frac{e^{\frac{1}{\Psi_0}}-C_1}{e^{\frac{1}{\Psi_0}}+C_1}\Bigg) = r\Gamma_0.
\end{equation}
We note here that considering an apparent horizon is not necessary for those cases where singularity forms only when $t \rightarrow \infty$, i.e. when there is no real singularity at all.
}
\item{{\bf Case $3$ : $m=-\frac{3}{2}$} \\
In a similar manner, for $V(\phi)= -\frac{2}{\phi^{1/2}}$, the condition for an apparent horizon may be written as
\begin{equation}
t_{ap} = r\Big(\frac{{\chi_0}^2}{2\tau}-z_0\Big),
\end{equation}
where $\chi_0$ and $\tau$ are constants of integration and are dependent on suitable choice of initial conditions.
}
\end{enumerate}

In all the cases we find that an apparent horizon is formed at a finite time. As there is no explicit expression for $t_{s}$ (the time of formation of the singularity) the visibility of the singularity cannot be ascertained clearly. But it is quite clear that the singularity can not be visible indefinitely. If $t_{app} > t_{s}$, it can be visible for a finite time only and for $t_{app} < t_{s}$, the singularity is never visible. In the next section, we shall discuss the possible visibility of singularity in another way.

\section{Nature of the singularity}
In a spherically symmetric gravitational collapse leading to a central singularity, if the centre gets trapped prior to the formation of singularity, then it is covered and a black hole results. Otherwise, it could be naked, when non-space like future directed trajectories escape from it. Therefore the important point is to determine whether there are any future directed non-spacelike geodesics emerging from the singularity.             \\
From this point of view, general relativistic solutions of self-similar collapse of an adiabatic perfect fluid was discussed by Ori and Piran \cite{ori1} where it was argued that a shell-focussing naked singularity may appear if the equation of state is soft enough. Marginally bound self-similar collapsing Tolman spacetimes were examined and the necessary conditions for the formation of a naked shell focussing singularity were discussed by Waugh and Lake \cite{waughlake1, waughlake2, waughlake3}. For a comprehensive description of the mathematical formulations on occurence of naked singularity in a spherically symmetric gravitational collapse, we refer to the works of Joshi and Dwivedi \cite{joshidwivedi1, joshidwivedi2, joshidwivedi3, joshidwivedi4}, Dwivedi and Dixit \cite{dwivedidixit}. Structure and visibility of central singularity with an arbitrary number of dimensions and with a general type $I$ matter field was discussed by Goswami and Joshi \cite{goswamijoshic5}. They showed that the space-time evolution goes to a final state which is either a black hole or a naked singularity, depending on the nature of initial data, and is also subject to validity of the weak energy condition. Following the work of Joshi and Dwivedi \cite{joshidwivedi1, joshidwivedi2, joshidwivedi3, joshidwivedi4}, a similar discussion was given on the occurrence of naked singularities in the gravitational collapse of an adiabatic perfect fluid in self-similar higher dimensional space–times, by Ghosh and Deshkar \cite{ghoshdeshkar}. It was shown that strong curvature naked singularities could occur if the weak energy condition holds.              \\
One can write a general spherically symmetric metric that admits a self-similarity, as
\begin{equation}
ds^2 = e^{\vartheta}dt^2 - e^{\chi}dr^2 - r^2 S^2 d \Omega^2,
\end{equation}
where $\vartheta$, $\chi$ and $S$ are functions of $z = \frac{t}{r}$. This space-time admits a homothetic killing vector $\xi^{a} = r\frac{\partial}{\partial r} + t\frac{\partial}{\partial t}$. For null geodesics one can write $K^{a}K_{a} = 0$, where $K^{a} = \frac{dx^{a}}{dk}$ are tangent vectors to null geodesics. Since $\xi^{a}$ is a homothetic killing vector, one can write 
\begin{equation}
{\mathcal{L}}_{\xi} g_{ab} = \xi_{a;b} + \xi_{b;a} = 2 g_{ab}
\end{equation}
where $\mathcal{L}$ denotes the Lie derivative. Then it is straightforward to prove that $\frac{d}{dk}(\xi^{a}K_a) = (\xi^{a}K_a)_{;b}K^b = 0$ (for mathematical details we refer to \cite{dwivedidixit}). Therefore one can write
\begin{equation}
\xi^{a}K_a = C,
\end{equation}
for null geodesics where $C$ is a constant.
From this algebraic equation and the null condition, one gets the following expressions for $K^t$ and $K^r$ as \cite{ghosh}
\begin{equation}\label{kt1c5}
K^t = \frac{C \left[z \pm e^{\chi} \Pi\right]}{r\left[ e^{ \chi} - e^{\vartheta} z^2 \right]},
\end{equation}

\begin{equation}\label{kr1c5}
K^r = \frac{C \left[1 \pm z e^{\vartheta} \Pi\right]}{r\left[ e^{\chi} - e^{\vartheta} z^2 \right]}, 
\end{equation}

where $\Pi =\sqrt{e^{-\chi - \vartheta}} > 0$. Radial null geodesics, by virtue of Eqs. (\ref{kt1c5}) and (\ref{kr1c5}), satisfy
\begin{equation}\label{de1c5}
\frac{dt}{dr} = \frac{z \pm e^{ \chi} \Pi}{1 \pm z e^{ \vartheta}\Pi}.
\end{equation}

The singularity that might have been there is at least locally naked if there exist radial null geodesics emerging from the singularity, and if no such geodesics exist it is a black hole. If the singularity is naked, then there exists a real and positive value of $z_{0}$ as a solution to the algebraic equation
\begin{equation}\label{lm1c5}
z_{0} = \lim_{t\rightarrow t_{s} \atop r\rightarrow 0} z = \lim_{t\rightarrow t_{s} \atop r\rightarrow 0} \frac{t}{r}=\lim_{t\rightarrow t_{s} \atop r\rightarrow 0} \frac{dt}{dr}.
\end{equation}

Waugh and Lake \cite{waughlake1, waughlake2, waughlake3} discussed shell-focussing naked singularities in self-similar spacetimes, considering a general radial homothetic killing trajectory. The lagrangian can be written in terms of $V(z) = (e^{\chi}-{z^2}e^{\vartheta})$, whose value determines the nature of the trajectory, for instance, if $V(z) = 0$, the trajectory is null, and for $V(z) > 0$ the trajectory is space-like. In the null case the trajectory can be shown to be geodesic. If $V(z) = 0$ has no real positive roots then the singularity is not naked. On the other hand the central shell focusing is at-least locally naked if $V(z_0) = 0$ admits one or more positive roots. The values of the roots give the tangents of the escaping geodesics near the singularity. \cite{ghoshdeshkar, joshidwivedi1, joshidwivedi2, joshidwivedi3, joshidwivedi4}.                \\

For a conformally flat space-time, the metric is defined as $ds^2=\frac{1}{{A(r,t)}^2}\Big[dt^2-dr^2-r^2d\Omega^2\Big]$, so we have $e^{\chi}=e^{\vartheta}= \frac{1}{A^2}$, $V(z)= \frac{1}{A^2}(1-z^2)$ and $\Pi = A^2$. So the equation for radial null geodesic simplifies considerably into $\frac{dt}{dr} = 1$ (where we have considered only the positive solution as we are looking for a radial outgoing ray). We consider the three cases  for which we have found exact collapsing solutions in section $(5)$.

\begin{enumerate}
\item{{\bf Case 1 : $m=-\frac{3}{5}$}  \\
Using the time evolution (\ref{exact-sol2c5}) and the condition that the central shell focusing is at-least locally naked for $V(z_0) = 0$ one finds 
\begin{equation}\label{zzc5}
z_0+(z_0^2-1)^\frac{1}{2} = e^{2C_1},
\end{equation}

for the conformally flat metric, from which it is straight-forward to write $z_0 = \frac{1}{2}\Big(e^{2C_1}+e^{-2C_1}\Big)$. The visibility of the singularity depends on the existence of positive roots to Eq. (\ref{zzc5}), therefore on the positivity of $\Big(e^{2C_1}+e^{-2C_1}\Big)$. Since $z_{0} = \lim_{t\rightarrow t_{s} \; r\rightarrow 0} \frac{dt}{dr}$, the equation for the null geodesic emerging from the singularity may be written as
\begin{equation}\label{ngeoc5}
t-t_{s}(0) = \frac{1}{2}\Big(e^{2C_1}+e^{-2C_1}\Big) r.
\end{equation}
For all values of $C_{1}$, $\Big(e^{2C_1}+e^{-2C_1}\Big)$ is always greater than zero. Therefore it is always possible to find a radially outward null geodesic emerging from the singularity, indicating a naked singularity.
}
\item{{\bf Case 2 : $m=3$} \\
Using (\ref{exact-sol3c5}) in $V(z_0) = 0$, one obtains 
\begin{equation}
\frac{C_{1}(1+z_0)}{(1-z_0)} = 1.
\end{equation}
As already discussed, for this particular model $C_1$ must be a negative number. Now, $z_0$ can be written as $z_0 = \frac{1-C_1}{1+C_1}$. Thus one can indeed have a naked singularity for $-1 < C_{1} < 0$. The singularity in this case is realised at a finite value of $z = \frac{t}{r}$, not at $r \rightarrow 0$ and therefore is not a central singularity. This kind of singularities are generally expected to be covered by a horizon \cite{joshiritudadhi}.
}
\item{{\bf Case 3 : $m=-\frac{3}{2}$}   \\
Equations (\ref{exact-sol4c5}) and the condition $V(z_0) = 0$ lead to
\begin{equation}
\frac{1}{4 C_{1} z_0} = 0,
\end{equation}
which can never have any finite positive solution for $z_0$. Therefore the central singularity in this particular case, defined at $\frac{1}{A(z)} \rightarrow 0$ at $r \rightarrow 0$, is always hidden from an exterior observer.
}
\end{enumerate}

However, $V(z_0) = 0$ admitting positive roots is necessary but not a sufficient condition to assert the existence of a naked singularity. The existence of a locally naked singularity also requires that the apparent horizon
curve must be an increasing function at the central singularity. If the tangent to the apparent horizon curve is negative, then even if there are positive roots of $V(z_0) = 0$, the outcome will still be a black hole. \\

For $m = -\frac{3}{2}$, one has a black hole solution as an end-state. However, for $m = -\frac{3}{5}$ and $m = 3$, i.e., for $V(\phi) = \frac{5\phi^{2/5}}{2}$ and $V(\phi) = \frac{\phi^4}{4}$, the spacetime sigularity can remain naked in principle. We can check the apparent horizon curve as calculated in section $5.5$. The apparent horizon curve for the case $m = -\frac{3}{5}$ is defined as $t_{ap} = \frac{r e^{-\gamma}}{2}(e^{2\gamma}-1) = r sinh{\gamma}$. Therefore the apparent horizon curve is an increasing function of $r$ at the central singularity if and only if $sinh{\gamma} > 0$. Therefore the visibility of the central singularity depends on the value of the parameter $\gamma$, defined as $\gamma = -2(C_1-\delta_0^{1/2})$ and $\delta_0$ is a constant of integration. Similarly, for $m = 3$, the apparent horizon curve is defined as $t_{ap} = r \Bigg(\frac{e^{\frac{1}{\Psi_0}}-C_1}{e^{\frac{1}{\Psi_0}}+C_1}\Bigg) = r \Gamma_0$ ($\Psi_0$ is a constant of integration) and the visibility of the ultimate spacetime singularity depends on the signature of $\Gamma_0$.    \\

Ii was shown by Hamid, Goswami and Maharaj \cite{hgmc4} that a spherically symmetric matter cloud evolving from a regular initial epoch, obeying physically reasonable energy conditions, is free of shell crossing singularities. A corollary of their work is that for a continued gravitational collapse of a spherically symmetric perfect fluid obeying the strong energy condition $\rho \geq 0$ and $(\rho + 3p) \geq 0$, the end state of the collapse is necessarily a black hole for a conformally flat spacetime. This idea was also confirmed by the massive scalar field analogue of the Oppenheimer-Snyder collapsing model in a conformally flat spacetime in the last chapter. This case, however, is quite different, as it deals with matter where the strong energy condition is not guaranteed. This can perhaps be related to the contribution of the scalar field to the energy momentum tensor and the dissipation part of the stress-energy tensor in the form of a heat flux. The choices of potential and the initial conditions in the form of the constants of integration can indeed conspire amongst themselves so that the energy conditions are violated which leads to the formation of a naked singularity. A violation of energy condition by massive scalar fields is quite usual and in fact forms the basis of its use as a dark energy. In the present case, as discussed, we find various possibilities, and a black hole is not at all the sole possibility, i.e. the central singularity may not always be covered, at least for some specific choices of the self-interacting potential, for instance, for $V(\phi)=\frac{5\phi^{2/5}}{2}$, as found out in the present work.

\section{Matching with an exterior Vaidya Solution}
Generally, in collapsing models, a spherically symmetric interior is matched with a suitable exterior solution; Vaidya metric or a Schwarzschild metric depending on the prevailing conditions \cite{joshi1}. This requires the continuity of both the metric and the extrinsic curvature on the boundary hypersurface. As the interior has a heat transport defined, the radiating Vaidya solution is chosen as a relevant exterior to be matched with the collapsing sphere. The interior metric is defined as
\begin{equation}
\label{metric2c5}
ds^2=\frac{1}{{A(r,t)}^2}\Bigg[dt^2-dr^2-r^2d\Omega^2\Bigg],
\end{equation}
and the Vaidya metric is given by
\begin{equation}\label{vaidyac5}
ds^2=\Bigg[1-\frac{2m(v)}{R}\Bigg]dv^2+2dvdR-R^2d\Omega^2.
\end{equation}
The quantity $m(v)$ represents the Newtonian mass of the gravitating body as measured by an observer at infinity. The metric (\ref{vaidyac5}) is the unique spherically symmetric solution of the Einstein field equations for radiation in the form of a null fluid.  The necessary conditions for the smooth matching of the interior spacetime to the exterior spacetime was presented by Santos \cite{santosc4} and also discussed in detail by Chan \cite{chanc4}, Maharaj and Govender \cite{maharajc5} in the context of a radiating gravitational collapse. Following their work, The relevant equations matching (\ref{metric2c5}) with (\ref{vaidyac5}) can be written as                         
\begin{equation}
\frac{r}{A(r,t)}_{\Sigma} = R,
\end{equation}

\begin{equation}
m(v)_{\Sigma} = \frac{r^3}{2A^3}\Bigg(\dot{A}^2-A'^2+\frac{A'A}{r}\Bigg),
\end{equation}
and
\begin{equation}\label{prqc5}
p_{r_{\small \Sigma}} = \frac{q}{A(r,t)},
\end{equation}                        
where $\Sigma$ is the boundary of the collapsing fluid.  \\
The relation between radial pressure and the heat flux as in equation (\ref{prqc5}) yields a nonlinear differential condition between the conformal factor and the scalar field to be satisfied on the boundary hypersurface $\Sigma$. In view of equations (\ref{radpressc5}) and (\ref{heatfc5}) the condition can be written as
\begin{equation}
\Bigg[\Bigg(2\frac{\ddot{A}}{A}-2\frac{\dot{A}^2}{A^2}+3\frac{A'^2}{A^2}-4\frac{A'}{Ar}+2\frac{\dot{A}'}{A}\Bigg)-\frac{1}{2}(\dot{\phi}+\phi')^2 +\frac{1}{A^2}\frac{\phi^{(m+1)}}{(m+1)}\Bigg]_{\Sigma} = 0.
\end{equation} 

It deserves mention that we cannot obtain any analytical expression from the matching conditions and thus their feedback to the interior solutions could not be discussed. Actually the condition of integrability of the scalar field equation has been utilized to find a solution for the metric but the evolution equation for the scalar field defined by (\ref{scalarfieldc5}) could not be explicitly integrated.

\section{Non-existence of shear}
For a general spherically symmetric metric
\begin{equation}
\label{metric-genc5}
ds^2 = S^2dt^2 - B^2dr^2 - R^2 d\Omega^2,
\end{equation}
where $S, B, R$ are functions of $r, t$, and for a comoving observer the velocity vector is defined as 
\begin{equation}
u^{\alpha}=S^{-1}\delta_0^{\alpha},
\end{equation} 
and
\begin{equation}
u^{\alpha}u_{\alpha}= 1.
\end{equation}
The shear tensor components can be easily calculated in this case. The acceleration $a_{\alpha}$ and the expansion $\Theta$ of the fluid are given by
\begin{equation}
a_{\alpha} = u_{\alpha ;\beta}u^{\beta}, \;\; \Theta = {u^{\alpha}}_{;\alpha}. \label{b}
\end{equation}
and the shear $\sigma_{\alpha\beta}$ by
\begin{equation}
\sigma_{\alpha\beta} = u_{(\alpha;\beta)}+a_{(\alpha}u_{\beta)} - \frac{1}{3} \Theta(g_{\alpha\beta} + u_{\alpha}u_{\beta}),
\label{c}
\end{equation}

From equations (\ref{b}) and (\ref{c}) one can calculate the non zero components of shear as
\begin{equation}
\sigma_{11}=\frac{2}{3}B^2\sigma, \;\;
\sigma_{22}=\frac{\sigma_{33}}{\sin^2\theta}=-\frac{1}{3}R^2\sigma,
 \label{shearcompo}
\end{equation}
and the scalar as
\begin{equation}
\sigma^{\alpha\beta}\sigma_{\alpha\beta}=\frac{2}{3}\sigma^2,
\label{shearscal}
\end{equation}
where the net shear scalar $\sigma$ can be found out as
\begin{equation}
\sigma=\frac{1}{S}\left(\frac{\dot{B}}{B}-\frac{\dot{R}}{R}\right).\label{shearc5}
\end{equation}

For the conformally flat metric chosen in the present work, $S = B = \frac{R}{r} = \frac{1}{A(r,t)}$. It is straightforward to see that $\sigma = 0$ in the present case.  In the present case, the existence of anisotropic pressure and dissipative processes might suggest the existence of shear in the spacetime, but the situation is actually shearfree. A shearfree motion is quite common in the discussion of gravitational collapse, and it is not unjustified either. But it should be noted that a shearfree condition, particularly in the presence of anisotropy of the pressure and dissipation, leads to instability. This result has been discussed in detail by Herrera, Prisco and Ospino \cite{herreraospinogrg}. 

\section{Combination of power-law potentials : $V(\phi)=\frac{\phi^2}{2}+\frac{\phi^{(m+1)}}{(m+1)}$}
The domain of the coordinate transformation restricted us not to choose a few values of $m$ ($m\neq{-3,-1,0,1}$). This excluded any chance of studying cases with a quadratic potential. In this section we assume a form for the potential such that $\frac{dV}{d\phi} = \phi + \phi^m$, i.e. $V(\phi)$ is a combination of two power-law terms, one of them being quadratic in $\phi$. Such interaction potentials also do carry additional motivations, for instance, considering a combination of $\phi^2$ and $\phi^4$, Lyth and Stewart \cite{lythstew} showed that it can lead to a cosmological history radically different from what is usually assumed to have occurred between the standard inflationary and nucleosynthesis epochs, which may solve the gravitino and Polonyi-moduli problems in a natural way. With this choice, the integrability criterion yields a non-linear second order differential equation for the conformal factor $B(z)$ as

\begin{equation} \label{evolu2c5}
\frac{B^{\circ\circ}}{B}-3\frac{(m+1)}{(m+3)}\Bigg(\frac{B^{\circ}}{B}\Bigg)^2-\frac{8(2m+3)}{6(m+3)}\frac{B^{\circ}z}{B(1-z^2)} -\frac{2(1+m)z^2-2(m+3)}{6(m+3)(1-z^2)^2}+\frac{(m+3)}{6B^2(1-z^2)}=0.               
\end{equation}                                   

It is very difficult to find an exact analytical form of $B(z)$ from (\ref{evolu2c5}) and thus any further analytical investigations regarding the collapsing geometry is quite restricted. However, we make use of a numerical method to analyse the evolution which of course depends heavily on choice of initial values and ranges of $z$, $B(z)$ and $\frac{dB}{dz}$. First the equation (\ref{evolu2c5}) is written in terms of $D(z) = \frac{1}{B(z)}$ as

\begin{equation} \label{devolu2c5}
\frac{D^{\circ\circ}}{D}= \Bigg[2-\frac{3(m+1)}{m+3}\Bigg]\Bigg(\frac{D^{\circ}}{D}\Bigg)^2+\frac{8(2m+3)}{6(m+3)}\frac{D^{\circ}z}{D(1-z^2)}-\frac{2(1+m)z^2-2(m+3)}{6(m+3)(1-z^2)^2}+\frac{(m+3)D^2}{6(1-z^2)}=0.
\end{equation}

Now equation (\ref{devolu2c5}) is solved numerically and studied graphically as $D(z)$ vs $z$ in the limit $z >>1$. Since we mean to study a collapsing model, $D^{\circ}(z)$ is always chosen as negative. The evolution of the collapsing sphere is sensitive to the factor $\mid\frac{D(z)}{D^{\circ}(z)}\mid$, at least for some choices of potential as will be shown in the subsequent analysis. We present the results obtained in three different categories.     

\begin{figure}[h]\label{RK41}
\begin{center}
\includegraphics[width=0.40\textwidth]{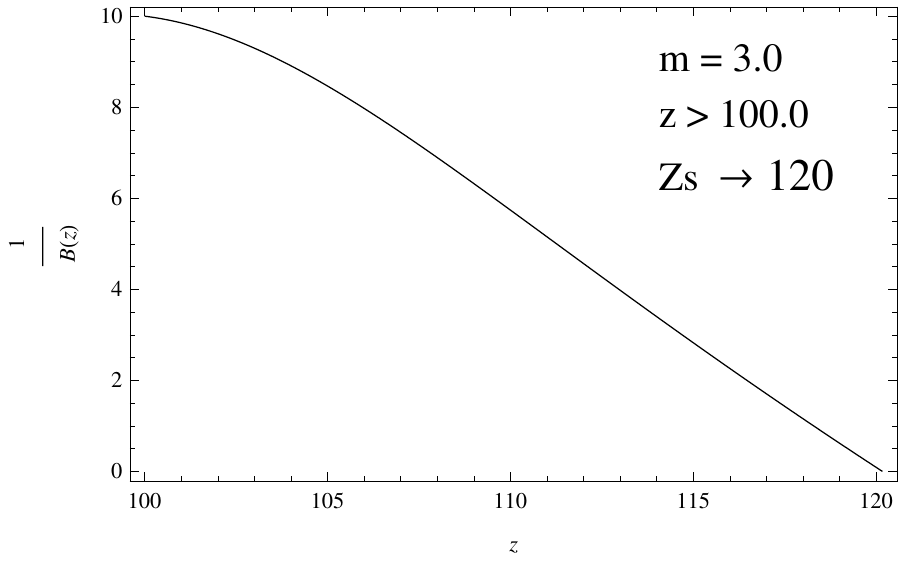}
\includegraphics[width=0.40\textwidth]{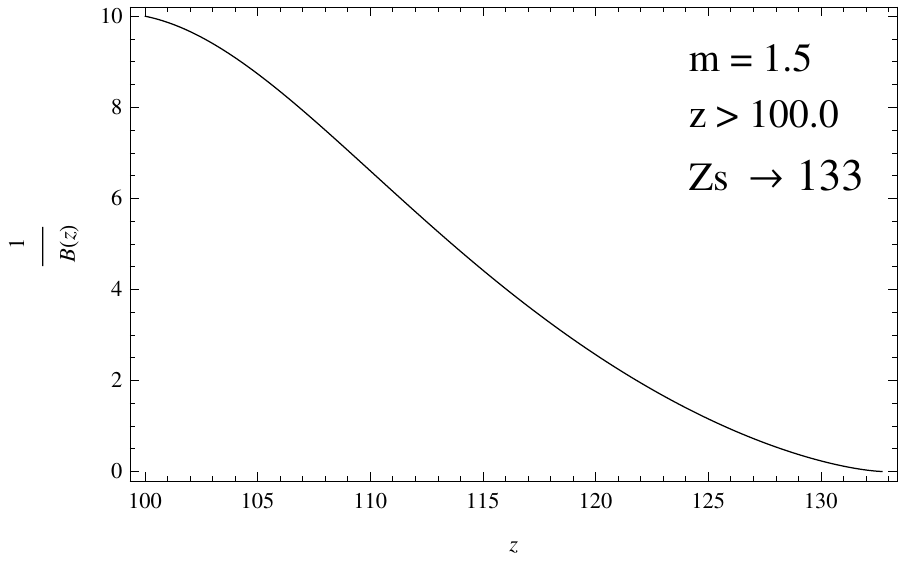}
\includegraphics[width=0.40\textwidth]{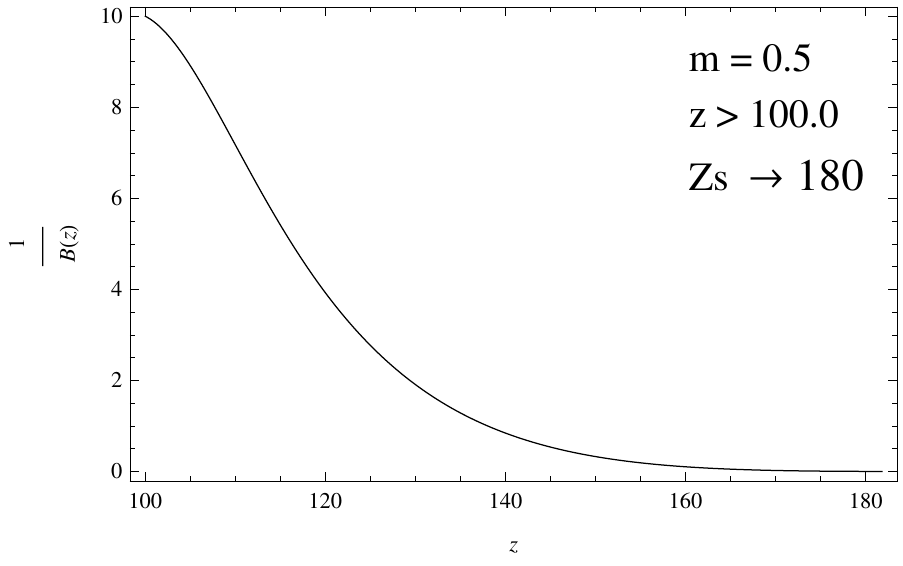}
\caption{Evolution of $\frac{1}{B(z)}$ with respect to $z$ for different choices of potential where $m > 0$: $m = 3.0, 1.5, 0.5$ i.e. $V(\phi) = \frac{\phi^2}{2} + 2\frac{\phi^{5/2}}{5}$, $V(\phi)= \frac{\phi^2}{2}+\frac{\phi^4}{4}$ and $V(\phi)= \frac{\phi^2}{2} + 2\frac{\phi^{3/2}}{3}$ respectively; for different initial conditions.}
\end{center}
\label{finfig}
\end{figure}

\begin{figure}[h]\label{RK42}
\begin{center}
\includegraphics[width=0.40\textwidth]{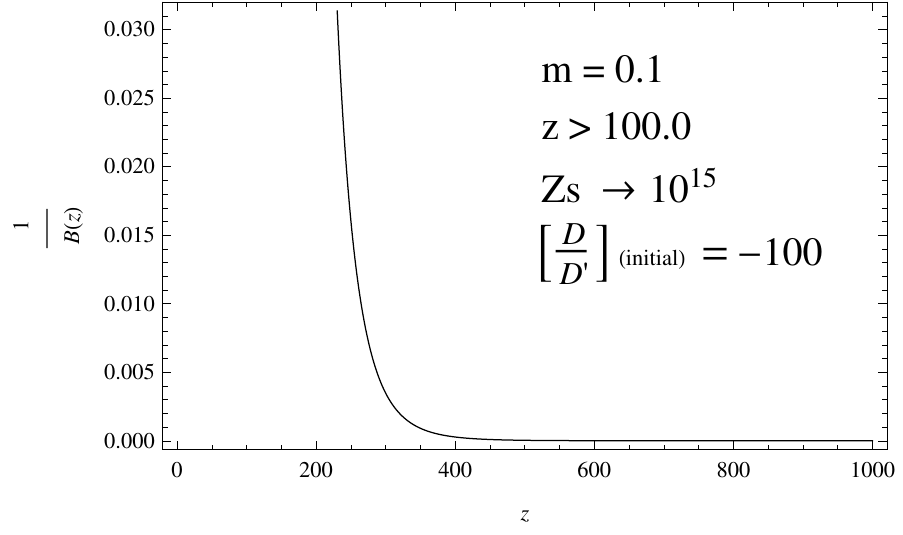}
\includegraphics[width=0.44\textwidth]{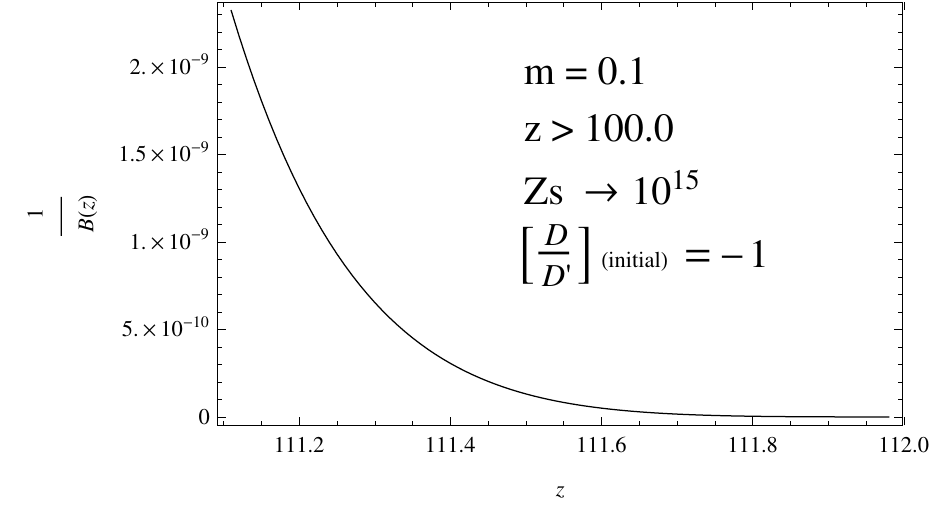}
\caption{Evolution of $\frac{1}{B(z)}$ with respect to $z$ for $m = \frac{1}{10}$ or $V(\phi)= \frac{\phi^2}{2} + 10\frac{\phi^{11/10}}{11}$. The case where $\mid D^{\circ}(z)\mid << D(z)$ is plotted on the LHS and the case where $\mid D^{\circ}(z)\mid \sim D(z)$ is on the RHS.}
\end{center}
\label{finfig}
\end{figure} 

\begin{enumerate}
\item{ For all $m > 0$, the spherical body collapses to a zero proper volume singularity at a finite future defined by $z = z_{s}$ as shown in figure $5.5$. For a large positive value of $m$, e.g $m \geq 3$, the singularity is reached more-or-less steadily. However, for choices of $m$ smaller in magnitude (for example, $m = 1.5, m = 0.5$), the system collapses at an increasingly larger value of $z_{s}$ and the evolution starts to look like an asymptotic curve. However, eventually the system attains a zero proper volume but at a very large but finite value of $z$, provided $m > 0$. This qualitative behaviour is independent of the initial condition defined by different choices of $\mid\frac{D(z)}{D^{\circ}(z)}\mid$.
}

\item{ When $m$ is a very small positive number (for example {\bf $m \sim \frac{1}{10}$}), the rate of collapse becomes increasingly sensitive to the initial condition $\mid\frac{D(z)}{D^{\circ}(z)}\mid$. The ultimate qualitative behaviour remains the same; the system reaches a zero proper volume at a finite but very large $z_{s}$. For $\frac{D(z)}{D^{\circ}(z)} \sim -100$, the system falls very rapidly as suggested by figure $5.6$, followed by a stage when the rate of collapse is slowed down significantly and the singularity is not reached until $z \sim 10^{15}$. The characteristic difference with the case when $\frac{D(z)}{D^{\circ}(z)} \sim -1$ is clear from figure $5.6$.
}

\item{ For $-3 < m < 0 (m \neq -3,-1)$, the collapsing system approaches a zero proper volume with respect to $z$ asymptotically, i.e., only when $z \rightarrow \infty$. In figure $5.7$, two examples are studied, for $m = -0.1$ and $m = -2.0$. The slope of the curves, i.e. the rate at which the spherical body approaches singularity may be different for different choices of $m$ in this domain, due to different signatures of the nonlinearities in equation (\ref{devolu2c5}). The collapse occurs more rapidly when $m = -0.1$ than $m = -2.0$. This nature is independent of the initial condition, i.e., whether $\mid D^{\circ}(z)\mid << D(z)$ or $\mid D^{\circ}(z)\mid \sim D(z)$.   
}

\item{ For $m < -3$, the evolution of the sphere is extremely sensitive to the choice of initial value of $D^{\circ}(z)$ with respect to $z$. An example is given in figure $5.8$, where we have studied the scenario for $m = -7.0$, or $V(\phi)= \frac{\phi^2}{2} - \frac{1}{6\phi^6}$ for different types of initial conditions. For some initial condition, when $\mid D^{\circ}(z)\mid \sim D(z)$, the sphere after an initial steady collapsing epoch, falls very sharply with respect to $z$; eventually hitting the zero proper volume singularity at a finite future. However, for $\mid D^{\circ}(z)\mid << D(z)$, the system can sometimes exhibit a somewhat oscillatory evolution as shown on the left panel of figure $5.8$. The oscillatory motion is followed by a very rapid, almost instanteneous drop to zero proper volume singularity.
}
\end{enumerate}

\begin{figure}[h]\label{RK43}
\begin{center}
\includegraphics[width=0.40\textwidth]{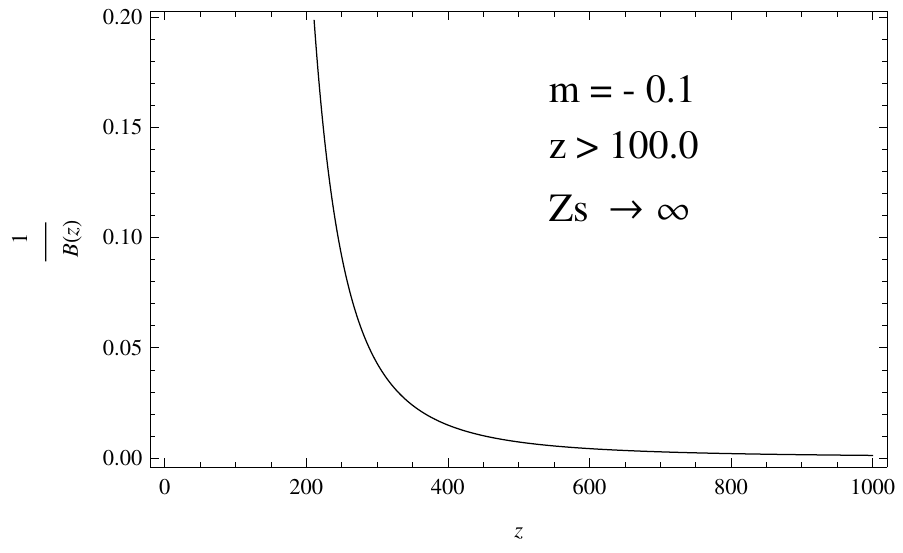}
\includegraphics[width=0.40\textwidth]{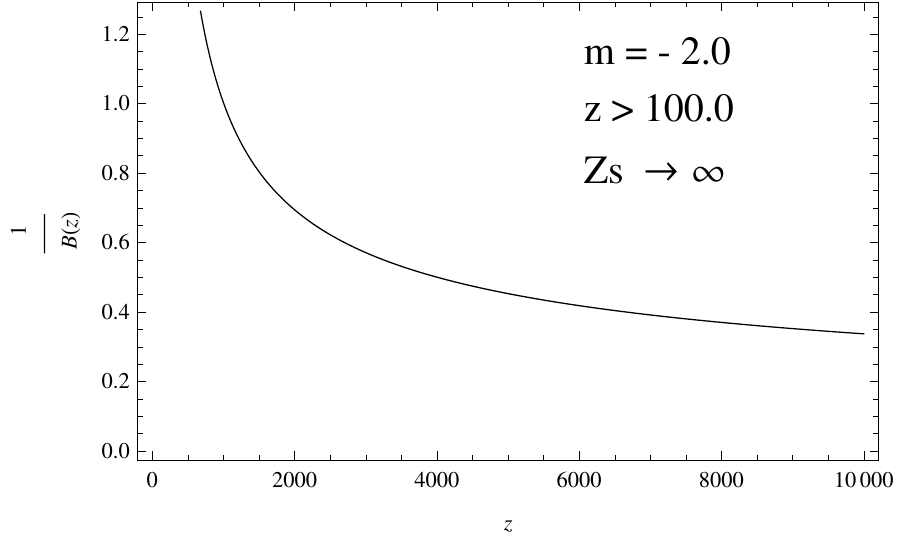}
\caption{Evolution of $\frac{1}{B(z)}$ with respect to $z$ for $m = -\frac{1}{10}$ or $V(\phi)= \frac{\phi^2}{2} + 10\frac{\phi^{9/10}}{9}$ and for $m = -2.0$ or $V(\phi)= \frac{\phi^2}{2} - \frac{1}{\phi}$ for different initial conditions defined by $\mid D^{\circ}(z)\mid << D(z)$ or $D^{\circ}(z) \sim D(z)$. The qualitative behaviour seems to be independent of the initial choice of parameters.}
\end{center}
\label{finfig}
\end{figure}

\begin{figure}[h]\label{RK44}
\begin{center}
\includegraphics[width=0.40\textwidth]{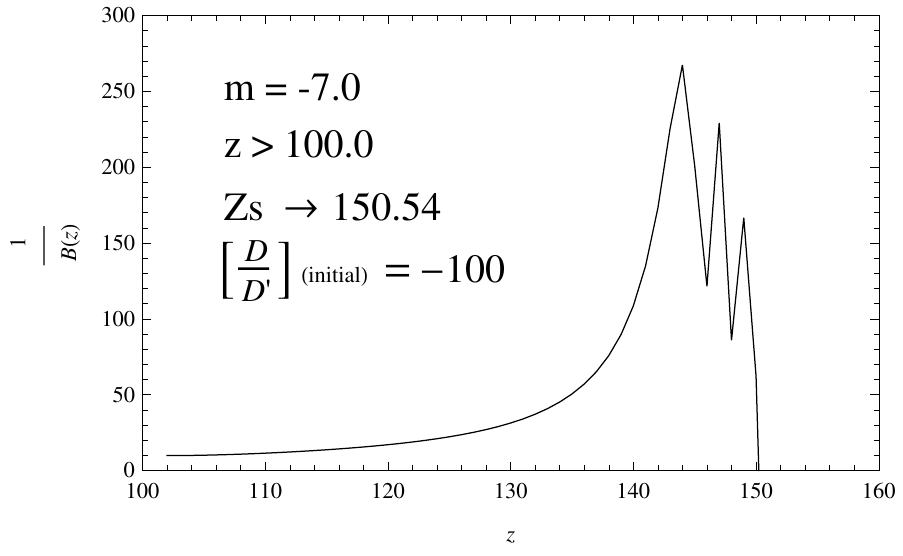}
\includegraphics[width=0.40\textwidth]{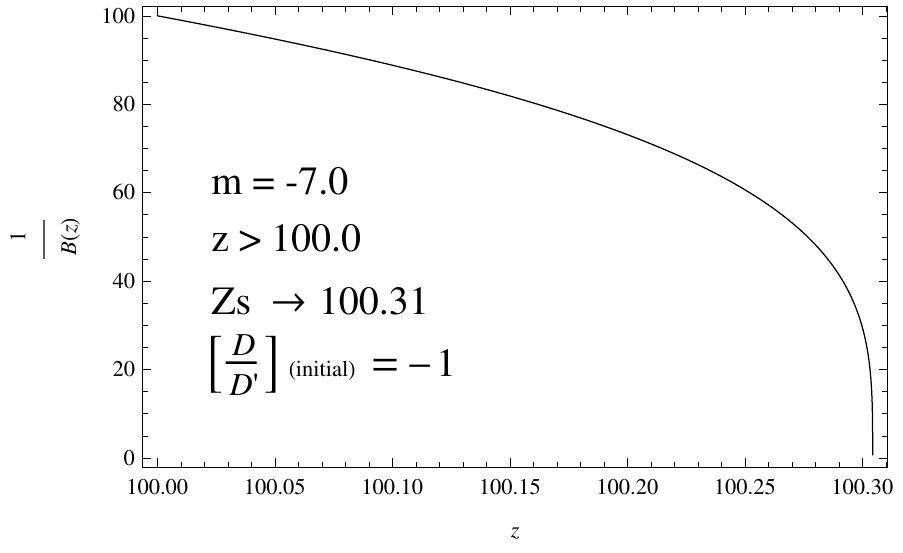}
\caption{Evolution of $B(z)$ with respect to $z$ for $m = -7.0$, or $V(\phi)= \frac{\phi^2}{2} - \frac{1}{6\phi^6}$ for different initial conditions.}
\end{center}
\label{finfig}
\end{figure}

\section{Discussion}
The collapse of a massive scalar field distribution is investigated in this chapter, with the basic method being a search for the integrability condition for the Klein-Gordon equation for the scalar field. The present chapter includes a lot more generalization in the matter content in the sense that it has an anisotropic pressure and a heat flux as well. For this generalization regarding the matter field, another symmetry requirement, that of a self similarity, has been imposed so as to overcome the difficulty in integrating the scalar field equation. Without actually solving the set of Einstein equations, a lot of informations regarding the metric can be obtained where the scalar field equation is integrable in principle.             \\
The scalar potential is assumed to be a power law one, which finds a lot of interest in cosmology, Except for a few powers, restricted by the domain of applicability of Eulers theorem, the method in fact applies for any other power law. We have picked up a few powers as examples, namely, ${\phi}^{4}$, ${\phi}^{\frac{2}{3}}$ and ${\phi}^{-\frac{1}{2}}$.      \\

Indeed, collapsing modes leading to a final singular state, are found with the help of numerical plots. Whether the singularity of a zero proper volume occurs at a finite future or the modes are for ever collapsing without practically hitting the singular state depends on the potential as well as the initial conditions. In order to facilitate further analysis like the formation of apparent horizon etc., we also do some analytical study with the help of some reasonable approximate solutions. The qualitative features of the analytical solutions are consistent with the numerical solutions. Some of the singularities are found to be not covered by an apparent horizon. This possibility of the collapse, ultimately leading to a naked singularity, can perhaps be related to the fact that the scalar field contribution can violate the energy conditions. Anisotropy of the fluid pressure and the heat flux (departure from the perfect fluid) can also contribute towards this existence of naked singularities.       \\

A ${\phi}^{2}$ potential is excluded from the purview of the investigation as this corresponds to $n=1$ which falls in the disallowed category. We have included an investigation of a potential of the form $V = a {\phi}^{2} + b {\phi}^n$. In this case we could obtain only some numerical plots for the scale factor. For various choices of $n$, collapsing modes are found, and all of them reaches the singularity sooner or later. \\

It also deserves mention that the particular potentials chosen as examples do have definite physical motivations but the condition for integrability stems from a mathematical interest, and no definite physics is associated with this. Still they represent interesting physical situations, and a study of scalar field collapse is possible.

\chapter[Scalar Field Collapse with an exponential potential]{Scalar Field Collapse with an exponential potential\footnote[1]{The results of this chapter are reported in General Relativity and Gravity (2017) {\bf 49}: 24}}

% **************************** Define Graphics Path **************************

\ifpdf
    \graphicspath{{Chapter6/Figs/Raster/}{Chapter6/Figs/PDF/}{Chapter6/Figs/}}
\else
    \graphicspath{{Chapter6/Figs/Vector/}{Chapter6/Figs/}}
\fi

Despite of exhaustive attempts over decades both in general relativity \cite{celer, magli2, harada, ritumala, maartens2, joshidwivedi1, joshidwivedi2, joshidwivedi3} and it's viable modifications \cite{ritu1, cembra, hwang, ghosh}, exact models for self-gravitating scalar field collapse remains rather limited. The aim of this chapter is to look at the collapse of a massive scalar field alongwith a very common functional form for the self-interaction potential, that of an exponential dependence upon the scalar field.   \\
Considerable importance and focus have been given to scalar field cosmology with an exponential interaction. A homogeneous isotropic cosmological model driven by a scalar field with an exponential potential was studied and a solution with power-law inflation was shown to be an attractor by Halliwell \cite{hall}. An exponential potential is predicted to be found in higher-order \cite{whit, barrowcot, wand} or higher-dimensional gravity theories \cite{greenbook}. The nature of the universe filled with a scalar field, with an exponential potential, has been studied for both homogeneous and inhomogeneous scalar fields. Inflationary Models with exponential Potentials were studied by Barrow and Burd \cite{barrowplb, burdbarrow}. Exact Bianchi type $I$ models for an exponential-potential scalar field were studied by Aguirregabiria and Chimento \cite{agui}. Exact general solution for cosmological models arising from the interaction of the gravitational field with two scalar fields in both flat FRW and the locally rotationally symmetric Bianchi $I$ spacetime filled with an exponential potential was given by Chimento \cite{chimento}. Rubano and Scudellaro \cite{rubano} presented general exact solutions for two classes of exponential potentials in scalar field models for quintessence. Using a particular type of exponential potential similar to that arising from the hyperbolic or flux compactification of higher-dimensional theories Neupane \cite{neupane} studied the four-dimensional flat and open FLRW cosmologies and gave both analytic and numerical solutions with exponential behavior of scale factors. Russo showed that the general solution of scalar field cosmology in $d$ dimensions with exponential potentials for flat FRW metric can be found in a straightforward way by introducing new variables which completely decouple the system \cite{russo}. Andrianov, Cannata and Kamenshchik studied in detail the general solution for a scalar field cosmology with an exponential potential. Piedipalumbo, Scudellaro, Esposito and Rubano studied dark energy models \cite{piedi} with a minimally-coupled scalar field and exponential potentials, admitting exact solutions for the cosmological equations and showed that for this class of potentials the Einstein field equations exhibit alternative Lagrangians, and are completely integrable and separable. These perhaps indicate the significance of investigations regarding a scalar field with an exponential potential in cosmology.         \\

However, it remains a challenge to write the highly non-linear Einstein field equations in an integrable form for different set up. The exact time-evolution here is studied analytically for a homogeneous scalar field in a flat FLRW spacetime. In a recent approach by Harko, Lobo and Mak \cite{harko1}, a new formalism for the analysis of scalar fields in flat isotropic and homogeneous cosmological models was presented. The basic evolution equations of the model were reduced to a first order non-linear differential equation. The transformation introduced therein, is used to simplify the evolution equations enough so that a collapsing model can be studied.       

\section{Mathematical formulaion}
The metric for a spherically symmetric spacetime with a spatial homogeneity and isotropy can be written as 
\begin{equation}
\label{metricltbc6}
ds^2=dt^2-T(t)^2(dr^2+r^2d\Omega^2).
\end{equation}

The time evolution is governed solely by the function $T(t)$. This indeed is a very simple case, but this would lead to some tractable solutions so that the possibility of collapse can be investigated. When a scalar field $\phi$ is minimally coupled to gravity, the relevant action is given by 
\begin{equation}\label{actionc6}
{\cal A} = \int{\sqrt{-g}d^4x[R+\frac{1}{2}\partial^{\mu}\phi \partial_{\mu}\phi - V(\phi) + L_{m}]},
\end{equation}

where $V(\phi)$ is the scalar potential and $L_{m}$ is the Lagrangian density for the fluid distribution. In this particular case, we assume that there is no fluid contribution in the action, thus, $L_{m}=0$. From this action, the contribution to the energy-momentum tensor from the scalar field $\phi$ can be  written as
\begin{equation}\label{minimallyscalarc6}
T^\phi_{\mu\nu}=\partial_\mu\phi\partial_\nu\phi-g_{\mu\nu}\Bigg[\frac{1}{2}g^{\alpha\beta}\partial_\alpha\phi\partial_\beta\phi-V(\phi)\Bigg]. 
\end{equation}

We assume the scalar field to be spatially homogeneous, i.e., $\phi=\phi(t)$. With this assumption, the Einstein field equations for the metric (\ref{metricltbc6}) can be written as (in units where $8 \pi G = 1$)
\begin{equation} \label{fe1ltbc6}
3\Bigg(\frac{\dot{T}}{T}\Bigg)^{2} = \frac{\dot{\phi}^{2}}{2}+V\left( \phi \right),
\end{equation}  
\begin{equation} \label{fe2ltbc6}
-2\frac{\ddot{T}}{T}-\Bigg(\frac{\dot{T}}{T}\Bigg)^{2} = \frac{\dot{\phi}^{2}}{2}-V\left( \phi \right).
\end{equation}
The evolution equation for the scalar field is given by
\begin{equation} \label{philtbc6}
\ddot{\phi}+3\frac{\dot{T}}{T}\dot{\phi}+\frac{dV(\phi)}{d\phi} = 0.  
\end{equation}                               

The overhead dot denotes the derivative with respect to the time-coordinate $t$. We will restrict our study to collapsing models, which satisfy the condition that radius of the two-sphere is a monotonically decreasing function of time. Therefore we discuss only those cases where $\frac{\dot{T}}{T} < 0$ is satisfied.        \\

By substituting $\frac{\dot{T}}{T}$ from Eq. (\ref{fe1ltbc6}) into Eq. (\ref{philtbc6}), one obtains the basic equation describing the scalar field evolution as
\begin{equation} \label{philtb1c6}
\ddot{\phi}-\sqrt{3}\dot{\phi}\sqrt{\frac{\dot{\phi}^{2}}{2}+V\left(\phi \right)}+\frac{dV}{d\phi }=0.  
\end{equation}

Defining a new function $f(\phi)$ so that $\dot{\phi}^{2}=f(\phi)$, and changing the independent variable from $t$ to $\phi$, eq. (\ref{philtb1c6}) can be written as
\begin{equation} \label{philtb2c6}
\frac{1}{2}\frac{df(\phi)}{d\phi}-\sqrt{3}\sqrt{\frac{f(\phi)}{2}+V(\phi)}\sqrt{f(\phi)}+\frac{dV}{d\phi}=0,  
\end{equation}
which may be reorganized into the following form
\begin{equation} \label{philtb3c6}
\frac{\frac{1}{2}\frac{df(\phi)}{d\phi}+\frac{dV}{d\phi}}{2\sqrt{\frac{f(\phi)}{2}+V(\phi)}}-\frac{\sqrt{3}}{2}\sqrt{f(\phi)}=0.  
\end{equation}                       

The transformations introduced by Harko et. al., (for the step by step systematic description, we refer to \cite{harko1}), defined by $F(\phi)=\sqrt{\frac{f(\phi)}{2}+V(\phi)}$, $F(\phi)=u(\phi)\sqrt{V(\phi)}$ and $u(\phi)=\cosh G(\phi)$ is considered here; such that one can simplify (\ref{philtb3c6}) enough to arrive at the basic equation governing the dynamics of the scalar field collapse as
\begin{equation} \label{finltbc6}
\frac{dG}{d\phi}+\frac{1}{2V}\frac{dV}{d\phi}\coth G-\sqrt{\frac{3}{2}}=0.
\end{equation}

For a flat FRW spacetime, it can be shown that the functions $f(\phi)$ and $F(\phi)$ are related to the Hubble function and its time derivative \cite{harko1} using the field equations (\ref{fe1ltbc6}) and (\ref{fe2ltbc6}). Another similar approach was considered by Salopek and Bond \cite{salop}, where the Hubble function was assumed to be a function of the scalar field $\phi$. The time evolution of the two-sphere governed by $T(t)$ can be written in terms of the scalar field (with $\frac{\dot{T(t)}}{T(t)}<0$, which corresponds to a collapsing mode), by the equation
\begin{equation} \label{Ttc6}
\frac{1}{T(\phi)}\frac{dT(\phi)}{d\phi}=-\frac{1}{\sqrt{6}}\coth{G(\phi)}.
\end{equation}                        

For a large number of choices of the functional form of the self-interacting potential, the first order evolution equation, Eq. (\ref{finltbc6}) can be solved exactly or parametrically and a collapsing model can be obtained, provided they satisfy a proper junction conditions. Here we present only a special case, a simple example of an exponential potential, such that a complete collapsing scenario can be investigated. Relevant possible choices of the potential and their dynamical scenario in case of scalar field cosmologies are discussed in \cite{harko1}.

\section{Exact solution}
If $V^{\prime}/V=\sqrt{6}\alpha_{0}$ where $\alpha_{0}$ is a constant, the scalar field self-interaction potential is of the exponential form given as
\begin{equation} \label{potentialc6}
V=V_{0}e^{\left(\sqrt{6}\alpha_{0}\phi\right)}.
\end{equation}

Taking into account Eq. (\ref{potentialc6}), Eq. (\ref{finltbc6}) takes the form
\begin{equation} \label{evoexpc6}
\frac{dG}{d\phi}+\sqrt{\frac{3}{2}}\left(\alpha_{0}\coth G-1\right)=0.
\end{equation}

A particular solution of the field equations corresponds to the case $G(\phi)=G_{0}=\mathrm{constant}$. In this case Eq.(\ref{evoexpc6}) is identically satisfied, with $G_{0}$ given by
\begin{equation}
G_{0}=\mathrm{arccoth}\left(\frac{1}{\alpha_{0}}\right).
\end{equation}

From Eq.(\ref{Ttc6}) it follows that the scale factor can be obtained as a function of the scalar field as
\begin{equation} \label{evoscalec6}
T(\phi)=T_{0}e^{-{\phi}/{\sqrt{6}\alpha_{0}}},  
\end{equation}
where $T_{0}$ is a constant of integration. The time variation of the scalar field is determined from the relation $G(\phi)=\mathrm{arccosh} \sqrt{1+\frac{\dot{\phi}^{2}}{2V(\phi)}}$ as
\begin{equation} \label{phidotc6}
\dot{\phi}=\pm \sqrt{2V_{0}}\Bigg(\frac{\alpha_{0}}{({\alpha_{0}}^2+1)^{\frac{1}{2}}}\Bigg) e^{\sqrt{3/2}\alpha_{0}\phi}.
\end{equation}
It is straightforward to integrate (\ref{phidotc6}) to write the exact evolution of scalar field with respect to time as
\begin{equation}
e^{-\sqrt{3/2}\alpha_{0}\phi}=\mp \sqrt{3V_{0}}\Bigg(\frac{\alpha_{0}^2}{({\alpha_{0}}^2+1)^{\frac{1}{2}}}\Bigg)(t-t_{0}),
\end{equation}
where $t_{0}$ is an arbitrary constant of integration.
With the help of Eq. (\ref{evoscalec6}), one can obtain the exact time evolution of the collapsing scalar field in the form
\begin{equation}
T(t)=T_{0}\Bigg[\mp \sqrt{3V_{0}}\Bigg(\frac{\alpha_{0}^2}{({\alpha_{0}}^2+1)^{\frac{1}{2}}}\Bigg)(t-t_0)\Bigg]^{\frac{1}{3\alpha_{0}^{2}}}.
\end{equation}                
Since for a collapsing scenario, $\dot{T}(t) < 0$, it is easy to check that one must choose the negative signature inside the parenthesis to write the time-evolution, thus giving the time evolution as
\begin{equation}\label{exactevoc6}
T(t)=T_{0}[N_{0}(t_0-t)]^{\frac{1}{3{\alpha_{0}}^2}}.
\end{equation}           

We have written $N_{0}=\sqrt{3V_{0}}\Big(\frac{\alpha_{0}^2}{({\alpha_{0}}^2+1)^{\frac{1}{2}}}\Big)$, which must always be greater than zero. It is straightforward to note that radius of the two-sphere $r T(t)$ goes to zero when $t \rightarrow t_{0}$, giving rise to a finite time zero proper volume singularity.

\begin{figure}[h]\label{fig:ltb}
\begin{center}
\includegraphics[width=0.5\textwidth]{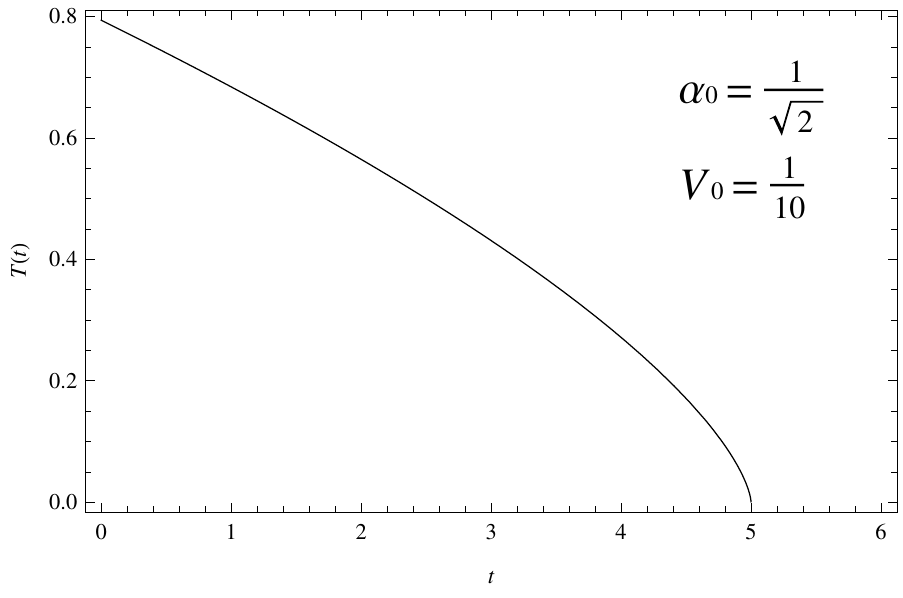}
\includegraphics[width=0.5\textwidth]{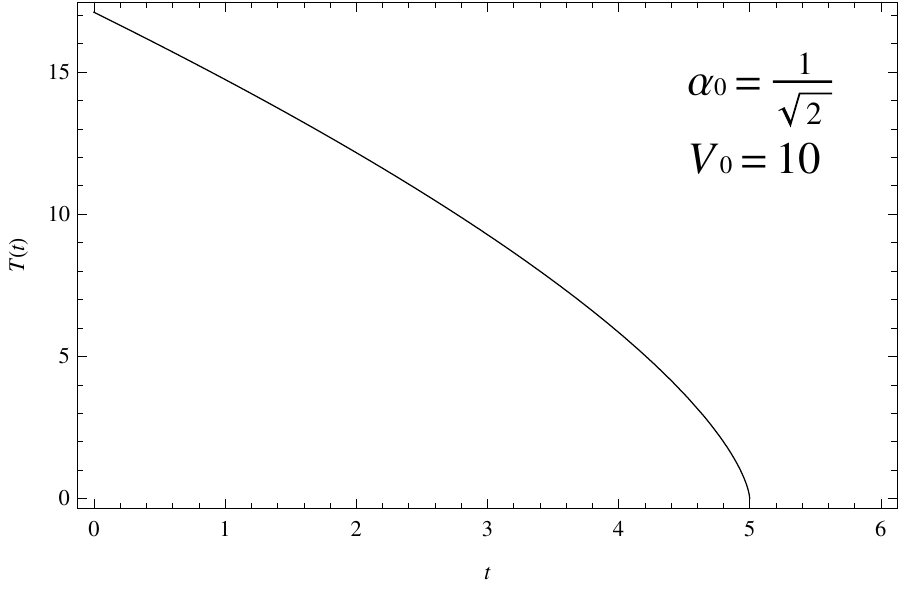}
\caption{\bf Time-evolution of the function $T(t)$ for different values of $V_{0}$}
\end{center}
\end{figure}

Since, $N_{0}$ must always be real and positive, it is obvious that one must always choose positive values of $V_{0}$. We plot the evolution of $T(t)$ with respect to $t$ in Figure $6.1$ for different values of $V_{0}$. Without any loss of generality, one can assume $T_{0}=1$. The choice of $\alpha_{0}$ is restricted by the constraints developed from the boundary matching discussed in the next section. The spherical body collapses with time almost uniformly until it reaches $t=t_{0}=t_{s}$, where it hurries towards a zero proper volume singularity. This behavior is not affected by different choices of $V_{0}$ as long as it is positive. We have presented here two specific examples with $V_{0}=10$ and $V_{0}=\frac{1}{10}$ in Figure $6.1$.        \\
One must look into the behavior of Ricci and Kretschmann curvature scalars to judge the nature of spacetime singularity. The curvature scalars can be written from the metric (\ref{metricltbc6}) as

\begin{equation}\label{riccic6}
R=-6\Bigg[\frac{\ddot{T}(t)}{T(t)}+\frac{{\dot{T}(t)}^2}{T(t)^2}\Bigg],
\end{equation}
and
\begin{equation}\label{kretc6}
K=6\Bigg[\frac{{\ddot{T}(t)}^2}{T(t)^2}+\frac{{\dot{T}(t)}^4}{T(t)^4}\Bigg].
\end{equation}

Using equation (\ref{exactevoc6}), the expressions for the scalars can be simplified into
\begin{equation}
R=\frac{2}{{\alpha_{0}}^2}\Bigg(1-\frac{2}{3{\alpha_{0}}^2}\Bigg)\frac{1}{(t-t_0)^2},
\end{equation}
and
\begin{equation}
K=\frac{2}{3{\alpha_{0}}^4}\Bigg[\frac{1}{9{\alpha_{0}}^4}+\Bigg(\frac{1}{3{\alpha_{0}}^2}-1\Bigg)^2\Bigg]\frac{1}{(t-t_0)^4}.
\end{equation}

One can clearly see, for all values of $\alpha_{0}$, both Ricci and Kretschmann scalar diverges to infinity when $t \rightarrow t_{0}$. For ${\alpha_{0}}^2 = \frac{2}{3}$, the Ricci scalar vanishes but the Kretschmann scalar diverges anyway. Therefore the collapsing sphere discussed here ends up in a curvature singularity.                         \\

Singularities formed in collapse can be shell focussing or shell crossing in nature. For a spherically symmetric collapse the shell focusing singularity occurs at $g_{\theta\theta} = 0$. It is evident from (\ref{metricltbc6}) and (\ref{exactevoc6}) that $g_{\theta\theta} = rT(t) \rightarrow 0$ when $t \rightarrow t_{0}$. Thus the curvature singularity indeed is a shell-focussing one.

\section{Matching of the interior space-time with an exterior geometry}
For a complete and consistent analysis of gravitational collapse, proper junction conditions are to be examined carefully which allow a smooth matching of an exterior geometry with the collapsing interior. Any astrophysical object is immersed in vacuum or almost vacuum spacetime, and hence the exterior spacetime around a spherically symmetric star is well described by the Schwarzschild geometry. Moreover it was extensively shown by Goncalves and Moss \cite{goncalves} that any sufficiently massive collapsing scalar field can be formally treated as collapsing dust ball. From the continuity of the first and second differential forms, the matching of the sphere to a Schwarzschild spacetime on the boundary surface, $\Sigma$, is extensively worked out in literature as already discussed in the earlier chapters (see also \cite{santosc4, chanc4, kolla, maharajc5}).         \\

However, conceptually this may lead to an inconsistency since the treatment allowed for a dust collapse may not be valid for a scalar field in general. For instance, since Schwarzschild has zero scalar field, such a matching would lead to a discontinuity in the scalar field, which means a delta function in the gradient of
the scalar field. As a consequence, there will appear square of a delta function in the stress-energy, which is definitely an inconsistency. In modified theories of gravity an alternative scenario is discussed sometimes where the exterior is non-static. However, the solar system experiments constrain heavily such a scenario. Another possible way to avoid such a scenario can perhaps be allowing jump in the curvature terms in the field equations. Such cases must result in surface stress energy terms, which in collapsing models must have observational signatures and can be established via experimental evidences \cite{ritu1}.        \\
Following the arguements of Goswami and Joshi \cite{massiveritu1, massiveritu11}, Ganguly and Banerjee \cite{massiveganguly}, we match the spherical ball of collapsing scalar field to a Vaidya exterior across a boundary hypersurface defined by $\Sigma$. The metric just inside $\Sigma$ is,
\begin{equation}\label{interiorc6}
d{s_-}^2=dt^2-T(t)^2dr^2-r^2 T(t)^2d{\Omega}^2,
\end{equation}
and the metric in the exterior of $\Sigma$ is given by
\begin{equation}\label{exteriorc6}
d{s_+}^2=(1-\frac{2M(r_v,v)}{r_v})dv^2+2dvdr_v-{r_v}^2d{\Omega}^2.
\end{equation}
Matching the first fundamental form on the hypersurface we get
\begin{equation}\label{cond1c6}
\Bigg({\frac{dv}{dt}}\Bigg)_{\Sigma} = \frac{1}{\sqrt{1-\frac{2M(r_v,v)}{r_v}+\frac{2dr_v}{dv}}}
\end{equation}
and
\begin{equation}\label{cond2c6}
r_{v_{\small \Sigma}} = r T(t) \frac{\small {\Sigma}}{-} rT_{0}[N_{0}(t_0-t)]^{\frac{1}{3{\alpha_{0}}^2}}.
\end{equation}
Matching the second fundamental form yields,
\begin{equation}\label{cond3c6}
rT(t)_{\Sigma} = r_v\left(\frac{1-\frac{2M(r_v,v)}{r_v}+\frac{dr_v}{dv}}{\sqrt{1-\frac{2M(r_v,v)}{r_v}+\frac{2dr_v}{dv}}}\right)
\end{equation}
Using equations (\ref{cond1c6}), (\ref{cond2c6}) and (\ref{cond3c6}) one can write
\begin{equation}\label{dvdt2c6}
\frac{dv}{dt}_{\Sigma} = \frac{T^2-\frac{r}{3}}{T^2-\frac{2M}{r}T}.
\end{equation}
From equation (\ref{cond3c6}) one obtains
\begin{equation}\label{Mc6}
M_{\Sigma} = \frac{r^{-1}T^{-1}+\frac{r}{9}T^{-5}+\sqrt{\frac{1}{r^{2}}T^{-2}+\frac{r^2}{81}T^{-10}-\frac{2}{9}T^{-6}}}{\frac{4}{r^2}T^{-2}}.
\end{equation}
Matching the second fundamental form we can also write the derivative of $M(v,r_v)$ as
\begin{equation}\label{dMc6}
M{(r_v,v)}_{,r_{v_{\small \Sigma}}} = \frac{M}{rT}-\frac{2r^2}{9T^{4}}.
\end{equation}
Equations (\ref{cond2c6}), (\ref{dvdt2c6}), (\ref{Mc6}) and (\ref{dMc6}) completely specify the matching conditions at the boundary of the collapsing scalar field.

\section{Visibility of singularity}
Whether the ultimate spacetime singularity is visible to an exterior observer depends on the formation of an apparent horizon. Such a surface is defined as
\begin{equation}
g^{\mu\nu}Y_{,\mu}Y_{,\nu}=0,
\end{equation}    
where $Y(r,t)$ is the proper radius of the collapsing sphere. Using (\ref{metricltbc6}) and (\ref{exactevoc6}), one can express this equation as
\begin{equation}
N_{0}(t_{0}-t_{ap})=\Bigg(\frac{9{{\alpha_{0}}^2}\delta}{{T_{0}}^2{N_{0}}^2}\Bigg)^{\frac{3{\alpha_{0}}^2}{2-6{\alpha_{0}}^2}}.
\end{equation}
Here, $\delta$ is a constant of separation. Thus the time of formation of apparent horizon can be expressed as
\begin{equation}\label{apparentc6}
t_{ap}=t_{0}-\frac{1}{N_{0}}\Bigg(\frac{9{\alpha_{0}}^2\delta}{{T_{0}}^2{N_{0}}^2}\Bigg)^{\frac{3{\alpha_{0}}^2}{2-6{\alpha_{0}}^2}}.
\end{equation}
Both $\alpha_{0}$ and $N_{0}$ depend on the choice of the potential given their definitions $V=V_{0}e^{\left(\sqrt{6}\alpha_{0}\phi\right)}$ and $N_{0}=\sqrt{3V_{0}}\Big(\frac{\alpha_{0}^2}{({\alpha_{0}}^2+1)^{\frac{1}{2}}}\Big)$. The arbitrary constants $T_{0}$ and $\delta$ can be estimated from proper matching of the collapsing scalar field with an appropriate exterior geometry and on initial conditions. From (\ref{exactevoc6}), one can write at $t=t_{ap}$,
\begin{equation}\label{tdotc6}
\frac{\dot{T}}{T}=\frac{1}{3{\alpha_0}^2(t_{ap}-t_0)} < 0.
\end{equation}
Since we are dealing with a geometry where the sphere must always decrease in volume with respect to time, this expression is consistent if and only if $t_{0} > t_{ap}$. This means that the apparent horizon, if any, must always form before the formation of singularity.        \\

In the present case the time of formation of singularity $t_{0}$ is independent of $r$ and therefore is not a central singularity. The entire collapsing body reaches the singularity simulteneously at $t = t_{0}$. This kind of singularity is always expected to be covered by the formation of an apparent horizon as already discussed by Joshi, Goswami and Dadhich \cite{joshiritudadhi}. The result that apparent horizon must always form before the formation of singularity, is therefore a consistent result.

\section{Discussion}
There are only very limited amount of work on massive scalar field collapse in the literature. We have discussed a spherically symmetric collapse of a massive scalar field where an exponential potential describes the self-interaction. The scalar field is chosen to be spatially homogeneous, representing the scalar field analogue of Oppenheimer-Snyder collapse.           

A systematic discussion is presented, assuming the matter contribution $L_{m}$ to be zero. The Klein-Gordon equation describing the dynamics of the scalar field is simplified considerably into a first order non-linear differential equation. A simple exact solution predicts the end state of the collapse to be a finite time shell-focussing singularity. The evolution of the system is found to be independent of different parameters defining the self-interacting potential $V=V_{0}e^{\left(\sqrt{6}\alpha_{0}\phi\right)}$. A proper boundary matching with an exterior Vaidya geometry is also discussed. The visibility of the end-state is sensitive to different choices of initial conditions. The collapse is simulteneous and results in a singularity which acts as a sink for all the curves of the collapsing congruence, and the volume elements shrink to zero along all the collapsing trajectories. An apparent horizon is always expected to form before the formation of zero proper volume singularity, which therefore remains hidden forever.      

The reduction of the klein-gordon equation into a first order differential equation can be fairly useful in many cosmological contexts. Exact and approximate solutions of massive scalar field collapse is an interesting subject and using this formalism the study can be extended, for example for cosine hyperbolic potentials.

\chapter{Conclusions}

Studies of exact solutions and their properties, symmetries, local geometries and singularities play a non-trivial role in general rel­ativity \cite{maccul}. Finding non-trivial solutions to the Einstein equations requires some reduction of the problem, which usually is done by exploiting symmetries or other properties. As a result, there is no single method preferred for finding solutions to the Einstein equations. In this thesis, we have used different analytical methods to study exact collapsing solutions of the highly non-linear field equations gravity. \\

Viable $f(R)$ models are quite successful in providing a geometrical origin of the dark energy sector of
the universe. However, they pose considerable problems in some other significant sectors, for instance, it is extremely difficult to find exact solutions of static or dynamic stellar objects as the field equations are fourth order differential equations in the metric components. Moreover, for any stellar object, the spacetime of the interior of the star has to be matched smoothly with the exterior spacetime. This is not a straight-forward task since the fourth order field equations generate extra matching conditions between two spacetimes beyond the usual Israel-Darmois conditions in General Relativity. The extra conditions arising from the matching of the Ricci scalar and it’s normal derivative across the matching surface, heavily
constrict the set of useful astrophysical solutions. We have addressed some inhomogeneous collapsing models in $f(R)$ gravity such that the collapsing stellar matter obeys all the energy conditions and at the comoving boundary of the collapsing star, the interior spacetime matches smoothly with an exterior spacetime. Under different symmetry assumptions the highly non-linear field equations are written in integrable forms to extract exact collapsing solutions. The presence and importance of spatial inhomogeneity is duely noted and discussed wherever relevant.   \\

We have also studied some models of gravitational collapse of a self-interacting scalar field minimally coupled to gravity. No equation of state for the fluid description is assumed at the outset for the models described.
\begin{itemize}
\item{ The field equations for a conformally flat spatially homogeneous system are investigated by exploiting the integrability of the scalar field evolution equation and the final fate of the collapse is discussed along with a smooth boundary matching of the collapsing interior spacetime with a proper exterior. The time evolution can be studied in a rigorus manner and it confirms a recent interesting finding that (Hamid, Goswami and Maharaj \cite{hgmc4}) for a continued gravitational collapse of a spherically symmetric perfect fluid obeying the strong energy condition ($\rho \geq 0$ and $(\rho + 3p) \geq 0$), the end state of the collapse is necessarily a black hole, for a conformally flat spacetime. 
}
\item{ A collapsing model admitting a Homothetic Killing vector implying a self-similarity in the spacetime is thoroughly addressed. The end-state of the collapse is investigated by analyzing radial null geodesics emanating from the spacetime singularity. This work is a lot more generalised and it deals with matter where the strong energy condition is not always guaranteed, which perhaps can be related to the contribution of the scalar field to the energy momentum tensor and the dissipation part of the stress-energy tensor in the form of a heat flux. It is proved that the choices of potential and the initial conditions in the form of the constants of integration can indeed conspire amongst themselves so that the energy conditions are violated which leads to the formation of a Naked Singularity. A violation of energy condition by massive scalar fields is quite usual and in fact forms the basis of its use as a dark energy. In the present case, we have found different possibilities, and a black hole is indeed, not the sole possibility, i.e. the central singularity may not always be covered, at least for some specific choices of the self-interacting potential.
}
\item{ A scalar field collapse for a Lemaitre-Tolman-Bondi type spacetime without matter contribution is studied by transforming the system of equations into a first order integrable ordinary differential equation.
}
\end{itemize}

% ********************************** Back Matter *******************************
% Backmatter should be commented out, if you are using appendices after References
%\backmatter

% ********************************** Bibliography ******************************
\begin{spacing}{0.9}

%\bibliographystyle{myapsrev}
%\bibliography{dissertation}
%\addcontentsline{toc}{chapter}{Bibliography}
% To use the conventional natbib style referencing
% Bibliography style previews: http://nodonn.tipido.net/bibstyle.php
% Reference styles: http://sites.stat.psu.edu/~surajit/present/bib.htm

\bibliographystyle{apsrev}
\cleardoublepage
\bibliography{thesis} % Path to your References.bib file

\begin{thebibliography}{200}

\bibitem{r1} S. Perlmutter et al, Bull. Am. Astron.Soc., {\bf 29}, 1351 (1997).\\
             S. Perlmutter et al, Astrophys. J., {\bf 517}, 565 (1999).\\
             J. L. Tonry et al, Astrophys. J., {\bf 594}, 1 (2003).\\
             S. Bridle, O. Lahav, J.P. Ostriker and P.J. Steihardt, Science, {\bf 299}. 1532 (2003).\\
             G. Hinshaw et al, Astrophys. J. Suppl., {\bf 148}, 135 (2003).\\
             A. Kogut et al, AstroAstrophys. J. Suppl., {\bf 148}, 161 (2003).\\
             D.N. Spergel et al, Astrophys. J. Suppl., {\bf 148}, 175 (2003).\\
             C.L. Bennet at al, Astrophys. J. Suppl., {\bf 148}, 1 (2003).

\bibitem{r20} A.G. Riess et al, Astrophys. J., {\bf 560}, 49 (2001).

\bibitem{r2} T. Padmanabhan and T. Roy Choudury, Mon. Not. R. Astron. Soc., {\bf 344}, 823 (2003).\\
             T. Roy Choudury and T. Padmanabhan, Astron. Astrophys., {\bf 823}, 807 (2005).

\bibitem{r3} V. Sahni and A. Starobinsky, Int. J. Mod. Phys. D, {\bf 9}, 373 (1000).\\
             T. Padmanabhan, Phys. Rep., {\bf 380}, 235 (2003).

\bibitem{r4} J. Martin, astro-ph/0803.4076.

\bibitem{r5} I. Zlatev and P.J. Steinhardt. Phys.Lett.B, {\bf 459}, 570 (1999).

\bibitem{r} N. Banerjee and S. Das, Mod. Phys. Lett. A, {\bf 21}, 2663 (2006).

\bibitem{r6} I. Zlatev, L. Wang and P.J. Steinhardt, Phys. Rev. Lett. {\bf 82}, 896 (1999).\\
             P.J. Steinhardt, L. Wang and I. Zlatev, Phys. Rev.D, {\bf 59}, 123504(1999).\\
             L. Wang, R.R. Caldwell, J.P. Ostriker and P.J. Steinhardt, Astrophys. J., {\bf 530}, 17 (2000).

\bibitem{r7} V.B.Johri Class.Quant.Grav {\bf 19}, 5959 (2002).

\bibitem{r8} L.A.Urena-Lopez and T.Matos Phys.Rev.D, {\bf 62}, 081302 (2000).

\bibitem{r9} M.Sahlen, AR.Liddle, D. Parkinson Phys.Rev.D, {\bf 75}, 023502 (2007).

\bibitem{r10} S.Dodelson, M.Kaplinghat and E. Stewart Phys.Rev.Lett, {\bf 85}, 5276(2000).

\bibitem{r11} P-Y. Wang, C.W Chen and P.Chen. JCAP 2012.


\bibitem{r12} S.C.C.Ng, N.J. Nunes, F.Rosati, Phys.Rev.D, {\bf64} ,083510(2001).

\bibitem{r13} \textit{Dynamical Systems in Cosmology}, J. Wainwright and G.F.R.Ellis (eds); Cambridge University Press, (1997).\\   \textit{Dynamical System and Cosmology}, A.A.Coley, Kluwer Academic Publishers (2003).

\bibitem{r14} L.Lara and M.Castagnins, Int.J.Theor. Phys. {\bf 44}, 1839(2005).

\bibitem{r15} E.Gunzig, V.Faraoni, A.Figeredo, T.M.Rocha Filho and L.Brenig, Class.Quant.Grav., {\bf 17}, 1783(2000).

\bibitem{r16} J.Carot and M.M.collinge, Class.Quanta.Grav., {\bf 20}, 707(2003).

\bibitem{r17} L.Arturo, Urena-Lopez, JCAP, {\bf 0509}, 013(2005).

\bibitem{r18} S.J.Kolitch and B.Hall, arxiv:[gr-qc /9410039].\\
              S.J. Kolitch and D.M. Eardley, Ann. Phys., {\bf 241}, 128, 1995.

\bibitem{r19} \textit{Nonlinear dynamics and chaos: With applications to Physics, Biology, Chemistry and Engineering}, S.H. Strogatz, Westview Press (2001).

\bibitem{r21} V. Sahni, arxiv:{astro-ph/0403324}.
%---------------------------------------------------------------------------------------CHAMELEON--------------------------------------
\bibitem{varun} V. Sahni and A. Starobinsky, Int. J Mod. Phys. D, {\bf 9}, 373 (2000).
\bibitem{paddy} T. Padmanabhan, Phys. Rep., {\bf 380}, 235 (2003).
\bibitem{sami} E.J. Copeland, M. Sami and S. Tsujikawa, Int. J Mod. Phys. D, {\bf 15}, 1753 (2006).
\bibitem{martin} J. Martin, Mod. Phys. Lett. A, {\bf 23}, 1252 (2008).
\bibitem{review} T.P. Sotiriou and V. Faraoni; arxiv:gr-qc/0805.1726;\\
                 A. De Felice  and S. Tsujikawa; arxiv:gr-qc/1002.4928.\\
                 S. Nojiri and S.D. Odintsov, Phys. Rep. {\bf 505}, 59 (2011).
\bibitem{nbdp}O. Bertolami and P.J. Martins, Phys. Rev. D, {\bf 61}, 064007 (2000).\\
              N. Banerjee and D. Pavon, Phys. Rev. D, {\bf 63}, 043504 (2001).\\
              S. Sen and A.A. Sen, Phys. Rev. D, {\bf 63}, 124006 (2001).\\
              E. Elizalde, S. Nojiri and S.D. Odintsov, Phys. Rev. D, {\bf 70}, 043539 (2004).\\
            
\bibitem{justin1} J. Khoury and A. Weltman, Phys. Rev D, {\bf 69}, 044026 (2004).
\bibitem{justin2} J. Khoury and A. Weltman, Phys. Rev. Lett., {\bf 93}, 171104 (2004).
\bibitem{mota1} D.F. Mota and D.J. Shaw; arxiv:0805.3430. 
\bibitem{mota2} D.F. Mota and D.J. Shaw Phys. Rev. Lett., {\bf 97}, 151102 (2006).
\bibitem{shaw1} C. Burrage, A.C. Davis and D.J. Shaw, Phys. Rev. D, {\bf 79}, 044028 (2009).
\bibitem{shaw2} A.C. Davis, C.A.O. Schelpe and D.J. Shaw, Phys.Rev. D, {\bf 80}, 064016 (2009).
\bibitem{buri} P. Burikham and S. Panpanich, Int. J. Mod. Phys. D, {\bf 21}, 1250041 (2012).
\bibitem{brax1} P. Brax, C. Burrage, A.C. Davis, D. Seery and A. Weltman, Phys. Lett. B, {\bf 699}, 5 (2011).
\bibitem{mota3} D.F. Mota and C.A.O. Schelpe, Phys. Rev. D, {\bf 86}, 123002 (2012).
\bibitem{khoury} J. Khoury, Class. Quant. Grav., {\bf 30}, 214004 (2013).
\bibitem{brax} P. Brax, C. van de Bruck, A.C. Davis, J. Khoury and A. Weltman, Phys. Rev. D, {\bf 70}, 123518 (2004).
\bibitem{das} S. Das, P.S. Corasaniti and J. Khoury, Phys. Rev. D, {\bf 73}, 083509 (2006).
\bibitem{nbsdkg} N. Banerjee, S. Das and K. Ganguly, Pramana, {\bf 74}, L481 (2010).
\bibitem{sdnb} S. Das and N. Banerjee, Phys. Rev.D, {\bf 78}, 043512 (2008).
\bibitem{jamil} M. Setare and M. Jamil. Phys. Lett. B, {\bf 690}, 1 (2014).
\bibitem{tavakol} S. Tsujikawa, T. Tamaki and R. Tavakol, JCAP, {\bf 05}, 020 (2009).
\bibitem{gunzig} E. Gunzig, V. Faraoni, A. Figeredo and L. Brenig, Class. Quantum Grav., {\bf 17}, 1783 (2000).\\
                 J. Carot and M.M. Collinge, Class. Quantum Grav., {\bf 20}, 707 (2003).
                 L.A. Urena-Lopez, JCAP, {\bf 0509}, 013 (2005).
                 S. Sen, A.A. Sen and M. Sami, Phys. Lett. B, {\bf 686}, 1 (2010).
                 S. Kumar, S. Panda and A.A. Sen, Class. Quantum Grav., {\bf 30}, 155011 (2013).
                 N. Roy and N. Banerjee, Gen. Rel. Grav., {\bf 46}, 1651 (2014).
\bibitem{coley}  A.A. Coley, {\it Dynamical Systems and Cosmology}, Cambridge University Press (2003).
\bibitem{ellis}  J. Wainwright and G.F.R. Ellis, {\it Dynamical Systems in Cosmology}, Springer (2005).
\bibitem{harko1} O. Minazzoli and T.Harko, Phys. Rev. D, {\bf 86}, 087502 (2012).
\bibitem{harko2} T.Harko, Phys. Rev. D, {\bf 81}, 044021 (2010).
\bibitem{strogatz}S.H. Strogatz, {\it Nonlinear dynamics and chaos: With Applications to Physics, Biology Chemistry and  		             Engineering}, Westview Press, Boulder (2001).
\bibitem{nrnb} N. Roy and N. Banerjee, Euro. Phys. J Plus., {\bf 129}, 162 (2014).
\bibitem{giostri}R. Giostri, M.V. dos Santos, I. Waga, R.R.R. Reis, M.O. Calvao and B.L. Lago, J. Cos. Astrophys., {\bf 027}, 
                 1203 (2012).
%------------------------------------Holo-----------------------------------------------------------------------------------
\bibitem{'thooft} G. 't Hooft, arxiv: gr-qc/9310026.
%\bibitem{susskind} L. Susskind, J. Math. Phys., {\bf 36}, 6377 (1995).
%\bibitem{diego} D. Pavon, J. Phys. A:Math. Theor., {\bf 40}, 6865 (2007).
%\bibitem{li} M.Li, Phys. Lett. B {\bf 603}, 1 (2004).
%\bibitem{diego1} D.Pavon and W. Zimdahl, Phys. Lett. B {\bf 628}, 206 (2005).
%\bibitem{diego2} D.Pavon and W. Zimdahl, Class. Quantum Grav. {\bf 24}, 5461 (2007).
%\bibitem{diego3} N. Banerjee and D. Pavon, Phy. Lett. B {\bf 647}, 477 (2007).
%\bibitem{setare} M.R. Setare and M. Jamil, Phys. Lett B {\bf 690}, 1 (2010).
%\bibitem{ellis} J. Wainwright and G. F. R. Ellis, \textit{Dynamical System in Cosmology} (Cambridge University Press, 2005).
%\bibitem{coley} A.A.Coley, \textit{Dynamical System and Cosmology} (Springer, 2003).
%\bibitem{gunzig} E. Gunzig, V. Faraoni, A. Figeredo and L. Brenig, Class. Quantum Grav. {\bf 17}, 1783 (2000).
%\bibitem{carot} J. Carot and M.M. Collinge, Class. Quantum Grav. {\bf 20}, 707 (2003).
%\bibitem{urena} L.A. Urena-Lopez, JCAP {\bf 0509}, 013 (2005).
%\bibitem{nandan2} N. Roy and N. Banerjee, Eur. Phys. J. Plus, {\bf 129}, 162 (2014).
%\bibitem{anjan} S. Kumar, S. Panda and A.A. Sen, Quantum Grav. {\bf 30}, 155011 (2013).
%\bibitem{soma} S. Sen, A.A. Sen and M. Sami, Phys. Lett B. {\bf 686}, 1 (2010).
%\bibitem{nandan1} N. Roy and N. Banerjee, Gen. Rel. Grav. {\bf 46}, 1651 (2014).
%\bibitem{fang} W. Fang, H. Tu, J. Huang and C. Shu, arxiv:[1402.4005].
%\bibitem{nairi1} N. Mazumder, R. Biswas and S. Chakraborty, arxiv:[1106.4627].
%\bibitem{nairi2} N. Mazumder, R. Biswas and S. Chakraborty, arxiv:[1106.4626].
%\bibitem{setare1} M.R. Setare and E.C. Vagenas, Int. J. Mod. Phys. D, {\bf 18}, 147 (2009).
%\bibitem{strog} S.H. Strogatz, \textit{Nonlinear Dynamics and Chaos: With Applications to Physics, Biology, Chemistry and     Engineering};                             Westview Press, Boulder (2001). 

%\bibitem{reza} R. Tavakol, `` Introduction to dynamical systems '' in ref \cite{ellis} .
                
\end{thebibliography}


\begin{thebibliography}{305}
\expandafter\ifx\csname natexlab\endcsname\relax\def\natexlab#1{#1}\fi
\expandafter\ifx\csname bibnamefont\endcsname\relax
  \def\bibnamefont#1{#1}\fi
\expandafter\ifx\csname bibfnamefont\endcsname\relax
  \def\bibfnamefont#1{#1}\fi
\expandafter\ifx\csname citenamefont\endcsname\relax
  \def\citenamefont#1{#1}\fi
\expandafter\ifx\csname url\endcsname\relax
  \def\url#1{\texttt{#1}}\fi
\expandafter\ifx\csname urlprefix\endcsname\relax\def\urlprefix{URL }\fi
\providecommand{\bibinfo}[2]{#2}
\providecommand{\eprint}[2][]{\url{#2}}

\bibitem[{\citenamefont{{Thorne}}(1965)}]{thorne}
\bibinfo{author}{\bibfnamefont{K.~S.} \bibnamefont{{Thorne}}},
  \bibinfo{journal}{Science} \textbf{\bibinfo{volume}{150}},
  \bibinfo{pages}{1671} (\bibinfo{year}{1965}).

\bibitem[{\citenamefont{Shapiro and Teukolsky}(1991{\natexlab{a}})}]{shapiro}
\bibinfo{author}{\bibfnamefont{S.~L.} \bibnamefont{Shapiro}} \bibnamefont{and}
  \bibinfo{author}{\bibfnamefont{S.~A.} \bibnamefont{Teukolsky}},
  \bibinfo{journal}{American Scientist} \textbf{\bibinfo{volume}{79}},
  \bibinfo{pages}{330} (\bibinfo{year}{1991}{\natexlab{a}}).

\bibitem[{\citenamefont{Einstein}(1915{\natexlab{a}})}]{einstein1}
\bibinfo{author}{\bibfnamefont{A.}~\bibnamefont{Einstein}},
  \bibinfo{journal}{Sitzungsber Preuss. Akad. Wiss.} p. \bibinfo{pages}{778}
  (\bibinfo{year}{1915}{\natexlab{a}}).

\bibitem[{\citenamefont{Einstein}(1915{\natexlab{b}})}]{einstein2}
\bibinfo{author}{\bibfnamefont{A.}~\bibnamefont{Einstein}},
  \bibinfo{journal}{Sitzungsber Preuss. Akad. Wiss.} p. \bibinfo{pages}{844}
  (\bibinfo{year}{1915}{\natexlab{b}}).

\bibitem[{\citenamefont{Einstein}(1916)}]{einstein3}
\bibinfo{author}{\bibfnamefont{A.}~\bibnamefont{Einstein}},
  \bibinfo{journal}{Ann. Phys. (Leipzig)} \textbf{\bibinfo{volume}{49}},
  \bibinfo{pages}{769} (\bibinfo{year}{1916}).

\bibitem[{\citenamefont{Schwarzschild}(1916)}]{sch}
\bibinfo{author}{\bibfnamefont{K.}~\bibnamefont{Schwarzschild}},
  \bibinfo{journal}{Sitzber. Deut. Akad. Wiss. Berlin, Kl. Math. Phys. Tech.,}
  p. \bibinfo{pages}{189} (\bibinfo{year}{1916}).

\bibitem[{\citenamefont{Chandrasekhar}(1931)}]{chandra}
\bibinfo{author}{\bibfnamefont{S.}~\bibnamefont{Chandrasekhar}},
  \bibinfo{journal}{Astrophys. J.} \textbf{\bibinfo{volume}{74}},
  \bibinfo{pages}{82} (\bibinfo{year}{1931}).

\bibitem[{\citenamefont{Landau}(1932)}]{landau}
\bibinfo{author}{\bibfnamefont{L.~D.} \bibnamefont{Landau}},
  \bibinfo{journal}{Phys. Z. Sowjetunion} \textbf{\bibinfo{volume}{1}},
  \bibinfo{pages}{285} (\bibinfo{year}{1932}).

\bibitem[{\citenamefont{Eddington}(1935)}]{eddi}
\bibinfo{author}{\bibfnamefont{A.~S.} \bibnamefont{Eddington}},
  \bibinfo{journal}{Mon. Not. Roy. Astron. Soc.} \textbf{\bibinfo{volume}{95}},
  \bibinfo{pages}{194} (\bibinfo{year}{1935}).

\bibitem[{\citenamefont{Oppenheimer and Snyder}(1939)}]{os}
\bibinfo{author}{\bibfnamefont{J.~R.} \bibnamefont{Oppenheimer}}
  \bibnamefont{and} \bibinfo{author}{\bibfnamefont{H.~S.}
  \bibnamefont{Snyder}}, \bibinfo{journal}{Phys. Rev.}
  \textbf{\bibinfo{volume}{56}}, \bibinfo{pages}{455} (\bibinfo{year}{1939}).

\bibitem[{\citenamefont{Wald}(1984)}]{wald}
\bibinfo{author}{\bibfnamefont{R.~M.} \bibnamefont{Wald}},
  \emph{\bibinfo{title}{General Relativity}} (\bibinfo{publisher}{The
  University of Chicago Press, Chicago}, \bibinfo{year}{1984}).

\bibitem[{\citenamefont{Penrose}(1965)}]{penrose}
\bibinfo{author}{\bibfnamefont{R.}~\bibnamefont{Penrose}},
  \bibinfo{journal}{Phys. Rev. Lett.} \textbf{\bibinfo{volume}{14}},
  \bibinfo{pages}{57} (\bibinfo{year}{1965}).

\bibitem[{\citenamefont{Hawking and Ellis}((1973))}]{hp}
\bibinfo{author}{\bibfnamefont{S.~W.} \bibnamefont{Hawking}} \bibnamefont{and}
  \bibinfo{author}{\bibfnamefont{G.~F.~R.} \bibnamefont{Ellis}},
  \emph{\bibinfo{title}{Large Scale Structure of Space-time}}
  (\bibinfo{publisher}{Cambridge University Press, Cambridge},
  \bibinfo{year}{(1973)}).

\bibitem[{\citenamefont{Oppenheimer and Volkoff}(1939)}]{ov}
\bibinfo{author}{\bibfnamefont{J.~R.} \bibnamefont{Oppenheimer}}
  \bibnamefont{and} \bibinfo{author}{\bibfnamefont{G.~M.}
  \bibnamefont{Volkoff}}, \bibinfo{journal}{Phys. Rev.}
  \textbf{\bibinfo{volume}{55}}, \bibinfo{pages}{374} (\bibinfo{year}{1939}).

\bibitem[{\citenamefont{Datt}(1938)}]{datt}
\bibinfo{author}{\bibfnamefont{B.}~\bibnamefont{Datt}}, \bibinfo{journal}{Z.
  Phys.} \textbf{\bibinfo{volume}{108}}, \bibinfo{pages}{314}
  (\bibinfo{year}{1938}).

\bibitem[{\citenamefont{{Penrose}}(1969)}]{penrose1}
\bibinfo{author}{\bibfnamefont{R.}~\bibnamefont{{Penrose}}},
  \bibinfo{journal}{Nuovo Cimento Rivista Serie} \textbf{\bibinfo{volume}{1}}
  (\bibinfo{year}{1969}).

\bibitem[{\citenamefont{Yodzis et~al.}(1973)\citenamefont{Yodzis, Seifert, and
  Muller~zum Hagen}}]{seif1}
\bibinfo{author}{\bibfnamefont{P.}~\bibnamefont{Yodzis}},
  \bibinfo{author}{\bibfnamefont{H.~J.} \bibnamefont{Seifert}},
  \bibnamefont{and} \bibinfo{author}{\bibfnamefont{H.}~\bibnamefont{Muller~zum
  Hagen}}, \bibinfo{journal}{Comm. Math. Phys.} \textbf{\bibinfo{volume}{34}},
  \bibinfo{pages}{135} (\bibinfo{year}{1973}).

\bibitem[{\citenamefont{Yodzis et~al.}(1974)\citenamefont{Yodzis, Seifert, and
  Muller~zum Hagen}}]{seif2}
\bibinfo{author}{\bibfnamefont{P.}~\bibnamefont{Yodzis}},
  \bibinfo{author}{\bibfnamefont{H.~J.} \bibnamefont{Seifert}},
  \bibnamefont{and} \bibinfo{author}{\bibfnamefont{H.}~\bibnamefont{Muller~zum
  Hagen}}, \bibinfo{journal}{Comm. Math. Phys.} \textbf{\bibinfo{volume}{37}},
  \bibinfo{pages}{29} (\bibinfo{year}{1974}).

\bibitem[{\citenamefont{Eardley and Smarr}(1979)}]{smarr}
\bibinfo{author}{\bibfnamefont{D.~M.} \bibnamefont{Eardley}} \bibnamefont{and}
  \bibinfo{author}{\bibfnamefont{L.}~\bibnamefont{Smarr}},
  \bibinfo{journal}{Phys. Rev. D.} \textbf{\bibinfo{volume}{19}},
  \bibinfo{pages}{2239} (\bibinfo{year}{1979}).

\bibitem[{\citenamefont{Christodoulou}(1984)}]{christozero}
\bibinfo{author}{\bibfnamefont{D.}~\bibnamefont{Christodoulou}},
  \bibinfo{journal}{Comm. Math. Phys.} \textbf{\bibinfo{volume}{93}},
  \bibinfo{pages}{171} (\bibinfo{year}{1984}).

\bibitem[{\citenamefont{Newman}(1986)}]{newman}
\bibinfo{author}{\bibfnamefont{R.~P. A.~C.} \bibnamefont{Newman}},
  \bibinfo{journal}{Class. Quant. Grav.} \textbf{\bibinfo{volume}{3}},
  \bibinfo{pages}{527} (\bibinfo{year}{1986}).

\bibitem[{\citenamefont{Joshi}(1993)}]{joshi1}
\bibinfo{author}{\bibfnamefont{P.~S.} \bibnamefont{Joshi}},
  \emph{\bibinfo{title}{Global Aspects in Gravitation and Cosmology.}}
  (\bibinfo{publisher}{Oxford: Clarendon Press, Oxford University Press.},
  \bibinfo{year}{1993}).

\bibitem[{\citenamefont{Joshi}(2000)}]{joshi2}
\bibinfo{author}{\bibfnamefont{P.~S.} \bibnamefont{Joshi}},
  \bibinfo{journal}{Pramana} \textbf{\bibinfo{volume}{55}},
  \bibinfo{pages}{529} (\bibinfo{year}{2000}).

\bibitem[{\citenamefont{Joshi and Dwivedi}(1992)}]{joshidwivedi1}
\bibinfo{author}{\bibfnamefont{P.~S.} \bibnamefont{Joshi}} \bibnamefont{and}
  \bibinfo{author}{\bibfnamefont{I.~H.} \bibnamefont{Dwivedi}},
  \bibinfo{journal}{Comm. Math. Phys.} \textbf{\bibinfo{volume}{146}},
  \bibinfo{pages}{333} (\bibinfo{year}{1992}).

\bibitem[{\citenamefont{Joshi and Dwivedi}(1993{\natexlab{a}})}]{joshidwivedi2}
\bibinfo{author}{\bibfnamefont{P.~S.} \bibnamefont{Joshi}} \bibnamefont{and}
  \bibinfo{author}{\bibfnamefont{I.~H.} \bibnamefont{Dwivedi}},
  \bibinfo{journal}{Phys. Rev. D.} \textbf{\bibinfo{volume}{47}},
  \bibinfo{pages}{5357} (\bibinfo{year}{1993}{\natexlab{a}}).

\bibitem[{\citenamefont{Joshi and Dwivedi}(1993{\natexlab{b}})}]{joshidwivedi3}
\bibinfo{author}{\bibfnamefont{P.~S.} \bibnamefont{Joshi}} \bibnamefont{and}
  \bibinfo{author}{\bibfnamefont{I.~H.} \bibnamefont{Dwivedi}},
  \bibinfo{journal}{Lett. Math. Phys.} \textbf{\bibinfo{volume}{27}},
  \bibinfo{pages}{235} (\bibinfo{year}{1993}{\natexlab{b}}).

\bibitem[{\citenamefont{Waugh and Lake}(1989)}]{waughlake1}
\bibinfo{author}{\bibfnamefont{B.}~\bibnamefont{Waugh}} \bibnamefont{and}
  \bibinfo{author}{\bibfnamefont{K.}~\bibnamefont{Lake}},
  \bibinfo{journal}{Phys. Rev. D.} \textbf{\bibinfo{volume}{40}},
  \bibinfo{pages}{2137} (\bibinfo{year}{1989}).

\bibitem[{\citenamefont{Waugh and Lake}(1988)}]{waughlake2}
\bibinfo{author}{\bibfnamefont{B.}~\bibnamefont{Waugh}} \bibnamefont{and}
  \bibinfo{author}{\bibfnamefont{K.}~\bibnamefont{Lake}},
  \bibinfo{journal}{Phys. Rev. D.} \textbf{\bibinfo{volume}{38}},
  \bibinfo{pages}{1315} (\bibinfo{year}{1988}).

\bibitem[{\citenamefont{Ori and Piran}(1987)}]{ori1}
\bibinfo{author}{\bibfnamefont{A.}~\bibnamefont{Ori}} \bibnamefont{and}
  \bibinfo{author}{\bibfnamefont{T.}~\bibnamefont{Piran}},
  \bibinfo{journal}{Phys. Rev. Lett.} \textbf{\bibinfo{volume}{59}},
  \bibinfo{pages}{2137} (\bibinfo{year}{1987}).

\bibitem[{\citenamefont{Ori and Piran}(1990)}]{ori2}
\bibinfo{author}{\bibfnamefont{A.}~\bibnamefont{Ori}} \bibnamefont{and}
  \bibinfo{author}{\bibfnamefont{T.}~\bibnamefont{Piran}},
  \bibinfo{journal}{Phys. Rev. D.} \textbf{\bibinfo{volume}{42}},
  \bibinfo{pages}{1068} (\bibinfo{year}{1990}).

\bibitem[{\citenamefont{Jhingan and Magli}(2000)}]{magli1}
\bibinfo{author}{\bibfnamefont{S.}~\bibnamefont{Jhingan}} \bibnamefont{and}
  \bibinfo{author}{\bibfnamefont{G.}~\bibnamefont{Magli}},
  \bibinfo{journal}{Phys. Rev. D.} \textbf{\bibinfo{volume}{61}},
  \bibinfo{pages}{124006} (\bibinfo{year}{2000}).

\bibitem[{\citenamefont{Giambo et~al.}(2003)\citenamefont{Giambo, Giannoni,
  Magli, and Piccione}}]{magli2}
\bibinfo{author}{\bibfnamefont{R.}~\bibnamefont{Giambo}},
  \bibinfo{author}{\bibfnamefont{F.}~\bibnamefont{Giannoni}},
  \bibinfo{author}{\bibfnamefont{G.}~\bibnamefont{Magli}}, \bibnamefont{and}
  \bibinfo{author}{\bibfnamefont{P.}~\bibnamefont{Piccione}},
  \bibinfo{journal}{Comm. Math. Phys.} \textbf{\bibinfo{volume}{235}},
  \bibinfo{pages}{545} (\bibinfo{year}{2003}).

\bibitem[{\citenamefont{Giambo et~al.}(2004)\citenamefont{Giambo, Giannoni,
  Magli, and Piccione}}]{magli3}
\bibinfo{author}{\bibfnamefont{R.}~\bibnamefont{Giambo}},
  \bibinfo{author}{\bibfnamefont{F.}~\bibnamefont{Giannoni}},
  \bibinfo{author}{\bibfnamefont{G.}~\bibnamefont{Magli}}, \bibnamefont{and}
  \bibinfo{author}{\bibfnamefont{P.}~\bibnamefont{Piccione}},
  \bibinfo{journal}{Gen. Rel. Grav.} \textbf{\bibinfo{volume}{36}},
  \bibinfo{pages}{1279} (\bibinfo{year}{2004}).

\bibitem[{\citenamefont{Iguchi et~al.}(1998)\citenamefont{Iguchi, Nakao, and
  Harada}}]{nakao}
\bibinfo{author}{\bibfnamefont{H.}~\bibnamefont{Iguchi}},
  \bibinfo{author}{\bibfnamefont{K.~I.} \bibnamefont{Nakao}}, \bibnamefont{and}
  \bibinfo{author}{\bibfnamefont{T.}~\bibnamefont{Harada}},
  \bibinfo{journal}{Phys. Rev. D.} \textbf{\bibinfo{volume}{57}},
  \bibinfo{pages}{7262} (\bibinfo{year}{1998}).

\bibitem[{\citenamefont{Shapiro and Teukolsky}(1991{\natexlab{b}})}]{shapiro1}
\bibinfo{author}{\bibfnamefont{S.~L.} \bibnamefont{Shapiro}} \bibnamefont{and}
  \bibinfo{author}{\bibfnamefont{S.~A.} \bibnamefont{Teukolsky}},
  \bibinfo{journal}{Phys. Rev. Lett.} \textbf{\bibinfo{volume}{66}},
  \bibinfo{pages}{994} (\bibinfo{year}{1991}{\natexlab{b}}).

\bibitem[{\citenamefont{Shapiro and Teukolsky}(1992)}]{shapiro2}
\bibinfo{author}{\bibfnamefont{S.~L.} \bibnamefont{Shapiro}} \bibnamefont{and}
  \bibinfo{author}{\bibfnamefont{S.~A.} \bibnamefont{Teukolsky}},
  \bibinfo{journal}{Phys. Rev. D.} \textbf{\bibinfo{volume}{45}},
  \bibinfo{pages}{2006} (\bibinfo{year}{1992}).

\bibitem[{\citenamefont{Joshi et~al.}(2000)\citenamefont{Joshi, Dadhich, and
  Maartens}}]{maartens1}
\bibinfo{author}{\bibfnamefont{P.~S.} \bibnamefont{Joshi}},
  \bibinfo{author}{\bibfnamefont{N.}~\bibnamefont{Dadhich}}, \bibnamefont{and}
  \bibinfo{author}{\bibfnamefont{R.}~\bibnamefont{Maartens}},
  \bibinfo{journal}{Mod. Phys. Lett. A.} \textbf{\bibinfo{volume}{15}},
  \bibinfo{pages}{991} (\bibinfo{year}{2000}).

\bibitem[{\citenamefont{Joshi et~al.}(2002)\citenamefont{Joshi, Dadhich, and
  Maartens}}]{maartens2}
\bibinfo{author}{\bibfnamefont{P.~S.} \bibnamefont{Joshi}},
  \bibinfo{author}{\bibfnamefont{N.}~\bibnamefont{Dadhich}}, \bibnamefont{and}
  \bibinfo{author}{\bibfnamefont{R.}~\bibnamefont{Maartens}},
  \bibinfo{journal}{Phys. Rev. D.} \textbf{\bibinfo{volume}{65}},
  \bibinfo{pages}{101501} (\bibinfo{year}{2002}).

\bibitem[{\citenamefont{Joshi et~al.}(2004)\citenamefont{Joshi, Goswami, and
  Dadhich}}]{joshiritudadhi}
\bibinfo{author}{\bibfnamefont{P.~S.} \bibnamefont{Joshi}},
  \bibinfo{author}{\bibfnamefont{R.}~\bibnamefont{Goswami}}, \bibnamefont{and}
  \bibinfo{author}{\bibfnamefont{N.}~\bibnamefont{Dadhich}},
  \bibinfo{journal}{Phys. Rev. D.} \textbf{\bibinfo{volume}{70}},
  \bibinfo{pages}{087502} (\bibinfo{year}{2004}).

\bibitem[{\citenamefont{Vaidya}(1943)}]{vaidya1}
\bibinfo{author}{\bibfnamefont{P.~C.} \bibnamefont{Vaidya}},
  \bibinfo{journal}{Curr. Sci.} \textbf{\bibinfo{volume}{12}},
  \bibinfo{pages}{183} (\bibinfo{year}{1943}).

\bibitem[{\citenamefont{Vaidya}(1951)}]{vaidya2}
\bibinfo{author}{\bibfnamefont{P.~C.} \bibnamefont{Vaidya}},
  \bibinfo{journal}{Proc. Ind. Acad. Sci. A.} \textbf{\bibinfo{volume}{33}},
  \bibinfo{pages}{264} (\bibinfo{year}{1951}).

\bibitem[{\citenamefont{Miller and Sciama}(1979)}]{millersciama}
\bibinfo{author}{\bibfnamefont{J.~C.} \bibnamefont{Miller}} \bibnamefont{and}
  \bibinfo{author}{\bibfnamefont{D.~W.} \bibnamefont{Sciama}},
  \emph{\bibinfo{title}{Gravitational collapse to black hole state: “General
  relativity and Gravitation: One hundred years after the birth of Albert
  Einstein”.}} (\bibinfo{publisher}{Plenum Press; New York and London},
  \bibinfo{year}{1979}).

\bibitem[{\citenamefont{Bronnikov and Kovalchuk}(1983)}]{bronni1}
\bibinfo{author}{\bibfnamefont{K.~A.} \bibnamefont{Bronnikov}}
  \bibnamefont{and}
  \bibinfo{author}{\bibfnamefont{M.}~\bibnamefont{Kovalchuk}},
  \bibinfo{journal}{Gen. Rel. Grav.} \textbf{\bibinfo{volume}{15}},
  \bibinfo{pages}{809} (\bibinfo{year}{1983}).

\bibitem[{\citenamefont{Bronnikov}(1983)}]{bronni2}
\bibinfo{author}{\bibfnamefont{K.~A.} \bibnamefont{Bronnikov}},
  \bibinfo{journal}{Gen. Rel. Grav.} \textbf{\bibinfo{volume}{15}},
  \bibinfo{pages}{823} (\bibinfo{year}{1983}).

\bibitem[{\citenamefont{Bronnikov and Kovalchuk}(1984)}]{bronni3}
\bibinfo{author}{\bibfnamefont{K.~A.} \bibnamefont{Bronnikov}}
  \bibnamefont{and}
  \bibinfo{author}{\bibfnamefont{M.}~\bibnamefont{Kovalchuk}},
  \bibinfo{journal}{Gen. Rel. Grav.} \textbf{\bibinfo{volume}{16}},
  \bibinfo{pages}{15} (\bibinfo{year}{1984}).

\bibitem[{\citenamefont{Ganguly and
  Banerjee}(2011{\natexlab{a}})}]{gangulybanerjee}
\bibinfo{author}{\bibfnamefont{K.}~\bibnamefont{Ganguly}} \bibnamefont{and}
  \bibinfo{author}{\bibfnamefont{N.}~\bibnamefont{Banerjee}},
  \bibinfo{journal}{Gen. Rel. Grav.} \textbf{\bibinfo{volume}{43}},
  \bibinfo{pages}{2141} (\bibinfo{year}{2011}{\natexlab{a}}).

\bibitem[{\citenamefont{{Joshi}}(2013)}]{joshi3}
\bibinfo{author}{\bibfnamefont{P.~S.} \bibnamefont{{Joshi}}},
  \bibinfo{journal}{ArXiv e-prints}  (\bibinfo{year}{2013}),
  \eprint{1305.1005}.

\bibitem[{\citenamefont{{Joshi} and {Malafarina}}(2011)}]{joshimala1}
\bibinfo{author}{\bibfnamefont{P.~S.} \bibnamefont{{Joshi}}} \bibnamefont{and}
  \bibinfo{author}{\bibfnamefont{D.}~\bibnamefont{{Malafarina}}},
  \bibinfo{journal}{Int. J. Mod. Phys. D.} \textbf{\bibinfo{volume}{20}},
  \bibinfo{pages}{2641} (\bibinfo{year}{2011}).

\bibitem[{\citenamefont{{Joshi} et~al.}(2011)\citenamefont{{Joshi},
  {Malafarina}, and {Narayan}}}]{joshimalanarayan1}
\bibinfo{author}{\bibfnamefont{P.~S.} \bibnamefont{{Joshi}}},
  \bibinfo{author}{\bibfnamefont{D.}~\bibnamefont{{Malafarina}}},
  \bibnamefont{and}
  \bibinfo{author}{\bibfnamefont{R.}~\bibnamefont{{Narayan}}},
  \bibinfo{journal}{Class. Quant. Grav.} \textbf{\bibinfo{volume}{28}},
  \bibinfo{pages}{235018} (\bibinfo{year}{2011}).

\bibitem[{\citenamefont{{Patil} and {Joshi}}(2010)}]{patiljoshi1}
\bibinfo{author}{\bibfnamefont{M.}~\bibnamefont{{Patil}}} \bibnamefont{and}
  \bibinfo{author}{\bibfnamefont{P.~S.} \bibnamefont{{Joshi}}},
  \bibinfo{journal}{Phys. Rev. D.} \textbf{\bibinfo{volume}{82}},
  \bibinfo{pages}{104049} (\bibinfo{year}{2010}).

\bibitem[{\citenamefont{{Patil} et~al.}(2011)\citenamefont{{Patil}, {Joshi},
  and {Malafarina}}}]{patiljoshi2}
\bibinfo{author}{\bibfnamefont{M.}~\bibnamefont{{Patil}}},
  \bibinfo{author}{\bibfnamefont{P.~S.} \bibnamefont{{Joshi}}},
  \bibnamefont{and}
  \bibinfo{author}{\bibfnamefont{D.}~\bibnamefont{{Malafarina}}},
  \bibinfo{journal}{Phys. Rev. D.} \textbf{\bibinfo{volume}{83}},
  \bibinfo{pages}{064007} (\bibinfo{year}{2011}).

\bibitem[{\citenamefont{{Sahu} et~al.}(2012)\citenamefont{{Sahu}, {Patil},
  {Narasimha}, and {Joshi}}}]{sahupatiljoshi}
\bibinfo{author}{\bibfnamefont{S.}~\bibnamefont{{Sahu}}},
  \bibinfo{author}{\bibfnamefont{M.}~\bibnamefont{{Patil}}},
  \bibinfo{author}{\bibfnamefont{D.}~\bibnamefont{{Narasimha}}},
  \bibnamefont{and} \bibinfo{author}{\bibfnamefont{P.~S.}
  \bibnamefont{{Joshi}}}, \bibinfo{journal}{Phys. Rev. D.}
  \textbf{\bibinfo{volume}{86}}, \bibinfo{pages}{063010}
  (\bibinfo{year}{2012}).

\bibitem[{\citenamefont{{Patil} and {Joshi}}(2012{\natexlab{a}})}]{patiljoshi3}
\bibinfo{author}{\bibfnamefont{M.}~\bibnamefont{{Patil}}} \bibnamefont{and}
  \bibinfo{author}{\bibfnamefont{P.~S.} \bibnamefont{{Joshi}}},
  \bibinfo{journal}{Phys. Rev. D.} \textbf{\bibinfo{volume}{86}},
  \bibinfo{pages}{044040} (\bibinfo{year}{2012}{\natexlab{a}}).

\bibitem[{\citenamefont{{Patil} and {Joshi}}(2012{\natexlab{b}})}]{patiljoshi4}
\bibinfo{author}{\bibfnamefont{M.}~\bibnamefont{{Patil}}} \bibnamefont{and}
  \bibinfo{author}{\bibfnamefont{P.~S.} \bibnamefont{{Joshi}}},
  \bibinfo{journal}{Phys. Rev. D.} \textbf{\bibinfo{volume}{85}},
  \bibinfo{pages}{104014} (\bibinfo{year}{2012}{\natexlab{b}}).

\bibitem[{\citenamefont{{Joshi} et~al.}(2012)\citenamefont{{Joshi},
  {Malafarina}, and {Saraykar}}}]{joshimala2}
\bibinfo{author}{\bibfnamefont{P.~S.} \bibnamefont{{Joshi}}},
  \bibinfo{author}{\bibfnamefont{D.}~\bibnamefont{{Malafarina}}},
  \bibnamefont{and} \bibinfo{author}{\bibfnamefont{R.~V.}
  \bibnamefont{{Saraykar}}}, \bibinfo{journal}{Int. J. Mod. Phys. D.}
  \textbf{\bibinfo{volume}{21}}, \bibinfo{pages}{1250066}
  (\bibinfo{year}{2012}).

\bibitem[{\citenamefont{{Joshi} et~al.}(2014)\citenamefont{{Joshi},
  {Malafarina}, and {Narayan}}}]{joshimalanarayan2}
\bibinfo{author}{\bibfnamefont{P.~S.} \bibnamefont{{Joshi}}},
  \bibinfo{author}{\bibfnamefont{D.}~\bibnamefont{{Malafarina}}},
  \bibnamefont{and}
  \bibinfo{author}{\bibfnamefont{R.}~\bibnamefont{{Narayan}}},
  \bibinfo{journal}{Class. Quant. Grav.} \textbf{\bibinfo{volume}{31}},
  \bibinfo{pages}{015002} (\bibinfo{year}{2014}).

\bibitem[{\citenamefont{{Faraoni}}(2008)}]{faraoni}
\bibinfo{author}{\bibfnamefont{V.}~\bibnamefont{{Faraoni}}},
  \bibinfo{journal}{ArXiv e-prints}  (\bibinfo{year}{2008}),
  \eprint{0810.2602}.

\bibitem[{\citenamefont{{Clifton} et~al.}(2012)\citenamefont{{Clifton},
  {Ferreira}, {Padilla}, and {Skordis}}}]{clifton}
\bibinfo{author}{\bibfnamefont{T.}~\bibnamefont{{Clifton}}},
  \bibinfo{author}{\bibfnamefont{P.~G.} \bibnamefont{{Ferreira}}},
  \bibinfo{author}{\bibfnamefont{A.}~\bibnamefont{{Padilla}}},
  \bibnamefont{and}
  \bibinfo{author}{\bibfnamefont{C.}~\bibnamefont{{Skordis}}},
  \bibinfo{journal}{Phys. Rep.} \textbf{\bibinfo{volume}{513}},
  \bibinfo{pages}{1} (\bibinfo{year}{2012}).

\bibitem[{\citenamefont{{Sotiriou}}(2007)}]{soti1}
\bibinfo{author}{\bibfnamefont{T.~P.} \bibnamefont{{Sotiriou}}}, Ph.D. thesis
  (\bibinfo{year}{2007}).

\bibitem[{\citenamefont{{Sotiriou} and {Faraoni}}(2010)}]{soti2}
\bibinfo{author}{\bibfnamefont{T.~P.} \bibnamefont{{Sotiriou}}}
  \bibnamefont{and}
  \bibinfo{author}{\bibfnamefont{V.}~\bibnamefont{{Faraoni}}},
  \bibinfo{journal}{Rev. Mod. Phys.} \textbf{\bibinfo{volume}{82}},
  \bibinfo{pages}{451} (\bibinfo{year}{2010}).

\bibitem[{\citenamefont{{Buchbinder} et~al.}(1992)\citenamefont{{Buchbinder},
  {Odintsov}, and {Shapiro}}}]{buchbinder}
\bibinfo{author}{\bibfnamefont{I.~L.} \bibnamefont{{Buchbinder}}},
  \bibinfo{author}{\bibfnamefont{S.~D.} \bibnamefont{{Odintsov}}},
  \bibnamefont{and} \bibinfo{author}{\bibfnamefont{I.~L.}
  \bibnamefont{{Shapiro}}}, \emph{\bibinfo{title}{{Effective action in quantum
  gravity.}}} (\bibinfo{year}{1992}).

\bibitem[{\citenamefont{{Bondi}}(1952)}]{bondibook}
\bibinfo{author}{\bibfnamefont{H.}~\bibnamefont{{Bondi}}},
  \emph{\bibinfo{title}{{Cosmology.}}} (\bibinfo{year}{1952}).

\bibitem[{\citenamefont{Birrell and Davies}(1984)}]{birrell}
\bibinfo{author}{\bibfnamefont{N.~D.} \bibnamefont{Birrell}} \bibnamefont{and}
  \bibinfo{author}{\bibfnamefont{P.~C.~W.} \bibnamefont{Davies}},
  \emph{\bibinfo{title}{{Quantum Fields in Curved Space}}}, Cambridge
  Monographs on Mathematical Physics (\bibinfo{publisher}{Cambridge Univ.
  Press}, \bibinfo{address}{Cambridge, UK}, \bibinfo{year}{1984}).

\bibitem[{\citenamefont{Vilkovisky}(1992)}]{vilkov}
\bibinfo{author}{\bibfnamefont{G.~A.} \bibnamefont{Vilkovisky}},
  \bibinfo{journal}{Classical and Quantum Gravity}
  \textbf{\bibinfo{volume}{9}}, \bibinfo{pages}{895} (\bibinfo{year}{1992}).

\bibitem[{\citenamefont{Gasperini and Veneziano}(1992)}]{gasperini}
\bibinfo{author}{\bibfnamefont{M.}~\bibnamefont{Gasperini}} \bibnamefont{and}
  \bibinfo{author}{\bibfnamefont{G.}~\bibnamefont{Veneziano}},
  \bibinfo{journal}{Phys. Lett.} \textbf{\bibinfo{volume}{B277}},
  \bibinfo{pages}{256} (\bibinfo{year}{1992}).

\bibitem[{\citenamefont{Capozziello
  et~al.}(2010{\natexlab{a}})\citenamefont{Capozziello, De~Laurentis, and
  Faraoni}}]{birdseye}
\bibinfo{author}{\bibfnamefont{S.}~\bibnamefont{Capozziello}},
  \bibinfo{author}{\bibfnamefont{M.}~\bibnamefont{De~Laurentis}},
  \bibnamefont{and} \bibinfo{author}{\bibfnamefont{V.}~\bibnamefont{Faraoni}},
  \bibinfo{journal}{Open Astron. J.} \textbf{\bibinfo{volume}{3}},
  \bibinfo{pages}{49} (\bibinfo{year}{2010}{\natexlab{a}}).

\bibitem[{\citenamefont{{Starobinsky, A. A.}}(1980)}]{starob}
\bibinfo{author}{\bibnamefont{{Starobinsky, A. A.}}}, \bibinfo{journal}{Phys.
  Lett. B.} \textbf{\bibinfo{volume}{91}}, \bibinfo{pages}{99}
  (\bibinfo{year}{1980}).

\bibitem[{\citenamefont{{Kazanas}}(1980)}]{kazanas}
\bibinfo{author}{\bibfnamefont{D.}~\bibnamefont{{Kazanas}}},
  \bibinfo{journal}{Astrophys. J.} \textbf{\bibinfo{volume}{241}},
  \bibinfo{pages}{L59} (\bibinfo{year}{1980}).

\bibitem[{\citenamefont{Guth}(1981)}]{guth}
\bibinfo{author}{\bibfnamefont{A.~H.} \bibnamefont{Guth}},
  \bibinfo{journal}{Phys. Rev. D.} \textbf{\bibinfo{volume}{23}},
  \bibinfo{pages}{347} (\bibinfo{year}{1981}).

\bibitem[{\citenamefont{{Sato}}(1981)}]{sato}
\bibinfo{author}{\bibfnamefont{K.}~\bibnamefont{{Sato}}},
  \bibinfo{journal}{Mon. Not. Roy. Astron. Soc.}
  \textbf{\bibinfo{volume}{195}}, \bibinfo{pages}{467} (\bibinfo{year}{1981}).

\bibitem[{\citenamefont{Liddle and Lyth}(2000)}]{liddle}
\bibinfo{author}{\bibfnamefont{A.~R.} \bibnamefont{Liddle}} \bibnamefont{and}
  \bibinfo{author}{\bibfnamefont{D.~H.} \bibnamefont{Lyth}},
  \emph{\bibinfo{title}{Cosmological inflation and Large-Scale Structure}}
  (\bibinfo{publisher}{Cambridge University Press; New York},
  \bibinfo{year}{2000}).

\bibitem[{\citenamefont{Lyth and Riotto}(1999)}]{lyth}
\bibinfo{author}{\bibfnamefont{D.~H.} \bibnamefont{Lyth}} \bibnamefont{and}
  \bibinfo{author}{\bibfnamefont{A.}~\bibnamefont{Riotto}},
  \bibinfo{journal}{Phys. Rep.} \textbf{\bibinfo{volume}{314}},
  \bibinfo{pages}{1} (\bibinfo{year}{1999}).

\bibitem[{\citenamefont{Bassett
  et~al.}(2006{\natexlab{a}})\citenamefont{Bassett, Tsujikawa, and
  Wands}}]{bassette}
\bibinfo{author}{\bibfnamefont{B.~A.} \bibnamefont{Bassett}},
  \bibinfo{author}{\bibfnamefont{S.}~\bibnamefont{Tsujikawa}},
  \bibnamefont{and} \bibinfo{author}{\bibfnamefont{D.}~\bibnamefont{Wands}},
  \bibinfo{journal}{Rev. Mod. Phys.} \textbf{\bibinfo{volume}{78}},
  \bibinfo{pages}{537} (\bibinfo{year}{2006}{\natexlab{a}}).

\bibitem[{\citenamefont{{Smoot} et~al.}(1992)\citenamefont{{Smoot}, {Bennett},
  {Kogut}, {Wright}, {Aymon}, {Boggess}, {Cheng}, {de Amici}, {Gulkis},
  {Hauser} et~al.}}]{smoot}
\bibinfo{author}{\bibfnamefont{G.~F.} \bibnamefont{{Smoot}}},
  \bibinfo{author}{\bibfnamefont{C.~L.} \bibnamefont{{Bennett}}},
  \bibinfo{author}{\bibfnamefont{A.}~\bibnamefont{{Kogut}}},
  \bibinfo{author}{\bibfnamefont{E.~L.} \bibnamefont{{Wright}}},
  \bibinfo{author}{\bibfnamefont{J.}~\bibnamefont{{Aymon}}},
  \bibinfo{author}{\bibfnamefont{N.~W.} \bibnamefont{{Boggess}}},
  \bibinfo{author}{\bibfnamefont{E.~S.} \bibnamefont{{Cheng}}},
  \bibinfo{author}{\bibfnamefont{G.}~\bibnamefont{{de Amici}}},
  \bibinfo{author}{\bibfnamefont{S.}~\bibnamefont{{Gulkis}}},
  \bibinfo{author}{\bibfnamefont{M.~G.} \bibnamefont{{Hauser}}},
  \bibnamefont{et~al.}, \bibinfo{journal}{Astrophys. J. Lett.}
  \textbf{\bibinfo{volume}{396}}, \bibinfo{pages}{L1} (\bibinfo{year}{1992}).

\bibitem[{\citenamefont{Huterer and Turner}(1999)}]{huterer}
\bibinfo{author}{\bibfnamefont{D.}~\bibnamefont{Huterer}} \bibnamefont{and}
  \bibinfo{author}{\bibfnamefont{M.~S.} \bibnamefont{Turner}},
  \bibinfo{journal}{Phys. Rev. D.} \textbf{\bibinfo{volume}{60}},
  \bibinfo{pages}{081301} (\bibinfo{year}{1999}).

\bibitem[{\citenamefont{Sahni and Starobinsky}(2000)}]{sahni}
\bibinfo{author}{\bibfnamefont{V.}~\bibnamefont{Sahni}} \bibnamefont{and}
  \bibinfo{author}{\bibfnamefont{A.}~\bibnamefont{Starobinsky}},
  \bibinfo{journal}{Int. J. Mod. Phys. D} \textbf{\bibinfo{volume}{09}},
  \bibinfo{pages}{373} (\bibinfo{year}{2000}).

\bibitem[{\citenamefont{Padmanabhan}(2003)}]{paddy}
\bibinfo{author}{\bibfnamefont{T.}~\bibnamefont{Padmanabhan}},
  \bibinfo{journal}{Phys. Rep.} \textbf{\bibinfo{volume}{380}},
  \bibinfo{pages}{235} (\bibinfo{year}{2003}).

\bibitem[{\citenamefont{Peebles and Ratra}(2003)}]{peebles}
\bibinfo{author}{\bibfnamefont{P.~J.~E.} \bibnamefont{Peebles}}
  \bibnamefont{and} \bibinfo{author}{\bibfnamefont{B.}~\bibnamefont{Ratra}},
  \bibinfo{journal}{Rev. Mod. Phys.} \textbf{\bibinfo{volume}{75}},
  \bibinfo{pages}{559} (\bibinfo{year}{2003}).

\bibitem[{\citenamefont{Copeland et~al.}(2006)\citenamefont{Copeland, Sami, and
  Tsujikawa}}]{copeland}
\bibinfo{author}{\bibfnamefont{E.~J.} \bibnamefont{Copeland}},
  \bibinfo{author}{\bibfnamefont{M.}~\bibnamefont{Sami}}, \bibnamefont{and}
  \bibinfo{author}{\bibfnamefont{S.}~\bibnamefont{Tsujikawa}},
  \bibinfo{journal}{Int. J. Mod. Phys. D.} \textbf{\bibinfo{volume}{15}},
  \bibinfo{pages}{1753} (\bibinfo{year}{2006}).

\bibitem[{\citenamefont{Weinberg}(1989)}]{weinberg}
\bibinfo{author}{\bibfnamefont{S.}~\bibnamefont{Weinberg}},
  \bibinfo{journal}{Rev. Mod. Phys.} \textbf{\bibinfo{volume}{61}},
  \bibinfo{pages}{1} (\bibinfo{year}{1989}).

\bibitem[{\citenamefont{{de la Cruz-Dombriz}}(2010)}]{cruz}
\bibinfo{author}{\bibfnamefont{A.}~\bibnamefont{{de la Cruz-Dombriz}}},
  \bibinfo{journal}{ArXiv e-prints}  (\bibinfo{year}{2010}),
  \eprint{1004.5052}.

\bibitem[{\citenamefont{Sotiriou and Liberati}(2007)}]{soti3}
\bibinfo{author}{\bibfnamefont{T.~P.} \bibnamefont{Sotiriou}} \bibnamefont{and}
  \bibinfo{author}{\bibfnamefont{S.}~\bibnamefont{Liberati}},
  \bibinfo{journal}{Ann. Phys.} \textbf{\bibinfo{volume}{322}},
  \bibinfo{pages}{935} (\bibinfo{year}{2007}).

\bibitem[{\citenamefont{Dolgov and Kawasaki}(2003)}]{dolgov}
\bibinfo{author}{\bibfnamefont{A.~D.} \bibnamefont{Dolgov}} \bibnamefont{and}
  \bibinfo{author}{\bibfnamefont{M.}~\bibnamefont{Kawasaki}},
  \bibinfo{journal}{Phys. Lett. B.} \textbf{\bibinfo{volume}{573}},
  \bibinfo{pages}{1} (\bibinfo{year}{2003}).

\bibitem[{\citenamefont{Nojiri and Odintsov}(2003)}]{nojiodi1}
\bibinfo{author}{\bibfnamefont{S.}~\bibnamefont{Nojiri}} \bibnamefont{and}
  \bibinfo{author}{\bibfnamefont{S.~D.} \bibnamefont{Odintsov}},
  \bibinfo{journal}{Phys. Rev. D.} \textbf{\bibinfo{volume}{68}},
  \bibinfo{pages}{123512} (\bibinfo{year}{2003}).

\bibitem[{\citenamefont{Nojiri and Odintsov}(2004)}]{nojiodi2}
\bibinfo{author}{\bibfnamefont{S.}~\bibnamefont{Nojiri}} \bibnamefont{and}
  \bibinfo{author}{\bibfnamefont{S.~D.} \bibnamefont{Odintsov}},
  \bibinfo{journal}{Gen. Rel. Grav.} \textbf{\bibinfo{volume}{36}},
  \bibinfo{pages}{1765} (\bibinfo{year}{2004}).

\bibitem[{\citenamefont{Baghram et~al.}(2007)\citenamefont{Baghram, Farhang,
  and Rahvar}}]{bagh}
\bibinfo{author}{\bibfnamefont{S.}~\bibnamefont{Baghram}},
  \bibinfo{author}{\bibfnamefont{M.}~\bibnamefont{Farhang}}, \bibnamefont{and}
  \bibinfo{author}{\bibfnamefont{S.}~\bibnamefont{Rahvar}},
  \bibinfo{journal}{Phys. Rev. D.} \textbf{\bibinfo{volume}{75}},
  \bibinfo{pages}{044024} (\bibinfo{year}{2007}).

\bibitem[{\citenamefont{Faraoni}(2006)}]{faraoni1}
\bibinfo{author}{\bibfnamefont{V.}~\bibnamefont{Faraoni}},
  \bibinfo{journal}{Phys. Rev. D.} \textbf{\bibinfo{volume}{74}},
  \bibinfo{pages}{104017} (\bibinfo{year}{2006}).

\bibitem[{\citenamefont{{Cognola} and {Zerbini}}(2008)}]{cognola}
\bibinfo{author}{\bibfnamefont{G.}~\bibnamefont{{Cognola}}} \bibnamefont{and}
  \bibinfo{author}{\bibfnamefont{S.}~\bibnamefont{{Zerbini}}},
  \bibinfo{journal}{Int. J. Theor. Phys.} \textbf{\bibinfo{volume}{47}},
  \bibinfo{pages}{3186} (\bibinfo{year}{2008}).

\bibitem[{\citenamefont{Faraoni}(2007)}]{faraoni2}
\bibinfo{author}{\bibfnamefont{V.}~\bibnamefont{Faraoni}},
  \bibinfo{journal}{Phys. Rev. D.} \textbf{\bibinfo{volume}{75}},
  \bibinfo{pages}{067302} (\bibinfo{year}{2007}).

\bibitem[{\citenamefont{Seifert}(2007)}]{mdseifert}
\bibinfo{author}{\bibfnamefont{M.~D.} \bibnamefont{Seifert}},
  \bibinfo{journal}{Phys. Rev. D.} \textbf{\bibinfo{volume}{76}},
  \bibinfo{pages}{064002} (\bibinfo{year}{2007}).

\bibitem[{\citenamefont{Sawicki and Hu}(2007)}]{husawi}
\bibinfo{author}{\bibfnamefont{I.}~\bibnamefont{Sawicki}} \bibnamefont{and}
  \bibinfo{author}{\bibfnamefont{W.}~\bibnamefont{Hu}}, \bibinfo{journal}{Phys.
  Rev. D.} \textbf{\bibinfo{volume}{75}}, \bibinfo{pages}{127502}
  (\bibinfo{year}{2007}).

\bibitem[{\citenamefont{Salgado}(2006)}]{salgado1}
\bibinfo{author}{\bibfnamefont{M.}~\bibnamefont{Salgado}},
  \bibinfo{journal}{Class. Quant. Grav.} \textbf{\bibinfo{volume}{23}},
  \bibinfo{pages}{4719} (\bibinfo{year}{2006}).

\bibitem[{\citenamefont{{Salgado} et~al.}(2008)\citenamefont{{Salgado},
  {Martinez}~del Rio, {Alcubierre}, and {Nunez}}}]{salgado2}
\bibinfo{author}{\bibfnamefont{M.}~\bibnamefont{{Salgado}}},
  \bibinfo{author}{\bibfnamefont{D.}~\bibnamefont{{Martinez}~del Rio}},
  \bibinfo{author}{\bibfnamefont{M.}~\bibnamefont{{Alcubierre}}},
  \bibnamefont{and} \bibinfo{author}{\bibfnamefont{D.}~\bibnamefont{{Nunez}}},
  \bibinfo{journal}{Phys. Rev. D.} \textbf{\bibinfo{volume}{77}},
  \bibinfo{pages}{104010} (\bibinfo{year}{2008}).

\bibitem[{\citenamefont{Noakes}(1983)}]{noakes}
\bibinfo{author}{\bibfnamefont{D.~R.} \bibnamefont{Noakes}},
  \bibinfo{journal}{J. Math. Phys.} \textbf{\bibinfo{volume}{24}},
  \bibinfo{pages}{1846} (\bibinfo{year}{1983}).

\bibitem[{\citenamefont{Cocke and Cohen}(1968)}]{cocke}
\bibinfo{author}{\bibfnamefont{W.~J.} \bibnamefont{Cocke}} \bibnamefont{and}
  \bibinfo{author}{\bibfnamefont{J.~M.} \bibnamefont{Cohen}},
  \bibinfo{journal}{J. Math. Phys.} \textbf{\bibinfo{volume}{9}},
  \bibinfo{pages}{971} (\bibinfo{year}{1968}).

\bibitem[{\citenamefont{Nunez and Solganik}(2004)}]{nunez}
\bibinfo{author}{\bibfnamefont{A.}~\bibnamefont{Nunez}} \bibnamefont{and}
  \bibinfo{author}{\bibfnamefont{S.}~\bibnamefont{Solganik}}
  (\bibinfo{year}{2004}), \eprint{hep-th/0403159}.

\bibitem[{\citenamefont{Amendola
  et~al.}(2007{\natexlab{a}})\citenamefont{Amendola, Polarski, and
  Tsujikawa}}]{amendola1}
\bibinfo{author}{\bibfnamefont{L.}~\bibnamefont{Amendola}},
  \bibinfo{author}{\bibfnamefont{D.}~\bibnamefont{Polarski}}, \bibnamefont{and}
  \bibinfo{author}{\bibfnamefont{S.}~\bibnamefont{Tsujikawa}},
  \bibinfo{journal}{Phys. Rev. Lett.} \textbf{\bibinfo{volume}{98}},
  \bibinfo{pages}{131302} (\bibinfo{year}{2007}{\natexlab{a}}).

\bibitem[{\citenamefont{Amendola
  et~al.}(2007{\natexlab{b}})\citenamefont{Amendola, Gannouji, Polarski, and
  Tsujikawa}}]{amendola2}
\bibinfo{author}{\bibfnamefont{L.}~\bibnamefont{Amendola}},
  \bibinfo{author}{\bibfnamefont{R.}~\bibnamefont{Gannouji}},
  \bibinfo{author}{\bibfnamefont{D.}~\bibnamefont{Polarski}}, \bibnamefont{and}
  \bibinfo{author}{\bibfnamefont{S.}~\bibnamefont{Tsujikawa}},
  \bibinfo{journal}{Phys. Rev. D.} \textbf{\bibinfo{volume}{75}},
  \bibinfo{pages}{083504} (\bibinfo{year}{2007}{\natexlab{b}}).

\bibitem[{\citenamefont{Capozziello et~al.}(2006)\citenamefont{Capozziello,
  Nojiri, Odintsov, and Troisi}}]{capo1}
\bibinfo{author}{\bibfnamefont{S.}~\bibnamefont{Capozziello}},
  \bibinfo{author}{\bibfnamefont{S.}~\bibnamefont{Nojiri}},
  \bibinfo{author}{\bibfnamefont{S.~D.} \bibnamefont{Odintsov}},
  \bibnamefont{and} \bibinfo{author}{\bibfnamefont{A.}~\bibnamefont{Troisi}},
  \bibinfo{journal}{Phys. Lett. B} \textbf{\bibinfo{volume}{639}},
  \bibinfo{pages}{135} (\bibinfo{year}{2006}).

\bibitem[{\citenamefont{Song et~al.}(2007)\citenamefont{Song, Hu, and
  Sawicki}}]{hu}
\bibinfo{author}{\bibfnamefont{Y.~S.} \bibnamefont{Song}},
  \bibinfo{author}{\bibfnamefont{W.}~\bibnamefont{Hu}}, \bibnamefont{and}
  \bibinfo{author}{\bibfnamefont{I.}~\bibnamefont{Sawicki}},
  \bibinfo{journal}{Phys. Rev. D.} \textbf{\bibinfo{volume}{75}},
  \bibinfo{pages}{044004} (\bibinfo{year}{2007}).

\bibitem[{\citenamefont{{Sokolowski}}(2007)}]{soko}
\bibinfo{author}{\bibfnamefont{L.~M.} \bibnamefont{{Sokolowski}}},
  \bibinfo{journal}{Class. Quant. Grav.} \textbf{\bibinfo{volume}{24}},
  \bibinfo{pages}{3391} (\bibinfo{year}{2007}).

\bibitem[{\citenamefont{Faulkner et~al.}(2007)\citenamefont{Faulkner, Tegmark,
  Bunn, and Mao}}]{faulk}
\bibinfo{author}{\bibfnamefont{T.}~\bibnamefont{Faulkner}},
  \bibinfo{author}{\bibfnamefont{M.}~\bibnamefont{Tegmark}},
  \bibinfo{author}{\bibfnamefont{E.~F.} \bibnamefont{Bunn}}, \bibnamefont{and}
  \bibinfo{author}{\bibfnamefont{Y.}~\bibnamefont{Mao}},
  \bibinfo{journal}{Phys. Rev. D.} \textbf{\bibinfo{volume}{76}},
  \bibinfo{pages}{063505} (\bibinfo{year}{2007}).

\bibitem[{\citenamefont{{Will}}(1993)}]{willbook}
\bibinfo{author}{\bibfnamefont{C.~M.} \bibnamefont{{Will}}},
  \emph{\bibinfo{title}{{Theory and Experiment in Gravitational Physics}}}
  (\bibinfo{year}{1993}).

\bibitem[{\citenamefont{Stelle}(1978)}]{stellegrg}
\bibinfo{author}{\bibfnamefont{K.~S.} \bibnamefont{Stelle}},
  \bibinfo{journal}{Gen. Rel. Grav.} \textbf{\bibinfo{volume}{9}},
  \bibinfo{pages}{353} (\bibinfo{year}{1978}).

\bibitem[{\citenamefont{Sanders and McGaugh}(2002)}]{sanders1}
\bibinfo{author}{\bibfnamefont{R.~H.} \bibnamefont{Sanders}} \bibnamefont{and}
  \bibinfo{author}{\bibfnamefont{S.~S.} \bibnamefont{McGaugh}},
  \bibinfo{journal}{Ann. Rev. Astron. Astrophys.}
  \textbf{\bibinfo{volume}{40}}, \bibinfo{pages}{263} (\bibinfo{year}{2002}).

\bibitem[{\citenamefont{Sanders}(2001{\natexlab{a}})}]{sanders2}
\bibinfo{author}{\bibfnamefont{R.~H.} \bibnamefont{Sanders}}
  (\bibinfo{year}{2001}{\natexlab{a}}), \eprint{astro-ph/0106558}.

\bibitem[{\citenamefont{Sanders}(2001{\natexlab{b}})}]{sanders3}
\bibinfo{author}{\bibfnamefont{R.~H.} \bibnamefont{Sanders}},
  \bibinfo{journal}{Astrophys. J.} \textbf{\bibinfo{volume}{560}},
  \bibinfo{pages}{1} (\bibinfo{year}{2001}{\natexlab{b}}).

\bibitem[{\citenamefont{Capozziello}(2004)}]{caponewton}
\bibinfo{author}{\bibfnamefont{S.}~\bibnamefont{Capozziello}},
  \bibinfo{journal}{Submitted to: GR-QC}  (\bibinfo{year}{2004}),
  \eprint{gr-qc/0412088}.

\bibitem[{\citenamefont{Chiba et~al.}(2007)\citenamefont{Chiba, Smith, and
  Erickcek}}]{chiba}
\bibinfo{author}{\bibfnamefont{T.}~\bibnamefont{Chiba}},
  \bibinfo{author}{\bibfnamefont{T.~L.} \bibnamefont{Smith}}, \bibnamefont{and}
  \bibinfo{author}{\bibfnamefont{A.~L.} \bibnamefont{Erickcek}},
  \bibinfo{journal}{Phys. Rev. D.} \textbf{\bibinfo{volume}{75}},
  \bibinfo{pages}{124014} (\bibinfo{year}{2007}).

\bibitem[{\citenamefont{Olmo}(2007)}]{olmo}
\bibinfo{author}{\bibfnamefont{G.~J.} \bibnamefont{Olmo}},
  \bibinfo{journal}{Phys. Rev. D.} \textbf{\bibinfo{volume}{75}},
  \bibinfo{pages}{023511} (\bibinfo{year}{2007}).

\bibitem[{\citenamefont{Clifton and Barrow}(2005)}]{cliftonbarrowweakfield}
\bibinfo{author}{\bibfnamefont{T.}~\bibnamefont{Clifton}} \bibnamefont{and}
  \bibinfo{author}{\bibfnamefont{J.~D.} \bibnamefont{Barrow}},
  \bibinfo{journal}{Phys. Rev. D.} \textbf{\bibinfo{volume}{72}},
  \bibinfo{pages}{103005} (\bibinfo{year}{2005}).

\bibitem[{\citenamefont{Paul et~al.}(2009)\citenamefont{Paul, Debnath, and
  Ghose}}]{paul}
\bibinfo{author}{\bibfnamefont{B.~C.} \bibnamefont{Paul}},
  \bibinfo{author}{\bibfnamefont{P.~S.} \bibnamefont{Debnath}},
  \bibnamefont{and} \bibinfo{author}{\bibfnamefont{S.}~\bibnamefont{Ghose}},
  \bibinfo{journal}{ArXiv e-prints}  (\bibinfo{year}{2009}),
  \eprint{0904.0345v1}.

\bibitem[{\citenamefont{Das et~al.}(2006)\citenamefont{Das, Banerjee, and
  Dadhich}}]{dasbanerjee}
\bibinfo{author}{\bibfnamefont{S.}~\bibnamefont{Das}},
  \bibinfo{author}{\bibfnamefont{N.}~\bibnamefont{Banerjee}}, \bibnamefont{and}
  \bibinfo{author}{\bibfnamefont{N.}~\bibnamefont{Dadhich}},
  \bibinfo{journal}{Class. Quant. Grav.} \textbf{\bibinfo{volume}{23}},
  \bibinfo{pages}{4159} (\bibinfo{year}{2006}).

\bibitem[{\citenamefont{Nzioki et~al.}(2010)\citenamefont{Nzioki, Carloni,
  Goswami, and Dunsby}}]{nzioki}
\bibinfo{author}{\bibfnamefont{A.~M.} \bibnamefont{Nzioki}},
  \bibinfo{author}{\bibfnamefont{S.}~\bibnamefont{Carloni}},
  \bibinfo{author}{\bibfnamefont{R.}~\bibnamefont{Goswami}}, \bibnamefont{and}
  \bibinfo{author}{\bibfnamefont{P.~K.~S.} \bibnamefont{Dunsby}},
  \bibinfo{journal}{Phys. Rev. D.} \textbf{\bibinfo{volume}{81}},
  \bibinfo{pages}{084028} (\bibinfo{year}{2010}).

\bibitem[{\citenamefont{Cembranos et~al.}(2012)\citenamefont{Cembranos,
  Cruz-Dombriz, and Montes-Nunez}}]{cembra}
\bibinfo{author}{\bibfnamefont{J.~A.~R.} \bibnamefont{Cembranos}},
  \bibinfo{author}{\bibfnamefont{A.~d.~l.} \bibnamefont{Cruz-Dombriz}},
  \bibnamefont{and}
  \bibinfo{author}{\bibfnamefont{B.}~\bibnamefont{Montes-Nunez}},
  \bibinfo{journal}{J.C.A.P} \textbf{\bibinfo{volume}{1204}},
  \bibinfo{pages}{021} (\bibinfo{year}{2012}).

\bibitem[{\citenamefont{Ghosh and Maharaj}(2012)}]{ghosh}
\bibinfo{author}{\bibfnamefont{S.~G.} \bibnamefont{Ghosh}} \bibnamefont{and}
  \bibinfo{author}{\bibfnamefont{S.~D.} \bibnamefont{Maharaj}},
  \bibinfo{journal}{Phys. Rev. D.} \textbf{\bibinfo{volume}{85}},
  \bibinfo{pages}{124064} (\bibinfo{year}{2012}).

\bibitem[{\citenamefont{Capozziello
  et~al.}(2010{\natexlab{b}})\citenamefont{Capozziello, De~Laurentis, and
  Stabile}}]{capo2}
\bibinfo{author}{\bibfnamefont{S.}~\bibnamefont{Capozziello}},
  \bibinfo{author}{\bibfnamefont{M.}~\bibnamefont{De~Laurentis}},
  \bibnamefont{and} \bibinfo{author}{\bibfnamefont{A.}~\bibnamefont{Stabile}},
  \bibinfo{journal}{Class. Quant. Grav.} \textbf{\bibinfo{volume}{27}},
  \bibinfo{pages}{165008} (\bibinfo{year}{2010}{\natexlab{b}}).

\bibitem[{\citenamefont{Multamaki and Vilja}(2006)}]{multamaki}
\bibinfo{author}{\bibfnamefont{T.}~\bibnamefont{Multamaki}} \bibnamefont{and}
  \bibinfo{author}{\bibfnamefont{I.}~\bibnamefont{Vilja}},
  \bibinfo{journal}{Phys. Rev. D.} \textbf{\bibinfo{volume}{74}},
  \bibinfo{pages}{064022} (\bibinfo{year}{2006}).

\bibitem[{\citenamefont{Cognola et~al.}(2005)\citenamefont{Cognola, Elizalde,
  Nojiri, Odintsov, and Zerbini}}]{cognola1}
\bibinfo{author}{\bibfnamefont{G.}~\bibnamefont{Cognola}},
  \bibinfo{author}{\bibfnamefont{E.}~\bibnamefont{Elizalde}},
  \bibinfo{author}{\bibfnamefont{S.}~\bibnamefont{Nojiri}},
  \bibinfo{author}{\bibfnamefont{S.~D.} \bibnamefont{Odintsov}},
  \bibnamefont{and} \bibinfo{author}{\bibfnamefont{S.}~\bibnamefont{Zerbini}},
  \bibinfo{journal}{J.C.A.P.} \textbf{\bibinfo{volume}{2005}},
  \bibinfo{pages}{010} (\bibinfo{year}{2005}).

\bibitem[{\citenamefont{de~la Cruz-Dombriz et~al.}(2009)\citenamefont{de~la
  Cruz-Dombriz, Dobado, and Maroto}}]{delacruz}
\bibinfo{author}{\bibfnamefont{A.}~\bibnamefont{de~la Cruz-Dombriz}},
  \bibinfo{author}{\bibfnamefont{A.}~\bibnamefont{Dobado}}, \bibnamefont{and}
  \bibinfo{author}{\bibfnamefont{A.~L.} \bibnamefont{Maroto}},
  \bibinfo{journal}{Phys. Rev. D} \textbf{\bibinfo{volume}{80}},
  \bibinfo{pages}{124011} (\bibinfo{year}{2009}).

\bibitem[{\citenamefont{Sebastiani and Zerbini}(2011)}]{sebastiani}
\bibinfo{author}{\bibfnamefont{L.}~\bibnamefont{Sebastiani}} \bibnamefont{and}
  \bibinfo{author}{\bibfnamefont{S.}~\bibnamefont{Zerbini}},
  \bibinfo{journal}{Eur. Phys. J. C.} \textbf{\bibinfo{volume}{71}},
  \bibinfo{pages}{1} (\bibinfo{year}{2011}).

\bibitem[{\citenamefont{Bergliaffa and de~Oliveira~Nunes}(2011)}]{berg}
\bibinfo{author}{\bibfnamefont{S.~E.~P.} \bibnamefont{Bergliaffa}}
  \bibnamefont{and} \bibinfo{author}{\bibfnamefont{Y.~E.~C.}
  \bibnamefont{de~Oliveira~Nunes}}, \bibinfo{journal}{Phys. Rev. D.}
  \textbf{\bibinfo{volume}{84}}, \bibinfo{pages}{084006}
  (\bibinfo{year}{2011}).

\bibitem[{\citenamefont{Aghmohammadi et~al.}(2010)\citenamefont{Aghmohammadi,
  Saaidi, Abolhassani, and Vajdi}}]{agha}
\bibinfo{author}{\bibfnamefont{A.}~\bibnamefont{Aghmohammadi}},
  \bibinfo{author}{\bibfnamefont{K.}~\bibnamefont{Saaidi}},
  \bibinfo{author}{\bibfnamefont{M.~R.} \bibnamefont{Abolhassani}},
  \bibnamefont{and} \bibinfo{author}{\bibfnamefont{A.}~\bibnamefont{Vajdi}},
  \bibinfo{journal}{Int. J. Theor. Phys.} \textbf{\bibinfo{volume}{49}},
  \bibinfo{pages}{709} (\bibinfo{year}{2010}).

\bibitem[{\citenamefont{Moon et~al.}(2011)\citenamefont{Moon, Myung, and
  Son}}]{moon}
\bibinfo{author}{\bibfnamefont{T.}~\bibnamefont{Moon}},
  \bibinfo{author}{\bibfnamefont{Y.~S.} \bibnamefont{Myung}}, \bibnamefont{and}
  \bibinfo{author}{\bibfnamefont{E.~J.} \bibnamefont{Son}},
  \bibinfo{journal}{Gen. Rel. Grav.} \textbf{\bibinfo{volume}{43}},
  \bibinfo{pages}{3079} (\bibinfo{year}{2011}).

\bibitem[{\citenamefont{Mazharimousavi and Halilsoy}(2011)}]{mazha1}
\bibinfo{author}{\bibfnamefont{S.~H.} \bibnamefont{Mazharimousavi}}
  \bibnamefont{and} \bibinfo{author}{\bibfnamefont{M.}~\bibnamefont{Halilsoy}},
  \bibinfo{journal}{Phys. Rev. D.} \textbf{\bibinfo{volume}{84}},
  \bibinfo{pages}{064032} (\bibinfo{year}{2011}).

\bibitem[{\citenamefont{Mazharimousavi
  et~al.}(2012)\citenamefont{Mazharimousavi, Halilsoy, and Tahamtan}}]{mazha2}
\bibinfo{author}{\bibfnamefont{S.~H.} \bibnamefont{Mazharimousavi}},
  \bibinfo{author}{\bibfnamefont{M.}~\bibnamefont{Halilsoy}}, \bibnamefont{and}
  \bibinfo{author}{\bibfnamefont{T.}~\bibnamefont{Tahamtan}},
  \bibinfo{journal}{Eur. Phys. J. C.} \textbf{\bibinfo{volume}{72}},
  \bibinfo{pages}{1} (\bibinfo{year}{2012}).

\bibitem[{\citenamefont{Hollenstein and Lobo}(2008)}]{lobo}
\bibinfo{author}{\bibfnamefont{L.}~\bibnamefont{Hollenstein}} \bibnamefont{and}
  \bibinfo{author}{\bibfnamefont{F.~S.~N.} \bibnamefont{Lobo}},
  \bibinfo{journal}{Phys. Rev. D.} \textbf{\bibinfo{volume}{78}},
  \bibinfo{pages}{124007} (\bibinfo{year}{2008}).

\bibitem[{\citenamefont{Hendi and Momeni}(2011)}]{hendi1}
\bibinfo{author}{\bibfnamefont{S.~H.} \bibnamefont{Hendi}} \bibnamefont{and}
  \bibinfo{author}{\bibfnamefont{D.}~\bibnamefont{Momeni}},
  \bibinfo{journal}{Eur. Phys. J. C.} \textbf{\bibinfo{volume}{71}},
  \bibinfo{pages}{1} (\bibinfo{year}{2011}).

\bibitem[{\citenamefont{Hendi et~al.}(2012)\citenamefont{Hendi, Panah, and
  Mousavi}}]{hendi2}
\bibinfo{author}{\bibfnamefont{S.~H.} \bibnamefont{Hendi}},
  \bibinfo{author}{\bibfnamefont{B.~E.} \bibnamefont{Panah}}, \bibnamefont{and}
  \bibinfo{author}{\bibfnamefont{S.~M.} \bibnamefont{Mousavi}},
  \bibinfo{journal}{Gen. Rel. Grav.} \textbf{\bibinfo{volume}{44}},
  \bibinfo{pages}{835} (\bibinfo{year}{2012}).

\bibitem[{\citenamefont{Capozziello et~al.}(2007)\citenamefont{Capozziello,
  Stabile, and Troisi}}]{troisi}
\bibinfo{author}{\bibfnamefont{S.}~\bibnamefont{Capozziello}},
  \bibinfo{author}{\bibfnamefont{A.}~\bibnamefont{Stabile}}, \bibnamefont{and}
  \bibinfo{author}{\bibfnamefont{A.}~\bibnamefont{Troisi}},
  \bibinfo{journal}{Phys. Rev. D.} \textbf{\bibinfo{volume}{76}},
  \bibinfo{pages}{104019} (\bibinfo{year}{2007}).

\bibitem[{\citenamefont{Faraoni}(2010)}]{vale}
\bibinfo{author}{\bibfnamefont{V.}~\bibnamefont{Faraoni}},
  \bibinfo{journal}{Phys. Rev. D.} \textbf{\bibinfo{volume}{81}},
  \bibinfo{pages}{044002} (\bibinfo{year}{2010}).

\bibitem[{\citenamefont{{Capozziello} and
  {S{\'a}ez-G{\'o}mez}}(2012)}]{saezgomez}
\bibinfo{author}{\bibfnamefont{S.}~\bibnamefont{{Capozziello}}}
  \bibnamefont{and}
  \bibinfo{author}{\bibfnamefont{D.}~\bibnamefont{{S{\'a}ez-G{\'o}mez}}},
  \bibinfo{journal}{Ann. der Phys.} \textbf{\bibinfo{volume}{524}},
  \bibinfo{pages}{279} (\bibinfo{year}{2012}).

\bibitem[{\citenamefont{{Bamba} et~al.}(2011)\citenamefont{{Bamba}, {Nojiri},
  and {Odintsov}}}]{bamba}
\bibinfo{author}{\bibfnamefont{K.}~\bibnamefont{{Bamba}}},
  \bibinfo{author}{\bibfnamefont{S.}~\bibnamefont{{Nojiri}}}, \bibnamefont{and}
  \bibinfo{author}{\bibfnamefont{S.~D.} \bibnamefont{{Odintsov}}},
  \bibinfo{journal}{Phys. Lett. B.} \textbf{\bibinfo{volume}{698}},
  \bibinfo{pages}{451} (\bibinfo{year}{2011}).

\bibitem[{\citenamefont{{Arbuzova} and {Dolgov}}(2011)}]{arbuzo}
\bibinfo{author}{\bibfnamefont{E.~V.} \bibnamefont{{Arbuzova}}}
  \bibnamefont{and} \bibinfo{author}{\bibfnamefont{A.~D.}
  \bibnamefont{{Dolgov}}}, \bibinfo{journal}{Phys. Lett. B.}
  \textbf{\bibinfo{volume}{700}}, \bibinfo{pages}{289} (\bibinfo{year}{2011}).

\bibitem[{\citenamefont{{Santos}}(2012)}]{esantos}
\bibinfo{author}{\bibfnamefont{E.}~\bibnamefont{{Santos}}},
  \bibinfo{journal}{ArXiv e-prints}  (\bibinfo{year}{2012}),
  \eprint{1104.2140}.

\bibitem[{\citenamefont{{Hwang} et~al.}(2011)\citenamefont{{Hwang}, {Lee}, and
  {Yeom}}}]{hwang}
\bibinfo{author}{\bibfnamefont{D.-i.} \bibnamefont{{Hwang}}},
  \bibinfo{author}{\bibfnamefont{B.-H.} \bibnamefont{{Lee}}}, \bibnamefont{and}
  \bibinfo{author}{\bibfnamefont{D.-h.} \bibnamefont{{Yeom}}},
  \bibinfo{journal}{J.C.A.P.} \textbf{\bibinfo{volume}{12}},
  \bibinfo{pages}{006} (\bibinfo{year}{2011}).

\bibitem[{\citenamefont{Borisov et~al.}(2012)\citenamefont{Borisov, Jain, and
  Zhang}}]{bori}
\bibinfo{author}{\bibfnamefont{A.}~\bibnamefont{Borisov}},
  \bibinfo{author}{\bibfnamefont{B.}~\bibnamefont{Jain}}, \bibnamefont{and}
  \bibinfo{author}{\bibfnamefont{P.}~\bibnamefont{Zhang}},
  \bibinfo{journal}{Phys. Rev. D.} \textbf{\bibinfo{volume}{85}},
  \bibinfo{pages}{063518} (\bibinfo{year}{2012}).

\bibitem[{\citenamefont{Guo et~al.}(2014)\citenamefont{Guo, Wang, and
  Frolov}}]{guo}
\bibinfo{author}{\bibfnamefont{J.}~\bibnamefont{Guo}},
  \bibinfo{author}{\bibfnamefont{D.}~\bibnamefont{Wang}}, \bibnamefont{and}
  \bibinfo{author}{\bibfnamefont{A.~V.} \bibnamefont{Frolov}},
  \bibinfo{journal}{Phys. Rev. D.} \textbf{\bibinfo{volume}{90}},
  \bibinfo{pages}{024017} (\bibinfo{year}{2014}).

\bibitem[{\citenamefont{Kausar and Noureen}(2014)}]{kaus}
\bibinfo{author}{\bibfnamefont{H.~R.} \bibnamefont{Kausar}} \bibnamefont{and}
  \bibinfo{author}{\bibfnamefont{I.}~\bibnamefont{Noureen}},
  \bibinfo{journal}{Eur. Phys. J. C.} \textbf{\bibinfo{volume}{74}},
  \bibinfo{pages}{2760} (\bibinfo{year}{2014}).

\bibitem[{\citenamefont{Sharif and Yousaf}(2013{\natexlab{a}})}]{sharifyousaf1}
\bibinfo{author}{\bibfnamefont{M.}~\bibnamefont{Sharif}} \bibnamefont{and}
  \bibinfo{author}{\bibfnamefont{Z.}~\bibnamefont{Yousaf}},
  \bibinfo{journal}{Eur. Phys. J. C.} \textbf{\bibinfo{volume}{73}},
  \bibinfo{pages}{2633} (\bibinfo{year}{2013}{\natexlab{a}}).

\bibitem[{\citenamefont{Sharif and Yousaf}(2013{\natexlab{b}})}]{sharifyousaf2}
\bibinfo{author}{\bibfnamefont{M.}~\bibnamefont{Sharif}} \bibnamefont{and}
  \bibinfo{author}{\bibfnamefont{Z.}~\bibnamefont{Yousaf}},
  \bibinfo{journal}{Phys. Rev. D.} \textbf{\bibinfo{volume}{88}},
  \bibinfo{pages}{024020} (\bibinfo{year}{2013}{\natexlab{b}}).

\bibitem[{\citenamefont{Darmois}(1927)}]{darmo}
\bibinfo{author}{\bibfnamefont{J.~G.} \bibnamefont{Darmois}},
  \bibinfo{journal}{Memorial des Sciences Mathematiques, Gauthier-Villars,
  Paris} \textbf{\bibinfo{volume}{25}}, \bibinfo{pages}{1}
  (\bibinfo{year}{1927}).

\bibitem[{\citenamefont{Israel}(1966)}]{israel}
\bibinfo{author}{\bibfnamefont{W.}~\bibnamefont{Israel}},
  \bibinfo{journal}{Nuovo Cim. B.} \textbf{\bibinfo{volume}{44}},
  \bibinfo{pages}{1} (\bibinfo{year}{1966}).

\bibitem[{\citenamefont{Clifton et~al.}(2013)\citenamefont{Clifton, Dunsby,
  Goswami, and Nzioki}}]{cliftondunsby}
\bibinfo{author}{\bibfnamefont{T.}~\bibnamefont{Clifton}},
  \bibinfo{author}{\bibfnamefont{P.~K.~S.} \bibnamefont{Dunsby}},
  \bibinfo{author}{\bibfnamefont{R.}~\bibnamefont{Goswami}}, \bibnamefont{and}
  \bibinfo{author}{\bibfnamefont{A.~M.} \bibnamefont{Nzioki}},
  \bibinfo{journal}{Phys. Rev. D.} \textbf{\bibinfo{volume}{87}},
  \bibinfo{pages}{063517} (\bibinfo{year}{2013}).

\bibitem[{\citenamefont{Ganguly et~al.}(2014)\citenamefont{Ganguly, Gannouji,
  Goswami, and Ray}}]{ganguly}
\bibinfo{author}{\bibfnamefont{A.}~\bibnamefont{Ganguly}},
  \bibinfo{author}{\bibfnamefont{R.}~\bibnamefont{Gannouji}},
  \bibinfo{author}{\bibfnamefont{R.}~\bibnamefont{Goswami}}, \bibnamefont{and}
  \bibinfo{author}{\bibfnamefont{S.}~\bibnamefont{Ray}},
  \bibinfo{journal}{Phys. Rev. D.} \textbf{\bibinfo{volume}{89}},
  \bibinfo{pages}{064019} (\bibinfo{year}{2014}).

\bibitem[{\citenamefont{Goswami et~al.}(2014)\citenamefont{Goswami, Nzioki,
  Maharaj, and Ghosh}}]{ritu1}
\bibinfo{author}{\bibfnamefont{R.}~\bibnamefont{Goswami}},
  \bibinfo{author}{\bibfnamefont{A.~M.} \bibnamefont{Nzioki}},
  \bibinfo{author}{\bibfnamefont{S.~D.} \bibnamefont{Maharaj}},
  \bibnamefont{and} \bibinfo{author}{\bibfnamefont{S.~G.} \bibnamefont{Ghosh}},
  \bibinfo{journal}{Phys. Rev. D.} \textbf{\bibinfo{volume}{90}},
  \bibinfo{pages}{084011} (\bibinfo{year}{2014}).

\bibitem[{\citenamefont{Lemaitre}(1933)}]{lemaitre}
\bibinfo{author}{\bibfnamefont{G.}~\bibnamefont{Lemaitre}},
  \bibinfo{journal}{Ann. Soc. Sci. Bruxelles} p.~\bibinfo{pages}{51}
  (\bibinfo{year}{1933}).

\bibitem[{\citenamefont{Tolman}(1934)}]{tolman}
\bibinfo{author}{\bibfnamefont{R.~C.} \bibnamefont{Tolman}},
  \bibinfo{journal}{Proc. Natl. Acad. Sci. USA.} \textbf{\bibinfo{volume}{20}},
  \bibinfo{pages}{169} (\bibinfo{year}{1934}).

\bibitem[{\citenamefont{Bondi}(1948)}]{bondi}
\bibinfo{author}{\bibfnamefont{H.}~\bibnamefont{Bondi}}, \bibinfo{journal}{Mon.
  Not. Roy. Astron. Soc.} \textbf{\bibinfo{volume}{107}}, \bibinfo{pages}{410}
  (\bibinfo{year}{1948}).

\bibitem[{\citenamefont{Senovilla}(2013)}]{seno}
\bibinfo{author}{\bibfnamefont{J.~M.~M.} \bibnamefont{Senovilla}},
  \bibinfo{journal}{Phys. Rev. D.} \textbf{\bibinfo{volume}{88}},
  \bibinfo{pages}{064015} (\bibinfo{year}{2013}).

\bibitem[{\citenamefont{Deruelle et~al.}(2008)\citenamefont{Deruelle, Sasaki,
  and Sendouda}}]{deru}
\bibinfo{author}{\bibfnamefont{N.}~\bibnamefont{Deruelle}},
  \bibinfo{author}{\bibfnamefont{M.}~\bibnamefont{Sasaki}}, \bibnamefont{and}
  \bibinfo{author}{\bibfnamefont{Y.}~\bibnamefont{Sendouda}},
  \bibinfo{journal}{{Prog. Theor. Phys.}} \textbf{\bibinfo{volume}{119}},
  \bibinfo{pages}{237} (\bibinfo{year}{2008}).

\bibitem[{\citenamefont{Nzioki}(2013)}]{nziokithesis}
\bibinfo{author}{\bibfnamefont{A.~M.} \bibnamefont{Nzioki}}, Ph.D. thesis,
  \bibinfo{school}{University of Cape Town} (\bibinfo{year}{2013}).

\bibitem[{\citenamefont{Clifton}(2006)}]{tclifton}
\bibinfo{author}{\bibfnamefont{T.}~\bibnamefont{Clifton}},
  \bibinfo{journal}{Class. Quant. Grav.} \textbf{\bibinfo{volume}{23}},
  \bibinfo{pages}{7445} (\bibinfo{year}{2006}).

\bibitem[{\citenamefont{Riess et~al.}(1998)}]{riess1}
\bibinfo{author}{\bibfnamefont{A.~G.} \bibnamefont{Riess}} \bibnamefont{et~al.}
  (\bibinfo{collaboration}{Supernova Search Team}), \bibinfo{journal}{Astron.
  J.} \textbf{\bibinfo{volume}{116}}, \bibinfo{pages}{1009}
  (\bibinfo{year}{1998}).

\bibitem[{\citenamefont{Riess et~al.}(1999)\citenamefont{Riess, Kirshner,
  Schmidt, Jha, Challis, Garnavich, Esin, Carpenter, Grashius, Schild
  et~al.}}]{riess2}
\bibinfo{author}{\bibfnamefont{A.~G.} \bibnamefont{Riess}},
  \bibinfo{author}{\bibfnamefont{R.~P.} \bibnamefont{Kirshner}},
  \bibinfo{author}{\bibfnamefont{B.~P.} \bibnamefont{Schmidt}},
  \bibinfo{author}{\bibfnamefont{S.}~\bibnamefont{Jha}},
  \bibinfo{author}{\bibfnamefont{P.}~\bibnamefont{Challis}},
  \bibinfo{author}{\bibfnamefont{P.~M.} \bibnamefont{Garnavich}},
  \bibinfo{author}{\bibfnamefont{A.~A.} \bibnamefont{Esin}},
  \bibinfo{author}{\bibfnamefont{C.}~\bibnamefont{Carpenter}},
  \bibinfo{author}{\bibfnamefont{R.}~\bibnamefont{Grashius}},
  \bibinfo{author}{\bibfnamefont{R.~E.} \bibnamefont{Schild}},
  \bibnamefont{et~al.}, \bibinfo{journal}{Astron. J.}
  \textbf{\bibinfo{volume}{117}}, \bibinfo{pages}{707} (\bibinfo{year}{1999}).

\bibitem[{\citenamefont{Perlmutter et~al.}(1999)}]{perl}
\bibinfo{author}{\bibfnamefont{S.}~\bibnamefont{Perlmutter}}
  \bibnamefont{et~al.} (\bibinfo{collaboration}{Supernova Cosmology Project}),
  \bibinfo{journal}{Astrophys. J.} \textbf{\bibinfo{volume}{517}},
  \bibinfo{pages}{565} (\bibinfo{year}{1999}).

\bibitem[{\citenamefont{Zlatev et~al.}(1999)\citenamefont{Zlatev, Wang, and
  Steinhardt}}]{zlatev1}
\bibinfo{author}{\bibfnamefont{I.}~\bibnamefont{Zlatev}},
  \bibinfo{author}{\bibfnamefont{L.}~\bibnamefont{Wang}}, \bibnamefont{and}
  \bibinfo{author}{\bibfnamefont{P.~J.} \bibnamefont{Steinhardt}},
  \bibinfo{journal}{Phys. Rev. Lett.} \textbf{\bibinfo{volume}{82}},
  \bibinfo{pages}{896} (\bibinfo{year}{1999}).

\bibitem[{\citenamefont{Steinhardt et~al.}(1999)\citenamefont{Steinhardt, Wang,
  and Zlatev}}]{zlatev2}
\bibinfo{author}{\bibfnamefont{P.~J.} \bibnamefont{Steinhardt}},
  \bibinfo{author}{\bibfnamefont{L.}~\bibnamefont{Wang}}, \bibnamefont{and}
  \bibinfo{author}{\bibfnamefont{I.}~\bibnamefont{Zlatev}},
  \bibinfo{journal}{Phys. Rev. D.} \textbf{\bibinfo{volume}{59}},
  \bibinfo{pages}{123504} (\bibinfo{year}{1999}).

\bibitem[{\citenamefont{Ratra and Peebles}(1988)}]{ratra}
\bibinfo{author}{\bibfnamefont{B.}~\bibnamefont{Ratra}} \bibnamefont{and}
  \bibinfo{author}{\bibfnamefont{P.~J.~E.} \bibnamefont{Peebles}},
  \bibinfo{journal}{Phys. Rev. D.} \textbf{\bibinfo{volume}{37}},
  \bibinfo{pages}{3406} (\bibinfo{year}{1988}).

\bibitem[{\citenamefont{Liddle and Scherrer}(1999)}]{liddlescher}
\bibinfo{author}{\bibfnamefont{A.~R.} \bibnamefont{Liddle}} \bibnamefont{and}
  \bibinfo{author}{\bibfnamefont{R.~J.} \bibnamefont{Scherrer}},
  \bibinfo{journal}{Phys. Rev. D.} \textbf{\bibinfo{volume}{59}},
  \bibinfo{pages}{023509} (\bibinfo{year}{1999}).

\bibitem[{\citenamefont{Barrow and Cotsakis}(1988)}]{barrowcot}
\bibinfo{author}{\bibfnamefont{J.~D.} \bibnamefont{Barrow}} \bibnamefont{and}
  \bibinfo{author}{\bibfnamefont{S.}~\bibnamefont{Cotsakis}},
  \bibinfo{journal}{Phys. Lett. B.} \textbf{\bibinfo{volume}{214}},
  \bibinfo{pages}{515} (\bibinfo{year}{1988}).

\bibitem[{\citenamefont{Whitt}(1984)}]{whit}
\bibinfo{author}{\bibfnamefont{B.}~\bibnamefont{Whitt}},
  \bibinfo{journal}{Phys. Lett. B.} \textbf{\bibinfo{volume}{145}},
  \bibinfo{pages}{176} (\bibinfo{year}{1984}).

\bibitem[{\citenamefont{{Wands}}(1994)}]{wand}
\bibinfo{author}{\bibfnamefont{D.}~\bibnamefont{{Wands}}},
  \bibinfo{journal}{Class. Quant. Grav.} \textbf{\bibinfo{volume}{11}},
  \bibinfo{pages}{269} (\bibinfo{year}{1994}).

\bibitem[{\citenamefont{{Liddle}}(1998)}]{liddle1}
\bibinfo{author}{\bibfnamefont{A.~R.} \bibnamefont{{Liddle}}},
  \bibinfo{journal}{Phys. Rep.} \textbf{\bibinfo{volume}{307}},
  \bibinfo{pages}{53} (\bibinfo{year}{1998}).

\bibitem[{\citenamefont{{Harko} et~al.}(2014)\citenamefont{{Harko}, {Lobo}, and
  {Mak}}}]{harko1}
\bibinfo{author}{\bibfnamefont{T.}~\bibnamefont{{Harko}}},
  \bibinfo{author}{\bibfnamefont{F.~S.~N.} \bibnamefont{{Lobo}}},
  \bibnamefont{and} \bibinfo{author}{\bibfnamefont{M.~K.} \bibnamefont{{Mak}}},
  \bibinfo{journal}{Eur. Phys. J. C.} \textbf{\bibinfo{volume}{74}},
  \bibinfo{pages}{2784} (\bibinfo{year}{2014}).

\bibitem[{\citenamefont{{Caldwell} et~al.}(1998)\citenamefont{{Caldwell},
  {Dave}, and {Steinhardt}}}]{cald}
\bibinfo{author}{\bibfnamefont{R.~R.} \bibnamefont{{Caldwell}}},
  \bibinfo{author}{\bibfnamefont{R.}~\bibnamefont{{Dave}}}, \bibnamefont{and}
  \bibinfo{author}{\bibfnamefont{P.~J.} \bibnamefont{{Steinhardt}}},
  \bibinfo{journal}{Phys. Rev. Lett.} \textbf{\bibinfo{volume}{80}},
  \bibinfo{pages}{1582} (\bibinfo{year}{1998}).

\bibitem[{\citenamefont{Bassett
  et~al.}(2006{\natexlab{b}})\citenamefont{Bassett, Tsujikawa, and
  Wands}}]{bas}
\bibinfo{author}{\bibfnamefont{B.~A.} \bibnamefont{Bassett}},
  \bibinfo{author}{\bibfnamefont{S.}~\bibnamefont{Tsujikawa}},
  \bibnamefont{and} \bibinfo{author}{\bibfnamefont{D.}~\bibnamefont{Wands}},
  \bibinfo{journal}{Rev. Mod. Phys.} \textbf{\bibinfo{volume}{78}},
  \bibinfo{pages}{537} (\bibinfo{year}{2006}{\natexlab{b}}).

\bibitem[{\citenamefont{{Caldwell} and {Linder}}(2005)}]{linder}
\bibinfo{author}{\bibfnamefont{R.~R.} \bibnamefont{{Caldwell}}}
  \bibnamefont{and} \bibinfo{author}{\bibfnamefont{E.~V.}
  \bibnamefont{{Linder}}}, \bibinfo{journal}{Phys. Rev. Lett.}
  \textbf{\bibinfo{volume}{95}}, \bibinfo{pages}{141301}
  (\bibinfo{year}{2005}).

\bibitem[{\citenamefont{Barreiro et~al.}(2000)\citenamefont{Barreiro, Copeland,
  and Nunes}}]{barre}
\bibinfo{author}{\bibfnamefont{T.}~\bibnamefont{Barreiro}},
  \bibinfo{author}{\bibfnamefont{E.~J.} \bibnamefont{Copeland}},
  \bibnamefont{and} \bibinfo{author}{\bibfnamefont{N.~J.} \bibnamefont{Nunes}},
  \bibinfo{journal}{Phys. Rev. D.} \textbf{\bibinfo{volume}{61}},
  \bibinfo{pages}{127301} (\bibinfo{year}{2000}).

\bibitem[{\citenamefont{Maleknejad et~al.}(2013)\citenamefont{Maleknejad,
  Sheikh-Jabbari, and Soda}}]{malek}
\bibinfo{author}{\bibfnamefont{A.}~\bibnamefont{Maleknejad}},
  \bibinfo{author}{\bibfnamefont{M.~M.} \bibnamefont{Sheikh-Jabbari}},
  \bibnamefont{and} \bibinfo{author}{\bibfnamefont{J.}~\bibnamefont{Soda}},
  \bibinfo{journal}{Phys. Rep.} \textbf{\bibinfo{volume}{528}},
  \bibinfo{pages}{161} (\bibinfo{year}{2013}), \bibinfo{note}{gauge Fields and
  Inflation}.

\bibitem[{\citenamefont{{Schunck} and {Mielke}}(2008)}]{schunk}
\bibinfo{author}{\bibfnamefont{F.~E.} \bibnamefont{{Schunck}}}
  \bibnamefont{and} \bibinfo{author}{\bibfnamefont{E.~W.}
  \bibnamefont{{Mielke}}}, \bibinfo{journal}{ArXiv e-prints}
  (\bibinfo{year}{2008}), \eprint{0809.4462}.

\bibitem[{\citenamefont{Linde}(1982)}]{linde}
\bibinfo{author}{\bibfnamefont{A.~D.} \bibnamefont{Linde}},
  \bibinfo{journal}{Phys. Lett. B.} \textbf{\bibinfo{volume}{108}},
  \bibinfo{pages}{389} (\bibinfo{year}{1982}).

\bibitem[{\citenamefont{Albrecht and Steinhardt}(1982)}]{alb}
\bibinfo{author}{\bibfnamefont{A.}~\bibnamefont{Albrecht}} \bibnamefont{and}
  \bibinfo{author}{\bibfnamefont{P.~J.} \bibnamefont{Steinhardt}},
  \bibinfo{journal}{Phys. Rev. Lett.} \textbf{\bibinfo{volume}{48}},
  \bibinfo{pages}{1220} (\bibinfo{year}{1982}).

\bibitem[{\citenamefont{{Linde}}(1983)}]{chaoticlinde}
\bibinfo{author}{\bibfnamefont{A.~D.} \bibnamefont{{Linde}}},
  \bibinfo{journal}{Phys. Lett. B.} \textbf{\bibinfo{volume}{129}},
  \bibinfo{pages}{177} (\bibinfo{year}{1983}).

\bibitem[{\citenamefont{Kolb}(1999)}]{kolb}
\bibinfo{author}{\bibfnamefont{E.~W.} \bibnamefont{Kolb}}, in
  \emph{\bibinfo{booktitle}{{Pritzker Symposium and Workshop on the Status of
  Inflationary Cosmology Chicago, Illinois, January 29-February 3, 1999}}}
  (\bibinfo{year}{1999}), \eprint{hep-ph/9910311}.

\bibitem[{\citenamefont{Linde}(1994)}]{hybridlinde}
\bibinfo{author}{\bibfnamefont{A.}~\bibnamefont{Linde}},
  \bibinfo{journal}{Phys. Rev. D.} \textbf{\bibinfo{volume}{49}},
  \bibinfo{pages}{748} (\bibinfo{year}{1994}).

\bibitem[{\citenamefont{Copeland et~al.}(1994)\citenamefont{Copeland, Liddle,
  Lyth, Stewart, and Wands}}]{wandetal}
\bibinfo{author}{\bibfnamefont{E.~J.} \bibnamefont{Copeland}},
  \bibinfo{author}{\bibfnamefont{A.~R.} \bibnamefont{Liddle}},
  \bibinfo{author}{\bibfnamefont{D.~H.} \bibnamefont{Lyth}},
  \bibinfo{author}{\bibfnamefont{E.~D.} \bibnamefont{Stewart}},
  \bibnamefont{and} \bibinfo{author}{\bibfnamefont{D.}~\bibnamefont{Wands}},
  \bibinfo{journal}{Phys. Rev. D.} \textbf{\bibinfo{volume}{49}},
  \bibinfo{pages}{6410} (\bibinfo{year}{1994}).

\bibitem[{\citenamefont{Sami}(2007)}]{booksami}
\bibinfo{author}{\bibfnamefont{M.}~\bibnamefont{Sami}},
  \emph{\bibinfo{title}{Models of dark energy}} (\bibinfo{publisher}{Springer
  Berlin Heidelberg}, \bibinfo{address}{Berlin, Heidelberg},
  \bibinfo{year}{2007}), p. \bibinfo{pages}{219}.

\bibitem[{\citenamefont{{Tsujikawa}}(2013)}]{tsuji}
\bibinfo{author}{\bibfnamefont{S.}~\bibnamefont{{Tsujikawa}}},
  \bibinfo{journal}{Class. Quant. Grav.} \textbf{\bibinfo{volume}{30}},
  \bibinfo{pages}{214003} (\bibinfo{year}{2013}).

\bibitem[{\citenamefont{Armendáriz-Picón
  et~al.}(1999)\citenamefont{Armendáriz-Picón, Damour, and Mukhanov}}]{arme1}
\bibinfo{author}{\bibfnamefont{C.}~\bibnamefont{Armendáriz-Picón}},
  \bibinfo{author}{\bibfnamefont{T.}~\bibnamefont{Damour}}, \bibnamefont{and}
  \bibinfo{author}{\bibfnamefont{V.}~\bibnamefont{Mukhanov}},
  \bibinfo{journal}{Phys. Lett. B.} \textbf{\bibinfo{volume}{458}},
  \bibinfo{pages}{209} (\bibinfo{year}{1999}).

\bibitem[{\citenamefont{Armendariz-Picon
  et~al.}(2000)\citenamefont{Armendariz-Picon, Mukhanov, and
  Steinhardt}}]{arme2}
\bibinfo{author}{\bibfnamefont{C.}~\bibnamefont{Armendariz-Picon}},
  \bibinfo{author}{\bibfnamefont{V.~F.} \bibnamefont{Mukhanov}},
  \bibnamefont{and} \bibinfo{author}{\bibfnamefont{P.~J.}
  \bibnamefont{Steinhardt}}, \bibinfo{journal}{Phys. Rev. Lett.}
  \textbf{\bibinfo{volume}{85}}, \bibinfo{pages}{4438} (\bibinfo{year}{2000}).

\bibitem[{\citenamefont{Armendariz-Picon
  et~al.}(2001)\citenamefont{Armendariz-Picon, Mukhanov, and
  Steinhardt}}]{arme3}
\bibinfo{author}{\bibfnamefont{C.}~\bibnamefont{Armendariz-Picon}},
  \bibinfo{author}{\bibfnamefont{V.~F.} \bibnamefont{Mukhanov}},
  \bibnamefont{and} \bibinfo{author}{\bibfnamefont{P.~J.}
  \bibnamefont{Steinhardt}}, \bibinfo{journal}{Phys. Rev. D.}
  \textbf{\bibinfo{volume}{63}}, \bibinfo{pages}{103510}
  (\bibinfo{year}{2001}).

\bibitem[{\citenamefont{Caldwell}(2002)}]{caldphantom}
\bibinfo{author}{\bibfnamefont{R.~R.} \bibnamefont{Caldwell}},
  \bibinfo{journal}{Phys. Lett. B.} \textbf{\bibinfo{volume}{545}},
  \bibinfo{pages}{23} (\bibinfo{year}{2002}).

\bibitem[{\citenamefont{Carroll et~al.}(2003)\citenamefont{Carroll, Hoffman,
  and Trodden}}]{carrol}
\bibinfo{author}{\bibfnamefont{S.~M.} \bibnamefont{Carroll}},
  \bibinfo{author}{\bibfnamefont{M.}~\bibnamefont{Hoffman}}, \bibnamefont{and}
  \bibinfo{author}{\bibfnamefont{M.}~\bibnamefont{Trodden}},
  \bibinfo{journal}{Phys. Rev. D.} \textbf{\bibinfo{volume}{68}},
  \bibinfo{pages}{023509} (\bibinfo{year}{2003}).

\bibitem[{\citenamefont{{Kamenshchik}}(2013)}]{kamen}
\bibinfo{author}{\bibfnamefont{A.~Y.} \bibnamefont{{Kamenshchik}}},
  \bibinfo{journal}{Class. Quant. Grav.} \textbf{\bibinfo{volume}{30}},
  \bibinfo{pages}{173001} (\bibinfo{year}{2013}).

\bibitem[{\citenamefont{Kamada et~al.}(2011)\citenamefont{Kamada, Kobayashi,
  Yamaguchi, and Yokoyama}}]{kamada}
\bibinfo{author}{\bibfnamefont{K.}~\bibnamefont{Kamada}},
  \bibinfo{author}{\bibfnamefont{T.}~\bibnamefont{Kobayashi}},
  \bibinfo{author}{\bibfnamefont{M.}~\bibnamefont{Yamaguchi}},
  \bibnamefont{and} \bibinfo{author}{\bibfnamefont{J.}~\bibnamefont{Yokoyama}},
  \bibinfo{journal}{Phys. Rev. D.} \textbf{\bibinfo{volume}{83}},
  \bibinfo{pages}{083515} (\bibinfo{year}{2011}).

\bibitem[{\citenamefont{Kobayashi et~al.}(2011)\citenamefont{Kobayashi,
  Yamaguchi, and Yokoyama}}]{kobaya}
\bibinfo{author}{\bibfnamefont{T.}~\bibnamefont{Kobayashi}},
  \bibinfo{author}{\bibfnamefont{M.}~\bibnamefont{Yamaguchi}},
  \bibnamefont{and} \bibinfo{author}{\bibfnamefont{J.}~\bibnamefont{Yokoyama}},
  \bibinfo{journal}{Prog. Theor. Phys.} \textbf{\bibinfo{volume}{126}},
  \bibinfo{pages}{511} (\bibinfo{year}{2011}).

\bibitem[{\citenamefont{{Gon{\c c}alves} and {Moss}}(1997)}]{goncalves}
\bibinfo{author}{\bibfnamefont{S.~M.~C.~V.} \bibnamefont{{Gon{\c c}alves}}}
  \bibnamefont{and} \bibinfo{author}{\bibfnamefont{I.~G.}
  \bibnamefont{{Moss}}}, \bibinfo{journal}{Class. Quant. Grav.}
  \textbf{\bibinfo{volume}{14}}, \bibinfo{pages}{2607} (\bibinfo{year}{1997}).

\bibitem[{\citenamefont{Christodoulou}(1987{\natexlab{a}})}]{christo1}
\bibinfo{author}{\bibfnamefont{D.}~\bibnamefont{Christodoulou}},
  \bibinfo{journal}{Comm. Math. Phys.} \textbf{\bibinfo{volume}{109}},
  \bibinfo{pages}{591} (\bibinfo{year}{1987}{\natexlab{a}}).

\bibitem[{\citenamefont{Christodoulou}(1987{\natexlab{b}})}]{christo2}
\bibinfo{author}{\bibfnamefont{D.}~\bibnamefont{Christodoulou}},
  \bibinfo{journal}{Comm. Math. Phys.} \textbf{\bibinfo{volume}{109}},
  \bibinfo{pages}{613} (\bibinfo{year}{1987}{\natexlab{b}}).

\bibitem[{\citenamefont{Christodoulou}(1994)}]{christo3}
\bibinfo{author}{\bibfnamefont{D.}~\bibnamefont{Christodoulou}},
  \bibinfo{journal}{Ann. Math.} \textbf{\bibinfo{volume}{140}},
  \bibinfo{pages}{607} (\bibinfo{year}{1994}).

\bibitem[{\citenamefont{Goldwirth and Piran}(1987)}]{piran}
\bibinfo{author}{\bibfnamefont{D.~S.} \bibnamefont{Goldwirth}}
  \bibnamefont{and} \bibinfo{author}{\bibfnamefont{T.}~\bibnamefont{Piran}},
  \bibinfo{journal}{Phys. Rev. D.} \textbf{\bibinfo{volume}{36}},
  \bibinfo{pages}{3575} (\bibinfo{year}{1987}).

\bibitem[{\citenamefont{Choptuik}(1993)}]{chop}
\bibinfo{author}{\bibfnamefont{M.~W.} \bibnamefont{Choptuik}},
  \bibinfo{journal}{Phys. Rev. Lett.} \textbf{\bibinfo{volume}{70}},
  \bibinfo{pages}{9} (\bibinfo{year}{1993}).

\bibitem[{\citenamefont{{Brady} et~al.}(1997)\citenamefont{{Brady}, {Chambers},
  and {Gon{\c c}alves}}}]{brady}
\bibinfo{author}{\bibfnamefont{P.~R.} \bibnamefont{{Brady}}},
  \bibinfo{author}{\bibfnamefont{C.~M.} \bibnamefont{{Chambers}}},
  \bibnamefont{and} \bibinfo{author}{\bibfnamefont{S.~M.~C.~V.}
  \bibnamefont{{Gon{\c c}alves}}}, \bibinfo{journal}{Phys. Rev. D.}
  \textbf{\bibinfo{volume}{56}}, \bibinfo{pages}{R6057} (\bibinfo{year}{1997}).

\bibitem[{\citenamefont{Gundlach}(1999)}]{gund1}
\bibinfo{author}{\bibfnamefont{C.}~\bibnamefont{Gundlach}},
  \bibinfo{journal}{Liv. Rev. Rel.} \textbf{\bibinfo{volume}{2}},
  \bibinfo{pages}{1} (\bibinfo{year}{1999}).

\bibitem[{\citenamefont{Gundlach}(1995)}]{gund2}
\bibinfo{author}{\bibfnamefont{C.}~\bibnamefont{Gundlach}},
  \bibinfo{journal}{Phys. Rev. Lett.} \textbf{\bibinfo{volume}{75}},
  \bibinfo{pages}{3214} (\bibinfo{year}{1995}).

\bibitem[{\citenamefont{Roberts}(1989)}]{roberts}
\bibinfo{author}{\bibfnamefont{M.~D.} \bibnamefont{Roberts}},
  \bibinfo{journal}{Gen. Rel. Grav.} \textbf{\bibinfo{volume}{21}},
  \bibinfo{pages}{907} (\bibinfo{year}{1989}).

\bibitem[{\citenamefont{{de Oliveira} and {Cheb-Terrab}}(1996)}]{oli}
\bibinfo{author}{\bibfnamefont{H.~P.} \bibnamefont{{de Oliveira}}}
  \bibnamefont{and} \bibinfo{author}{\bibfnamefont{E.~S.}
  \bibnamefont{{Cheb-Terrab}}}, \bibinfo{journal}{Class. Quant. Grav.}
  \textbf{\bibinfo{volume}{13}}, \bibinfo{pages}{425} (\bibinfo{year}{1996}).

\bibitem[{\citenamefont{Frolov}(1997)}]{frolov}
\bibinfo{author}{\bibfnamefont{A.~V.} \bibnamefont{Frolov}},
  \bibinfo{journal}{Phys. Rev. D.} \textbf{\bibinfo{volume}{56}},
  \bibinfo{pages}{6433} (\bibinfo{year}{1997}).

\bibitem[{\citenamefont{{Soda} and {Hirata}}(1996)}]{soda}
\bibinfo{author}{\bibfnamefont{J.}~\bibnamefont{{Soda}}} \bibnamefont{and}
  \bibinfo{author}{\bibfnamefont{K.}~\bibnamefont{{Hirata}}},
  \bibinfo{journal}{Phys. Lett. B.} \textbf{\bibinfo{volume}{387}},
  \bibinfo{pages}{271} (\bibinfo{year}{1996}).

\bibitem[{\citenamefont{Green and Liddle}(1999)}]{green}
\bibinfo{author}{\bibfnamefont{A.~M.} \bibnamefont{Green}} \bibnamefont{and}
  \bibinfo{author}{\bibfnamefont{A.~R.} \bibnamefont{Liddle}},
  \bibinfo{journal}{Phys. Rev. D.} \textbf{\bibinfo{volume}{60}},
  \bibinfo{pages}{063509} (\bibinfo{year}{1999}).

\bibitem[{\citenamefont{Hara et~al.}(1996)\citenamefont{Hara, Koike, and
  Adachi}}]{hara}
\bibinfo{author}{\bibfnamefont{T.}~\bibnamefont{Hara}},
  \bibinfo{author}{\bibfnamefont{T.}~\bibnamefont{Koike}}, \bibnamefont{and}
  \bibinfo{author}{\bibfnamefont{S.}~\bibnamefont{Adachi}}
  (\bibinfo{year}{1996}), \eprint{gr-qc/9607010}.

\bibitem[{\citenamefont{Olabarrieta et~al.}(2007)\citenamefont{Olabarrieta,
  Ventrella, Choptuik, and Unruh}}]{olabari}
\bibinfo{author}{\bibfnamefont{I.}~\bibnamefont{Olabarrieta}},
  \bibinfo{author}{\bibfnamefont{J.~F.} \bibnamefont{Ventrella}},
  \bibinfo{author}{\bibfnamefont{M.~W.} \bibnamefont{Choptuik}},
  \bibnamefont{and} \bibinfo{author}{\bibfnamefont{W.~G.} \bibnamefont{Unruh}},
  \bibinfo{journal}{Phys. Rev. D.} \textbf{\bibinfo{volume}{76}},
  \bibinfo{pages}{124014} (\bibinfo{year}{2007}).

\bibitem[{\citenamefont{Hawley and Choptuik}(2000)}]{hawley}
\bibinfo{author}{\bibfnamefont{S.~H.} \bibnamefont{Hawley}} \bibnamefont{and}
  \bibinfo{author}{\bibfnamefont{M.~W.} \bibnamefont{Choptuik}}, p.
  \bibinfo{pages}{751} (\bibinfo{year}{2000}), \bibinfo{note}{[AIP Conf.
  Proc.586,751(2001)]}.

\bibitem[{\citenamefont{Olabarrieta and Choptuik}(2002)}]{olabari1}
\bibinfo{author}{\bibfnamefont{I.}~\bibnamefont{Olabarrieta}} \bibnamefont{and}
  \bibinfo{author}{\bibfnamefont{M.~W.} \bibnamefont{Choptuik}},
  \bibinfo{journal}{Phys. Rev. D.} \textbf{\bibinfo{volume}{65}},
  \bibinfo{pages}{024007} (\bibinfo{year}{2002}).

\bibitem[{\citenamefont{Ventrella and Choptuik}(2003)}]{ventrella}
\bibinfo{author}{\bibfnamefont{J.~F.} \bibnamefont{Ventrella}}
  \bibnamefont{and} \bibinfo{author}{\bibfnamefont{M.~W.}
  \bibnamefont{Choptuik}}, \bibinfo{journal}{Phys. Rev. D.}
  \textbf{\bibinfo{volume}{68}}, \bibinfo{pages}{044020}
  (\bibinfo{year}{2003}).

\bibitem[{\citenamefont{Choptuik et~al.}(2003)\citenamefont{Choptuik,
  Hirschmann, Liebling, and Pretorius}}]{pretorius}
\bibinfo{author}{\bibfnamefont{M.~W.} \bibnamefont{Choptuik}},
  \bibinfo{author}{\bibfnamefont{E.~W.} \bibnamefont{Hirschmann}},
  \bibinfo{author}{\bibfnamefont{S.~L.} \bibnamefont{Liebling}},
  \bibnamefont{and}
  \bibinfo{author}{\bibfnamefont{F.}~\bibnamefont{Pretorius}},
  \bibinfo{journal}{Phys. Rev. D.} \textbf{\bibinfo{volume}{68}},
  \bibinfo{pages}{044007} (\bibinfo{year}{2003}).

\bibitem[{\citenamefont{Noble and Choptuik}(2008)}]{noble1}
\bibinfo{author}{\bibfnamefont{S.~C.} \bibnamefont{Noble}} \bibnamefont{and}
  \bibinfo{author}{\bibfnamefont{M.~W.} \bibnamefont{Choptuik}},
  \bibinfo{journal}{Phys. Rev. D.} \textbf{\bibinfo{volume}{78}},
  \bibinfo{pages}{064059} (\bibinfo{year}{2008}).

\bibitem[{\citenamefont{Radice et~al.}(2010)\citenamefont{Radice, Rezzolla, and
  Kellermann}}]{radice1}
\bibinfo{author}{\bibfnamefont{D.}~\bibnamefont{Radice}},
  \bibinfo{author}{\bibfnamefont{L.}~\bibnamefont{Rezzolla}}, \bibnamefont{and}
  \bibinfo{author}{\bibfnamefont{T.}~\bibnamefont{Kellermann}},
  \bibinfo{journal}{Class. Quant. Grav.} \textbf{\bibinfo{volume}{27}},
  \bibinfo{pages}{235015} (\bibinfo{year}{2010}).

\bibitem[{\citenamefont{Kellermann et~al.}(2010)\citenamefont{Kellermann,
  Rezzolla, and Radice}}]{radice2}
\bibinfo{author}{\bibfnamefont{T.}~\bibnamefont{Kellermann}},
  \bibinfo{author}{\bibfnamefont{L.}~\bibnamefont{Rezzolla}}, \bibnamefont{and}
  \bibinfo{author}{\bibfnamefont{D.}~\bibnamefont{Radice}},
  \bibinfo{journal}{Class. Quant. Grav.} \textbf{\bibinfo{volume}{27}},
  \bibinfo{pages}{235016} (\bibinfo{year}{2010}).

\bibitem[{\citenamefont{Noble and Choptuik}(2016)}]{noble2}
\bibinfo{author}{\bibfnamefont{S.~C.} \bibnamefont{Noble}} \bibnamefont{and}
  \bibinfo{author}{\bibfnamefont{M.~W.} \bibnamefont{Choptuik}},
  \bibinfo{journal}{Phys. Rev. D.} \textbf{\bibinfo{volume}{93}},
  \bibinfo{pages}{024015} (\bibinfo{year}{2016}).

\bibitem[{\citenamefont{Akbarian and Choptuik}(2015)}]{akbarchop}
\bibinfo{author}{\bibfnamefont{A.}~\bibnamefont{Akbarian}} \bibnamefont{and}
  \bibinfo{author}{\bibfnamefont{M.~W.} \bibnamefont{Choptuik}},
  \bibinfo{journal}{Phys. Rev. D.} \textbf{\bibinfo{volume}{92}},
  \bibinfo{pages}{084037} (\bibinfo{year}{2015}).

\bibitem[{\citenamefont{Giambo}(2005)}]{massivegiambo}
\bibinfo{author}{\bibfnamefont{R.}~\bibnamefont{Giambo}},
  \bibinfo{journal}{Class. Quant. Grav.} \textbf{\bibinfo{volume}{22}},
  \bibinfo{pages}{2295} (\bibinfo{year}{2005}).

\bibitem[{\citenamefont{Gon\ifmmode~\mbox{\c{c}}\else
  \c{c}\fi{}alves}(2000)}]{massivegoncalves}
\bibinfo{author}{\bibfnamefont{S.~M. C.~V.}
  \bibnamefont{Gon\ifmmode~\mbox{\c{c}}\else \c{c}\fi{}alves}},
  \bibinfo{journal}{Phys. Rev. D.} \textbf{\bibinfo{volume}{62}},
  \bibinfo{pages}{124006} (\bibinfo{year}{2000}).

\bibitem[{\citenamefont{Goswami and Joshi}(2004{\natexlab{a}})}]{massiveritu1}
\bibinfo{author}{\bibfnamefont{R.}~\bibnamefont{Goswami}} \bibnamefont{and}
  \bibinfo{author}{\bibfnamefont{P.~S.} \bibnamefont{Joshi}}
  (\bibinfo{year}{2004}{\natexlab{a}}), \eprint{gr-qc/0410144}.

\bibitem[{\citenamefont{Goswami and Joshi}(2007{\natexlab{a}})}]{massiveritu2}
\bibinfo{author}{\bibfnamefont{R.}~\bibnamefont{Goswami}} \bibnamefont{and}
  \bibinfo{author}{\bibfnamefont{P.~S.} \bibnamefont{Joshi}},
  \bibinfo{journal}{Mod. Phys. Lett. A.} \textbf{\bibinfo{volume}{22}},
  \bibinfo{pages}{65} (\bibinfo{year}{2007}{\natexlab{a}}).

\bibitem[{\citenamefont{Ganguly and Banerjee}(2013)}]{massiveganguly}
\bibinfo{author}{\bibfnamefont{K.}~\bibnamefont{Ganguly}} \bibnamefont{and}
  \bibinfo{author}{\bibfnamefont{N.}~\bibnamefont{Banerjee}},
  \bibinfo{journal}{Pramana} \textbf{\bibinfo{volume}{80}},
  \bibinfo{pages}{439} (\bibinfo{year}{2013}).

\bibitem[{\citenamefont{Baier et~al.}(2015)\citenamefont{Baier, Nishimura, and
  Stricker}}]{baiernishi}
\bibinfo{author}{\bibfnamefont{R.}~\bibnamefont{Baier}},
  \bibinfo{author}{\bibfnamefont{H.}~\bibnamefont{Nishimura}},
  \bibnamefont{and} \bibinfo{author}{\bibfnamefont{S.~A.}
  \bibnamefont{Stricker}}, \bibinfo{journal}{Class. Quant. Grav.}
  \textbf{\bibinfo{volume}{32}}, \bibinfo{pages}{135021}
  (\bibinfo{year}{2015}).

\bibitem[{\citenamefont{Condron and Nolan}(2013)}]{condron1}
\bibinfo{author}{\bibfnamefont{E.}~\bibnamefont{Condron}} \bibnamefont{and}
  \bibinfo{author}{\bibfnamefont{B.~C.} \bibnamefont{Nolan}}
  (\bibinfo{year}{2013}), \eprint{1305.4866}.

\bibitem[{\citenamefont{Condron and Nolan}(2014)}]{condron2}
\bibinfo{author}{\bibfnamefont{E.}~\bibnamefont{Condron}} \bibnamefont{and}
  \bibinfo{author}{\bibfnamefont{B.~C.} \bibnamefont{Nolan}},
  \bibinfo{journal}{Class. Quant. Grav.} \textbf{\bibinfo{volume}{31}},
  \bibinfo{pages}{165018} (\bibinfo{year}{2014}).

\bibitem[{\citenamefont{Ganguly and
  Banerjee}(2011{\natexlab{b}})}]{nonsphericalganguly}
\bibinfo{author}{\bibfnamefont{K.}~\bibnamefont{Ganguly}} \bibnamefont{and}
  \bibinfo{author}{\bibfnamefont{N.}~\bibnamefont{Banerjee}},
  \bibinfo{journal}{Gen. Rel. Grav.} \textbf{\bibinfo{volume}{43}},
  \bibinfo{pages}{2141} (\bibinfo{year}{2011}{\natexlab{b}}).

\bibitem[{\citenamefont{Wagh and Govinder}(2006{\natexlab{a}})}]{wagh1}
\bibinfo{author}{\bibfnamefont{S.~M.} \bibnamefont{Wagh}} \bibnamefont{and}
  \bibinfo{author}{\bibfnamefont{K.~S.} \bibnamefont{Govinder}},
  \bibinfo{journal}{Gen. Rel. Grav.} \textbf{\bibinfo{volume}{38}},
  \bibinfo{pages}{1253} (\bibinfo{year}{2006}{\natexlab{a}}).

\bibitem[{\citenamefont{Wagh et~al.}(2001)\citenamefont{Wagh, Saraykar,
  Muktibodh, and Govinder}}]{wagh2}
\bibinfo{author}{\bibfnamefont{S.~M.} \bibnamefont{Wagh}},
  \bibinfo{author}{\bibfnamefont{R.~V.} \bibnamefont{Saraykar}},
  \bibinfo{author}{\bibfnamefont{P.~S.} \bibnamefont{Muktibodh}},
  \bibnamefont{and} \bibinfo{author}{\bibfnamefont{K.~S.}
  \bibnamefont{Govinder}}, \bibinfo{journal}{Submitted to: Class. Quant. Grav.}
   (\bibinfo{year}{2001}), \eprint{gr-qc/0112033}.

\bibitem[{\citenamefont{Kerner}(1982)}]{kerner}
\bibinfo{author}{\bibfnamefont{R.}~\bibnamefont{Kerner}},
  \bibinfo{journal}{Gen. Rel. Grav.} \textbf{\bibinfo{volume}{14}},
  \bibinfo{pages}{453} (\bibinfo{year}{1982}).

\bibitem[{\citenamefont{Barrow and Ottewill}(1983)}]{barrow1}
\bibinfo{author}{\bibfnamefont{J.~D.} \bibnamefont{Barrow}} \bibnamefont{and}
  \bibinfo{author}{\bibfnamefont{A.~C.} \bibnamefont{Ottewill}},
  \bibinfo{journal}{Journ. Phys. A : Mathematical and General}
  \textbf{\bibinfo{volume}{16}}, \bibinfo{pages}{2757} (\bibinfo{year}{1983}).

\bibitem[{\citenamefont{Capozziello et~al.}(2003)\citenamefont{Capozziello,
  Cardone, Carloni, and Troisi}}]{capo}
\bibinfo{author}{\bibfnamefont{S.}~\bibnamefont{Capozziello}},
  \bibinfo{author}{\bibfnamefont{V.~F.} \bibnamefont{Cardone}},
  \bibinfo{author}{\bibfnamefont{S.}~\bibnamefont{Carloni}}, \bibnamefont{and}
  \bibinfo{author}{\bibfnamefont{A.}~\bibnamefont{Troisi}},
  \bibinfo{journal}{Int. J. Mod. Phys. D.} \textbf{\bibinfo{volume}{12}},
  \bibinfo{pages}{1969} (\bibinfo{year}{2003}).

\bibitem[{\citenamefont{{Carroll} et~al.}(2004)\citenamefont{{Carroll},
  {Duvvuri}, {Trodden}, and {Turner}}}]{carroll1}
\bibinfo{author}{\bibfnamefont{S.~M.} \bibnamefont{{Carroll}}},
  \bibinfo{author}{\bibfnamefont{V.}~\bibnamefont{{Duvvuri}}},
  \bibinfo{author}{\bibfnamefont{M.}~\bibnamefont{{Trodden}}},
  \bibnamefont{and} \bibinfo{author}{\bibfnamefont{M.~S.}
  \bibnamefont{{Turner}}}, \bibinfo{journal}{Phys. Rev. D.}
  \textbf{\bibinfo{volume}{70}}, \bibinfo{pages}{043528}
  (\bibinfo{year}{2004}).

\bibitem[{\citenamefont{De~Felice and Tsujikawa}(2010)}]{felice}
\bibinfo{author}{\bibfnamefont{A.}~\bibnamefont{De~Felice}} \bibnamefont{and}
  \bibinfo{author}{\bibfnamefont{S.}~\bibnamefont{Tsujikawa}},
  \bibinfo{journal}{Liv. Rev. Rel.} \textbf{\bibinfo{volume}{13}}
  (\bibinfo{year}{2010}).

\bibitem[{\citenamefont{Nojiri and Odintsov}(2011)}]{nojiriphysrep}
\bibinfo{author}{\bibfnamefont{S.}~\bibnamefont{Nojiri}} \bibnamefont{and}
  \bibinfo{author}{\bibfnamefont{S.~D.} \bibnamefont{Odintsov}},
  \bibinfo{journal}{Phys. Rep.} \textbf{\bibinfo{volume}{505}},
  \bibinfo{pages}{59} (\bibinfo{year}{2011}).

\bibitem[{\citenamefont{Misner and Sharp}(1964)}]{sharp}
\bibinfo{author}{\bibfnamefont{C.~W.} \bibnamefont{Misner}} \bibnamefont{and}
  \bibinfo{author}{\bibfnamefont{D.~H.} \bibnamefont{Sharp}},
  \bibinfo{journal}{Phys. Rev.} \textbf{\bibinfo{volume}{136}},
  \bibinfo{pages}{B571} (\bibinfo{year}{1964}).

\bibitem[{\citenamefont{Waugh and Lake}(1986)}]{waughlake3}
\bibinfo{author}{\bibfnamefont{B.}~\bibnamefont{Waugh}} \bibnamefont{and}
  \bibinfo{author}{\bibfnamefont{K.}~\bibnamefont{Lake}},
  \bibinfo{journal}{Phys. Rev. D} \textbf{\bibinfo{volume}{34}},
  \bibinfo{pages}{2978} (\bibinfo{year}{1986}).

\bibitem[{\citenamefont{Lukash and Strokov}(2013)}]{lukash}
\bibinfo{author}{\bibfnamefont{V.~N.} \bibnamefont{Lukash}} \bibnamefont{and}
  \bibinfo{author}{\bibfnamefont{V.~N.} \bibnamefont{Strokov}},
  \bibinfo{journal}{Int. J. Mod. Phys.} \textbf{\bibinfo{volume}{A28}},
  \bibinfo{pages}{1350007} (\bibinfo{year}{2013}).

\bibitem[{\citenamefont{Nzioki et~al.}(2014)\citenamefont{Nzioki, Goswami, and
  Dunsby}}]{nzioki2}
\bibinfo{author}{\bibfnamefont{A.~M.} \bibnamefont{Nzioki}},
  \bibinfo{author}{\bibfnamefont{R.}~\bibnamefont{Goswami}}, \bibnamefont{and}
  \bibinfo{author}{\bibfnamefont{P.~K.~S.} \bibnamefont{Dunsby}},
  \bibinfo{journal}{Phys. Rev. D.} \textbf{\bibinfo{volume}{89}},
  \bibinfo{pages}{064050} (\bibinfo{year}{2014}).

\bibitem[{\citenamefont{Goswami and Joshi}(2004{\natexlab{b}})}]{gosjoshi}
\bibinfo{author}{\bibfnamefont{R.}~\bibnamefont{Goswami}} \bibnamefont{and}
  \bibinfo{author}{\bibfnamefont{P.~S.} \bibnamefont{Joshi}},
  \bibinfo{journal}{Phys. Rev. D} \textbf{\bibinfo{volume}{69}},
  \bibinfo{pages}{027502} (\bibinfo{year}{2004}{\natexlab{b}}).

\bibitem[{\citenamefont{Choquet-Bruhat
  et~al.}(1982)\citenamefont{Choquet-Bruhat, Dewitt-Morette, and
  Dillard-Bleick}}]{bruhat}
\bibinfo{author}{\bibfnamefont{Y.}~\bibnamefont{Choquet-Bruhat}},
  \bibinfo{author}{\bibfnamefont{C.}~\bibnamefont{Dewitt-Morette}},
  \bibnamefont{and}
  \bibinfo{author}{\bibfnamefont{M.}~\bibnamefont{Dillard-Bleick}},
  \emph{\bibinfo{title}{Analysis, Manifolds and Physics}}
  (\bibinfo{publisher}{Amsterdam: North-Holland}, \bibinfo{year}{1982}).

\bibitem[{\citenamefont{Stephani et~al.}(2003)\citenamefont{Stephani, Kramer,
  MacCallum, Hoenselaers, and Herlt}}]{stephani}
\bibinfo{author}{\bibfnamefont{H.}~\bibnamefont{Stephani}},
  \bibinfo{author}{\bibfnamefont{D.}~\bibnamefont{Kramer}},
  \bibinfo{author}{\bibfnamefont{M.~A.~H.} \bibnamefont{MacCallum}},
  \bibinfo{author}{\bibfnamefont{C.}~\bibnamefont{Hoenselaers}},
  \bibnamefont{and} \bibinfo{author}{\bibfnamefont{E.}~\bibnamefont{Herlt}},
  \emph{\bibinfo{title}{Exact Solutions of Einstein's Field Equations}}
  (\bibinfo{publisher}{Cambridge: Cambridge University Press.},
  \bibinfo{year}{2003}).

\bibitem[{\citenamefont{Hall}(2004)}]{hall}
\bibinfo{author}{\bibfnamefont{G.~S.} \bibnamefont{Hall}},
  \emph{\bibinfo{title}{Symmetries and Curvature Structure in General
  Relativity}} (\bibinfo{publisher}{Singapore: World Scientific.},
  \bibinfo{year}{2004}).

\bibitem[{\citenamefont{Maartens and Maharaj}(1986)}]{maartens1c3}
\bibinfo{author}{\bibfnamefont{R.}~\bibnamefont{Maartens}} \bibnamefont{and}
  \bibinfo{author}{\bibfnamefont{S.~D.} \bibnamefont{Maharaj}},
  \bibinfo{journal}{Class. Quant. Grav.} \textbf{\bibinfo{volume}{3}},
  \bibinfo{pages}{1005} (\bibinfo{year}{1986}).

\bibitem[{\citenamefont{Keane and Barrett}(2000)}]{keane1}
\bibinfo{author}{\bibfnamefont{A.~J.} \bibnamefont{Keane}} \bibnamefont{and}
  \bibinfo{author}{\bibfnamefont{R.~K.} \bibnamefont{Barrett}},
  \bibinfo{journal}{Class. Quant. Grav.} \textbf{\bibinfo{volume}{17}},
  \bibinfo{pages}{201} (\bibinfo{year}{2000}).

\bibitem[{\citenamefont{Maartens and Maharaj}(1991)}]{maartens2c3}
\bibinfo{author}{\bibfnamefont{R.}~\bibnamefont{Maartens}} \bibnamefont{and}
  \bibinfo{author}{\bibfnamefont{S.~D.} \bibnamefont{Maharaj}},
  \bibinfo{journal}{Class. Quant. Grav.} \textbf{\bibinfo{volume}{8}},
  \bibinfo{pages}{503} (\bibinfo{year}{1991}).

\bibitem[{\citenamefont{Keane and Tupper}(2004)}]{keane2}
\bibinfo{author}{\bibfnamefont{A.~J.} \bibnamefont{Keane}} \bibnamefont{and}
  \bibinfo{author}{\bibfnamefont{B.~O.~J.} \bibnamefont{Tupper}},
  \bibinfo{journal}{Class. Quant. Grav.} \textbf{\bibinfo{volume}{21}},
  \bibinfo{pages}{2037} (\bibinfo{year}{2004}).

\bibitem[{\citenamefont{Tupper et~al.}(2003)\citenamefont{Tupper, Keane, Hall,
  Coley, and Carot}}]{tupper}
\bibinfo{author}{\bibfnamefont{B.~O.~J.} \bibnamefont{Tupper}},
  \bibinfo{author}{\bibfnamefont{A.~J.} \bibnamefont{Keane}},
  \bibinfo{author}{\bibfnamefont{G.~S.} \bibnamefont{Hall}},
  \bibinfo{author}{\bibfnamefont{A.~A.} \bibnamefont{Coley}}, \bibnamefont{and}
  \bibinfo{author}{\bibfnamefont{J.}~\bibnamefont{Carot}},
  \bibinfo{journal}{Class. Quant. Grav.} \textbf{\bibinfo{volume}{20}},
  \bibinfo{pages}{801} (\bibinfo{year}{2003}).

\bibitem[{\citenamefont{Saifullah and Yazdan}(2009)}]{saifullah}
\bibinfo{author}{\bibfnamefont{K.}~\bibnamefont{Saifullah}} \bibnamefont{and}
  \bibinfo{author}{\bibfnamefont{S.}~\bibnamefont{Yazdan}},
  \bibinfo{journal}{Int. J. Mod. Phys. D.} \textbf{\bibinfo{volume}{18}},
  \bibinfo{pages}{71} (\bibinfo{year}{2009}).

\bibitem[{\citenamefont{Maartens et~al.}(1995)\citenamefont{Maartens, Maharaj,
  and Tupper}}]{maartens3c3}
\bibinfo{author}{\bibfnamefont{R.}~\bibnamefont{Maartens}},
  \bibinfo{author}{\bibfnamefont{S.~D.} \bibnamefont{Maharaj}},
  \bibnamefont{and} \bibinfo{author}{\bibfnamefont{B.~O.~J.}
  \bibnamefont{Tupper}}, \bibinfo{journal}{Class. Quant. Grav.}
  \textbf{\bibinfo{volume}{12}}, \bibinfo{pages}{2577} (\bibinfo{year}{1995}).

\bibitem[{\citenamefont{Moopanar and Maharaj}(2010)}]{moopanar}
\bibinfo{author}{\bibfnamefont{S.}~\bibnamefont{Moopanar}} \bibnamefont{and}
  \bibinfo{author}{\bibfnamefont{S.~D.} \bibnamefont{Maharaj}},
  \bibinfo{journal}{Int. J. Theor. Phys.} \textbf{\bibinfo{volume}{49}},
  \bibinfo{pages}{1878} (\bibinfo{year}{2010}).

\bibitem[{\citenamefont{Chrobok and Borzeszkowski}(2006)}]{chrobok}
\bibinfo{author}{\bibfnamefont{T.}~\bibnamefont{Chrobok}} \bibnamefont{and}
  \bibinfo{author}{\bibfnamefont{H.~H.} \bibnamefont{Borzeszkowski}},
  \bibinfo{journal}{Gen. Rel. Grav.} \textbf{\bibinfo{volume}{38}},
  \bibinfo{pages}{397} (\bibinfo{year}{2006}).

\bibitem[{\citenamefont{Bohmer et~al.}(2008)\citenamefont{Bohmer, Harko, and
  Lobo}}]{bohmer}
\bibinfo{author}{\bibfnamefont{C.~G.} \bibnamefont{Bohmer}},
  \bibinfo{author}{\bibfnamefont{T.}~\bibnamefont{Harko}}, \bibnamefont{and}
  \bibinfo{author}{\bibfnamefont{F.~S.~N.} \bibnamefont{Lobo}},
  \bibinfo{journal}{Class. Quant. Grav.} \textbf{\bibinfo{volume}{25}},
  \bibinfo{pages}{075016} (\bibinfo{year}{2008}).

\bibitem[{\citenamefont{Mak and Harko}(2004)}]{mak}
\bibinfo{author}{\bibfnamefont{M.~K.} \bibnamefont{Mak}} \bibnamefont{and}
  \bibinfo{author}{\bibfnamefont{T.}~\bibnamefont{Harko}},
  \bibinfo{journal}{Int. J. Mod. Phys. D.} \textbf{\bibinfo{volume}{13}},
  \bibinfo{pages}{149} (\bibinfo{year}{2004}).

\bibitem[{\citenamefont{Esculpi and Aloma}(2010)}]{esculpi}
\bibinfo{author}{\bibfnamefont{M.}~\bibnamefont{Esculpi}} \bibnamefont{and}
  \bibinfo{author}{\bibfnamefont{E.}~\bibnamefont{Aloma}},
  \bibinfo{journal}{Eur. Phys. J. C.} \textbf{\bibinfo{volume}{67}},
  \bibinfo{pages}{521} (\bibinfo{year}{2010}).

\bibitem[{\citenamefont{Herrera et~al.}(2012)\citenamefont{Herrera, Di~Prisco,
  and Ibanez}}]{herrera1}
\bibinfo{author}{\bibfnamefont{L.}~\bibnamefont{Herrera}},
  \bibinfo{author}{\bibfnamefont{A.}~\bibnamefont{Di~Prisco}},
  \bibnamefont{and} \bibinfo{author}{\bibfnamefont{J.}~\bibnamefont{Ibanez}},
  \bibinfo{journal}{Phys. Lett. A.} \textbf{\bibinfo{volume}{376}},
  \bibinfo{pages}{899} (\bibinfo{year}{2012}).

\bibitem[{\citenamefont{Krasinski}(1997)}]{krasinski}
\bibinfo{author}{\bibfnamefont{A.}~\bibnamefont{Krasinski}},
  \emph{\bibinfo{title}{Inhomogeneous Cosmological Models}}
  (\bibinfo{publisher}{Cambridge: Cambridge University Press.},
  \bibinfo{year}{1997}).

\bibitem[{\citenamefont{Herrera and Santos}(2010)}]{herrera2}
\bibinfo{author}{\bibfnamefont{L.}~\bibnamefont{Herrera}} \bibnamefont{and}
  \bibinfo{author}{\bibfnamefont{N.~O.} \bibnamefont{Santos}},
  \bibinfo{journal}{Gen. Rel. Grav.} \textbf{\bibinfo{volume}{42}},
  \bibinfo{pages}{2383} (\bibinfo{year}{2010}).

\bibitem[{\citenamefont{Moopanar and Maharaj}(2013)}]{moopanar1}
\bibinfo{author}{\bibfnamefont{S.}~\bibnamefont{Moopanar}} \bibnamefont{and}
  \bibinfo{author}{\bibfnamefont{S.~D.} \bibnamefont{Maharaj}},
  \bibinfo{journal}{J. Eng. Math.} \textbf{\bibinfo{volume}{82}},
  \bibinfo{pages}{125} (\bibinfo{year}{2013}).

\bibitem[{\citenamefont{Cahill and Taub}(1971)}]{cahilltaub}
\bibinfo{author}{\bibfnamefont{M.~E.} \bibnamefont{Cahill}} \bibnamefont{and}
  \bibinfo{author}{\bibfnamefont{A.~H.} \bibnamefont{Taub}},
  \bibinfo{journal}{Comm. Math. Phys.} \textbf{\bibinfo{volume}{21}},
  \bibinfo{pages}{1} (\bibinfo{year}{1971}).

\bibitem[{\citenamefont{Carr and Coley}(1999)}]{carrcoley}
\bibinfo{author}{\bibfnamefont{B.~J.} \bibnamefont{Carr}} \bibnamefont{and}
  \bibinfo{author}{\bibfnamefont{A.~A.} \bibnamefont{Coley}},
  \bibinfo{journal}{Class. Quant. Grav.} \textbf{\bibinfo{volume}{16}},
  \bibinfo{pages}{R31} (\bibinfo{year}{1999}).

\bibitem[{\citenamefont{Wagh and Govinder}(2006{\natexlab{b}})}]{waghgovi}
\bibinfo{author}{\bibfnamefont{S.~M.} \bibnamefont{Wagh}} \bibnamefont{and}
  \bibinfo{author}{\bibfnamefont{K.~S.} \bibnamefont{Govinder}},
  \bibinfo{journal}{Gen. Rel. Grav.} \textbf{\bibinfo{volume}{38}},
  \bibinfo{pages}{1253} (\bibinfo{year}{2006}{\natexlab{b}}).

\bibitem[{\citenamefont{{Wagh} et~al.}(2001)\citenamefont{{Wagh}, {Govender},
  {Govinder}, {Maharaj}, {Muktibodh}, and {Moodley}}}]{waghh}
\bibinfo{author}{\bibfnamefont{S.~M.} \bibnamefont{{Wagh}}},
  \bibinfo{author}{\bibfnamefont{M.}~\bibnamefont{{Govender}}},
  \bibinfo{author}{\bibfnamefont{K.~S.} \bibnamefont{{Govinder}}},
  \bibinfo{author}{\bibfnamefont{S.~D.} \bibnamefont{{Maharaj}}},
  \bibinfo{author}{\bibfnamefont{P.~S.} \bibnamefont{{Muktibodh}}},
  \bibnamefont{and}
  \bibinfo{author}{\bibfnamefont{M.}~\bibnamefont{{Moodley}}},
  \bibinfo{journal}{Class. Quant. Grav.} \textbf{\bibinfo{volume}{18}},
  \bibinfo{pages}{2147} (\bibinfo{year}{2001}).

\bibitem[{\citenamefont{Goheer et~al.}(2009)\citenamefont{Goheer, Larena, and
  Dunsby}}]{goheer}
\bibinfo{author}{\bibfnamefont{N.}~\bibnamefont{Goheer}},
  \bibinfo{author}{\bibfnamefont{J.}~\bibnamefont{Larena}}, \bibnamefont{and}
  \bibinfo{author}{\bibfnamefont{P.~K.~S.} \bibnamefont{Dunsby}},
  \bibinfo{journal}{Phys. Rev. D.} \textbf{\bibinfo{volume}{80}},
  \bibinfo{pages}{061301} (\bibinfo{year}{2009}).

\bibitem[{\citenamefont{Sharma et~al.}(2015)\citenamefont{Sharma, Das, and
  Tikekar}}]{sharmadastike}
\bibinfo{author}{\bibfnamefont{R.}~\bibnamefont{Sharma}},
  \bibinfo{author}{\bibfnamefont{S.}~\bibnamefont{Das}}, \bibnamefont{and}
  \bibinfo{author}{\bibfnamefont{R.}~\bibnamefont{Tikekar}},
  \bibinfo{journal}{Gen. Rel. Grav.} \textbf{\bibinfo{volume}{47}},
  \bibinfo{pages}{25} (\bibinfo{year}{2015}).

\bibitem[{\citenamefont{Euler}(1997)}]{euler1}
\bibinfo{author}{\bibfnamefont{N.}~\bibnamefont{Euler}}, \bibinfo{journal}{J.
  Nonlin. Math. Phys.} \textbf{\bibinfo{volume}{4}}, \bibinfo{pages}{310}
  (\bibinfo{year}{1997}).

\bibitem[{\citenamefont{Euler et~al.}(1989)\citenamefont{Euler, Steeb, and
  Cyrus}}]{euler2}
\bibinfo{author}{\bibfnamefont{N.}~\bibnamefont{Euler}},
  \bibinfo{author}{\bibfnamefont{W.~H.} \bibnamefont{Steeb}}, \bibnamefont{and}
  \bibinfo{author}{\bibfnamefont{K.}~\bibnamefont{Cyrus}}, \bibinfo{journal}{J.
  Phys. A: Mathematical and General} \textbf{\bibinfo{volume}{22}},
  \bibinfo{pages}{L195} (\bibinfo{year}{1989}).

\bibitem[{\citenamefont{{Harko} et~al.}(2013)\citenamefont{{Harko}, {Lobo}, and
  {Mak}}}]{harkolobomak}
\bibinfo{author}{\bibfnamefont{T.}~\bibnamefont{{Harko}}},
  \bibinfo{author}{\bibfnamefont{F.~S.~N.} \bibnamefont{{Lobo}}},
  \bibnamefont{and} \bibinfo{author}{\bibfnamefont{M.~K.} \bibnamefont{{Mak}}},
  \bibinfo{journal}{ArXiv e-prints}  (\bibinfo{year}{2013}),
  \eprint{1304.1468}.

\bibitem[{\citenamefont{{Santos}}(1985)}]{santosc4}
\bibinfo{author}{\bibfnamefont{N.~O.} \bibnamefont{{Santos}}},
  \bibinfo{journal}{Mon. Not. Roy. Astron. Soc.}
  \textbf{\bibinfo{volume}{216}}, \bibinfo{pages}{403} (\bibinfo{year}{1985}).

\bibitem[{\citenamefont{Som and Santos}(1981)}]{som}
\bibinfo{author}{\bibfnamefont{M.~M.} \bibnamefont{Som}} \bibnamefont{and}
  \bibinfo{author}{\bibfnamefont{N.~O.} \bibnamefont{Santos}},
  \bibinfo{journal}{Phys. Lett. A.} \textbf{\bibinfo{volume}{87}},
  \bibinfo{pages}{89} (\bibinfo{year}{1981}).

\bibitem[{\citenamefont{Maiti}(1982)}]{maiti}
\bibinfo{author}{\bibfnamefont{S.~R.} \bibnamefont{Maiti}},
  \bibinfo{journal}{Phys. Rev. D.} \textbf{\bibinfo{volume}{25}},
  \bibinfo{pages}{2518} (\bibinfo{year}{1982}).

\bibitem[{\citenamefont{Modak}(1984)}]{modak}
\bibinfo{author}{\bibfnamefont{B.}~\bibnamefont{Modak}}, \bibinfo{journal}{J.
  Astrophys. Astron.} \textbf{\bibinfo{volume}{5}}, \bibinfo{pages}{317}
  (\bibinfo{year}{1984}).

\bibitem[{\citenamefont{Banerjee et~al.}(1989)\citenamefont{Banerjee,
  Choudhury, and Bhui}}]{bhui}
\bibinfo{author}{\bibfnamefont{A.}~\bibnamefont{Banerjee}},
  \bibinfo{author}{\bibfnamefont{S.~B.~D.} \bibnamefont{Choudhury}},
  \bibnamefont{and} \bibinfo{author}{\bibfnamefont{B.~K.} \bibnamefont{Bhui}},
  \bibinfo{journal}{Phys. Rev. D.} \textbf{\bibinfo{volume}{40}},
  \bibinfo{pages}{670} (\bibinfo{year}{1989}).

\bibitem[{\citenamefont{Patel and Tikekar}(1991)}]{patel}
\bibinfo{author}{\bibfnamefont{L.~K.} \bibnamefont{Patel}} \bibnamefont{and}
  \bibinfo{author}{\bibfnamefont{R.}~\bibnamefont{Tikekar}},
  \bibinfo{journal}{Mathematics Today} \textbf{\bibinfo{volume}{IX}},
  \bibinfo{pages}{19} (\bibinfo{year}{1991}).

\bibitem[{\citenamefont{Schafer and Goenner}(2000)}]{schafer}
\bibinfo{author}{\bibfnamefont{D.}~\bibnamefont{Schafer}} \bibnamefont{and}
  \bibinfo{author}{\bibfnamefont{H.~F.} \bibnamefont{Goenner}},
  \bibinfo{journal}{Gen. Rel. Grav.} \textbf{\bibinfo{volume}{42}},
  \bibinfo{pages}{2119} (\bibinfo{year}{2000}).

\bibitem[{\citenamefont{Ivanov}(2012)}]{ivanov}
\bibinfo{author}{\bibfnamefont{B.~V.} \bibnamefont{Ivanov}},
  \bibinfo{journal}{Gen. Rel. Grav.} \textbf{\bibinfo{volume}{44}},
  \bibinfo{pages}{1835} (\bibinfo{year}{2012}).

\bibitem[{\citenamefont{Herrera
  et~al.}(2004{\natexlab{a}})\citenamefont{Herrera, Le~Denmat, Santos, and
  Wang}}]{herreraetal}
\bibinfo{author}{\bibfnamefont{L.}~\bibnamefont{Herrera}},
  \bibinfo{author}{\bibfnamefont{G.}~\bibnamefont{Le~Denmat}},
  \bibinfo{author}{\bibfnamefont{N.~O.} \bibnamefont{Santos}},
  \bibnamefont{and} \bibinfo{author}{\bibfnamefont{A.}~\bibnamefont{Wang}},
  \bibinfo{journal}{Int. J. Mod. Phys. D.} \textbf{\bibinfo{volume}{13}},
  \bibinfo{pages}{583} (\bibinfo{year}{2004}{\natexlab{a}}).

\bibitem[{\citenamefont{Peebles and Ratra}(1988)}]{ratrac4}
\bibinfo{author}{\bibfnamefont{P.~J.~E.} \bibnamefont{Peebles}}
  \bibnamefont{and} \bibinfo{author}{\bibfnamefont{B.}~\bibnamefont{Ratra}},
  \bibinfo{journal}{Astrophys. J.} \textbf{\bibinfo{volume}{325}},
  \bibinfo{pages}{L17} (\bibinfo{year}{1988}).

\bibitem[{\citenamefont{Roman}(1986)}]{romanc4}
\bibinfo{author}{\bibfnamefont{T.~A.} \bibnamefont{Roman}},
  \bibinfo{journal}{Phys. Rev. D} \textbf{\bibinfo{volume}{33}},
  \bibinfo{pages}{3526} (\bibinfo{year}{1986}).

\bibitem[{\citenamefont{{Chan}}(2000)}]{chanc4}
\bibinfo{author}{\bibfnamefont{R.}~\bibnamefont{{Chan}}},
  \bibinfo{journal}{Mon. Not. Roy. Astron. Soc.}
  \textbf{\bibinfo{volume}{316}}, \bibinfo{pages}{588} (\bibinfo{year}{2000}).

\bibitem[{\citenamefont{{Kolassis} et~al.}(1988)\citenamefont{{Kolassis},
  {Santos}, and {Tsoubelis}}}]{kolla}
\bibinfo{author}{\bibfnamefont{C.~A.} \bibnamefont{{Kolassis}}},
  \bibinfo{author}{\bibfnamefont{N.~O.} \bibnamefont{{Santos}}},
  \bibnamefont{and}
  \bibinfo{author}{\bibfnamefont{D.}~\bibnamefont{{Tsoubelis}}},
  \bibinfo{journal}{Astrophys. J.} \textbf{\bibinfo{volume}{327}},
  \bibinfo{pages}{755} (\bibinfo{year}{1988}).

\bibitem[{\citenamefont{Maharaj and Govender}(2005)}]{maharajc5}
\bibinfo{author}{\bibfnamefont{S.~D.} \bibnamefont{Maharaj}} \bibnamefont{and}
  \bibinfo{author}{\bibfnamefont{M.}~\bibnamefont{Govender}},
  \bibinfo{journal}{Int. J. Mod. Phys. D.} \textbf{\bibinfo{volume}{14}},
  \bibinfo{pages}{667} (\bibinfo{year}{2005}).

\bibitem[{\citenamefont{Joshi and Goswami}(2004)}]{massiveritu11}
\bibinfo{author}{\bibfnamefont{P.~S.} \bibnamefont{Joshi}} \bibnamefont{and}
  \bibinfo{author}{\bibfnamefont{R.}~\bibnamefont{Goswami}},
  \bibinfo{journal}{Phys. Rev. D.} \textbf{\bibinfo{volume}{69}},
  \bibinfo{pages}{064027} (\bibinfo{year}{2004}).

\bibitem[{\citenamefont{Oliveira et~al.}(1988)\citenamefont{Oliveira, Kolassis,
  and Santos}}]{olic4}
\bibinfo{author}{\bibfnamefont{A.~K.~G.} \bibnamefont{Oliveira}},
  \bibinfo{author}{\bibfnamefont{C.~A.} \bibnamefont{Kolassis}},
  \bibnamefont{and} \bibinfo{author}{\bibfnamefont{N.~O.}
  \bibnamefont{Santos}}, \bibinfo{journal}{Mon. Not. Roy. Astron. Soc.}
  \textbf{\bibinfo{volume}{231}}, \bibinfo{pages}{1011} (\bibinfo{year}{1988}).

\bibitem[{\citenamefont{Hamid et~al.}(2014)\citenamefont{Hamid, Goswami, and
  Maharaj}}]{hgmc4}
\bibinfo{author}{\bibfnamefont{A.~I.~M.} \bibnamefont{Hamid}},
  \bibinfo{author}{\bibfnamefont{R.}~\bibnamefont{Goswami}}, \bibnamefont{and}
  \bibinfo{author}{\bibfnamefont{S.~D.} \bibnamefont{Maharaj}},
  \bibinfo{journal}{Class. Quant. Grav.} \textbf{\bibinfo{volume}{31}},
  \bibinfo{pages}{135010} (\bibinfo{year}{2014}).

\bibitem[{\citenamefont{Johri}(2001)}]{johri1}
\bibinfo{author}{\bibfnamefont{V.~B.} \bibnamefont{Johri}},
  \bibinfo{journal}{Phys. Rev. D} \textbf{\bibinfo{volume}{63}},
  \bibinfo{pages}{103504} (\bibinfo{year}{2001}).

\bibitem[{\citenamefont{Johri}(2002)}]{johri2}
\bibinfo{author}{\bibfnamefont{V.~B.} \bibnamefont{Johri}},
  \bibinfo{journal}{Class. Quant. Grav.} \textbf{\bibinfo{volume}{19}},
  \bibinfo{pages}{5959} (\bibinfo{year}{2002}).

\bibitem[{\citenamefont{{Herrera} and {Santos}}(1997)}]{herresantoc5}
\bibinfo{author}{\bibfnamefont{L.}~\bibnamefont{{Herrera}}} \bibnamefont{and}
  \bibinfo{author}{\bibfnamefont{N.~O.} \bibnamefont{{Santos}}},
  \bibinfo{journal}{Phys. Rep.} \textbf{\bibinfo{volume}{286}},
  \bibinfo{pages}{53} (\bibinfo{year}{1997}).

\bibitem[{\citenamefont{Herrera
  et~al.}(2004{\natexlab{b}})\citenamefont{Herrera, Di~Prisco, Martin, Ospino,
  Santos, and Troconis}}]{herreetal2}
\bibinfo{author}{\bibfnamefont{L.}~\bibnamefont{Herrera}},
  \bibinfo{author}{\bibfnamefont{A.}~\bibnamefont{Di~Prisco}},
  \bibinfo{author}{\bibfnamefont{J.}~\bibnamefont{Martin}},
  \bibinfo{author}{\bibfnamefont{J.}~\bibnamefont{Ospino}},
  \bibinfo{author}{\bibfnamefont{N.~O.} \bibnamefont{Santos}},
  \bibnamefont{and} \bibinfo{author}{\bibfnamefont{O.}~\bibnamefont{Troconis}},
  \bibinfo{journal}{Phys. Rev. D} \textbf{\bibinfo{volume}{69}},
  \bibinfo{pages}{084026} (\bibinfo{year}{2004}{\natexlab{b}}).

\bibitem[{\citenamefont{Herrera and Ponce~de León}(1985)}]{herreraleon}
\bibinfo{author}{\bibfnamefont{L.}~\bibnamefont{Herrera}} \bibnamefont{and}
  \bibinfo{author}{\bibfnamefont{J.}~\bibnamefont{Ponce~de León}},
  \bibinfo{journal}{J. Math. Phys.} \textbf{\bibinfo{volume}{26}},
  \bibinfo{pages}{2018} (\bibinfo{year}{1985}).

\bibitem[{\citenamefont{Kazanas and Schramm}(1979)}]{kaza}
\bibinfo{author}{\bibfnamefont{D.}~\bibnamefont{Kazanas}} \bibnamefont{and}
  \bibinfo{author}{\bibfnamefont{D.}~\bibnamefont{Schramm}},
  \emph{\bibinfo{title}{Sources of Gravitational Radiation}}
  (\bibinfo{publisher}{L. Smarr ed., (Cambridge University Press, Cambridge,
  1979).}, \bibinfo{year}{1979}).

\bibitem[{\citenamefont{Dwivedi and Joshi}(1994)}]{joshidwivedi4}
\bibinfo{author}{\bibfnamefont{I.~H.} \bibnamefont{Dwivedi}} \bibnamefont{and}
  \bibinfo{author}{\bibfnamefont{P.~S.} \bibnamefont{Joshi}},
  \bibinfo{journal}{Commun. Math. Phys.} \textbf{\bibinfo{volume}{166}},
  \bibinfo{pages}{117} (\bibinfo{year}{1994}).

\bibitem[{\citenamefont{Dwivedi and Dixit}(1991)}]{dwivedidixit}
\bibinfo{author}{\bibfnamefont{I.~H.} \bibnamefont{Dwivedi}} \bibnamefont{and}
  \bibinfo{author}{\bibfnamefont{S.}~\bibnamefont{Dixit}},
  \bibinfo{journal}{Prog. Theor. Phys.} \textbf{\bibinfo{volume}{85}},
  \bibinfo{pages}{433} (\bibinfo{year}{1991}).

\bibitem[{\citenamefont{Goswami and
  Joshi}(2007{\natexlab{b}})}]{goswamijoshic5}
\bibinfo{author}{\bibfnamefont{R.}~\bibnamefont{Goswami}} \bibnamefont{and}
  \bibinfo{author}{\bibfnamefont{P.~S.} \bibnamefont{Joshi}},
  \bibinfo{journal}{Phys. Rev. D.} \textbf{\bibinfo{volume}{76}},
  \bibinfo{pages}{084026} (\bibinfo{year}{2007}{\natexlab{b}}).

\bibitem[{\citenamefont{Ghosh and Deshkar}(2003)}]{ghoshdeshkar}
\bibinfo{author}{\bibfnamefont{S.~G.} \bibnamefont{Ghosh}} \bibnamefont{and}
  \bibinfo{author}{\bibfnamefont{D.~W.} \bibnamefont{Deshkar}},
  \bibinfo{journal}{Int. J. Mod. Phys. D.} \textbf{\bibinfo{volume}{12}},
  \bibinfo{pages}{913} (\bibinfo{year}{2003}).

\bibitem[{\citenamefont{Herrera et~al.}(2010)\citenamefont{Herrera, Di~Prisco,
  and Ospino}}]{herreraospinogrg}
\bibinfo{author}{\bibfnamefont{L.}~\bibnamefont{Herrera}},
  \bibinfo{author}{\bibfnamefont{A.}~\bibnamefont{Di~Prisco}},
  \bibnamefont{and} \bibinfo{author}{\bibfnamefont{J.}~\bibnamefont{Ospino}},
  \bibinfo{journal}{Gen. Rel. Grav.} \textbf{\bibinfo{volume}{42}},
  \bibinfo{pages}{1585} (\bibinfo{year}{2010}).

\bibitem[{\citenamefont{Lyth and Stewart}(1995)}]{lythstew}
\bibinfo{author}{\bibfnamefont{D.~H.} \bibnamefont{Lyth}} \bibnamefont{and}
  \bibinfo{author}{\bibfnamefont{E.~D.} \bibnamefont{Stewart}},
  \bibinfo{journal}{Phys. Rev. Lett.} \textbf{\bibinfo{volume}{75}},
  \bibinfo{pages}{201} (\bibinfo{year}{1995}).

\bibitem[{\citenamefont{Celerier and Szekeres}(2002)}]{celer}
\bibinfo{author}{\bibfnamefont{M.}~\bibnamefont{Celerier}} \bibnamefont{and}
  \bibinfo{author}{\bibfnamefont{P.}~\bibnamefont{Szekeres}},
  \bibinfo{journal}{Phys. Rev. D.} \textbf{\bibinfo{volume}{65}},
  \bibinfo{pages}{123516} (\bibinfo{year}{2002}).

\bibitem[{\citenamefont{Harada et~al.}(2002)\citenamefont{Harada, Iguchi, and
  Nakao}}]{harada}
\bibinfo{author}{\bibfnamefont{T.}~\bibnamefont{Harada}},
  \bibinfo{author}{\bibfnamefont{H.}~\bibnamefont{Iguchi}}, \bibnamefont{and}
  \bibinfo{author}{\bibfnamefont{K.~I.} \bibnamefont{Nakao}},
  \bibinfo{journal}{Prog. Theor. Phys.} \textbf{\bibinfo{volume}{107}},
  \bibinfo{pages}{449} (\bibinfo{year}{2002}).

\bibitem[{\citenamefont{Goswami et~al.}(2012)\citenamefont{Goswami, Joshi, and
  Malafarina}}]{ritumala}
\bibinfo{author}{\bibfnamefont{R.}~\bibnamefont{Goswami}},
  \bibinfo{author}{\bibfnamefont{P.~S.} \bibnamefont{Joshi}}, \bibnamefont{and}
  \bibinfo{author}{\bibfnamefont{D.}~\bibnamefont{Malafarina}},
  \bibinfo{journal}{ArXiv e-prints}  (\bibinfo{year}{2012}),
  \eprint{1202.6218}.

\bibitem[{\citenamefont{Mobius}(1988)}]{greenbook}
\bibinfo{author}{\bibfnamefont{P.}~\bibnamefont{Mobius}},
  \bibinfo{journal}{Jour. App. Math. Mech.} \textbf{\bibinfo{volume}{68}},
  \bibinfo{pages}{258} (\bibinfo{year}{1988}).

\bibitem[{\citenamefont{Barrow}(1987)}]{barrowplb}
\bibinfo{author}{\bibfnamefont{J.~D.} \bibnamefont{Barrow}},
  \bibinfo{journal}{Phys. Lett. B.} \textbf{\bibinfo{volume}{187}},
  \bibinfo{pages}{12} (\bibinfo{year}{1987}).

\bibitem[{\citenamefont{Burd and Barrow}(1988)}]{burdbarrow}
\bibinfo{author}{\bibfnamefont{A.~B.} \bibnamefont{Burd}} \bibnamefont{and}
  \bibinfo{author}{\bibfnamefont{J.~D.} \bibnamefont{Barrow}},
  \bibinfo{journal}{Nucl. Phys.} \textbf{\bibinfo{volume}{B308}},
  \bibinfo{pages}{929} (\bibinfo{year}{1988}).

\bibitem[{\citenamefont{Aguirregabiria and Chimento}(1996)}]{agui}
\bibinfo{author}{\bibfnamefont{J.~M.} \bibnamefont{Aguirregabiria}}
  \bibnamefont{and} \bibinfo{author}{\bibfnamefont{L.~P.}
  \bibnamefont{Chimento}}, \bibinfo{journal}{Class. Quant. Grav.}
  \textbf{\bibinfo{volume}{13}}, \bibinfo{pages}{3197} (\bibinfo{year}{1996}).

\bibitem[{\citenamefont{Chimento}(1998)}]{chimento}
\bibinfo{author}{\bibfnamefont{L.~P.} \bibnamefont{Chimento}},
  \bibinfo{journal}{Class. Quant. Grav.} \textbf{\bibinfo{volume}{15}},
  \bibinfo{pages}{965} (\bibinfo{year}{1998}).

\bibitem[{\citenamefont{Rubano and Scudellaro}(2002)}]{rubano}
\bibinfo{author}{\bibfnamefont{C.}~\bibnamefont{Rubano}} \bibnamefont{and}
  \bibinfo{author}{\bibfnamefont{P.}~\bibnamefont{Scudellaro}},
  \bibinfo{journal}{Gen. Rel. Grav.} \textbf{\bibinfo{volume}{34}},
  \bibinfo{pages}{307} (\bibinfo{year}{2002}).

\bibitem[{\citenamefont{Neupane}(2004)}]{neupane}
\bibinfo{author}{\bibfnamefont{I.~P.} \bibnamefont{Neupane}},
  \bibinfo{journal}{Class. Quant. Grav.} \textbf{\bibinfo{volume}{21}},
  \bibinfo{pages}{4383} (\bibinfo{year}{2004}).

\bibitem[{\citenamefont{Russo}(2004)}]{russo}
\bibinfo{author}{\bibfnamefont{J.~G.} \bibnamefont{Russo}},
  \bibinfo{journal}{Phys. Lett. B.} \textbf{\bibinfo{volume}{600}},
  \bibinfo{pages}{185} (\bibinfo{year}{2004}).

\bibitem[{\citenamefont{Piedipalumbo et~al.}(2012)\citenamefont{Piedipalumbo,
  Scudellaro, Esposito, and Rubano}}]{piedi}
\bibinfo{author}{\bibfnamefont{E.}~\bibnamefont{Piedipalumbo}},
  \bibinfo{author}{\bibfnamefont{P.}~\bibnamefont{Scudellaro}},
  \bibinfo{author}{\bibfnamefont{G.}~\bibnamefont{Esposito}}, \bibnamefont{and}
  \bibinfo{author}{\bibfnamefont{C.}~\bibnamefont{Rubano}},
  \bibinfo{journal}{Gen. Rel. Grav.} \textbf{\bibinfo{volume}{44}},
  \bibinfo{pages}{2611} (\bibinfo{year}{2012}).

\bibitem[{\citenamefont{Salopek and Bond}(1990)}]{salop}
\bibinfo{author}{\bibfnamefont{D.~S.} \bibnamefont{Salopek}} \bibnamefont{and}
  \bibinfo{author}{\bibfnamefont{J.~R.} \bibnamefont{Bond}},
  \bibinfo{journal}{Phys. Rev. D.} \textbf{\bibinfo{volume}{42}},
  \bibinfo{pages}{3936} (\bibinfo{year}{1990}).

\bibitem[{\citenamefont{MacCallum}(2006)}]{maccul}
\bibinfo{author}{\bibfnamefont{M.~A.~H.} \bibnamefont{MacCallum}},
  \bibinfo{journal}{A. I. P. Conf. Proc.} \textbf{\bibinfo{volume}{841}},
  \bibinfo{pages}{129} (\bibinfo{year}{2006}).

\end{thebibliography}

% If you would like to use BibLaTeX for your references, pass `custombib' as
% an option in the document class. The location of 'reference.bib' should be
% specified in the preamble.tex file in the custombib section.
% Comment out the lines related to natbib above and uncomment the following line.

%\printbibliography[heading=bibintoc, title={References}]

\end{spacing}

% ********************************** Appendices ********************************

%\begin{appendices} % Using appendices environment for more functunality

%\include{Appendix1/appendix1}
%\include{Appendix2/appendix2} 

%\end{appendices}

% *************************************** Index ********************************
%\printthesisindex % If index is present

\end{document}